\documentclass{emulateapj}

\usepackage{amsmath}
\usepackage{amssymb}
\usepackage{graphicx}
\usepackage{appendix}
\usepackage{color}

\newcommand{\lSect}[1]{{\label{sec:#1}}}
\newcommand{\lFig}[1]{{\label{fig:#1}}}
\newcommand{\lEq}[1]{{\label{eq:#1}}}
\newcommand{\lTab}[1]{{\label{tab:#1}}}
\newcommand{\FIGFF}[2]{{\ref{fig:#2}{#1}}}

\newcommand{\FIG}[2]{{Fig.~\FIGFF{#1}{#2}}}
\newcommand{\Fig}[1]{{\FIG{}{#1}}}

\newcommand{\Sectff}[1]{{\ref{sec:#1}}}
\newcommand{\Sect}[1]{{\S~\Sectff{#1}}}

\newcommand{\Eqref}[1]{{\ref{eq:#1}}}
\newcommand{\Eqff}[1]{{(\Eqref{#1})}}

\newcommand{\Eq}[1]{{eq.~\Eqff{#1}}}

\newcommand{\Msun}{\ensuremath{\mathrm{M}_\odot}}
\newcommand{\Rsun}{\ensuremath{\mathrm{R}_\odot}}
\newcommand{\Zsun}{\ensuremath{\mathrm{Z}_{\odot}}}
\newcommand{\Tab}[1]{{Table \ref{tab:#1}}}
\newcommand{\cp}{\ensuremath{\mathrm{\xi}_{2.5}}}

\def\gtaprx {\lower .1ex\hbox{\rlap{\raise .6ex\hbox{\hskip .3ex
	{\ifmmode{\scriptscriptstyle >}\else
		{$\scriptscriptstyle >$}\fi}}}
	\kern -.4ex{\ifmmode{\scriptscriptstyle \sim}\else
		{$\scriptscriptstyle\sim$}\fi}}}
\def\ltaprx {\lower .1ex\hbox{\rlap{\raise .6ex\hbox{\hskip .3ex
	{\ifmmode{\scriptscriptstyle <}\else
		{$\scriptscriptstyle <$}\fi}}}
	\kern -.4ex{\ifmmode{\scriptscriptstyle \sim}\else
		{$\scriptscriptstyle\sim$}\fi}}}

\graphicspath{{tmp_figs/}}

\begin{document}

\submitted{16 October, 2015}
\accepted{26 January, 2016}

\title{Core-Collapse Supernovae from 9 to 120 Solar Masses\\
Based on Neutrino-powered Explosions}

\author{Tuguldur Sukhbold\altaffilmark{1}, 
	             T.\ Ertl\altaffilmark{2,3}, 
          S.\ E.\ Woosley\altaffilmark{1}, 
        Justin\ M.\ Brown\altaffilmark{1}, 
          and H.-T. Janka\altaffilmark{2}}
\altaffiltext{1}{Department of Astronomy and Astrophysics, 
                 University of California, Santa Cruz, CA 
                 95064}
\altaffiltext{2}{Max-Planck-Institut f{\"u}r Astrophysik, 
                 Postfach 1317, 85741 Garching, Germany}
\altaffiltext{3}{Physik Department, Technische 
                 Universit\"at M\"unchen, 
                 James-Franck-Str.~1, 85748 Garching, 
                 Germany}
                 
\begin{abstract}
Nucleosynthesis, light curves, explosion energies, and remnant 
masses are calculated for a grid of supernovae resulting from 
massive stars with solar metallicity and masses from 9.0 to 
120 \Msun. The full evolution is followed using an adaptive 
reaction network of up to 2000 nuclei.  A novel aspect of the 
survey is the use of a one-dimensional neutrino transport model 
for the explosion.  This explosion model has been calibrated to 
give the observed energy for SN 1987A, using five standard 
progenitors, and for the Crab supernova using a 9.6 \Msun\ 
progenitor. As a result of using a calibrated central engine, 
the final kinetic energy of the supernova is variable and 
sensitive to the structure of each presupernova star. Many 
progenitors with extended core structures do not explode, but 
become black holes, and the masses of exploding stars do not 
form a simply connected set. The resulting nucleosynthesis 
agrees reasonably well with the sun provided that a reasonable 
contribution from Type Ia supernovae is also allowed, but with 
a deficiency of light s-process isotopes. The resulting neutron 
star IMF has a mean gravitational mass near 1.4 \Msun. The 
average black hole mass is about 9 \Msun \ if only the helium 
core implodes, and 14 \Msun, if the entire presupernova star 
collapses. Only $\sim$10\% of supernovae come from stars over 
20 \Msun\ and some of these are Type Ib or Ic. Some useful 
systematics of Type IIp light curves are explored.
\end{abstract}

\keywords{stars: supernovae: general, nucleosynthesis}

\section{Introduction}
\lSect{intro}

The study of nucleosynthesis in massive stars has a rich 
history \citep[e.g.][]{Bur57,Woo95,Thi96,Woo02,Nom13}. These 
studies have frequently taken the form of a detailed analysis 
for individual events, e.g., SN 1987A (87A), or surveys for 
stars of just a few different masses. In some cases where 
broader surveys were done, the nucleosynthesis was calculated 
separately from any consideration of the explosion physics, and 
the central engine was linked in as a parametrized inner 
boundary condition for a calculation that included only the 
matter outside of the collapsing core. The explosion was 
parametrized either by dumping a prescribed amount of energy 
into inner zones, or by the motion of a piston, generally 
constrained to give a fixed value of kinetic energy for the 
final supernova. Sometimes this energy was varied over a 
limited range and the sensitivity of outcomes determined.

Here we report the results of a different approach. The 
starting point is an ensemble of 200 presupernova models that 
together span the range of masses expected for common 
supernovae (9.0 - 120 \Msun). For the most part, these models 
or very similar ones have been published before 
\citep{Woo07,Suk14,Woo15}. All have solar metallicity. Each was 
calculated using the KEPLER code and identical stellar physics. 
The major innovation here is the treatment of the explosion 
mechanism. Two sorts of calculations of the explosion are 
carried out for each mass. Iron-core collapse and bounce, 
neutrino transport, and the propagation of the outgoing shock 
are followed in one simulation until sufficiently late that the 
final mass cut and explosion energy have been well determined. 
This calculation, which includes a high density equation of 
state and neutrino transport that captures the essential 
effects of neutrinos, also gives an estimate of nucleosynthesis, 
especially for $^{56}$Ni. The second calculation tracks the 
results of the first in a study of detailed nucleosynthesis and 
radiation transport that gives the bolometric light curve. A 
consistent mapping between the two codes gives, for each 
presupernova star, very nearly the same remnant mass, kinetic 
energy at infinity, and $^{56}$Ni mass.  Nucleosynthesis is 
calculated for all isotopes from hydrogen through lead, except 
that the nucleosynthesis in the neutrino wind is not included.

Both calculations are one-dimensional (1D). While the actual 
physics of neutrino-powered supernovae cannot be fully 
represented in a such a limited study 
\citep{Jan12a,Bur13,Fog15}, multi-dimensional models are 
expensive, scarce, and often lack community consensus. Of 
necessity, an expedient track is adopted that should capture 
many characteristics of a real explosion.  The 1D core-collapse 
code is used to calculate the explosion of five different 
progenitors for SN 1987A \citep{Ert15} and one progenitor for 
SN 1054 (the ``Crab''; \Sect{crab}). Adjustable parameters in 
the central engine model are varied to give the well-determined 
$^{56}$Ni mass, kinetic energy and neutrino burst time scale 
for SN 1987A, and the kinetic energy of the Crab. Based upon 
the systematics of presupernova core structure 
(\Sect{calibrate}), supernovae below 12.0 \Msun \ are deemed to 
be ``Crab-like'', while heavier stars are ``87A-like''. Further 
partition of the mass space might be practical given more 
observational constraints, especially supernovae with 
well-determined masses, but was not attempted here.

These central engines are then used to explore the outcomes 
when placed in the 200 presupernova stars. The use of these 
``calibrated central engines'', which are really just 
standardized descriptions of the inner 1.1 \Msun \ of the 
proto-neutron star, allows a calculation of the neutrino and 
accretion physics outside the inner core that depends 
sensitively upon the structure of individual presupernova stars. 
For some models, the standard central engine is unable to 
provide enough energy for explosion and a black hole forms. For 
others, the star explodes with a variable amount of kinetic 
energy. For the stars that do explode, the resulting patterns 
of supernova energetics, nucleosynthesis, remnant masses, and 
light curves are reasonable for four out of five of the 
calibrations for SN 1987A. One set of calculations that used a 
model for SN 1987A that exploded too easily, gave explosions 
that, on the average, were deemed too weak. In this case an 
inadequate number of stars exploded to make the abundances we 
see in the sun (\Sect{explosions}). Thus nucleosynthesis was not 
studied in detail using this model as a basis.

We are particularly interested in how the ``explodability'' of 
the presupernova models and their observable properties 
correlate with their ``compactness'' 
\citep[\Fig{cp_all};][]{Oco11}

\begin{equation}
\xi_{M}=\frac{M/\Msun}{R(M)/1000\, {\rm
    km}}\Big|_{t_{{\rm bounce}}}, 
    \lEq{compact}
\end{equation}

and other measures of presupernova core structure
(\Sect{calibrate};\citet{Ert15}).  Using a standard central 
engine in presupernova models of variable compactness, a 
significant correlation in outcome is found (\Sect{explosions}). 
As previously suggested, the resulting mass spectrum of 
successful supernovae is not simply connected 
\citep{Oco11,Ugl12,Suk14}.  That is, there is no single mass 
above which stars make black holes and below which they explode. 
Rather there are mass ranges that ``tend'' to explode, albeit 
with significant variations, even in narrow ranges.  The model 
stars below 15 \Msun\ always explode easily, while those from 
22 - 25 \ and 27 -30 \Msun, rarely explode. The outcome for 
other masses is either variable (15 - 22 \Msun \ and 
25 - 27 \Msun), or sensitive to the treatment of mass loss 
(M $>$ 30 \Msun).

The input physics and the systematics for all of the 
progenitors, including the models used for the explosion 
calibrations have been discussed in previous publications, but 
are summarized in \Sect{physandmods}. Details of the explosion 
modelling for both the P-HOTB code, which is used to study core 
collapse, and the KEPLER code, which is used to calculate 
nucleosynthesis and light curves, and how agreement is enforced 
between the two codes are discussed in \Sect{expl}. The 
properties of the explosions in relation to the progenitor 
compactness parameter, especially their explosion energies and 
$^{56}$Ni masses, are addressed in \Sect{explosions}, and the 
mass distributions of neutron stars and black holes from the 
explosions are discussed in \Sect{remnants}. The resulting 
nucleosynthesis from different mass ranges and the integrated 
yields for different element groups are examined in 
\Sect{nucleo}, and the light curves of various types emerging 
from the explosions and their systematics are analysed in 
\Sect{lite}. In \Sect{conclude}, we offer some conclusions.

\section{Presupernova Physics and Models}
\lSect{physandmods}


\begin{figure}[h] 
\centering 
\includegraphics[width=\columnwidth]{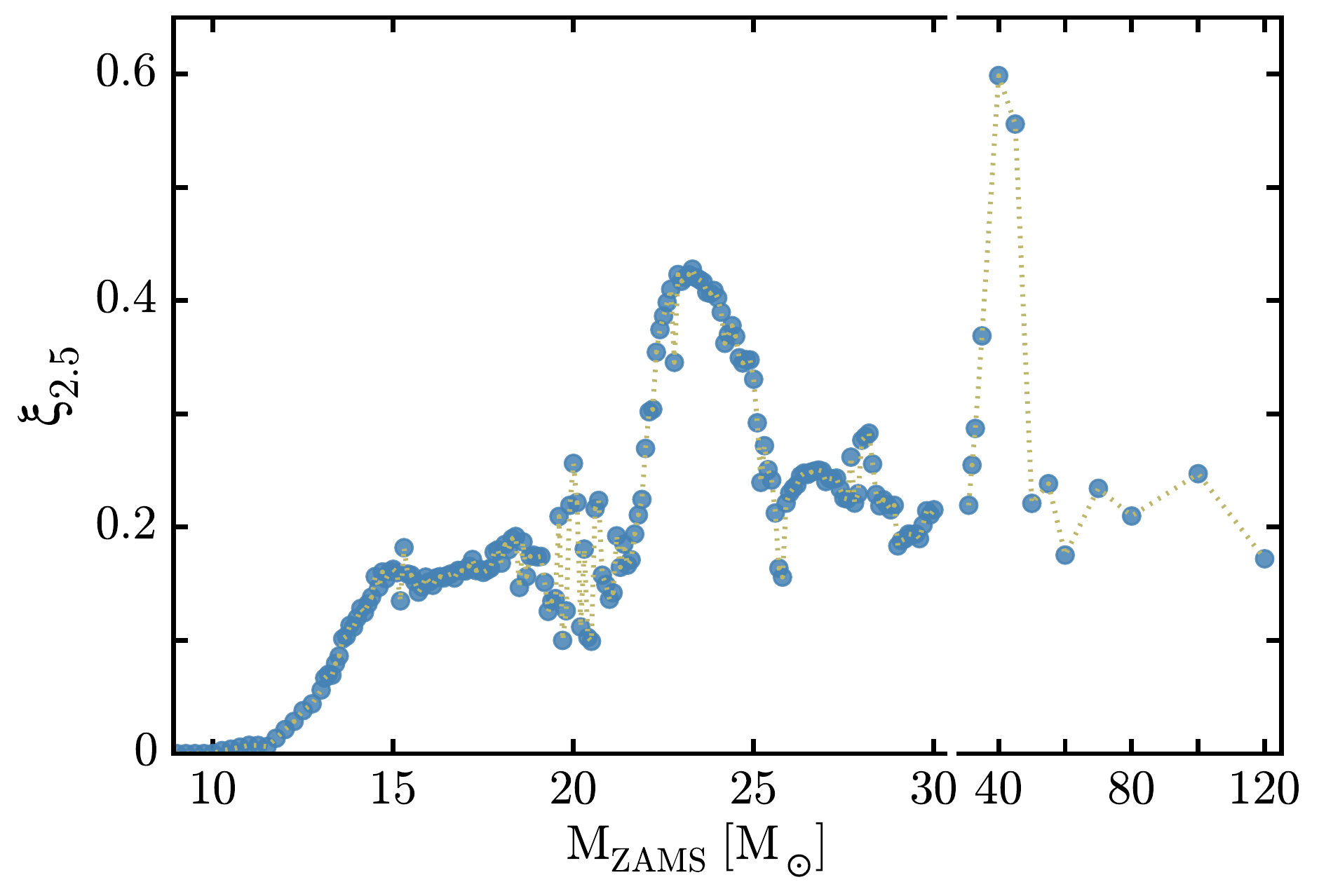}
\caption{The compactness parameter, \cp, 
         \citep[\Eq{compact};][]{Oco11} characterizing the 
         inner 2.5 \Msun\ of the presupernova star is shown as 
         a function of zero-age main sequence (ZAMS) mass for 
         all 200 models between 9.0 and 120 \Msun. The 
         compactness was evaluated when the collapse speed 
         anywhere in the core first exceeded 1000 km s$^{-1}$. 
         Studies have shown little difference for $\xi_{2.5}$ 
         evaluated at this time and at core bounce \citep{Suk14}. 
         Note the scale break at 32 \Msun. Above about 40 \Msun\ 
         the compactness parameter declines to a nearly constant 
         value due to the reduction of the helium core by mass 
         loss. Stars below 12 \Msun \ have an especially dilute 
         density structure around the iron core, due to the 
         lingering effects of degeneracy on the late stages of 
         evolution. Stars below 15 \Msun\ (helium core mass 
         below 4 \Msun) explode comparatively easily, while 
         stars from 22 to 26 \Msun\ are hard to explode. 
         \lFig{cp_all}}
\end{figure}

\subsection{Stellar Physics}
\lSect{presnphys}

All presupernova models were calculated using the KEPLER code 
\citep{Wea78}, the physics of which has been previously 
discussed \citep[e.g.,][]{Woo02}, and are included in the electronic 
edition in a tar.gz package. Indeed, most of the models, 
or very similar ones, have been published previously  \citep{Woo07,Suk14,Woo15}. All stars, except for one model 
used for the calibration of 87A, were non-rotating. Except 
for the calibration models for 87A and the Crab 
(\Sect{calibratepsn}), each had an initially solar 
\citep{Lod03} composition. All were evolved from the main 
sequence until the onset of iron-core collapse. The lower 
bound to the survey, 9.0 \Msun, is set by the lightest star to 
experience iron-core collapse in the KEPLER code \citep{Woo15}. 
The evolution of still lighter stars in the 7 to 9 \Msun\ range 
was not calculated. Some uncertain fraction of these will 
evolve to electron-capture supernovae and contribute to the 
supernova rate and remnant distribution. Those contributions 
are neglected here. Their contribution to nucleosynthesis, 
other than the $r$- and $s$-processes is likely to be small, 
however.

Full stars, including their hydrogenic envelopes, were 
calculated and mass loss was included \citep{Nie90,Wel99,Woo02}. 
The nuclear physics varied (though not the individual reaction 
rates). A rate for $^{12}$C($\alpha,\gamma)^{16}$O equal to 1.2 
times that of \citet{Buc96} was employed. For stars 13.0 \Msun\ 
and heavier, including the models for SN 1987A, the evolution 
up to core oxygen depletion was followed using a 19 isotope 
approximation network and, after central oxygen decreased to 
below 4\% by mass, using a 128 isotope quasi-equilibrium network 
\citep{Wea78}. Those calculations were ``co-processed'' using an 
adaptive network \citep{Rau02} that, near the end included, 
typically, 1500 isotopes and sometimes over 2000. In addition to 
following the nucleosynthesis zone by zone and step by step, the 
evolution of the (slowly evolving) electron mole number, $Y_e$, 
calculated by the large network was fed back into the structure 
calculation. Energy generation within an iterative loop, however, 
came from the 19 isotope network or the quasi-equilibrium 
network. 

For stars lighter than 13 \Msun, the tracking of off-center 
burning and convectively bounded flames requires a more careful 
treatment of the nuclear physics \citep{Woo15}. For these stars, 
the large network and all neutrino losses from weak reactions on 
individual nuclei, was coupled directly to the structure 
calculation. The 19 isotope network and quasi-equilibrium 
networks were not employed. Several stars in the 12 to 13 \Msun\ 
range were computed using both the direct coupling of the big 
network and coprocessing. The results were nearly identical.

During the explosion, all shock-induced transformations were 
followed as well as nucleosynthesis by the neutrino-process 
\citep{Woo90,Heg05}.  The r-, s-, and $\gamma$-processes were 
all followed in matter outside the iron core, but 
nucleosynthesis in the neutrino-powered wind was not followed 
in any detail.

Formally, only single stars were included. The key quantity 
for nucleosynthesis and remnant masses is the mass of the core 
of helium and heavy elements at the time of the explosion. To 
the extent that the Salpeter IMF properly reflects this 
distribution of helium core masses, our results would not be 
changed if a substantial fraction of the stars were in binary 
systems. If mass exchange proceeds so far as to appreciably 
reduce the mass of the helium core from what it would have been 
for a single star, including radiatively-driven mass loss, then 
the answer will change. For example, it seems most likely that 
common SN Ib and Ic result from binary systems containing stars 
whose presupernova mass is only 3 or 4 \Msun\ 
\citep{Des12,Ens88}. Many of these may come from 12 to 15 \Msun\ 
main sequence stars that lost their envelopes and little else to 
a companion. Treating these as single stars does not greatly 
distort the nucleosynthetic outcome, though it obviously changes 
the rate of Type Ib and Ic supernovae a lot. If the progenitors 
in binaries were much heavier, some of their nucleosynthesis, 
especially in the helium shell, might have been lost to a 
companion and it becomes an issue whether {\sl that} star 
blows up returning what it accreted to the interstellar medium. 
If more massive stars in binaries are involved in making SN Ic, 
then our results would be altered. The explosion of stripped 
down 4 \Msun\ cores of oxygen and heavy elements is different 
from the explosion of a 4 \Msun\ helium core, both in mechanism 
and consequence \citep{Woo93}. However, such supernovae are 
relatively rare and we do not expect substantial changes from 
what we report here. Including the contributions of SN Ib and 
Ic models explicitly would be a worthwhile project for the future.

\subsection{Calibration Models}
\lSect{calibratepsn}

\subsubsection{SN1987A}
\lSect{87a}

The presupernova evolution of Sk -69$^o$ 202 and its explosion 
as SN 1987A have been studied extensively for almost 30 years 
\citep[e.g.][and references therein]{Arn89,Utr14}, yet a 
comprehensive, self-consistent model for the progenitor star, 
its ring structure, and the explosion is still lacking. The 
observational constraints are tighter, though, than for any 
other supernova, so it is worthwhile to try to use SN 1987A as 
a calibration point.

It is known that the progenitor star was a Type B3-I blue 
supergiant, Sk -69$^o$ 202 \citep{Wal87}, with an effective 
temperature 14,000 - 17,000 K \citep{Hum84,Cro06,Smi07}, a 
luminosity $3 - 6 \times 10^{38}$ erg s$^{-1}$ 
\citep{Hum84,Arn89}, and thus a radius $2 - 4 \times 10^{12}$ 
cm. Since the presupernova luminosity is determined chiefly by 
the helium core mass, a core mass of $6 \pm 1$ \Msun\ has been 
inferred theoretically \citep[e.g.][]{Woo88a}. Based upon 
analysis of the supernova light curve, this core was surrounded 
by about 10 \Msun\ of envelope in  which helium and nitrogen 
were enriched, but hydrogen was still a major constituent \citep{Woo88a,Shi88}. The star had been a red supergiant before 
dying as a blue one in order to produce the colliding wind 
structure seen shortly after the explosion. Spectral analysis, 
as well as the light curve, give an explosion energy near 
$1.5 \times 10^{51}$ erg \citep{Arn89,Utr14}, and the late time 
light curve shows unambiguously that the explosion produced 0.07 
\Msun\ of $^{56}$Ni \citep{Bou91,Sun92}. The explosion also 
ejected approximately $1.4 \pm 0.7$ \Msun\ of oxygen 
\citep{Fra02}.

\begin{deluxetable*}{cccccccccc}
\tablewidth{0pt}
\tablecaption{SN 1987A Models}
   \tablehead{\colhead{model}                     & 
              \colhead{M$_{\rm preSN}$/M$_\odot$} & 
              \colhead{M$_{\rm He}$/M$_\odot$}    & 
              \colhead{M$_{\rm CO}$/M$_\odot$}    & 
              \colhead{L/$10^{38}$ erg s$^{-1}$}  & 
              \colhead{T$_{\rm eff}$}             & 
              \colhead{$\zeta_{2.5}$}             & 
              \colhead{Z/\Zsun}                   & 
              \colhead{Rotation}
              }
\startdata
W18   & 16.93 & 7.39 & 3.06 & 8.04 & 18000 & 0.10  & 1/3 & Yes \\
N20   & 16.3  & 6    & 3.76 & 5.0  & 15500 & 0.12  & low & No  \\
S19.8 & 15.85 & 6.09 & 4.49 & 5.65 & 3520  & 0.13  & 1   & No  \\
W15   & 15    & 4.15 & 2.02 & 2.0  & 15300 &  -    & 1/4 & No  \\
W20   & 19.38 & 5.78 & 2.32 & 5.16 & 13800 & 0.059 & 1/3 & No  \\
      &       &      &      &      &       &       &     &     \\
W16   & 15.37 & 6.55 & 2.57 & 6.35 & 21700 & 0.11  & 1/3 & Yes \\
W17   & 16.27 & 7.04 & 2.82 & 7.31 & 20900 & 0.11  & 1/3 & Yes \\
W18x  & 17.56 & 5.12 & 2.12 & 4.11 & 19000 & 0.10  & 1/3 & Yes \\
S18   & 14.82 & 5.39 & 3.87 & 4.83 & 3520  & 0.19  & 1   & No  \\
\lTab{87aprog}
\end{deluxetable*}

\paragraph{Models for Sk -69$^o$ 202}

Numerous attempts have been made to model the presupernova 
evolution of Sk -69$^o$ 202.  The final stages are complicated 
by the widespread belief that SN 1987A may have been a merger 
event \citep{Pod92,Mor07}, though see \citet{Smi07}. In any 
case, the star that exploded was single at the time, in 
hydrostatic and, given the necessary timescale for the 
red-to-blue transition, probably in approximate thermal 
equilibrium. Its helium core mass is thus constrained by the 
presupernova luminosity and so single star models are relevant.

Five principal models for SN 1987A are considered here, the ones 
above the blank line in \Tab{87aprog}, and span the space of 
possible helium and heavy element core structures (\Fig{hrd}). 
The 3D explosion and observable properties of four of them have 
been explored previously \citep{Utr14} and agree well with 
experimental data. These models are N20, W15, W18, and W20.


\begin{figure}[h] 
\centering 
\includegraphics[width=\columnwidth]{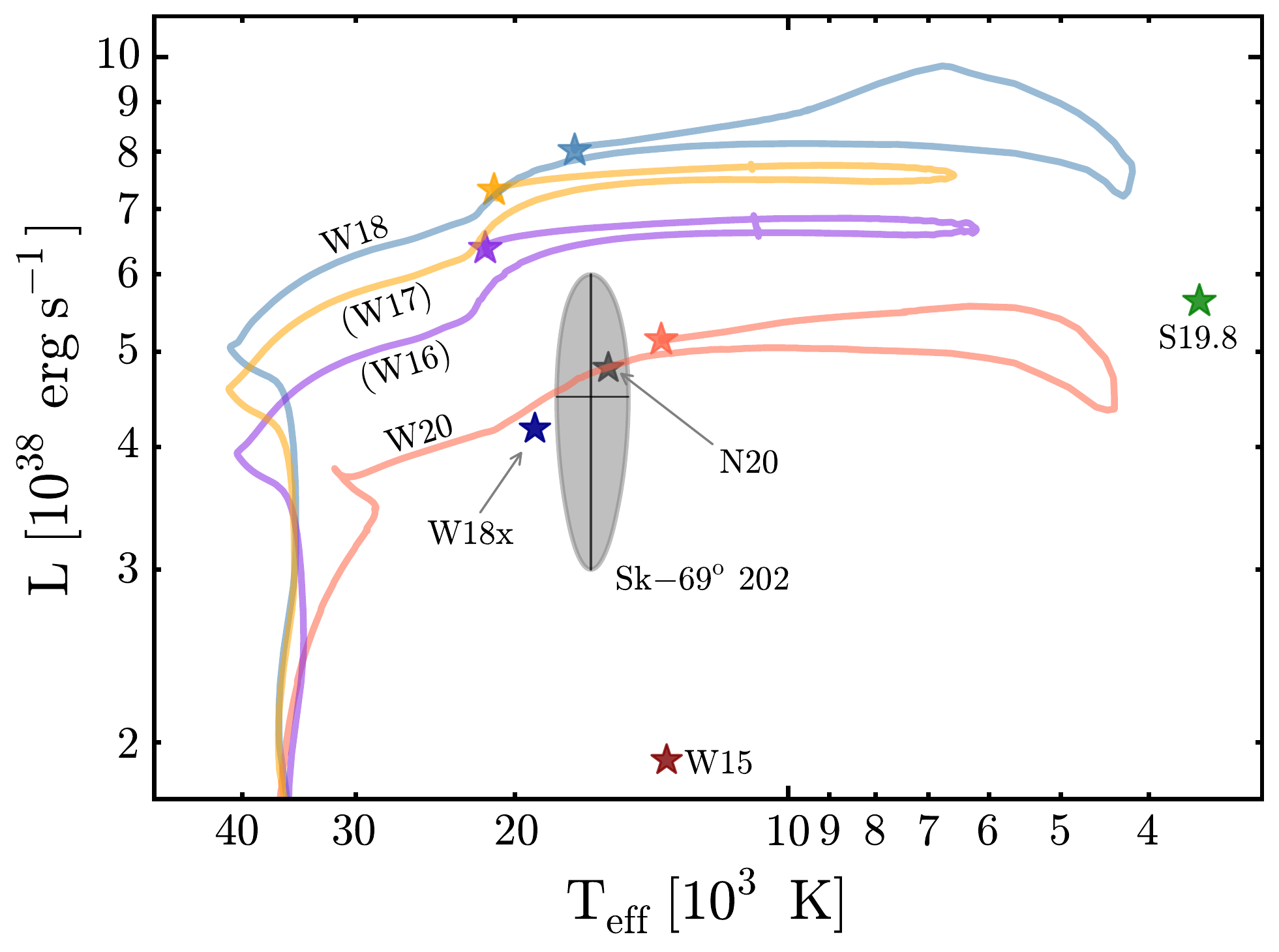}
\caption{Hertzsprung-Russell diagram for the SN 1987A progenitors 
         used in this study. The shaded region shows the observed 
         properties of Sk -69$^o$ 202 (see text). Model S19.8 
         produced a red supergiant; all the rest gave blue 
         supergiants at the time of core collapse, but were 
         preceded by a phase as a red supergiant. Models W16, W17, 
         W18, and W18x included rotation; the rest were 
         non-rotating. Model N20 was a construct obtained by 
         joining 6 \Msun\ helium star to an envelope. All models 
         beginning with a ``W'' were calculated using the KEPLER 
         code assuming reduced metallicity and restricted 
         semiconvection. Model W18x had 3/4 the angular momentum 
         of Model W18, but was otherwise identical. See also 
         \Tab{87aprog}. 
         \lFig{hrd}}
\end{figure}

Model N20 \citep{Shi90,Nom88,Sai88}, a historical model, was 
``prepared'' by artificially combining the presupernova 
structure of an evolved 6 \Msun\ helium core from one calculation 
with the hydrogen envelope from another.  This model was thus 
defined to satisfy observational constraints on helium core mass, 
hydrogen envelope mass, radius, and luminosity, but lacks the 
physical basis of a presupernova model evolved intact from the 
main sequence onwards. Since the helium core of Model N20 was 
evolved using the Schwarzschild criterion for convection 
\citep{Nom88}, its carbon-oxygen (CO) core is large compared with 
some of the other models (\Tab{87aprog}), which used restricted 
semiconvection in order to produce a blue supergiant progenitor. 
It is not clear that, had the 6 \Msun\ core been included in a 
self-consistent calculation with an envelope always in place, a 
blue supergiant would have resulted rather than a red one.

Model W15 \citep{Woo88}, another early model calculated shortly 
after SN 1987A occurred, was produced by evolving a single 15 
\Msun\ star from beginning to end, but relied upon opacities that 
are now outdated, reduced metallicity, and greatly restricted 
semiconvection to produce a blue supergiant. Mass loss was not 
included. The small ratio (less than about 0.6) of the CO core 
mass to helium core mass in W15 is characteristic of all models 
in \Tab{87aprog} that produce blue supergiant progenitors. So far 
as we know, this is a hallmark of all solutions that use Ledoux 
convection or restricted semiconvection to achieve this goal. The 
helium core mass and luminosity of W15 are too small compared 
with more accurate observations of Sk -69$^o$ 202. Still W15 is 
included here because it has been studied in the literature and 
it brackets, on the low side, the observed luminosity and helium 
core mass of Sk -69$^o$ 202 (\Fig{hrd}).

Model W20 \citep{Woo97} used more modern opacities 
\citep{Rog92,Igl96} and included mass loss.  Its luminosity, 
radius, and helium core mass agree well with observations of 
Sk -69$^o$ 202 (\Fig{hrd}). Its presupernova oxygen mass, 0.68 
\Msun, is on the small side compared with observations 
\citet{Fra02}, especially when allowance is made for the 
processing of some oxygen into heavier elements during the 
explosion. The envelope is also not enriched in helium. The 
surface mass fraction of helium is 0.26.

Model W18 is a more recent unpublished model that will be 
presented in greater detail later in this section. It achieves 
the desired goals of making a blue supergiant progenitor, 
producing a relatively large mass of oxygen, and displaying 
large enhancements in surface helium and nitrogen abundances, 
by invoking reduced metallicity, restricted semiconvection, and 
a substantial amount of rotation. Rotation makes the helium core 
larger and also stirs more helium into the envelope, raising the 
mean atomic weight there, and reducing the opacity. A larger 
helium core implies a larger luminosity, somewhat above the 
observed limit for SN 1987A (\Fig{hrd}), but the increased 
helium fraction in the envelope facilitates a blue solution 
\citep{Sai88,Lan92}.

Three other models similar to W18 are also given in \Tab{87aprog} 
and \Fig{hrd} to illustrate some sensitivities, but were not used 
in the remainder of the paper. W16 and W17 are identical to W18 
except for their different masses and rotation rates. The angular 
momentum on the zero age main sequence for Models W16, W17, and 
W18 was $J_{\rm ZAMS} =$ 2.7, 2.9 and 3.1 $\times 10^{52}$ erg s. 
The presupernova core properties for W16 and W17 are similar to 
W18 and one would expect similar results were they to be employed 
as standard central engines. W18x on the other hand is slightly 
different in input physics, yet quite different in outcome. Like 
W18, W18x includes rotation, but the total angular momentum on 
the main sequence was reduced by $\sim$25\% to J$_{\rm ZAMS} =
2.3 \times 10^{52}$ erg s. This small reduction in $J$ resulted 
in a dramatic decrease in the helium core mass and presupernova 
luminosity (\Tab{87aprog}; \Fig{hrd}). Another model, not shown, 
with an angular momentum 50\% of W18 gave almost identical 
results to W18x, showing that W18 itself was on the verge of 
becoming a chemically homogeneous model \citep{Mae87,Lan92,Woo06}. 
Nevertheless, we chose to use Model W18 itself as one of our 
standard models because it gave reasonable agreement with the 
observed properties of Sk -69$^o$ 202, had a core structure 
similar to the non-rotating, solar metallicity stars in the 
present survey, was available a few years ago when this study 
began, and had been previously studied in the literature \citep{Utr14,Ert15}.

In addition to the above models with restricted semi-convection 
and low metallicity, a non-rotating, solar metallicity red 
supergiant progenitor was also examined. ``S19.8'' \citep{Woo02}, 
was not considered by Utrobin et al., but has a similar helium 
core structure to W18 and N20 (\Tab{87aprog}). Because the 
surface boundary pressure is small in both cases, the structure 
of the helium core is insensitive to whether the envelope is that 
of a blue or red supergiant. Despite its vintage, Model S19.8 was 
calculated using essentially the same initial composition and 
stellar physics as in the present survey - including opacities, 
mass loss, semiconvection, and metallicity. The luminosity, 
helium core mass, and hydrogen envelope mass of presupernova 
Model S19.8 are all what one would want for SN 1987A, but the 
presupernova star was a red supergiant. Also given in 
\Tab{87aprog} is an 18 \Msun \ presupernova model from the 
present survey. The luminosity is very similar to Sk -69$^o$ 202 
and the structure similar to S19.8.

The point of these calculations was not to generate the best 
possible model for SN 1987A itself, but to use its well studied 
properties to bracket and calibrate a central engine capable of 
exploding helium cores with similar characteristics. It was 
important to study a range of models to test the sensitivity of 
the results to the choice of a model for SN 1987A (see 
\Sect{explosions} for results).

\paragraph{The Presupernova Evolution of Model W18}

Because it will be used to generate one of the standard engines 
and is not documented in the refereed literature, Model W18 
warrants some additional description.  The star had an initial 
mass of 18.0 \Msun\ and a composition of 74.6\% H, 25.0 \% He, 
0.052\% C, 0.012\% N, and\ 0.25\% O, with traces of heavier 
elements. Its iron mass fraction was $7.6 \times 10^{-4}$. These abundances, which for CNO are roughly 1/3 of solar, may reflect 
the smaller metallicity of the Large Magellanic Cloud. The star 
was given sufficient angular momentum that, while burning 
hydrogen on the main sequence, its equatorial rotational speed 
was 240 km s$^{-1}$.  On the main sequence, half way through 
hydrogen burning, its luminosity and effective temperature were 
$2.3 \times 10^{38}$ erg s$^{-1}$ and 36,100 K. The star spent 
most of its helium burning lifetime as a blue supergiant with 
radius near $2.5 \times 10^{12}$ cm, a luminosity of 
$7.7 \times 10^{38}$ erg s$^{-1}$, and an effective temperature 
of 20,000 K. Only towards the end of its life, when the central 
helium mass fraction declined below 7\%, did the radius increase 
above $5 \times 10^{12}$ cm.  When the radius reached 
$1 \times 10^{13}$ cm the helium mass fraction was only 1.5\%. 
The radius continued to increase reaching a maximum of $6.2 \times
10^{13}$ cm after helium had already been depleted and the central
temperature was $3.6 \times 10^8$ K.  The luminosity at that 
point was $8.1 \times 10^{38}$ erg s$^{-1}$.

Over the next 15,000 years, the radius of the star shrank back 
to less than 10$^{13}$ cm, and the star remained a blue supergiant 
for another 5000 years before dying with a radius of 
$3.3 \times 10^{12}$ cm and luminosity $8.04 \times 10^{38}$ erg 
s$^{-1}$. At that time, its helium core mass (where hydrogen mass 
fraction equals 1\%) was 7.39 \Msun.  Nucleosynthesis in W18, 
prior to any explosion, included 1.25 \Msun\ of $^{16}$O, in 
agreement with observations suggesting an oxygen mass of 
$1.4 \pm 0.7$ \Msun \ \citep{Fra02}. The helium mass fraction at 
the surface of the presupernova star was 53\%, in agreement with 
evidence for rotationally-induced mixing 
\citep[e.g.]{Wei88,Sai88,Arn89}. The final total star mass for 
W18 was 16.92 \Msun \ and its iron-core mass was 1.47 \Msun, 
which was also the location of the entropy jump $S/N_Ak$ = 4 
associated with the oxygen shell. Some further details of 
presupernova composition and structure of W18 and our other 
models for SN 1987A are given in \citet{Utr14} and \citet{Ert15}.

Because the model included both rotation and the transport of angular
momentum by magnetic torques \citep{Heg05}, it was possible to estimate
the importance of rotation and magnetic fields to the explosion.
The angular momentum interior to 1.47 \Msun \ in the presupernova
model W18 was $6.1 \times 10^{47}$ erg s, implying a rotation
rate of the resultant pulsar of about 12 ms.  No pulsar has been
discovered yet in SN 1987A, but this amount of rotation would be
inadequate to provide the observed $\sim 1.5 \times 10^{51}$ erg explosion.
It seems likely therefore that the explosion was neutrino powered.

In summary, W18 is the heaviest possible blue supergiant model for SN
1987A in terms of its helium core mass. While a bit overluminous, it
fits many other observed characteristics of the supernova well,
especially the oxygen mass synthesized, and its central core structure
is similar to other lighter models, W16 and W17. Table 1 of
\citet{Ert15} shows that, at least in terms of compactness, W18 is
quite similar to W20, N20, and S19.8. 

\subsubsection{SN 1054 - The Crab}
\lSect{crab}

While the Crab supernova was not studied spectroscopically, modern
observations of its remnant suggest that the kinetic energy was much
lower than that of SN 1987A and other common supernovae from more
massive stars \citep{Arn89,Kas09,Yan15}. The event has been
successfully and repeatedly modelled as the explosion of a star near
10 \Msun \ \citep{Nom82,Nom87,Smi13,Tom13}. Such stars are known to
have compact structures that explode easily by the neutrino-heating
mechanism, even in spherical symmetry \citep{Kit06,Fis10,Mel15}.  Use
of an 87A-like central engine in stars of this sort thus overestimates
the power of the supernova and gives an unphysically large kinetic
energy (\Sect{calibrate}). Systematics of the presupernova core
structure, to be discussed later, suggest that, for KEPLER models,
this low energy branch of supernovae should characterize events from
9.0 (or less) to 12.0 \Msun.

The fiducial model employed to represent the core physics in this low
mass range was a 9.6 \Msun\ zero-metallicity model (Z9.6) provided
by A.~Heger (2012, private communication). This star was also
calculated using the KEPLER code using physics described in
\citet{Heg10} and, while a star of zero metallicity, it has a very
similar core structure to the lightest solar metallicity models
employed here (\Fig{9to13rho}). The model has the merit of having been
successfully exploded in 1D and multi-dimensional calculations
\citep{Jan12,Mel15}, developing an explosion dynamically similar to
collapsing O-Ne-Mg core progenitors \citep{Kit06,Jan08}. It also gives
similar nucleosynthesis to the electron-capture supernovae
\citep{Wan15} often mentioned as candidates for the Crab
\citep{Wan11,Jer15}, but has the advantage of being an iron-core
collapse supernova similar to the ones we are presently modelling.


\begin{figure}
	\centering
	\includegraphics[width=\columnwidth]{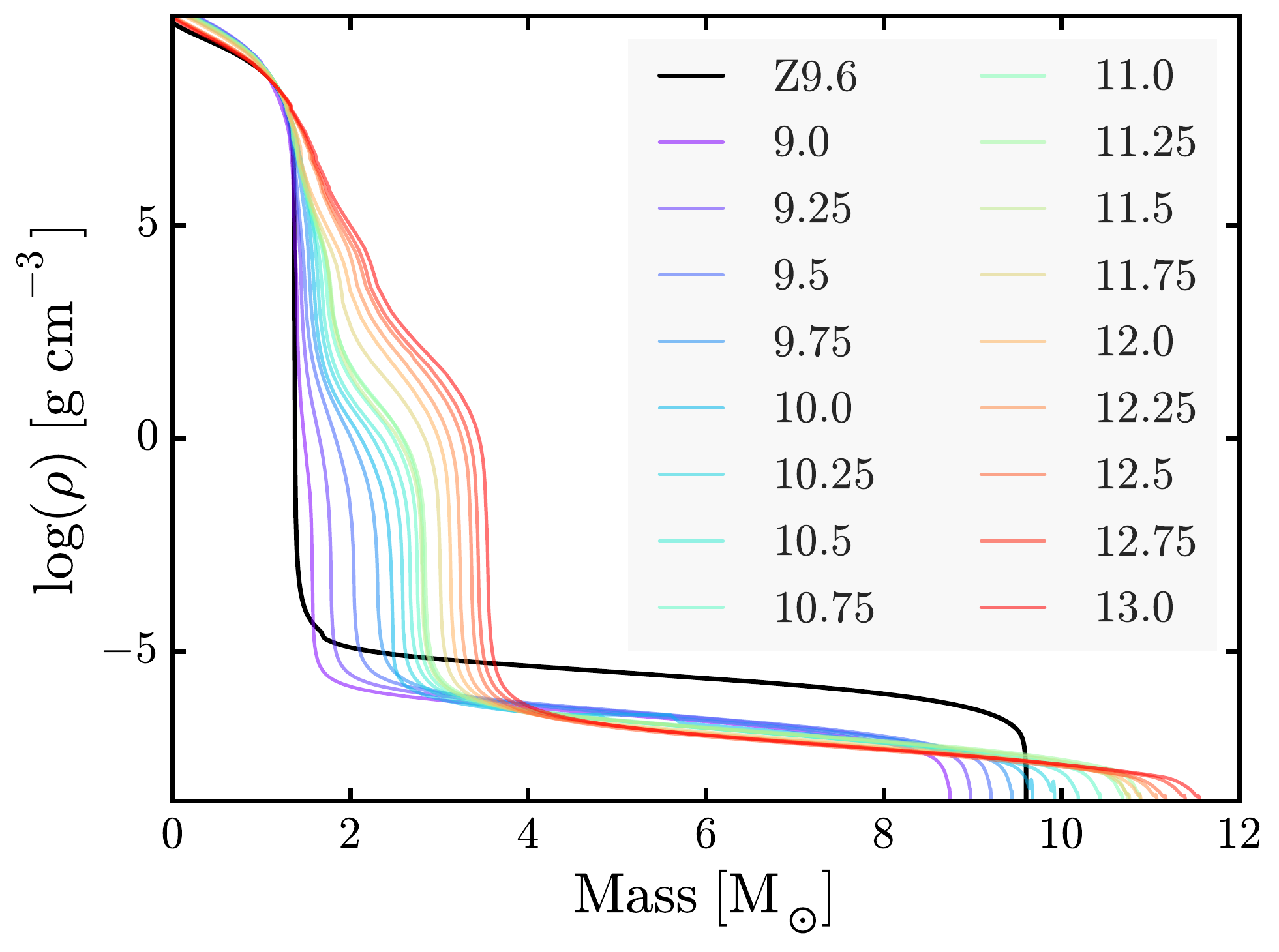}
	\includegraphics[width=\columnwidth]{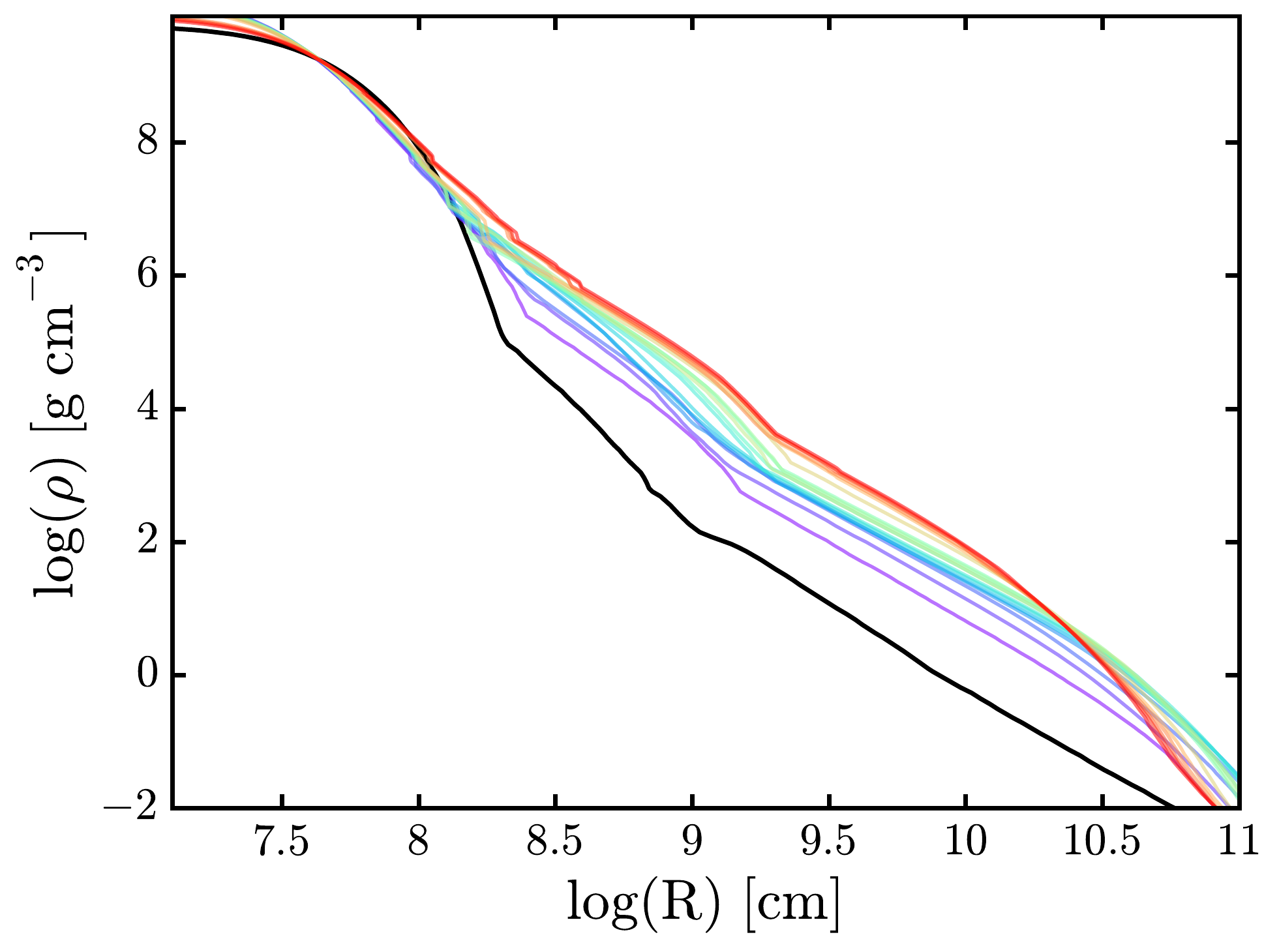}
	\caption{Density profiles for the 17 new presupernova models
          used for stars from 9.0 to 13.0 \Msun. These have a high degree
          of central concentration and are relatively easy to
          explode. Also shown for comparison is the zero metallicity
          calibration model used for Crab, Z9.6, which is similar in
          structure to the 9.0 \Msun \ solar metallicity models
          studied here.}  \lFig{9to13rho}
\end{figure}

\subsection{Presupernova Models Used For This Study}
\lSect{solgrid}

The presupernova models into which these calibrated central engines
are placed are all the outcomes of non-rotating stars with initially
solar metallicity. Mass loss was included in all calculations.
Qualitatively, the models reflect four partitions of expected behaviour
\citep[e.g.][]{Woo15}:

\begin{itemize}

\item{9 -- 12 \Msun. These are stars with degenerate compact cores,
  essentially tightly bound white dwarf-like cores inside loosely
  bound envelopes. They are easy to explode. Starting at 9 \Msun, an
  iron core is produced before the star dies. Slightly below 9 \Msun,
  an electron-capture supernova results or an oxygen-neon white dwarf
  is made. The presence of an active oxygen burning shell at the edge
  of the effective Chandrasekhar mass in the presupernova star causes
  a rapid fall off in density near the edge of a dense central core
  for models below 12 \Msun.}

\item{12 -- 20 \Msun. Starting at about 10.5 \Msun, all fuels ignite
  in the center of the star \citep{Woo15}. The compactness parameter,
  \Eq{compact}, which is a measure of the core's density structure,
  increases rapidly from 12 to 15 \Msun \ and then remains roughly
  constant (\Fig{cp_all}).  Carbon ignites exoergically and powers
  convection in all stars up to about 20 \Msun, and models near the
  transition to radiative central carbon burning end up creating
  highly variable presupernova core structures \ \citep{Suk14}. The
  central core is degenerate and is surrounded by helium and heavy
  element shells of increasing thickness. Most of these stars are also
  relatively easy to explode.}

\item{20 -- 30 \Msun. These stars ignite central carbon burning
  endoergically and are thus not convective during a critical phase
  when excess entropy might have been radiated by neutrinos. Thick
  shells of oxygen, neon and carbon contribute appreciably to
  nucleosynthesis, but, depending upon the location of carbon and
  oxygen burning shells during the later stages of evolution, lead to
  a greater compactness parameter. These stars may be difficult to
  explode \citep{Suk14}.}

\item{Above 30 \Msun, presupernova stars have increasingly thick
  shells of heavy elements and shallow density gradients around large
  iron cores making them difficult to explode with neutrinos. Mass
  loss plays a very important role. For the assumed prescription, the
  presupernova mass actually decreases with increasing initial mass
  above 23.5 \Msun. The maximum presupernova mass there is $\sim16.5$ 
  \Msun. Rotation may be increasingly important to the explosion 
  mechanism of stars in this mass range \citep{Heg05}. If mass loss 
  succeeds in removing the entire hydrogen envelope, as it did here 
  for main sequence stars above 40 \Msun, these stars may explode as 
  a Type Ib or Ic supernova.}

\end{itemize}

Because of their different characteristics, these mass ranges are
often modelled separately. Our models for 9 -- 13 \Msun \ stars, 13 --
30 \Msun \ stars, and 30 -- 120 \Msun \ stars come from different
studies, but all were calculated using the KEPLER code. All
presupernova models, along with data from \Sect{expl}, \Sect{nucleo} 
and \Sect{lite}, are available at the MPA-Garching archive\footnote{http://doi.org/10.17617/1.b \label{mpa_db}}.

\subsubsection{9 to 13 \Msun}
\lSect{9to13}

The 17 models in this mass range, 9.0 - 13.0 \Msun, in intervals of
0.25 \Msun, are based on the recent study of \citet{Woo15}. They use
the same stellar physics, especially for the propagation of
convectively-bounded oxygen and silicon-burning flames, as in that work
\citep[see also][]{Tim94}, but differ by carrying a much larger
network throughout the whole evolution. They are new models.

Stars from 9.0 to 10.25 \Msun \ ignite silicon in a strong flash that,
for the 10.0 and 10.25 \Msun \ models, could be violent enough to
eject the hydrogen envelope and helium shell prior to iron-core
collapse. What actually happens is uncertain though, and depends upon
how much silicon burns in the flash \citep{Woo15}. For the present
study, silicon burning in the flash was constrained to be small by
using a small convective transport coefficient, and envelope ejection
did not occur. For models 10.5 \Msun and heavier, all fuels ignited
stably in the center of the star.

The nuclear reaction network used was complete up to Bi (Z = 93) and
contained roughly 1000 isotopes during helium burning and over 1200
during oxygen and silicon burning. Presupernova stars contained
approximately 1900 isotopes. \citet{Woo15}, who were interested
chiefly in stellar structure and not nucleosynthesis, truncated their
network at Ge. The extra nuclei included here allowed a better
representation of the s-process in the presupernova star and its
modification during the explosion, but did not greatly alter the
structure.

\subsubsection{13.1 to 30 \Msun}
\lSect{12to30}

Models in this mass range were calculated using the same nuclear and
stellar physics as in \citet{Woo07}. The survey includes 151 models
from 15 to 30 \Msun\ taken from the "S-series" in \citet{Suk14}, and
19 additional models from 13.1 to 14.9 \Msun\ calculated the same way.
In order to demonstrate the agreement among models from two different
surveys, several additional models from 12 to 13 \Msun \ were
calculated for comparison with models from \Sect{9to13}, but were not
used in the survey. The presupernova structures of stars calculated both
ways were in good agreement, having e.g., nearly identical values of
the mass enclosed by inner 3000 km (\Fig{overlap}).


\begin{figure}[h]
	\centering
	\includegraphics[width=\columnwidth]{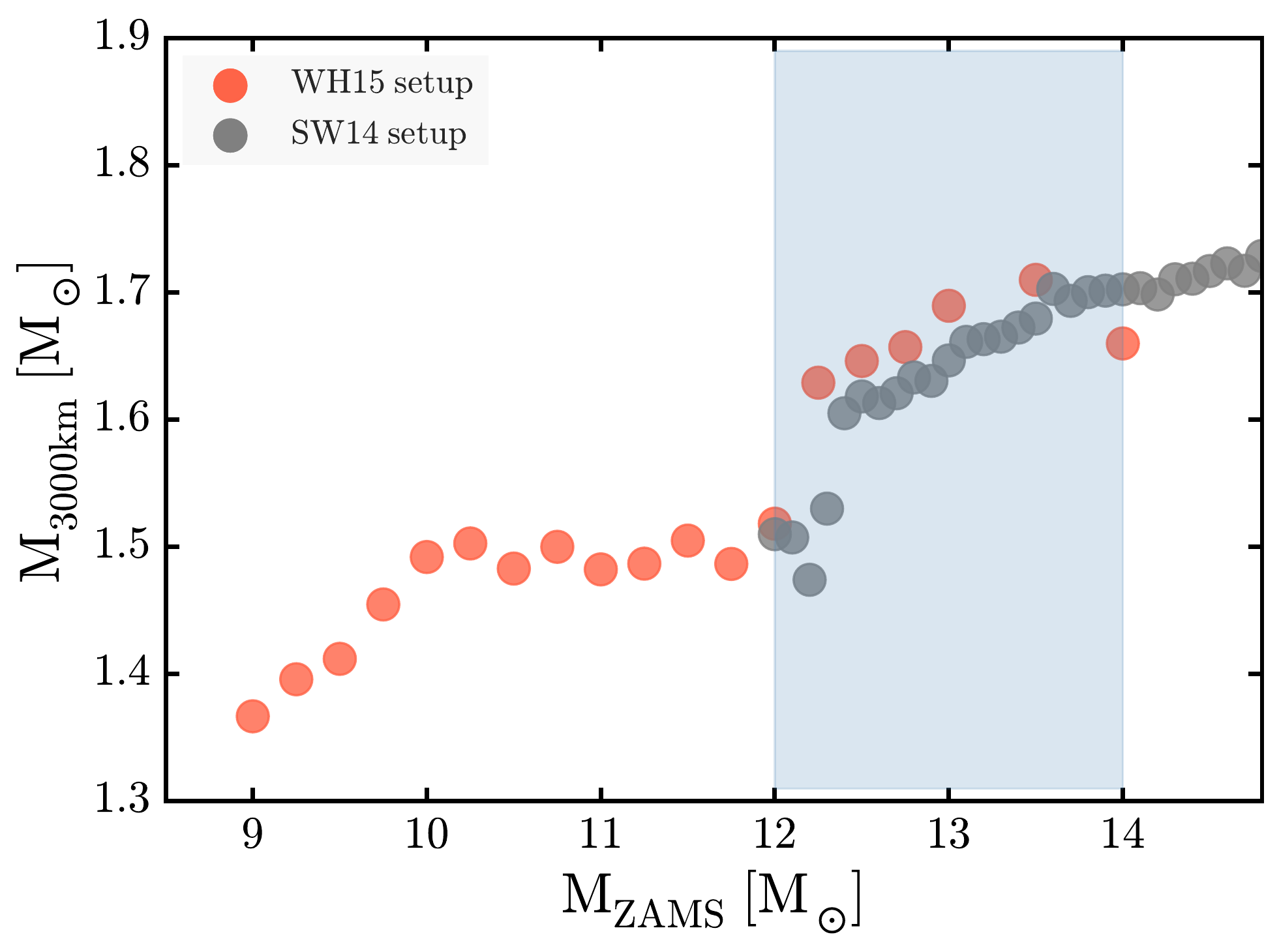}
	\caption{The presupernova core mass inside of 3000 km is shown
          for the 17 new models \citep[see also][]{Woo15} up to 13.0
          \Msun, along with the 19 new models down to 13.1
          \Msun\ calculated with the same input physics as in
          \citet{Suk14}. Though the models come from different
          studies, they are in good agreement. Several models between
          12 and 14 \Msun\ were calculated both ways to highlight the
          agreement (gray region).}  \lFig{overlap}
\end{figure}

The mesh of stellar masses was relatively fine, 0.1 \Msun, as is 
necessary to calculate the rich variation of stellar structure in this 
mass range. 

\subsubsection{31 to 120 \Msun}
\lSect{30to120}

A total of 13 models with varying mass increments were studied in this
range - 31, 32, 33, 35, 40, 45, 50, 55, 60, 70, 80, 100, 120 \Msun\ -
all coming from \citet{Woo07}. Starting at about 40 \Msun, these
models lost their entire envelopes and ended their lives as Wolf-Rayet
stars.  The final masses are thus sensitive to the treatment of mass
loss both for red supergiant stars and stripped cores. The final
presupernova masses for the 60, 70, 80, 100, and 120 \Msun \ stars
were 7.29, 6.41, 6.37, 6.04 and 6.16 \Msun\ respectively. Some
additional properties of all the presupernova models are given in
\Tab{progenitor}.
$\quad$\\
$\quad$\\
$\quad$\\
$\quad$\\

\LongTables
\begin{deluxetable}{ccccc}
\tablecaption{Progenitor Data}
\tablehead{\colhead{M$_{\rm i}\ [\Msun]$} & \colhead{M$_{\rm f}\ [\Msun]$} & \colhead{M$_\alpha\ [\Msun]$} & \colhead{M$_{\rm CO}\ [\Msun]$} & \colhead{log(R$_{\rm f}$) [cm]}}
\startdata
       9.0 & 8.746 & 1.571 & 1.402 & 13.46  \\ 
      9.25 & 8.978 & 1.783 & 1.452 & 13.45  \\ 
       9.5 & 9.208 & 2.041 & 1.501 & 13.46  \\ 
      9.75 & 9.445 & 2.305 & 1.555 & 13.49  \\ 
      10.0 & 9.675 & 2.476 & 1.608 & 13.55  \\ 
     10.25 & 9.927 & 2.590 & 1.648 & 13.58  \\ 
      10.5 & 10.194 & 2.672 & 1.689 & 13.58  \\ 
     10.75 & 10.440 & 2.752 & 1.741 & 13.59  \\ 
      11.0 & 10.688 & 2.819 & 1.781 & 13.60  \\ 
     11.25 & 10.889 & 2.842 & 1.793 & 13.60  \\ 
      11.5 & 10.754 & 2.806 & 1.781 & 13.60  \\ 
     11.75 & 10.775 & 2.999 & 1.910 & 13.63  \\ 
      12.0 & 10.902 & 3.116 & 2.086 & 13.65  \\ 
     12.25 & 11.083 & 3.220 & 2.166 & 13.66  \\ 
      12.5 & 11.188 & 3.342 & 2.254 & 13.68  \\ 
     12.75 & 11.404 & 3.418 & 2.332 & 13.68  \\ 
      13.0 & 11.567 & 3.516 & 2.416 & 13.69  \\ 
      13.1 & 11.469 & 3.636 & 2.496 & 13.71  \\ 
      13.2 & 11.538 & 3.660 & 2.515 & 13.71  \\ 
      13.3 & 11.703 & 3.656 & 2.509 & 13.70  \\ 
      13.4 & 11.733 & 3.708 & 2.566 & 13.71  \\ 
      13.5 & 11.777 & 3.762 & 2.610 & 13.72  \\ 
      13.6 & 11.766 & 3.824 & 2.654 & 13.73  \\ 
      13.7 & 11.885 & 3.835 & 2.680 & 13.73  \\ 
      13.8 & 11.864 & 3.905 & 2.726 & 13.73  \\ 
      13.9 & 12.124 & 3.876 & 2.707 & 13.73  \\ 
      14.0 & 12.079 & 3.950 & 2.765 & 13.74  \\ 
      14.1 & 12.084 & 4.012 & 2.831 & 13.74  \\ 
      14.2 & 12.315 & 3.994 & 2.805 & 13.74  \\ 
      14.3 & 12.347 & 4.038 & 2.864 & 13.74  \\ 
      14.4 & 12.384 & 4.082 & 2.904 & 13.75  \\ 
      14.5 & 12.201 & 4.196 & 3.006 & 13.76  \\ 
      14.6 & 12.599 & 4.131 & 2.952 & 13.75  \\ 
      14.7 & 12.531 & 4.214 & 2.999 & 13.76  \\ 
      14.8 & 12.780 & 4.200 & 3.013 & 13.75  \\ 
      14.9 & 12.894 & 4.223 & 3.040 & 13.76  \\ 
      15.0 & 12.603 & 4.352 & 3.144 & 13.77  \\ 
      15.1 & 12.933 & 4.328 & 3.130 & 13.76  \\ 
      15.2 & 12.584 & 4.459 & 3.257 & 13.78  \\ 
      15.3 & 13.121 & 4.378 & 3.193 & 13.76  \\ 
      15.4 & 13.226 & 4.419 & 3.221 & 13.77  \\ 
      15.5 & 13.279 & 4.451 & 3.227 & 13.77  \\ 
      15.6 & 13.230 & 4.518 & 3.292 & 13.77  \\ 
      15.7 & 13.187 & 4.579 & 3.342 & 13.78  \\ 
      15.8 & 13.414 & 4.579 & 3.378 & 13.78  \\ 
      15.9 & 13.563 & 4.600 & 3.370 & 13.78  \\ 
      16.0 & 13.145 & 4.749 & 3.510 & 13.79  \\ 
      16.1 & 13.741 & 4.672 & 3.435 & 13.78  \\ 
      16.2 & 13.551 & 4.765 & 3.524 & 13.79  \\ 
      16.3 & 13.861 & 4.756 & 3.538 & 13.78  \\ 
      16.4 & 13.974 & 4.787 & 3.552 & 13.79  \\ 
      16.5 & 13.992 & 4.827 & 3.581 & 13.79  \\ 
      16.6 & 13.863 & 4.904 & 3.657 & 13.79  \\ 
      16.7 & 14.067 & 4.915 & 3.679 & 13.79  \\ 
      16.8 & 14.190 & 4.942 & 3.701 & 13.79  \\ 
      16.9 & 14.244 & 4.983 & 3.739 & 13.80  \\ 
      17.0 & 14.301 & 5.018 & 3.801 & 13.80  \\ 
      17.1 & 14.362 & 5.058 & 3.831 & 13.80  \\ 
      17.2 & 14.348 & 5.107 & 3.853 & 13.80  \\ 
      17.3 & 14.514 & 5.141 & 3.900 & 13.81  \\ 
      17.4 & 14.297 & 5.216 & 3.955 & 13.82  \\ 
      17.5 & 14.573 & 5.220 & 3.978 & 13.81  \\ 
      17.6 & 14.689 & 5.254 & 4.000 & 13.82  \\ 
      17.7 & 14.654 & 5.308 & 4.056 & 13.83  \\ 
      17.8 & 14.794 & 5.338 & 4.087 & 13.83  \\ 
      17.9 & 14.882 & 5.372 & 4.135 & 13.83  \\ 
      18.0 & 14.936 & 5.401 & 4.176 & 13.83  \\ 
      18.1 & 15.065 & 5.441 & 4.200 & 13.83  \\ 
      18.2 & 15.024 & 5.485 & 4.223 & 13.84  \\ 
      18.3 & 15.156 & 5.527 & 4.301 & 13.84  \\ 
      18.4 & 15.151 & 5.566 & 4.315 & 13.84  \\ 
      18.5 & 14.943 & 5.651 & 4.385 & 13.85  \\ 
      18.6 & 15.247 & 5.654 & 4.409 & 13.84  \\ 
      18.7 & 15.289 & 5.684 & 4.414 & 13.84  \\ 
      18.8 & 14.977 & 5.799 & 4.522 & 13.85  \\ 
      18.9 & 15.042 & 5.838 & 4.555 & 13.86  \\ 
      19.0 & 15.503 & 5.804 & 4.541 & 13.85  \\ 
      19.1 & 15.576 & 5.838 & 4.584 & 13.85  \\ 
      19.2 & 15.545 & 5.884 & 4.608 & 13.86  \\ 
      19.3 & 15.536 & 5.929 & 4.642 & 13.86  \\ 
      19.4 & 15.537 & 5.975 & 4.715 & 13.86  \\ 
      19.5 & 15.670 & 6.002 & 4.710 & 13.86  \\ 
      19.6 & 15.650 & 6.051 & 4.773 & 13.87  \\ 
      19.7 & 15.757 & 6.087 & 4.817 & 13.87  \\ 
      19.8 & 15.851 & 6.123 & 4.851 & 13.87  \\ 
      19.9 & 15.857 & 6.169 & 4.886 & 13.87  \\ 
      20.0 & 15.848 & 6.207 & 4.970 & 13.87  \\ 
      20.1 & 15.900 & 6.256 & 4.955 & 13.87  \\ 
      20.2 & 15.835 & 6.292 & 5.020 & 13.87  \\ 
      20.3 & 16.057 & 6.316 & 5.035 & 13.87  \\ 
      20.4 & 15.935 & 6.376 & 5.101 & 13.88  \\ 
      20.5 & 16.093 & 6.402 & 5.126 & 13.87  \\ 
      20.6 & 15.660 & 6.533 & 5.243 & 13.89  \\ 
      20.7 & 15.941 & 6.505 & 5.217 & 13.89  \\ 
      20.8 & 16.017 & 6.539 & 5.232 & 13.89  \\ 
      20.9 & 16.232 & 6.558 & 5.288 & 13.89  \\ 
      21.0 & 16.119 & 6.617 & 5.335 & 13.89  \\ 
      21.1 & 16.183 & 6.651 & 5.360 & 13.89  \\ 
      21.2 & 16.372 & 6.682 & 5.373 & 13.90  \\ 
      21.3 & 16.306 & 6.720 & 5.436 & 13.90  \\ 
      21.4 & 16.163 & 6.772 & 5.475 & 13.90  \\ 
      21.5 & 16.044 & 6.842 & 5.537 & 13.91  \\ 
      21.6 & 16.331 & 6.848 & 5.565 & 13.90  \\ 
      21.7 & 16.390 & 6.880 & 5.630 & 13.91  \\ 
      21.8 & 16.336 & 6.924 & 5.629 & 13.91  \\ 
      21.9 & 16.514 & 6.969 & 5.708 & 13.91  \\ 
      22.0 & 16.218 & 7.026 & 5.776 & 13.91  \\ 
      22.1 & 16.297 & 7.045 & 5.786 & 13.92  \\ 
      22.2 & 16.520 & 7.091 & 5.839 & 13.92  \\ 
      22.3 & 16.368 & 7.142 & 5.892 & 13.92  \\ 
      22.4 & 16.385 & 7.174 & 5.932 & 13.92  \\ 
      22.5 & 16.364 & 7.201 & 5.959 & 13.93  \\ 
      22.6 & 16.489 & 7.252 & 6.012 & 13.93  \\ 
      22.7 & 16.386 & 7.307 & 6.066 & 13.93  \\ 
      22.8 & 16.260 & 7.330 & 6.093 & 13.94  \\ 
      22.9 & 16.433 & 7.385 & 6.147 & 13.94  \\ 
      23.0 & 16.468 & 7.419 & 6.174 & 13.94  \\ 
      23.1 & 16.464 & 7.450 & 6.215 & 13.94  \\ 
      23.2 & 16.444 & 7.497 & 6.269 & 13.95  \\ 
      23.3 & 16.533 & 7.525 & 6.282 & 13.95  \\ 
      23.4 & 16.475 & 7.594 & 6.351 & 13.95  \\ 
      23.5 & 16.564 & 7.626 & 6.379 & 13.95  \\ 
      23.6 & 16.131 & 7.668 & 6.420 & 13.96  \\ 
      23.7 & 16.274 & 7.720 & 6.475 & 13.96  \\ 
      23.8 & 16.053 & 7.788 & 6.546 & 13.96  \\ 
      23.9 & 16.460 & 7.799 & 6.573 & 13.96  \\ 
      24.0 & 16.062 & 7.848 & 6.601 & 13.96  \\ 
      24.1 & 16.366 & 7.853 & 6.614 & 13.96  \\ 
      24.2 & 15.236 & 8.077 & 6.830 & 13.98  \\ 
      24.3 & 16.357 & 7.951 & 6.712 & 13.97  \\ 
      24.4 & 15.981 & 8.018 & 6.769 & 13.97  \\ 
      24.5 & 15.988 & 8.038 & 6.782 & 13.97  \\ 
      24.6 & 15.967 & 8.101 & 6.869 & 13.98  \\ 
      24.7 & 15.743 & 8.152 & 6.897 & 13.98  \\ 
      24.8 & 16.215 & 8.161 & 6.910 & 13.98  \\ 
      24.9 & 16.121 & 8.165 & 6.908 & 13.98  \\ 
      25.0 & 15.849 & 8.251 & 7.011 & 13.99  \\ 
      25.1 & 15.827 & 8.302 & 7.039 & 13.99  \\ 
      25.2 & 15.076 & 8.414 & 7.127 & 13.99  \\ 
      25.3 & 16.264 & 8.355 & 7.080 & 13.99  \\ 
      25.4 & 15.621 & 8.432 & 7.169 & 13.99  \\ 
      25.5 & 15.445 & 8.460 & 7.197 & 13.99  \\ 
      25.6 & 15.599 & 8.525 & 7.256 & 14.00  \\ 
      25.7 & 15.867 & 8.569 & 7.315 & 14.00  \\ 
      25.8 & 15.954 & 8.572 & 7.312 & 14.00  \\ 
      25.9 & 15.461 & 8.649 & 7.387 & 14.00  \\ 
      26.0 & 15.462 & 8.688 & 7.385 & 14.00  \\ 
      26.1 & 15.473 & 8.723 & 7.460 & 14.00  \\ 
      26.2 & 15.400 & 8.775 & 7.489 & 14.01  \\ 
      26.3 & 15.276 & 8.826 & 7.549 & 14.00  \\ 
      26.4 & 15.147 & 8.886 & 7.593 & 14.01  \\ 
      26.5 & 15.616 & 8.896 & 7.638 & 14.01  \\ 
      26.6 & 15.317 & 8.959 & 7.683 & 14.01  \\ 
      26.7 & 15.073 & 8.988 & 7.712 & 14.01  \\ 
      26.8 & 15.268 & 9.027 & 7.757 & 14.02  \\ 
      26.9 & 15.449 & 9.031 & 7.770 & 14.01  \\ 
      27.0 & 15.397 & 9.159 & 7.896 & 14.02  \\ 
      27.1 & 15.133 & 9.170 & 7.909 & 14.02  \\ 
      27.2 & 15.281 & 9.184 & 7.921 & 14.02  \\ 
      27.3 & 15.068 & 9.238 & 7.983 & 14.02  \\ 
      27.4 & 15.019 & 9.316 & 8.029 & 14.02  \\ 
      27.5 & 15.008 & 9.306 & 8.058 & 14.02  \\ 
      27.6 & 14.862 & 9.392 & 8.137 & 14.02  \\ 
      27.7 & 14.831 & 9.443 & 8.183 & 14.03  \\ 
      27.8 & 14.846 & 9.493 & 8.263 & 14.03  \\ 
      27.9 & 15.169 & 9.441 & 8.226 & 14.03  \\ 
      28.0 & 15.120 & 9.521 & 8.289 & 14.03  \\ 
      28.1 & 14.747 & 9.599 & 8.352 & 14.03  \\ 
      28.2 & 14.561 & 9.655 & 8.399 & 14.03  \\ 
      28.3 & 14.532 & 9.668 & 8.412 & 14.03  \\ 
      28.4 & 14.235 & 9.776 & 8.527 & 14.03  \\ 
      28.5 & 14.578 & 9.697 & 8.456 & 14.03  \\ 
      28.6 & 14.310 & 9.794 & 8.535 & 14.03  \\ 
      28.7 & 14.363 & 9.838 & 8.582 & 14.03  \\ 
      28.8 & 14.442 & 9.908 & 8.664 & 14.03  \\ 
      28.9 & 14.292 & 9.941 & 8.694 & 14.03  \\ 
      29.0 & 15.087 & 9.867 & 8.662 & 14.03  \\ 
      29.1 & 14.264 & 10.013 & 8.807 & 14.04  \\ 
      29.2 & 14.609 & 10.021 & 8.802 & 14.03  \\ 
      29.3 & 14.181 & 10.080 & 8.849 & 14.03  \\ 
      29.4 & 14.030 & 10.138 & 8.897 & 14.03  \\ 
      29.5 & 13.674 & 10.195 & 8.963 & 14.03  \\ 
      29.6 & 14.094 & 10.215 & 8.976 & 14.03  \\ 
      29.7 & 14.113 & 10.264 & 9.024 & 14.04  \\ 
      29.8 & 13.803 & 10.278 & 9.036 & 14.03  \\ 
      29.9 & 13.882 & 10.379 & 8.277 & 14.03  \\ 
      30.0 & 13.805 & 10.359 & 9.101 & 14.03  \\ 
        31 & 13.634 & 10.821 & 9.611 & 14.01  \\ 
        32 & 13.414 & 11.270 & 10.070 & 13.99  \\ 
        33 & 13.239 & 11.599 & 10.396 & 13.96  \\ 
        35 & 13.663 & 12.561 & 11.426 & 12.16  \\ 
        40 & 15.336 & 14.682 & 13.581 & 11.91  \\ 
        45 & 13.018 & 13.018 & 13.018 & 10.83  \\ 
        50 & 9.817 & 9.817 & 9.817 & 11.58  \\ 
        55 & 9.380 & 9.380 & 9.380 & 11.72  \\ 
        60 & 7.289 & 7.289 & 7.289 & 10.49  \\ 
        70 & 6.408 & 6.408 & 6.408 & 10.63  \\ 
        80 & 6.368 & 6.368 & 6.368 & 10.62  \\ 
       100 & 6.036 & 6.036 & 6.036 & 10.58  \\ 
       120 & 6.160 & 6.160 & 6.160 & 11.55  \\ 
\enddata
\tablecomments{M$_{\rm i}$ is the ZAMS mass, M$_{\rm f}$ is the 
               final presupernova mass and R$_{\rm f}$ is the 
               corresponding final radius. He and CO core masses 
               have been measured using 20\% mass fraction point 
               X(H)$\leq$0.2 for M$_\alpha$ and X(He)$\leq$0.2 
               for M$_{\rm CO}$.}  
               \lTab{progenitor}
\end{deluxetable}


\section{Explosion Modeling}
\lSect{expl}

The presupernova models were exploded using two different
one-dimensional hydrodynamics codes, Prometheus-Hot Bubble (henceforth
P-HOTB; \citealt{Jan96,Kif2003}) and KEPLER. For central engines Z9.6,
W18, and N20, all models were exploded using both codes. For W15, W20,
and S19.8, models were exploded only with P-HOTB. All of our explosion 
results from the P-HOTB calculations are included in the electronic 
edition in a tar.gz package.

P-HOTB includes the necessary neutrino and high-density physics to
follow iron-core collapse and neutrino energy and lepton-number
transport. It also carries a small reaction network with 14 alpha
nuclei \citep{Mul86}, which is capable of tracking bulk
nucleosynthesis \citep{Ugl12,Ert15}. The neutrino transport is
approximated by a fast, gray scheme that solves the coupled set of
neutrino number and energy equations for all flavours as described in
\citet{Sch06}. The implementation of gravity, including general
relativistic corrections, has been detailed in \citet{Arc07}, and the
choice of the boundary conditions for hydrodynamics and transport
applied in our present work follows \citet{Ugl12} and
\citet{Ert15}. Extending the code used by \citet{Ugl12}, \citet{Ert15}
added a simple deleptonization scheme \citep{Lie05} for a fast
treatment of the core-collapse phase until shortly after core
bounce. They also improved the microphysics treatment in several
respects, e.g., by implementing a high-density equation of state (EoS)
and consistently coupling the burning network with the low-density EoS
and hydrodynamics.

KEPLER does not include neutrino transport and its equation of state
becomes unreliable above about 10$^{11}$ g cm$^{-3}$, but it does
carry a much larger nuclear reaction network of roughly 1500 isotopes,
capable of calculating detailed nucleosynthesis, and includes
flux-limited radiative diffusion so that bolometric light curves can
be determined.  For cases where both codes were used, the procedure
that was followed was to first calculate the explosion using P-HOTB,
continued to a sufficiently late time to determine the remnant mass,
and then to post-process the explosion using KEPLER.

\subsection{Explosion Simulations with P-HOTB}
\lSect{coremodel}

\subsubsection{Modeling Methodology}

A detailed discussion of our methodology for neutrino-driven
explosions for large sets of progenitor models can be found in
\citet{Ert15}, and more technical information is also given in 
\citet{Ugl12}. Here we summarize the essential characteristics.

The calculations with P-HOTB assume spherical symmetry and are 1D.
They follow the evolution from the onset of stellar collapse as
determined by the KEPLER code, when the collapse speed anywhere in the
core first exceeds 1000 km s$^{-1}$, through core bounce and
post-bounce accretion to either black-hole (BH) formation or
successful explosion. The fall back of matter that does not become
gravitationally unbound in the explosion is determined by running the
calculation for days to weeks. Proto-neutron star (PNS) cooling and
neutrino emission are followed for typically 15\,s, at which time the
shock has crossed $\sim10^{10}$\,cm and is outside of the CO core. 
By that time the power of the neutrino-driven wind, which
pushes the dense ejecta shell enclosed by the expanding supernova
shock and the wind-termination shock, has decayed to a dynamically
insignificant level. The models are then mapped to a grid extending to
larger radii, and the inner boundary is moved from below the
neutrinosphere to 10$^9$ cm in order to ease the numerical time-step
constraint.

Spherically symmetric, first-principles simulations do not yield
explosions by the neutrino-driven mechanism except for low mass
progenitors with oxygen-neon-magnesium cores \citep{Kit06,Jan08} or
small iron cores \citep{Mel15}. These stars are characterized by an
extremely dilute circum-core density and small compactness values,
$\xi_M$ for $M\ge 1.5$. Ignoring the beneficial effects of
hydrodynamic instabilities in initiating the revival of the stalled
bounce shock, explosions require a suitable tuning of the neutrino
source to revive the shock by neutrino heating.

In order to do so, we exclude the inner 1.1 \Msun \ of the
high-density core of the PNS at densities significantly above the
neutrinospheric region from the hydrodynamical modelling, and replace
them by an inner grid boundary that is contracted to mimic the
shrinking of the core. At this domain boundary, time-dependent
neutrino luminosities and mean energies \citep{Sch06,Arc07,Ugl12} are
imposed. Outside of this boundary (between neutrino optical depths
below $\sim$10, initially, and up to several 1000 after seconds of PNS
cooling), the lepton number and energy transport of all
species of neutrinos is solved using a fast, analytical, gray transport
approximation \citep{Sch06}.  The time-dependent neutrino emission
from PNS accretion, which is determined by the progenitor structure of
the collapsing star and by the dynamics of the shock front, is thus
included in our models, and ensures that corresponding feedback effects
are taken into account, at least as far as they can be represented in
spherical symmetry.

The excised PNS core is represented as a time-dependent neutrino
source using a simple one-zone model whose behaviour is
determined by applying fundamental physics relations
\citep{Ugl12}. The free parameters of this core model are calibrated
by comparing the results of explosion simulations for suitable
progenitor models to observations of well-studied supernovae. For this
purpose we consider important diagnostic parameters of the explosion,
i.e., the explosion energy and the mass of $^{56}$Ni produced, as well
as important properties of the PNS as a neutrino source, i.e., the
total release of neutrino energy and the time scale of this neutrino
emission.  For the neutrino properties, the detected neutrino
signal from SN~1987A is used as a benchmark, but we also make sure that
the neutrino-energy loss of our PNS models is consistent with binding
energies of neutron stars based on microphysical equations of state.

Excluding the high-density core of the PNS from direct modelling can be
justified by the still incomplete knowledge of the equation of state
in the supernuclear regime and significant uncertainties in the
neutrino opacities in dense, correlated nuclear matter, which
overshadow all fully self-consistent modelling. Of course, modelling supernova
explosions in spherical symmetry can only be an intermediate step, and
ultimately three-dimensional simulations will be needed.  Below we
will argue that 1D simulations in the case of tuned explosions exhibit
some aspects that are more similar to current 3D models than 2D
simulations.  According to current multi-dimensional calculations the
onset of the explosion differs between 2D and 3D with the former
favouring earlier explosions because of differences in the behaviour of
turbulence in the post-shock region (e.g., \citealt{Cou13,Han12}). Moreover, 
the imposed constraint of axisymmetry in 2D leads to artificial explosion geometries with pronounced axis features. Also the long-time phase of simultaneous accretion and ejection of neutrino-heated matter seems to 
differ considerably between 2D and 3D \citep{Mul15}, which can lead to 
more energetic explosions in 3D \citep{Mel15,Mul15}.

In the following, we present essential information on how the PNS is
described as neutrino source in our simulations and how the
calibration of this neutrino source is done by referring to
observational data.  Since results of nearly the same set of
presupernova models using P-HOTB have been recently discussed by
\citet{Ert15}, our description here can be terse and focused mainly on
newly introduced aspects.

\subsubsection{The PNS as Neutrino Source}

The central 1.1 \Msun \ core of the PNS resulting from the 
collapse of each presupernova star is excised from the computational
volume and replaced by a contracting
inner grid boundary a few milliseconds after core bounce, shortly
after the expanding shock has converted to a stalled accretion
shock at an enclosed mass of more than 1.1 \Msun. 
For the hydrodynamics we apply the conditions that are needed to
maintain hydrostatic equilibrium at this Lagrangian grid boundary
\citep{Jan96}. The shrinking of the cooling and deleptonizing
PNS is mimicked by the contraction of the inner
grid boundary, $R_\mathrm{ib}$, which is described by an exponential 
function (Eq.~1 in \citealt{Arc07}) with the parameter values given
in Sect.~2 of \citet{Ugl12}. In order to avoid too severe time-step
restrictions that could hamper the tracking of the combined evolution
of PNS and ejecta over more than 20\,s, we choose the case of
``standard'' instead of ``rapid'' contraction shown in Fig.~1
of \citet{Sch06}. This may lead to an underestimation of the 
accretion luminosity and of the gradual hardening of the radiated 
neutrino spectra compared to self-consistent supernova
simulations, but in our approach this can well be compensated by
a corresponding adjustment of the calibration of the inner
core as neutrino source. Since the accretion luminosity computed
by the transport adds to core luminosity imposed at the inner
grid boundary, the calibration sets constraints to the combined
value of these two contributions. 

As detailed in \citet{Ugl12}, we describe the high-density PNS
core of mass $M_\mathrm{c} = 1.1 \Msun$ by a simple one-zone
model under the constraints of energy conservation and the Virial
theorem including the effects associated with the growing pressure
of the accretion layer, whose accumulation around the PNS core is
followed by our hydrodynamic simulations. The one-zone core model
provides a time-dependent total core-neutrino luminosity,
$L_{\nu,\mathrm{c}}(t)$, which determines the boundary luminosity
imposed at the inner grid boundary (at $R_\mathrm{ib}$) of the
computational grid, suitably split between $\nu_e$, $\bar\nu_e$,
and heavy-lepton neutrinos (see \citealt{Ugl12} and \citealt{Sch06}).
Besides the luminosities of the different neutrino species also
the mean energies (or spectral temperatures) of the inflowing
neutrinos are prescribed at $R_\mathrm{ib}$ (cf.\ \citealt{Ugl12}).

The total core-neutrino luminosity $L_{\nu,\mathrm{c}}(t)$ is 
determined by the core mass, $M_\mathrm{c}$, the core radius,
$R_\mathrm{c}(t)$, the rate of contraction of this radius,
$\dot R_\mathrm{c}(t)$, the mass of the PNS accretion mantle around
the core, $m_\mathrm{acc}$ (taken to be the mass between the inner 
grid boundary and a lower density of $10^{10}\,$g\,cm$^{-3}$),
and the mass-accretion rate of the PNS, $\dot m_\mathrm{acc}$,
as computed in our hydrodynamical runs.
Equation~(4) of \citet{Ugl12} for the core luminosity can be 
rewritten as
\begin{equation}
L_{\nu,\mathrm{c}}(t) = \frac{1}{3(\Gamma-1)}
\left[(3\Gamma-4)(E_\mathrm{g}+S)\,\frac{\dot R_\mathrm{c}}{R_\mathrm{c}}
\,+\,S\,\frac{\dot m_\mathrm{acc}}{m_\mathrm{acc}}\right]\,,
\label{eq:corelum}
\end{equation}
with the factors
\begin{eqnarray}
E_\mathrm{g}+S &=& -\,\frac{2}{5}\,\frac{GM_\mathrm{c}}{R_\mathrm{c}}\,
\left(M_\mathrm{c} + \frac{5}{2}\,\zeta\,m_\mathrm{acc}\right)\,,
\label{eq:lterm1} \\
S &=& -\,\zeta\,\frac{G M_\mathrm{c}m_\mathrm{acc}}{R_\mathrm{c}}\,,
\label{eq:lterm2}
\end{eqnarray}
and with the adiabatic index $\Gamma$ and coefficient $\zeta$ 
($={\cal O}(1)$) being free parameters of the model. We note that
through the $S$-dependent terms the PNS-core luminosity depends on 
the structure (e.g., compactness) of the progenitor star, which 
determines the mass-accretion rate $\dot m_\mathrm{acc}$ and thus
the mass $m_\mathrm{acc}$ of the accretion mantle of the PNS.
The core radius as function of time is prescribed by
\begin{equation}
R_\mathrm{c}(t) \ = \ R_\mathrm{c,f} + 
\frac{R_\mathrm{c,i} - R_\mathrm{c,f}}{(1 + t)^n}\,,
\label{eq:rc}
\end{equation}
where the time $t$ is measured in seconds. The initial PNS core
radius, $R_\mathrm{c,i}$, is set equal to the initial radius of the
inner boundary of the hydrodynamic grid, $R_\mathrm{ib}^\mathrm{i}$,
and the final PNS core radius, $R_\mathrm{c,f}$, as well as the
exponent $n$ are also free parameters of our model.
Note that $R_\mathrm{c}$ should be considered as a representative
radius of the high-density core. Since we grossly simplify the
radial structure of the PNS core by a one-zone description, the 
numerical value of $R_\mathrm{c}$ has no exact physical meaning
except that is represents a rough measure of the size of the 
PNS core. However, neither its time evolution
nor its final value need to be equal to the inner boundary radius
of the hydrodynamic grid or to the radius of the 1.1 \Msun \ 
core of a self-consistently computed PNS model using a nuclear 
EoS. Nevertheless, with a suitably chosen (calibrated)
time dependence of $R_\mathrm{c}$, our model
is able to reproduce basic features of the neutrino emission of
cooling PNSs, and the choice of $n$ and $R_\mathrm{c,f}$ can 
be guided by the temporal behaviour of $R_\mathrm{c}(t)$ in
sophisticated PNS cooling calculations (see below).

Since both of the expressions of Eqs.~(\ref{eq:lterm1}) and
(\ref{eq:lterm2}) are negative and $\dot R_\mathrm{c} < 0$
and $\dot m_\mathrm{acc} > 0$ for the contracting and accreting
PNS, the first term in the brackets
on the r.h.s.\ of Eq.~(\ref{eq:corelum}) is
positive while the second term is negative. The former term 
represents the gravitational binding energy release as a 
consequence of the PNS-core settling in the gravitational 
potential including the compressional work delivered by the
surrounding accretion layer. In contrast, the latter (negative)
term accounts for the additional internal energy (and pressure) 
of the core that is needed for gravitational stability
when the overlying mantle grows in
mass. This second term thus reduces the energy that can be 
radiated in neutrinos.

\begin{deluxetable}{ccccccc}
\tablewidth{0pt}
\tablecaption{pns core-model parameters in p-hotb}
\tablehead{ \colhead{Model} & \colhead{$R_\mathrm{c,f}$ [km]} &
\colhead{$\Gamma$} & \colhead{$\zeta$} & \colhead{$n$} &
  \colhead{E$_{51}$} & \colhead{M($^{56}$Ni+ 1/2 Tr)}  }
\startdata
Z9.6   &     7.0    &  3.0     & 0.65 & 1.55 & 0.16 & $\:$0.0087 \\
S19.8  &     6.5    &  3.0     & 0.90 & 2.96 & 1.30 & 0.089  \\
W15    &     6.0    &  3.0     & 0.60 & 3.10 & 1.41 & 0.068  \\
W18    &     6.0    &  3.0     & 0.65 & 3.06 & 1.25 & 0.074  \\
W20    &     6.0    &  3.0     & 0.70 & 2.84 & 1.24 & 0.076  \\
N20    &     6.0    &  3.0     & 0.60 & $\:$3.23 & 1.49 & $\:$0.062 
\lTab{P-HOTBparams}
\end{deluxetable}

\begin{figure}[t]
\centering
\includegraphics[width=.48\textwidth]{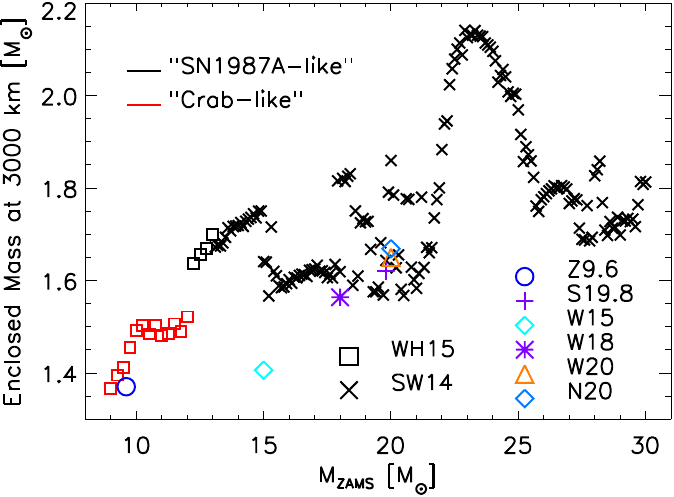}
\caption{ Mass inside a radius of 3000\,km for all progenitors up to
  30 \Msun \ at the time when the central density reaches the same
  value of $3\times 10^{10}$\,g\,cm$^{-3}$. The black crosses and
  squares mark indicate ``SN1987-like'' cases; the red squares,
  ``Crab-like'' ones. Squares are models similar to \citet{Woo15};
  crosses, those of the \citet{Suk14} progenitor set.  The locations of
  the calibration models are indicated by colored symbols as
  indicated in the figure legend. While N20, W20, and S19.8 cluster
  closely, W15 lies far off because this model has a very small
  compactness, $\xi_{2.5}$ (see \citealt{Ert15}).
\label{fig:mass3000}}
\end{figure}

\begin{figure}[h]
\centering
\includegraphics[width=.48\textwidth]{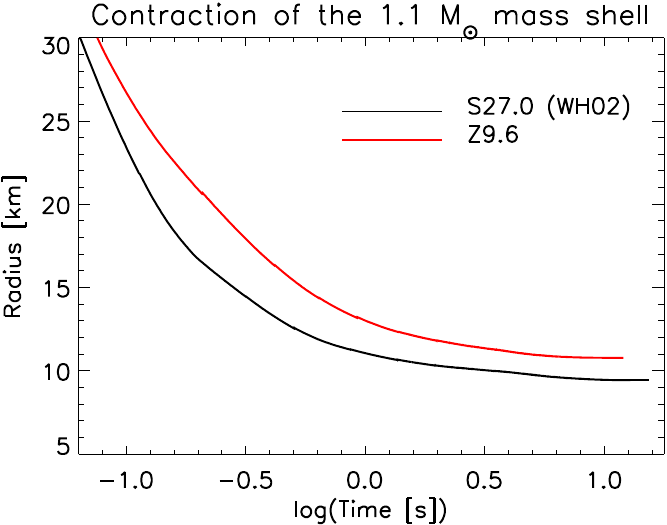}
\caption{ Radius of the 1.1 \Msun\ core as a function of post-bounce
  time for a low-mass neutron star (baryonic mass 1.36 \Msun;
  gravitational mass 1.25 \Msun) characteristic of the Z9.6 \Msun
  \ progenitor (red line), and for a high-mass neutron star (baryonic
  mass 1.78 \Msun; gravitational mass 1.59 \Msun) characteristic of a
  27 \Msun \ progenitor (black line). Both cores were evolved in a
  self-consistent PNS cooling simulation with detailed neutrino
  transport and the LS220 high-density EoS of \citet{Lat91}.  The
  same simulations were recently reported by \citet{Mir15}.  The different
  contraction behaviour of the high-density cores of low-mass and
  high-mass PNSs is obvious.
\label{fig:pnscontraction}}
\end{figure}

\subsubsection{Calibration to SN~1987A and SN~1054}
\lSect{calibrate}

The four free parameters of the PNS-core model, $\Gamma$, $\zeta$, $n$,
and $R_\mathrm{c,f}$, determine the core-neutrino emission as given by
Eq.~(\ref{eq:corelum}) and thus the solution to a given supernova explosion.
The observational cases that serve our needs for calibrating these
free parameters are SN~1987A for high-mass stars and the Crab supernova,
SN~1054, for low-mass stars. The progenitors and explosion properties
of these supernovae are distinctly different and require different
values for our model parameters as we shall argue below. For both 
SN~1987A and SN~1054 the explosion energy and $^{56}$Ni mass are fairly
well determined and reasonably good guesses for the progenitors exist.

For the SN~1987A progenitor, \citet{Ert15} considered five different
15--20 \Msun\ models, W18, N20, S19.8, W15, and W20.  The parameter
values that were found to approximately reproduce the known explosion
energy and $^{56}$Ni production of SN1987A, 1.3--$1.5\times
10^{51}$\,erg and 0.0723--0.0772 \Msun, respectively
\citep{Utr05,Utr11,Utr14}, are given in \Tab{P-HOTBparams}. In the
table, $E_{51}$ is the explosion energy in units of 10$^{51}$
erg=1 B=1 bethe, and M($^{56}$Ni + 1/2 Tr) is the mass of
shock-produced $^{56}$Ni plus one-half of the iron group species in
the neutrino-powered wind. Because $Y_e$ in the wind is not precisely
determined, the actual $^{56}$Ni synthesis is between M($^{56}$Ni) and
M($^{56}$Ni + Tr) \citep{Ert15}.  In practice, $\Gamma$, and for most
cases $R_\mathrm{c,f}$, are held constant ($\Gamma =3$ and
$R_\mathrm{c,f} = 6.0$\,km), while $n$ and $\zeta$ are adjusted in an
iterative process.

Once the parameter values of the PNS core-model are determined 
for a given SN~1987A progenitor, that same core history is used
in all the other presupernova stars of different masses. In this way
a set of supernovae calibrated to a given SN 1987A progenitor is
generated. This set of explosions is named according to the 
calibration model, so we have the ``W18 series'', the ``N20 series'', 
etc. In this paper we mainly focus on the sets of explosions for the
W18 and N20 calibrations. 

A problematic point in the results of \citet{Ugl12} was an
overestimation of the explosion energies for lower-mass
supernovae. Because of the calibration of the PNS-core parameters with
SN~1987A models, a high core-neutrino luminosity caused a strong
neutrino-driven wind for low-mass neutron stars. This led to explosion
energies in excess of 1.5\,B for most stars between 10 \Msun \ and
15 \Msun. In particular for progenitors below $\sim$12 \Msun \
such high energies are, on the one hand, not compatible with recent
self-consistent 2D and 3D models of the explosion, which obtain much
lower explosion energies in the range of $\lesssim$0.1\,B to at most a
few 0.1\,B for progenitors from $\sim$\,8.8 \Msun \ to
$\sim$12 \Msun \ with oxygen-neon-magnesium or iron cores
\citep{Kit06,Jan08,Jan12,Mel15,Mul15}.  Given the relatively dilute
layers and steep density gradients that surround the cores of these
progenitors, the hydrodynamic explosion models also suggest that it
will be very difficult to reach higher energies by the neutrino-driven
mechanism and our present knowledge of the involved physics. On the
other hand, high explosion energies for low-mass progenitors are also
disfavoured by observations, which seem to suggest some correlation
between the progenitor (or ejecta) mass and the explosion energy
(e.g., \citealt{Poz13,Chu14}), although the correlation appears
weakened in the light of a more detailed statistical analysis
\citep{Pej15}.

\begin{figure*}[!]
\centering
\includegraphics[width=.98\textwidth]{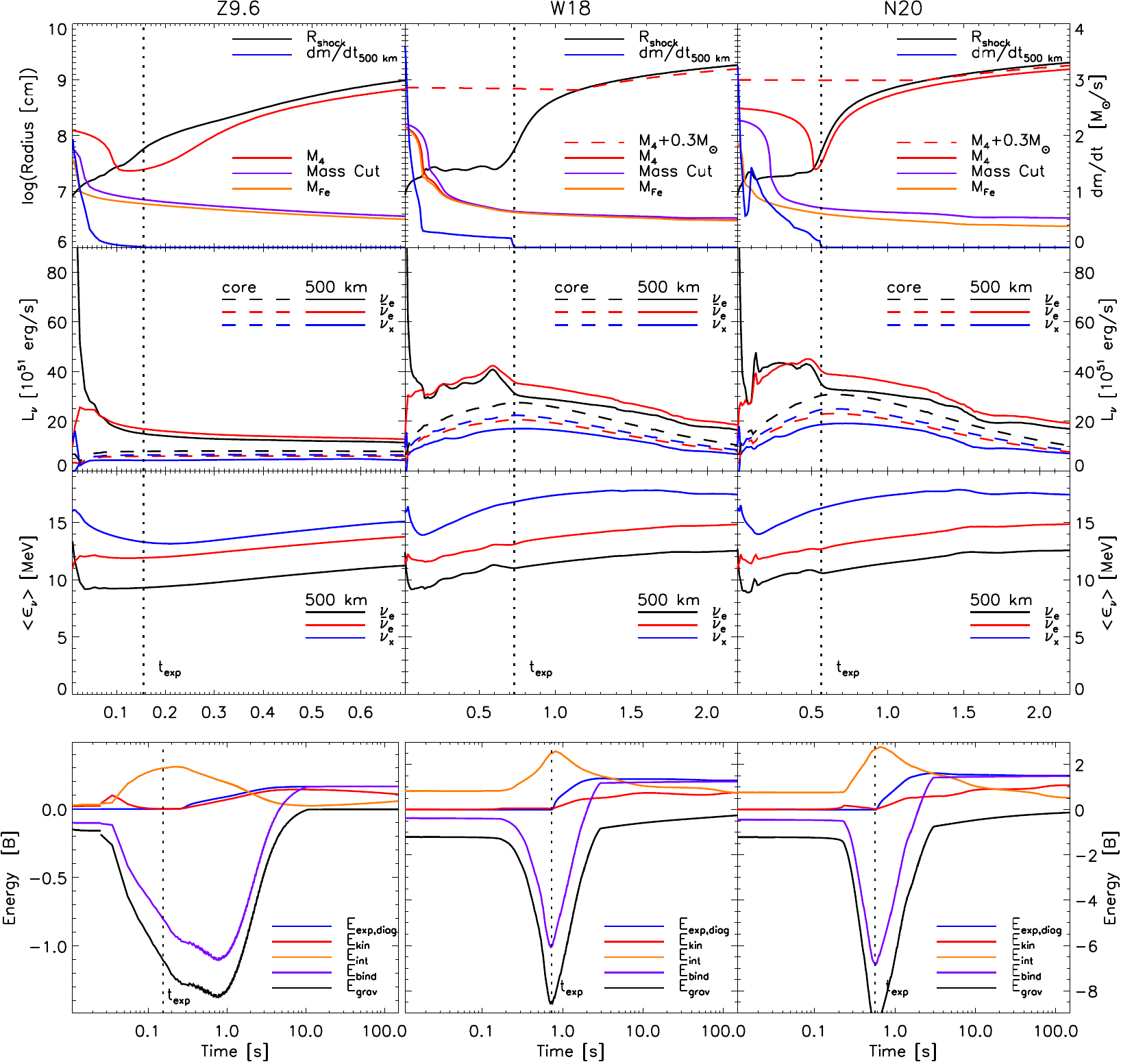}
\caption{
Time evolution of the Crab-like calibration model Z9.6 ({\em left}), and of
the SN~1987A calibration models W18 ({\em middle}) and N20 ({\em right}).
The {\em top panels} display as functions of post-bounce time the radius of
the outgoing shock (black line), the mass accretion rate measured at
500\,km (blue line; scale on the right side), and the radii of iron core
(orange), $M_4=m(s=4)$ (red; $s$ is the entropy per nucleon),
$M_4+0.3$ \Msun \ (red dashed) and trajectory
of the final mass cut (after completion of fallback; purple).
The {\em second panels from top} show the luminosities
of $\nu_e$, $\bar\nu_e$ and a single species of heavy-lepton neutrinos $\nu_x$
as labelled in the plot, measured at 500\,km (solid lines) and at the inner
grid boundary (dashed lines). The {\em third panels from top} show the mean
energies of all neutrino species as radiated at 500\,km. The vertical dotted
lines indicate the onset time of the explosion defined by the moment when
the outgoing shock passes the radius of 500\,km.
The {\em bottom panels} display the diagnostic energy
of the explosion (integrated energy of all post-shock zones with positive
total energy; blue line). Also shown are the kinetic energy (red),
gravitational energy (black), and internal energy (orange) as integrals
over the whole, final SN ejecta between the final mass cut (after fallback)
on the one side and the stellar surface on the other. The total (binding)
energy (purple) as the sum of these energies ultimately converges to the
diagnostic energy and both of these energies asymptote to the final
explosion energy. While this convergence is essentially reached after
$\sim$10\,s in the case of Z9.6, the convergence of total energy and
diagnostic energy takes tens of seconds in the other two cases.
\label{fig:z96w18n20evol}}
\end{figure*}

\begin{figure*}[!]
\centering
\includegraphics[width=.98\textwidth]{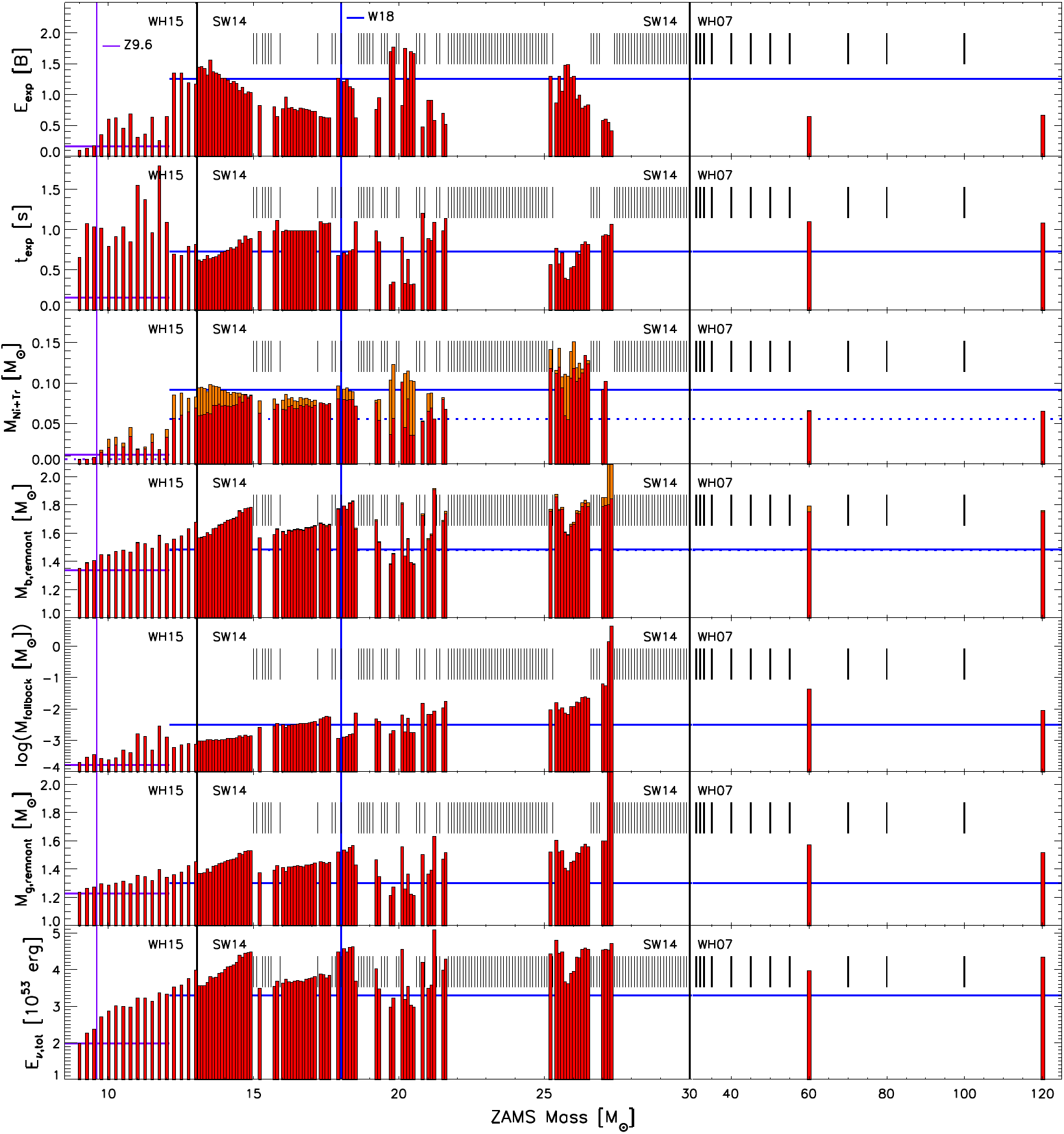}
\caption{Explosion properties for all models exploded with the Z9.6
  and W18 calibrations.  Black vertical lines mark the boundaries
  between the different progenitor sets our model sample is composed
  of. The panels, {\em from top to bottom}, show the final explosion
  energy, E, in units of 1\,B\,=\,1\,Bethe\,=\,$10^{51}$\,erg, the
  time of the onset of the explosion, $t_\mathrm{exp}$, the mass of
  finally ejected, explosively produced $^{56}$Ni (red bars) and
  tracer element (orange bars), the baryonic mass of the compact
  remnant with the fallback mass indicated by orange sections on the
  bars, the fallback mass, the gravitational mass of the compact
  remnant, and the total energy radiated in neutrinos,
  $E_{\nu,\mathrm{tot}}$. The masses of the calibration models are
  indicated by vertical blue lines, and the corresponding results by
  horizontal solid or dashed blue lines, which extend over the mass
  ranges that are considered to have Crab-like or SN1987A-like
  progenitor properties, respectively. Non-exploding cases are marked
  by short vertical black bars in the upper half of each panel.
\label{fig:w18expls}}
\end{figure*}

\begin{figure*}[!]
\centering
\includegraphics[width=.98\textwidth]{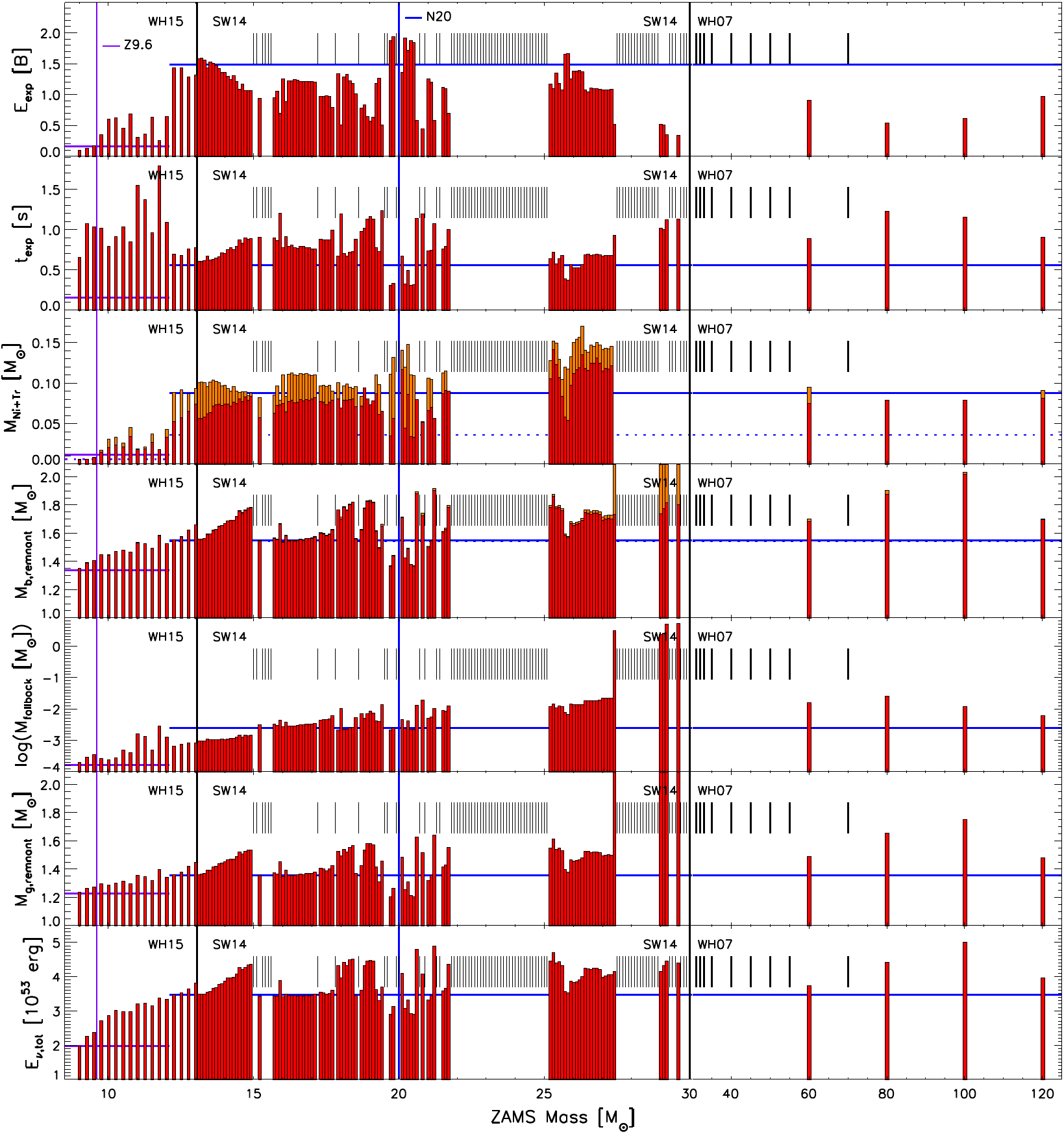}
\caption{
Same as Fig.~\ref{fig:w18expls}, but for P-HOTB simulations with the
Z9.6 and N20 calibrations.
\label{fig:n20expls}}
\end{figure*}

\citet{Ert15} applied a ``quick fix'' to cure the problem of
over-energetic low-mass explosions by using a linear scaling of the
compression parameter $\zeta$ of the PNS core model, coupled to the
decreasing compactness $\xi_{1.75}$ for progenitors with $M\le 13.5$
\Msun.  Here, we elaborate on the background of this measure and
introduce an alternative, similarly effective but physics-wise more
transparent approach than the $\zeta$ modification.

Figure~\ref{fig:mass3000} displays the mass inside of a radius
of 3000\,km for all considered progenitor stars up to 30 \Msun \
at the onset of core collapse. For most appropriate comparison
we evolved all models until they had contracted
to the same central density of $3\times 10^{10}$\,g\,cm$^{-3}$. 
The low-mass stars with $M\le 12.0$ \Msun \ are clearly separated
from the more massive progenitors by smaller enclosed masses
$M_{3000} = m(r\le 3000\,\mathrm{km})$. Since the Crab remnant is
thought to have originated from the explosion of a star in the
$\sim$9--10 \Msun \  
range (e.g., \citealt{Nom82,Nom87,Smi13,Tom13}), we 
call the models with small $M_{3000}$ ``Crab-like'', whereas we
consider the progenitors with $M> 12.0$ \Msun \ and considerably
larger values of $M_{3000}$ as ``SN1987A-like''. Our calibration
models for SN~1987A possess $M_{3000}$ cores that join with the 
SN~1987A-like cases (see in Fig.~\ref{fig:mass3000} the purple 
star and blue diamond for the W18 and N20 progenitors, respectively,
and the colored symbols for the other calibration models).

Since the value of $M_{3000}$ provides the mass that is located
within 3000\,km from the center, it correlates with the
mass of the PNS that forms in the collapse of the stellar core.
It is important to realize that the contraction behaviour and
final radius of the innermost 1.1 \Msun \ (which we describe
here by our analytic one-zone model) depend in a 
microphysical model on the mass of the newly formed NS. This
is shown in Fig.~\ref{fig:pnscontraction}, where we plot the
time evolution of the 1.1 \Msun \ cores for two self-consistent
PNS cooling simulations with all relevant microphysics
\citep{Mir15}, carried out with the nuclear EoS of \citet{Lat91}.
The slower and less extreme contraction in the case of the 
low-mass PNS compared to the high-mass PNS suggest that less
neutrino energy is more gradually released in the low-mass case.
This difference needs to be taken into account when modelling the
explosions of the low-mass progenitor stars of our ``Crab-like''
sample in order to avoid the overestimation of the explosion
energy in this subset of the progenitors.

Because the simple one-zone core model does not allow to just 
adopt the core-radius evolution from a self-consistent simulation,
we calibrate the ``Crab-like'' explosions by reproducing the
observationally inferred explosion energy of SN~1054 
($\sim$0.05--0.15\,B; \citealt{Tom13,Smi13,Yan15}) and
$^{56}$Ni mass ($\lesssim 10^{-2}$\Msun; \citealt{Tom13,Smi13})
with a suitable progenitor model.

\Tab{P-HOTBparams} lists the parameter values obtained in
our Crab calibration of the Z9.6 model. We again use $\Gamma=3.0$
and $\zeta=0.65$ as in the W18 series, but now employ a reduced
value of $n=1.55$ and a larger value of $R_\mathrm{c,f} = 7$\,km,
because these values describe a slower contraction of the 
PNS core to a larger final radius (cf.\ Eq.~\ref{eq:rc}) as found
for a low-mass PNS in the self-consistent PNS cooling simulation
(Fig.~\ref{fig:pnscontraction}).

Since model Z9.6 for the Crab progenitor is in the extreme corner 
of the Crab-like sample in terms of mass and $M_{3000}$
(see the blue circle in Fig.~\ref{fig:mass3000}) and the PNS
masses in this sample vary considerably (see \Fig{w18expls} and 
\Fig{n20expls}), we determine
the values of $n$ and $R_\mathrm{c,f}$ for the other Crab-like
progenitors (with masses $M$) by
interpolating as function of $M_{3000}$ between the Crab model 
and the SN1987-like progenitor with the smallest value of 
$M_{3000}$, which is the case at 15.2 \Msun \ in 
Fig.~\ref{fig:mass3000}, model SW14-15.2. On the SN~1987A-like
side we use here the SN~1987A calibration values of $n$ and
$R_\mathrm{c,f}$ for the W18 series. Our interpolation functions
for $y \equiv n,\,R_\mathrm{c,f}$ therefore are:
\begin{equation}
y(x) = y_0 + \frac{y_1-y_0}{x_1-x_0}\,(x-x_0)\,,
\label{eq:interpol}
\end{equation}
with $x = M_{3000}(M)$, $x_0 = M_{3000}(\mathrm{Z9.6})$,
$x_1 = M_{3000}$(SW14-15.2), $y(x) = X[M_{3000}(M)]$,
$y_0 = X[M_{3000}(\mathrm{Z9.6})]$, and
$y_1 = X[M_{3000}$(SW14-15.2)].
For all SN1987A-like progenitors, we still apply the same values
of the model parameters as obtained for the SN~1987A calibration
of each series and listed in \Tab{P-HOTBparams}.


For the 9.0 \Msun \ progenitor, which has a lower mass $M_{3000}$ than
Z9.6, we use the same values of all core-model parameters as for the
9.6 \Msun \ case. Although we employ Z9.6 as our template case for Crab
and can also refer to a recent 3D explosion simulation of this model
\citep{Mel15} for its parameter calibration, our 9.0 \Msun \ explosion
is equally well compatible with Crab. The computed properties of this
explosion (energy around 0.1\,B and $^{56}$Ni mass below 0.01 \Msun;
see following section) turn out to be fully consistent with
the observational limits for Crab.

\subsubsection{Explosion Results with P-HOTB}
\lSect{sec:explosionresults}

The basic explosion properties of all SN~1987A calibration
models considered here can be found in Table~1 of \citet{Ert15}.
Figure~\ref{fig:z96w18n20evol} shows the evolution towards explosion
for our calibration models Z9.6, W18, and N20. The shock radius and
characteristic mass trajectories as well as neutrino luminosities and
mean neutrino energies are given in the upper panels, and the
different contributions to the ejecta energy in the lower panels.
 
The expansion of the shock radius in model Z9.6 reproduces the results
of multi-dimensional simulations \citep{Mel15} fairly well, and the
diagnostic explosion energy also reaches $\sim$0.1\,B after roughly
0.5\,s to asymptote to a final energy of about 0.16\,B after several
seconds. This energy is in the upper range of estimates for SN~1054
but still compatible with them \citep{Tom13,Smi13,Yan15}. Also the
luminosities and mean energies of electron neutrinos ($\nu_e$) and
antineutrinos ($\bar\nu_e$), which are the crucial agents for driving
and powering the explosions, are in the ballpark of results of
sophisticated transport calculations for this stellar model
(cf.\ Fig.~12 of \citealt{Mir15}).  In all displayed runs the escaping
luminosities (solid lines in the second panel from top) for $\nu_e$
and $\bar\nu_e$ are enhanced compared to the core luminosities (dashed
lines) due to the neutrino emission produced by the accretion and
mantle cooling of the PNS.

While the Z9.6 progenitor explodes relatively quickly (at
$t_\mathrm{exp}\sim\,$0.15\,s after core bounce, defined as the time
when the outgoing shock passes 500\,km), it takes more than 0.5\,s for
the shocks in W18 and N20 to expand to 500\,km and to accelerate
outwards (vertical dashed lines in Fig.~\ref{fig:z96w18n20evol}).  In
all three cases the explosion sets in after the mass shell $M_4 =
m(s=4)$, where the progenitor entropy reaches a value of
4\,$k_\mathrm{B}$ per baryon, has fallen through the shock (see also
\citealt{Ert15}).

Figures~\ref{fig:w18expls} and \ref{fig:n20expls} provide an overview
of the explosion and remnant properties of all simulated models for
our Z9.6, W18 and N20 series. The general features are similar to the
results obtained by \citet{Ugl12} for a different progenitor set, and
they are identical to those discussed by \citet{Ert15} for stars with
ZAMS masses $M$ greater than 13 \Msun. In the mass range $M\le
13$ \Msun \ our present work does not only employ updated progenitor
models, but in contrast to the (seemingly ad hoc) $\zeta$ reduction
applied by \citet{Ert15} we also introduced the more elaborate
treatment of the Crab-like models (i.e., of cases $M\le
12.0$ \Msun) described in Sect.~\ref{sec:calibrate} to cure the
problem of over-energetic low-mass explosions in \citet{Ugl12}.

Figures~\ref{fig:w18expls} and \ref{fig:n20expls} demonstrate that the
$\zeta$ modification used by \citet{Ert15} (see Fig.~3 there) leads to
basically the same overall explosion results as our present Crab-like
calibration for the low-mass progenitors.  This can be understood by
the mathematical structure of the core luminosity as given by
Eq.~(\ref{eq:corelum}), where a similar effect on $L_{\nu,\mathrm{c}}$
can be achieved by either a reduction of the absolute value of $S$ (by
decreasing $\zeta$) or an increase of $R_\mathrm{c}$ through a larger
value of $R_\mathrm{c,f}$.  As discussed in detail by \citet{Ert15},
the explosions of these stars set in fairly late after an extended
phase of accretion, and the explosion is powered by a relatively
massive and energetic neutrino-driven wind. This seems compatible with
the recent result of multi-dimensional simulations by \citet{Mul15},
who found a long-lasting phase of simultaneous accretion and mass
ejection after the onset of the explosion for stars in the mass range
between 11 \Msun \ and 12 \Msun.

While supernovae near the low-mass end of the investigated mass range
tend to explode with the lowest energies and smallest production of
$^{56}$Ni, our results exhibit a moderately strong correlation of
these explosion parameters with ZAMS mass. Overall, our results appear
compatible with the observational data collected by \citet{Poz13}, in
particular in view of the weakness of the $M$-$E_\mathrm{exp}$
correlation concluded from a critical assessment of the observational
analysis by \citet{Pej15}.

Although the remainder of the paper focuses on models based on the W18
and N20 calibrations, some results from P-HOTB are also given for the
other three series, W15, W20, and S19.8. Among the SN~1987A models,
W15 has both an unusually low compactness parameter (Table~1 in
\citealt{Ert15}) and an extremely low value of $M_{3000}$
(Fig.~\ref{fig:mass3000}), more Crab-like than
SN1987A-like. Nevertheless, we found that W15, calibrated as engine as
described in Sect.~\ref{sec:calibrate}, was almost as successful
a central engine as W18, N20, and S19.8. In contrast, W20 is fairly
inconspicuous with respect to compactness and $M_{3000}$, but turned
out to be the clearly weakest engine with the largest number of failed
explosions.

The behavior of W15 can be understood in terms of the two parameters
$M_4$ and $\mu_4$ that determine neutrino-driven explosions as
discussed by \citet{Ert15}. $M_4$ is the stellar mass enclosed by the
radius where the entropy per baryon is $s = 4\,k_\mathrm{B}$ and
$\mu_4$, the gradient, $dM/dr$, just outside of this location.  W15
has much lower values of both $M_4$ and $\mu_4$ than all the other
engine models. For this reason W15 has a low accretion component of
the neutrino luminosity. Exploding SN~1987A with sufficiently high
energy using W15 therefore requires that the parameters of the high-density
core model provide a sufficiently powerful explosion, on their
own, without much assistance from accretion. With this
strong core component of the neutrino emission W15 also acts as strong
engine for other progenitor stars.

W20 is weak for other reasons. Its values of $M_4$ and $\mu_4$ are
actually quite similar to those of N20, which is a considerably
stronger engine. The main reason for the different strengths of these
two cases is the fact that we tuned the N20 parameters such that
SN~1987A exploded with about 1.5\,B (\Tab{P-HOTBparams}), whereas for
W20 we accepted a SN~1987A explosion energy of only 1.24\,B (see
Table~1 in \citealt{Ert15}). The higher explosion energy for N20 was
necessary to obtain enough $^{56}$Ni from this explosion and is
responsible for N20 being the more successful engine.


\subsection{Simulating the Explosion in KEPLER}
\lSect{kepler}

Linking the successful explosions back into the KEPLER code required
care in order to preserve the energetics, remnant masses, and bulk
nucleosynthesis determined in the more accurate simulation with 
neutrino physics. KEPLER is an implicit Lagrangian hydrodynamics code, 
and thus it can take longer time steps and carry a reaction network of
arbitrary size. Since the presupernova models used for the neutrino
transport calculation were calculated using the KEPLER code, it was
most natural to make the link at that time. Within the Lagrangian
code, one then has two options for simulating the explosion - dumping
a prescribed amount of energy in one or several zones, or moving a
specified inner boundary, a ``piston'', along a specified trajectory,
radius as a function of time.


\begin{figure}[h]
\centering
\includegraphics[width=.48\textwidth]{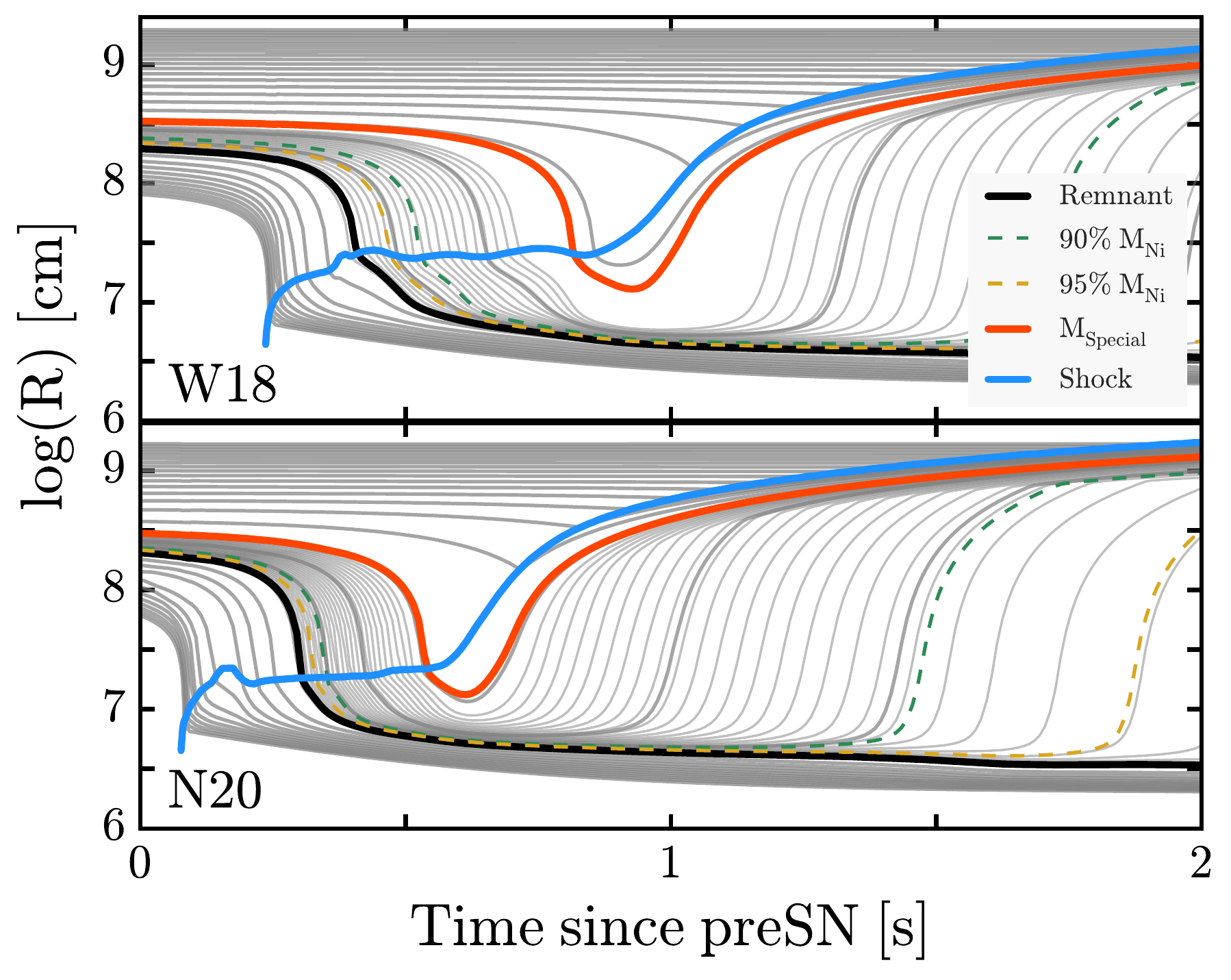}
\caption{ Evolution of baryonic mass shells from the explosion of
  two calibration models, W18 and N20. Thick gray lines are contours
  of constant enclosed masses in steps of 0.04 \Msun\ starting at 1.1
  \Msun.  The thin gray lines are mass shells in steps of 0.004
  \Msun\ resolving the neutrino-driven wind, which is defined as the
  difference between the proto-neutron star mass at the onset of the
  explosion and its mass after $\sim$15\,s.  The solid blue line shows
  the shock radius and solid black curve shows the mass-shell of the
  final mass cut. The solid red line shows the ``special trajectory''
  used in the KEPLER code. Dashed yellow and green lines show the
  location of mass shells internal to the initial ejection of 95\% and
  90\% of the iron-group elements in the P-HOTB calculations.
  \lFig{trajectory}}
\end{figure}

Here the piston approach was adopted, but considerable experimentation
was devoted to determining the optimal trajectory. Major recoding
would have been necessary to include neutrino energy transport in
KEPLER, so one requirement was that the piston trajectory be located
outside of the radius where, even at bounce, neutrino heating and
modification of the composition were negligible. On the other hand,
the piston needed to be deep enough in the star to give an accurate
estimate of the explosion energy and iron synthesis. Experience showed
that a piston situated too deep in the star accreted too much matter
while waiting for neutrinos to reverse the accretion. The sudden
outward motion of the large artificially accreted mass resulted in an
overly powerful explosion and too much $^{56}$Ni production. Similarly, a
piston situated too far out experienced an inadequate peak temperature to
make iron. The density also was too low to acquire enough momentum to
make a strong shock unless the piston was moved for a very long time.


\begin{figure}[h]
\centering
\includegraphics[width=.48\textwidth]{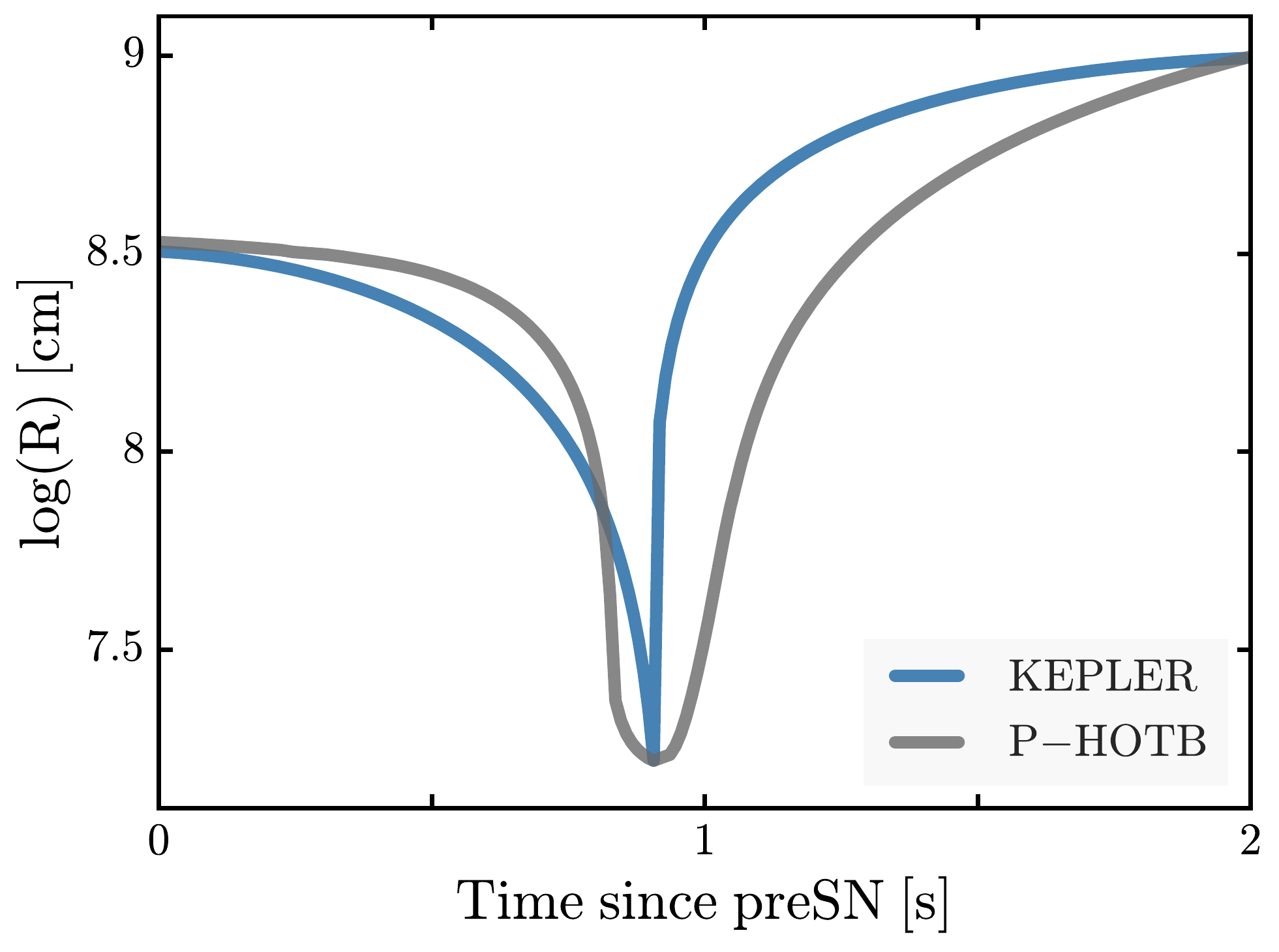}
\caption{The trajectory from the neutrino-driven explosion with P-HOTB
  (gray) is compared against the corresponding trajectory from KEPLER
  (blue) for the explosion of the W18 engine itself. In all cases, the
  trajectories from the two codes have a common starting radius and
  time and the same minimum radius and time.  \lFig{trajs}}
\end{figure}

After some experimentation with various trajectories, the ``special
trajectories'' were chosen (\Fig{trajectory}). These are Lagrangian
mass coordinates that, in the neutrino transport calculation, bounce
between 80 and 140 km (most between 120 and 140 km). They mark the
first mass shell to be accelerated outwards when the stalled shock
revives.  Several such trajectories were followed in KEPLER and gave
reasonably good agreement, typically within 20\%, with the results
from P-HOTB for both supernova energy and iron-group synthesis.
Better agreement could be obtained, however, if the significant
stagnation at small radius for the ``special trajectory'' used by
KEPLER was shortened. The procedure that was eventually adopted used a
piston with following velocity evolution for inward ($t<t_{\rm min}$)
and outward ($t>t_{\rm min}$) motions, where $t_{\rm min}$ is the time
when the piston reaches its smallest radius, $r_{\rm min}$:

\begin{equation}
\begin{aligned}
v_{\rm in} &= (r_{\rm min}-r_z-v_zt_{\rm min})t/t_{\rm min}^2+v_z & \rm for\ t<t_{\rm min} \\
v_{\rm out} &= \sqrt{\frac{2G\alpha m_z(r_{\rm max}-r)}{(r_{\rm max}r)}} & \rm for\ t>t_{\rm max}
\end{aligned}
\end{equation}

\noindent here subscript $z$ denotes the initial piston location 
in the presupernova model according to the special trajectory 
from the P-HOTB calculation, i.e. $m_z, r_z, v_z$ are the
initial Lagrangian location of the trajectory, the corresponding
initial radius and collapse velocity. The inward path of the piston
follows a parabola that connects this initial location to $r_{\rm
  min}$ at $t_{\rm min}$. The outward path from $r_{\rm min}$ to
$r_{\rm max}$ (taken as $10^9$cm), is determined by the time-dependent
velocity that is equal in speed with the free fall velocity that
corresponds to the gravitational potential from the piston mass $m_z$
times the multiplier $\alpha$ (\Fig{trajs}).


\begin{figure}[h]
\centering
\includegraphics[width=\columnwidth]{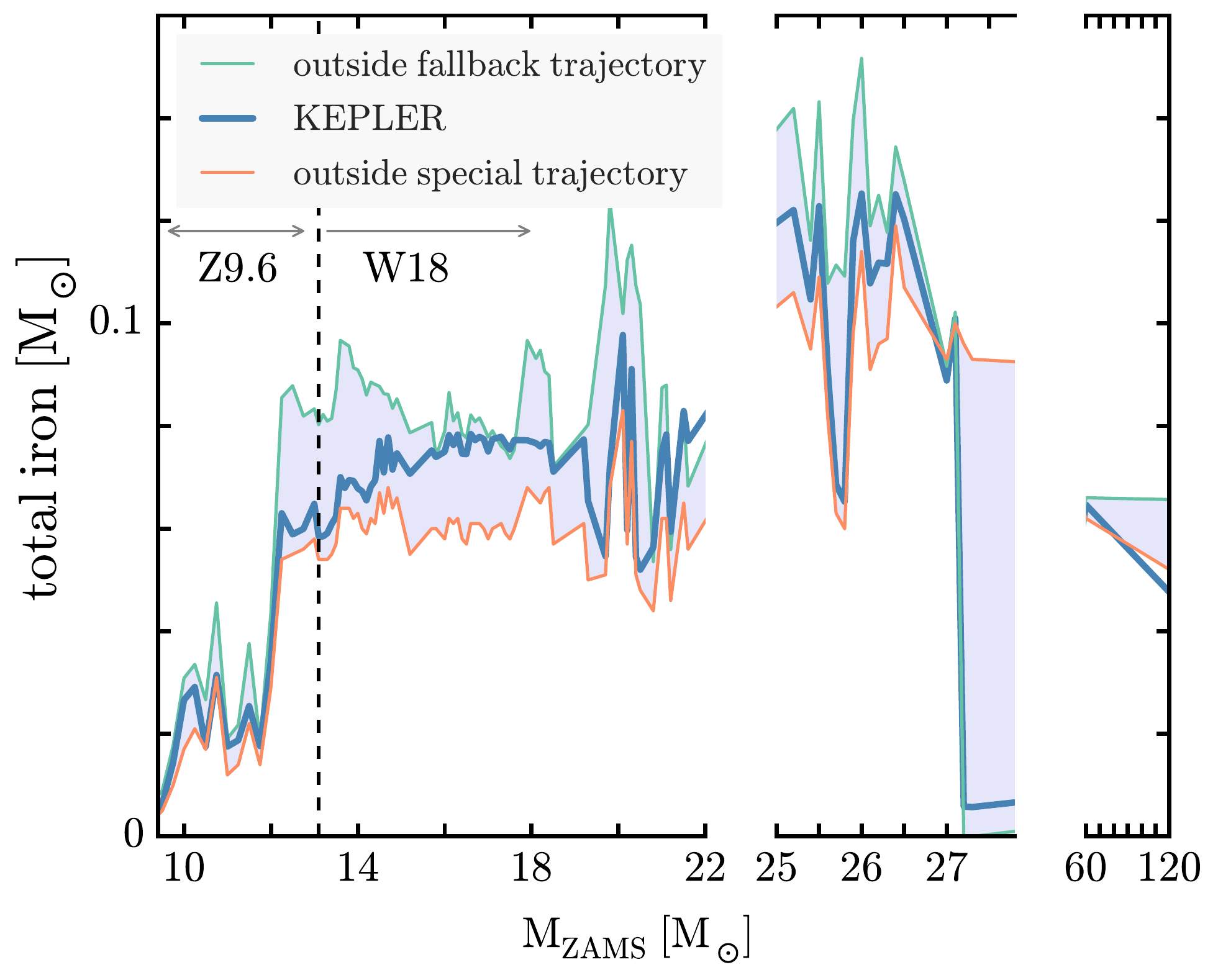}
\caption{Iron production is compared between the KEPLER and P-HOTB
  calculations for all models that exploded using the Z9.6 and W18
  engines. The shaded gray region is bounded on the bottom by the total
  iron produced by P-HOTB outside the 'special' trajectory (orange),
  and on the top by the total iron ejected (green). The thick blue
  curve represents total iron production in the converged KEPLER
  explosions.  \lFig{ni_W18}}
\end{figure}

For the stars that exploded in P-HOTB, the desired kinetic energy at
infinity in KEPLER is obtained by iterating on $\alpha$. This required
an earlier, more rapid motion outwards for the adopted piston. By
design, the two explosion models thus agreed almost exactly in
explosion energy and piston mass. They also agreed to typically better
than 10\% in the mass of iron-group nuclei that were synthesized
(\Tab{eni_Z9.6} and \Tab{eni_N20_W18}). Here the total iron in P-HOTB
calculation is taken as the amount outside the final fallback mass.

To make the agreement in $^{56}$Ni mass even better, first, for few
models the starting location in mass of the special trajectory, $m_z$,
was slightly varied, usually by $\sim 0.01$ \Msun , so that the KEPLER
total iron mass lies roughly in between the special and fallback
trajectories (\Fig{ni_W18}). Then using the innermost zone abundances,
most models were scaled up slightly until the fallback trajectory
value, so that the final disagreement of iron-group synthesis was a few
percent at most. The full tabulated list of all piston parameters for
all explosion calculations is available at the MPA-Garching archive 
\textsuperscript{\ref{mpa_db}}.

In the remainder of the paper, the baryonic remnant masses, the
kinetic energies at infinity of the ejecta, and the total iron-group
synthesis are based upon the 1D neutrino-powered explosions using
P-HOTB. Only the isotopic nucleosynthesis (of all elements including
presupernova mass loss) and the light curves are taken from KEPLER.


\section{Explosion Properties}
\lSect{explosions}

Inserting the standard ``central engines'' described in \Sect{expl} in
the various presupernova stars resulted in a wide variety of outcomes
depending upon the properties of each progenitor, especially its mass
and compactness, and the choice of 87A model used for the engine's
calibration (\Fig{mass_map}). Generally speaking, weaker central
engines  like W20 gave fewer supernova
than stronger engines like N20.

This is an interesting point that warrants elaboration. Not every
model for 87A will give equivalent, or even necessarily valid results
when its central engine is inserted in other stars.  SN 1987A was a
blue supergiant in a galaxy with lower metallicity than the sun. All
presupernova models considered here, except those that lost their
envelopes before exploding, are red supergiants with an initially
solar composition. The SN 1987A models, at least those that made blue
supergiant progenitors (Table 1), also used a different value for
semi-convective mixing that affected the size of the carbon-oxygen
core for that mass (made the core smaller). One of the models, W18,
included rotation while the present survey does not.  Our calculations
are 1D not 3D. Finally, one expects significant variations in
presupernova core structure even for two stars of very similar initial
mass and presupernova luminosity \citep{Suk14}.

The very similar results for ``explodability'' for models N20, W18,
W15 and S19.8 are thus welcome and suggest a robustness to the answer
than might not necessarily have existed.  They also justify the
neglect of model set W20 in the surveys of nucleosynthesis carried out
in \Sect{nucleo}. Use of such a weak engine would grossly underproduce
the heavy elements, especially the light $s$-process \citep{Bro13}. A
much larger supernova rate would be required to make even abundant
elements like silicon and oxygen.  The results obtained here for solar
metallicity stars are also very similar to those of \citet[][their
figure 12]{Pej15b}, who used a very different
approach. Qualitatively, the outcome seems more influenced by
presupernova structure than details of the central engine, provided
that engine is sufficiently powerful to explode many stars.

Models are normalized to SN 1987A here because it was a well studied
event with precise determinations for its explosion energy and
$^{56}$Ni mass as well as its progenitor properties.  One could take a
different tack and use an even more powerful central engine than N20
in order to achieve optimal agreement with the solar abundances. That
was not done here.


\begin{figure}[h]
\centering
\includegraphics[width=\columnwidth]{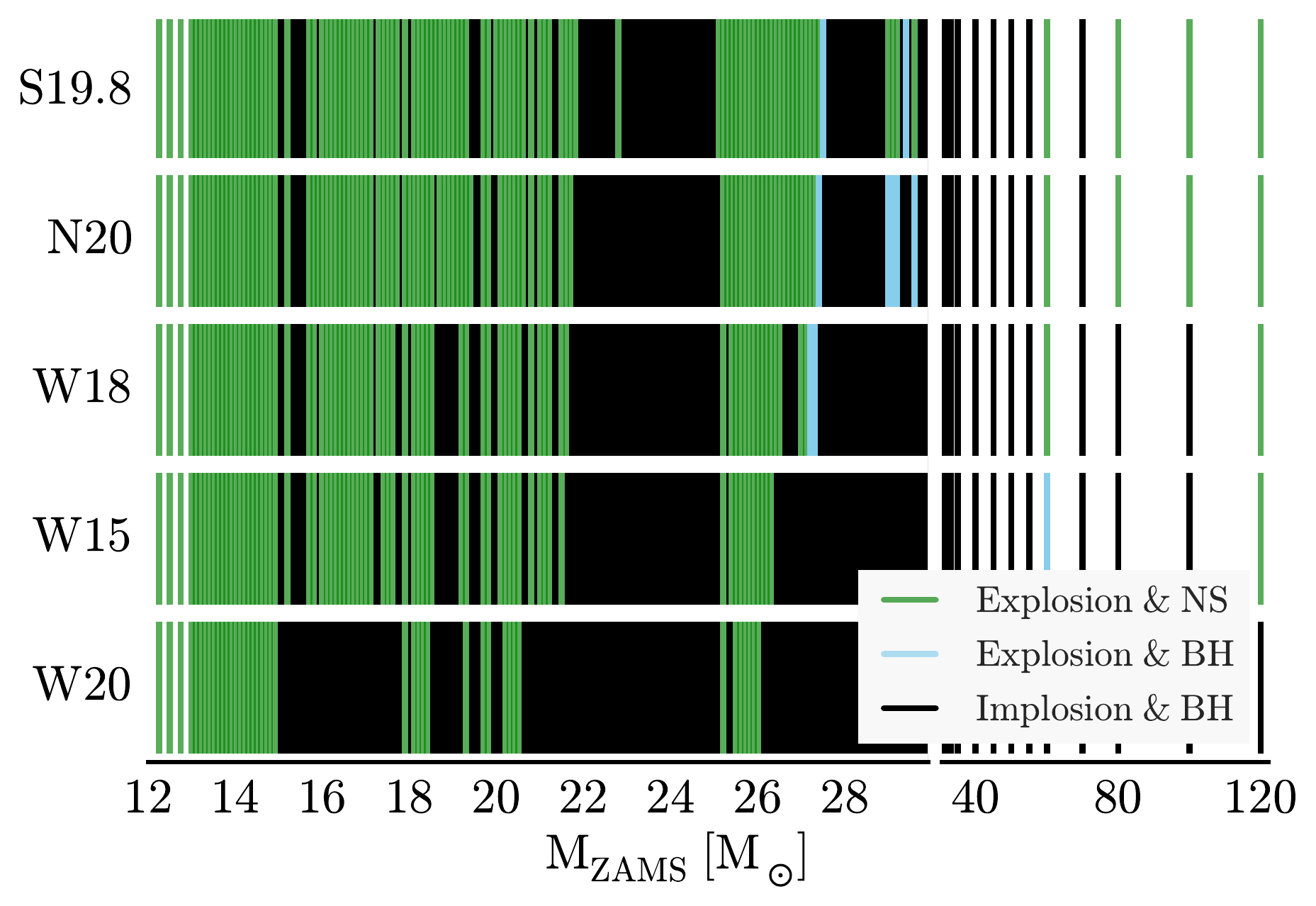}
\caption{The explosion outcomes from the five different central
  engines for SN 1987A (\Tab{87aprog} and \Tab{P-HOTBparams}) are
    shown in comparison. Successful explosions that make neutron stars
    are green, the explosions that make black holes through fall back
    are light blue, and the failures, which make black holes, are
    black lines. The calibrators are listed by the engine strength,
    weakest at the bottom. Models heavier than 12.25\Msun\ were
    covered by these five engines, all lighter models produced
    successful explosions by the Z9.6 engine calibrated to Crab supernova.
    \lFig{mass_map}}
\end{figure}

Also given in \Fig{SNperc} is the fraction of successful supernovae
above a certain main sequence mass, but below 30 \Msun. Since heavier
stars either fail to explode or explode after losing their hydrogen
envelopes, this would be the fraction of Type IIp supernovae.


\begin{figure}[h]
\centering
\includegraphics[width=.48\textwidth]{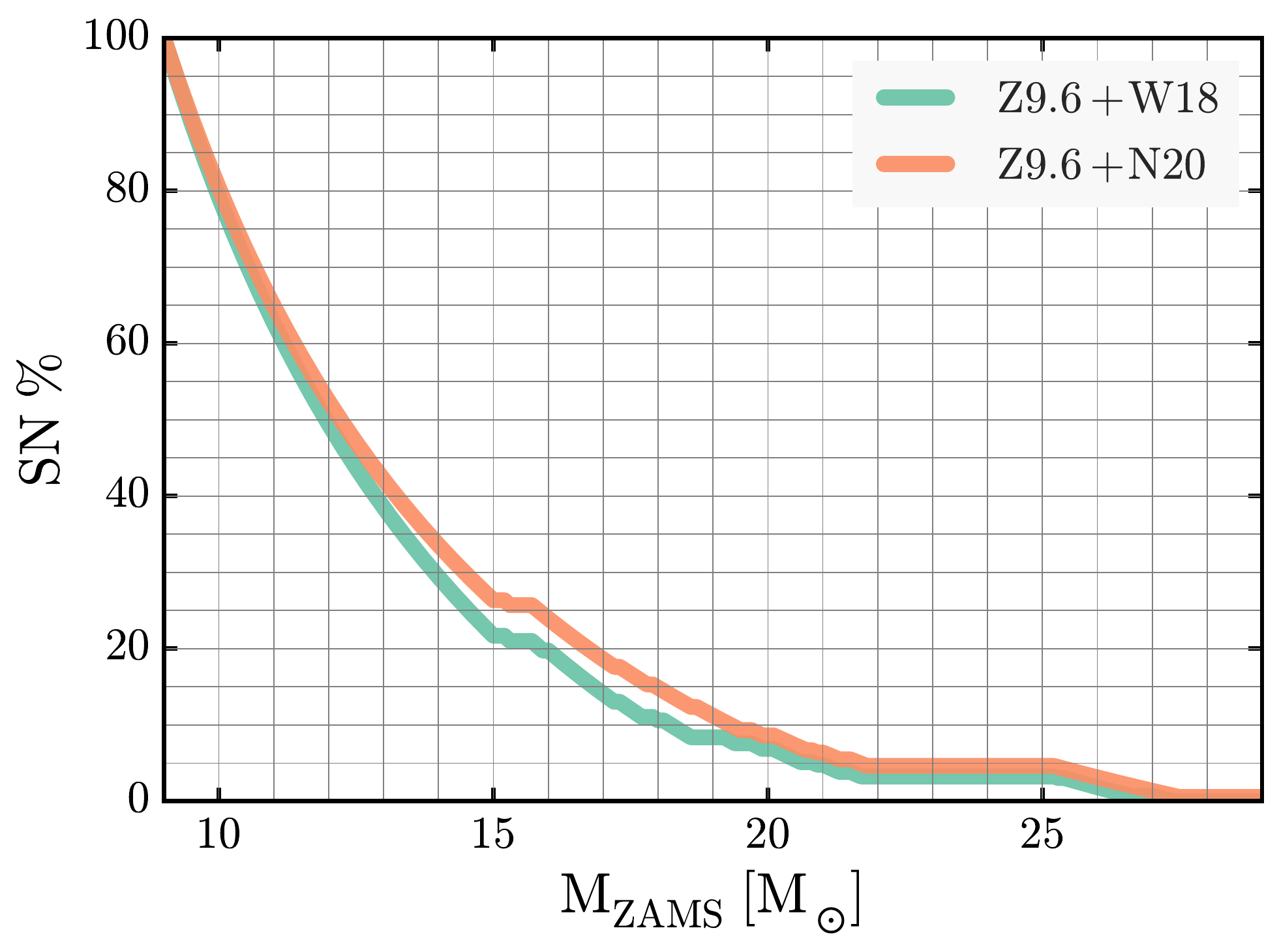}
\caption{The percentage of Type II supernovae above a given main
  sequence mass for explosions using the Z9.6 and W18 or N20
  engines. A Salpeter IMF has been assumed. Successful explosions
  above 30 \Msun \ do not make Type II supernovae. \lFig{SNperc}}
\end{figure}

\subsection{Relation to ``Compactness''}
\lSect{compact}

The distribution of successful explosions in \Fig{mass_map} is not a
simply connected set. The compactness parameter as originally defined
in \citet{Oco11} was innovative for its emphasis on the non-monotonic
outcome expected for the deaths of stars of different masses, but is
by no means a unique descriptor. Any parameter that samples the
density gradient outside the iron core will correlate with
explodability. Other measures could be for example, the free fall time
from a particular mass shell, the mass enclosed by a fiducial radius,
binding energy outside a fiducial mass, d(BE)/dr, the mass where the
dimensionless entropy equals 4, etc.  Recently \citet{Ert15} have
shown that a physically based two-parameter description of
``compactness'' can predict explodability for the present set of
models presented with almost 100\% accuracy (see Figure 6 of that
paper). No known single parameter criterion works as well. A less
accurate, but perhaps more physically intuitive predictor of explosion
is the mass derivative of the binding energy at $\sim$2\Msun, shown in
\Fig{dBEdm}.

As expected, the successful explosions are, by whatever measure, the
outcome of core collapse in stars with steep density gradients around
their iron cores. As shown in \Fig{cp_all}, $\xi_{2.5}$ has a very
small value for stars below 12.5 \Msun\, and it slowly increases to
about 0.2 until 15\Msun. All of these small models are particularly
easy to explode because they are essentially degenerate cores inside
of loosely bound hydrogenic envelopes, especially at the lower mass
end. All versions of the central engine explode stars lighter than
this limit. Between 15 and 22 \Msun\ stars become more difficult to
explode and the outcome can be highly variable for even small changes
in the initial mass, especially near 20\Msun\ , where central carbon
burning transitions from convective to radiative. From 22 to 25 \Msun,
very few or no successful explosions were found for all central
engines. There then comes an island around 25 to 27 \Msun \ that, once
again explodes. The non-monotonic nature of compactness is due to the
migration of the location of carbon and oxygen-burning shells with
changing mass \citep{Suk14}. From 30 \Msun \ on up to about 60 \Msun,
nothing explodes, except for the strongest engines. Eventually
however, the large mass loss appropriate for such large solar
metallicity stars removes the hydrogen envelope and whittles away at
the helium core making it once again compact and easier to explode, at
least for the stronger engines. The results are therefore sensitive 
to the mass loss prescription employed. 


\begin{figure}[h]
\centering
\includegraphics[width=\columnwidth]{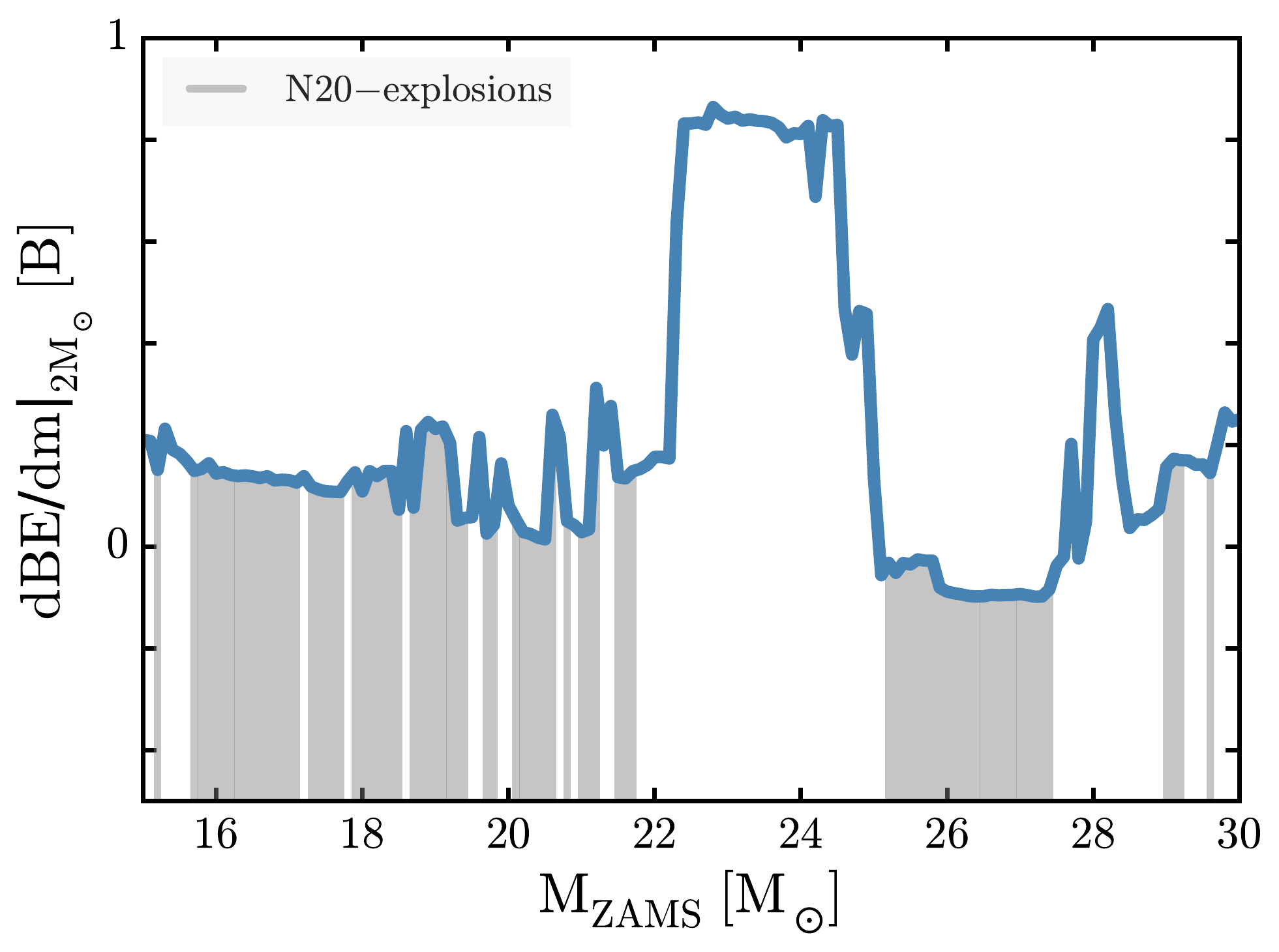}
\caption{The mass derivative of the binding energy (1 Bethe =
  10$^{51}$ erg) outside 2\Msun as a function of ZAMS mass is compared
  against the explosion outcome for the N20 engine. The successful
  explosions are noted with gray vertical lines. Though this doesn't
  work as well as the 2-parameter method of \citet{Ert15}, it is one
  of the better single-parameter criteria for an explosion.
  \lFig{dBEdm}}
\end{figure}

\subsection{Systematics of Explosion Energy and $^{56}$Ni Mass}
\lSect{system}

On general principles, one expects correlations to exist among the
explosion energy, $^{56}$Ni production, and compactness parameter in
successful explosions. The compactness parameter is a surrogate for
the density gradient outside of the iron core. The larger $\xi_{2.5}$
(\Fig{cp_all}), the shallower the density gradient and the greater the
mass closer to the origin where high temperature is attained in the
shock.  A frequently used approximation that takes advantage of the
fact that, during the epoch of nucleosynthesis, most of the energy
behind the shock is in the form of nearly isothermal radiation is
\citep{Wea80}
\begin{equation}
E \ \approx \ \frac{4}{3} \pi a T_s^4 R_s^3,
\end{equation}
where E is the total internal energy in the shocked
region, $T_s$, the temperature at the shock, and $R_s$, its
radius. After a short time, this internal energy converts into kinetic
energy and becomes nearly equal to the final kinetic energy of the
supernova. A shock temperature in excess of about $5 \times 10^9$ K is
required for the production of $^{56}$Ni, so for an explosion energy
of 10$^{51}$ erg, most of the matter between the final mass cut and a
point located at 3600 km in the presupernova star will end up as
$^{56}$Ni. This is provided, of course, that the 3600 km point does
not move a lot closer to the origin as the explosion develops, and the
final mass cut lies inside of the initial 3600 km mass
coordinate. Both assumptions are generally valid, although fall back
can occasionally reduce $^{56}$Ni synthesis to zero.


\begin{figure}[h]
\centering
\includegraphics[width=.48\textwidth]{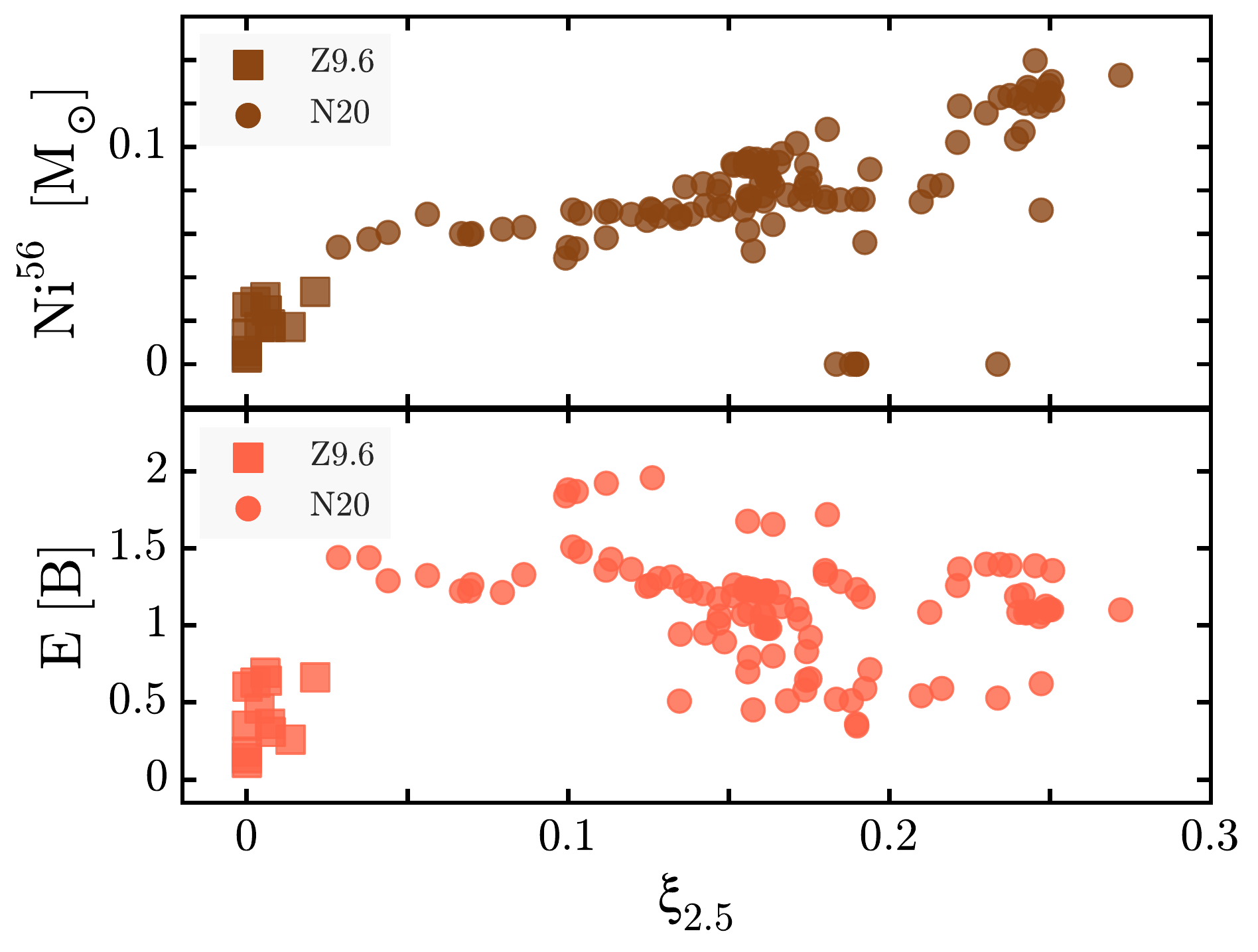}
\caption{The amount of nickel and explosion energies resulting from
  use of the Z9.6 and N20 engines are shown as functions of the
  compactness parameter (\Eq{compact}). The top panel shows a positive
  correlation of $^{56}$Ni production with compactness.  More matter
  is heated by the supernova shock for models with high
  $\xi_{2.5}$. The explosion energy is low for stars with very small
  compactness parameter because their thin shells are inefficient at
  trapping neutrino energy and there is very little luminosity from
  accreting matter. These effects saturate, however, around $10^{51}$
  ergs, since the energy provided by the neutrino source is limited
  and the the binding energy of the overlying shells is harder to
  overcome.  The results for the W18 engine are not plotted, but
  closely follow the N20 points plotted here.  \lFig{niandevscomp}}
\end{figure}

One expects then, for stars of similar initial compactness and final
remnant mass, a weak positive correlation between explosion energy and
$^{56}$Ni production. A greater explosion energy increases $R_s(5
\times 10^9 \rm K)$, and this larger radius encompasses a greater
mass.  This correlation can be obscured, or at least rendered
``noisy'' by variations in the compactness, remnant masses and
explosion energies. In particular, the compactness of presupernova
stars below 12 \Msun \ is very small, i.e., the density gradients at
the edges of their iron cores are very steep. These stars are also
easy to explode and have substantially lower final energies than the
heavier stars.  The radius that reaches $5 \times 10^9$ K is small and
the density gradient is also steep there. Thus, as has been known for
some time, stars below 12 \Msun \ are not prolific sources of
iron. These low mass supernovae, in fact, separate rather cleanly, in
theory at least, into a separate class with low energy and low
$^{56}$Ni yield (see \Fig{w18expls} and \Fig{n20expls}) - and as we
shall see in \Sect{lite}, shorter, fainter light curves.


\begin{figure}[h]
\centering
\includegraphics[width=.48\textwidth]{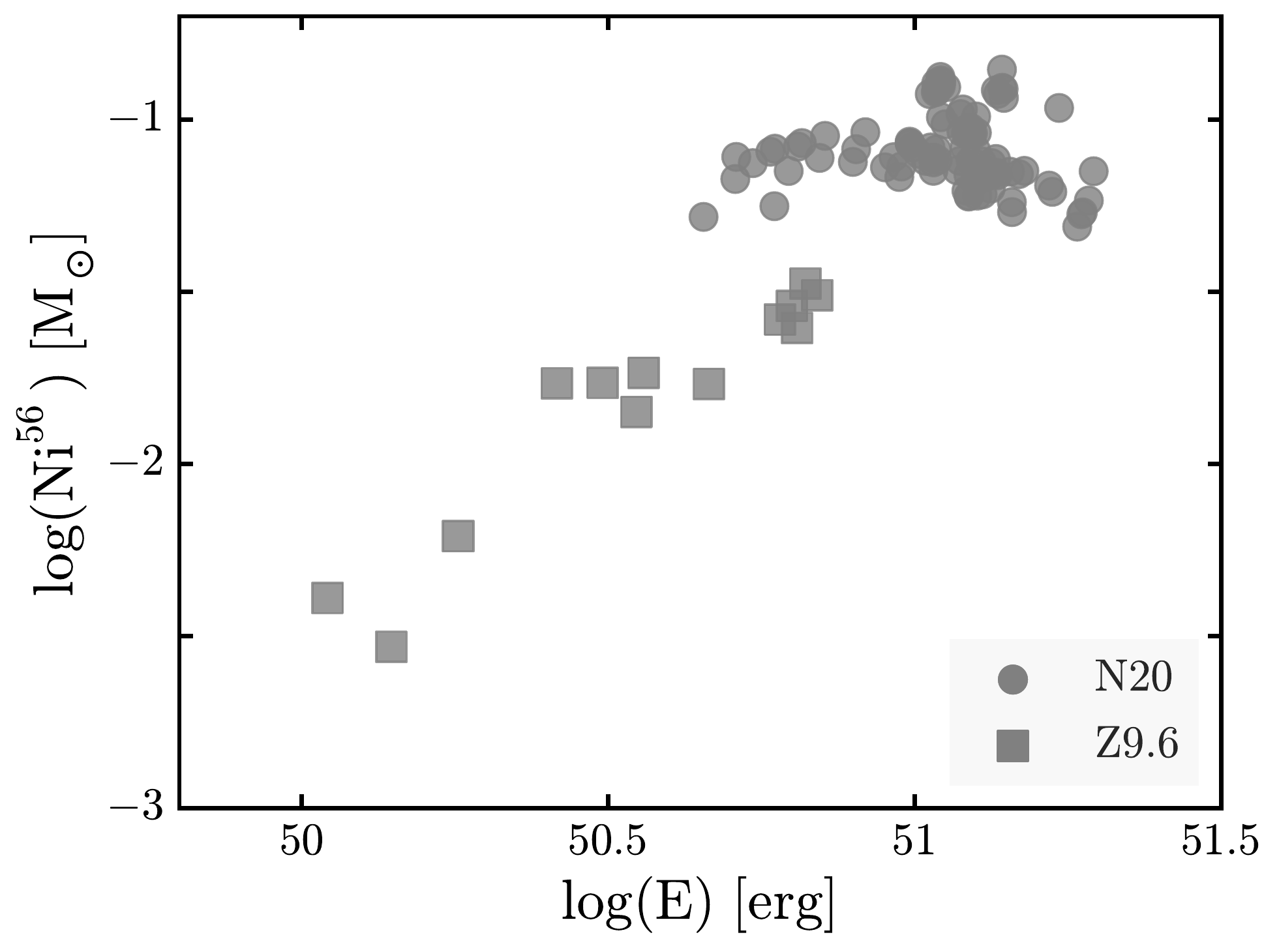}
\caption{The amount of $^{56}$Ni nucleosynthesis versus the logarithm
  of the final kinetic energy of the explosion in units of 10$^{51}$
  erg for the Z9.6 (squares) and N20 (circles) series. Models below an
  initial mass of 12 \Msun\ (half of all supernovae) explode
  easily and produce small amounts of nickel that correlate with their
  explosion energy. Larger mass models on the other hand, produce a
  nearly constant 0.07 \Msun\ nickel, except for a few cases with
  large fallback. The results for the W18 engine, though not plotted
  here for clarity, look very similar to those of N20.  \lFig{nike}}
\end{figure}

A correlation is also expected between explosion energy and
compactness, but which way does it go? Stars with more compact cores
(low values of $\xi_{2.5}$) are easier to explode and thus explode
with lower energy, but the stars with extended cores (larger values of
$\xi_{2.5}$) might also have lower final energy simply because they are
harder to explode.  The neutrinos have to do more work against infall
and the explosion may be delayed. The mantle has greater binding
energy that must be subtracted, but the additional accretion might
increase the neutrino luminosity and may give a larger explosion
energy.

\Fig{niandevscomp} shows the $^{56}$Ni mass vs. compactness parameter
for the model series Z9.6 and N20. The clustering of points around
$\xi_{2.5} = 0$ with low $^{56}$Ni and energy is expected, as is a
transition region to higher values of both. Over most of the
compactness parameter range, however, the explosion energy is roughly
constant with some small variation due to the effects just
mentioned. With a constant explosion energy, the $^{56}$Ni synthesis
is slightly greater for the stars with shallower density gradients,
i.e., larger $\xi_{2.5}$.

It is also interesting to compare the correlation between $^{56}$Ni
production and the explosion energy, especially since both can
potentially be measured. Two classes of events are expected - the
stars above and below 12 \Msun \ and above - and a slight positive
correlation of $^{56}$Ni with explosion energy is also expected in the
more massive stars. \Fig{nike} supports these expectations, but shows
that the variation of $^{56}$Ni production in stars above 13 \Msun
\ is really quite small. While it may be tempting to draw a straight
line through the full data set, this obscures what is really two
different sorts of behavior.  It is important to note that about half
of all observable supernovae in the current survey have masses below 12
\Msun \ (\Tab{stats}).

A similar correlation between $^{56}$Ni and kinetic energy has been
discussed by \citet{Pej15b} and compared with observational data
(their figure 20). The observations \citep{Ham03,Spi14} show a
particularly strong correlation of plateau luminosity with the
inferred mass of $^{56}$Ni, extending all the way down to 10$^{-3}$
\Msun \ for the latter. Our lowest $^{56}$Ni synthesis, in a
successful supernova that left a neutron star remnant, was 0.003 \Msun
\ for the $1.4 \times 10^{50}$ erg explosion calculated with KEPLER
for the 9.25 \Msun \ model using the Z9.6 central
engine. \Tab{eni_Z9.6} gives a substantially different value for
$^{56}$Ni synthesis in the 9.0 to 9.5 \Msun \ models using P-HOTB and
the discrepancy highlights some uncertain physics that warrants
discussion. Because the shock-produced $^{56}$Ni is very small in
these low mass stars, the neutrino wind contribution is
non-negligible. For the 9.25 \Msun \ model, P-HOTB calculates that
0.00659 \Msun \ of $^{56}$Ni is ejected, but about half of this is
made inside the ``special trajectory'' (\Fig{trajectory}) used by
KEPLER to represent the mass cut (\Sect{kepler}), and most of this
comes from the approximately-simulated wind in P-HOTB.
In that calculation all iron-group and trans-iron species in the
neutrino-powered wind are represented by $^{56}$Ni if neutrino
interactions lead to $Y_e \ge 0.49$. Otherwise, $^{56}$Ni is replaced
by a ``tracer nucleus'' for neutron-rich species. The network employed
does not therefore track the composition in detail.

For heavier stars that make most of the elements, this small
difference is negligible, but for these very light stars it is
not. The actual $^{56}$Ni synthesis in these very light stars is
probably between the KEPLER and P-HOTB values in
\Tab{eni_Z9.6}. Indeed, KEPLER was taken to be ``converged'' when its
$^{56}$Ni synthesis lay between these two values - the actual ejected
value in P-HOTB and the value outside of the ``special
trajectory''. This means that the very low values of $^{56}$Ni
synthesis calculated for the 9.0 to 9.5 \Msun \ models by KEPLER are
more uncertain than for other masses and possibly a lower bound to the
actual value. For nucleosynthesis purposes (\Sect{nucleo}), the KEPLER
iron-group synthesis was renormalized to agree exactly with the full
P-HOTB value. It is also possible to get still smaller values for
$^{56}$Ni in heavier stars that experience appreciable fallback.

We note in passing that the $^{56}$Ni produced by electron-capture
supernovae is mostly made in their neutrino-powered
winds. \citet{Wan11}, using 1D and 2D explosion models that
incorporated realistic neutrino transport, obtained about 0.003
$M_\odot$ of $^{56}$Ni for an electron-capture supernova, which is
very similar to the wind component predicted with P-HOTB for the
present explosions of our low-mass progenitors.

On the upper end, one might be tempted to extend this correlation of
$^{56}$Ni production and kinetic energy to still more energetic
events, including gamma-ray bursts and ultra-luminous supernovae
\citep{Kush15}, but these other events likely have other explosion
mechanisms, and the paper here is focused on non-rotating,
neutrino-powered models.


\section{Bound Remnants}
\lSect{remnants}

\subsection{Neutron Stars}
\lSect{nstar}

The P-HOTB calculations continue to sufficiently late times
($\sim10^6$ s) to include most fallback and accurately determine the
baryonic mass that collapses to a neutron star. They also give an
estimate of the gravitational mass based upon the neutrino loss of the
neutron-star cooling model employed in the simulations
(cf. \Fig{w18expls} and \Fig{n20expls}.) These results are in very
good agreement with a simple, radius-dependent correction
\citep{Lat01} that gives the neutron star's gravitational mass, $M_g$,
from its calculated baryonic mass, $M_b$:
\begin{align}
  \beta&=\frac{G M_{g}}{c^212\rm km}\,,\\
  \frac{M_{b}-M_{g}}{M_{g}}&=0.6\frac{\beta}{1-0.5\beta}\,,
\lEq{mgrav}
\end{align}
where $G$ is the gravitational constant and $c$ the speed of light.

A distribution of gravitational masses can then be constructed by
weighting the occurrence of each of our successful models according to
a Salpeter IMF.  The resulting frequencies are plotted as a function
of neutron star mass in \Fig{nstarmass} and are seen to be in
reasonably good agreement with the observed values compiled by 
\citet{Oze12} and \citet{Oze16} in terms of overall spread and mean
value. Our maximum neutron star mass for the W18 series is 1.68 
\Msun\ and the minimum is 1.23 \Msun. Use of 10 km for the neutron 
star radius reduces these numbers to 1.64 \Msun \ and 1.21 \Msun, and
reduces the average mass by about 0.02 \Msun. Here we note that, due 
to the complications of evolutionary effects, we did not build a PDF 
from the observed measurements, that reflects proper weights of 
different populations. Instead, the aim here is purely to provide a 
visual guide to the currently known measurements.

Different mass neutron stars, for the most part, come from different
ranges of main sequence mass with lower-mass progenitors producing
low-mass neutron stars. There is some overlap however. Remnants from
stars between 15 and 18 solar masses have slightly lower masses than
some below 15 solar masses. Stars above 18 \Msun\ contribute little to
the distribution, but do so over a broad range of neutron star
masses. This reflects both the variable mass of the very massive stars
at death, due to extensive mass loss, and inherent variability in
their compactness. While we have not carried out a statistical
analysis, \Fig{nstarmass} does suggest some bimodality in the
distribution, with a separate peak around 1.25 \Msun\ \citep[see
  also][]{Sch10}. It is interesting to note that these low mass
neutron stars are produced without invoking electron-capture
supernovae \citep[e.g.,][]{Sch10}, none of which are in our
sample. Indeed, heavier stars that experience iron-core collapse
supernovae are capable, in principle, of producing lighter neutron
stars than electron-capture supernovae since the values of $Y_e$ in
the cores of the latter are close to 0.5 at the onset of collapse,
while they are substantially less after oxygen and silicon burning in
the former. In quite heavy stars, this destabilizing effect is more
than compensated for by the high entropy and mild degeneracy, but in
lighter 9 - 11 \Msun stars, $Y_e$ becomes a major determinant for the
onset of core collapse.


\begin{figure}
\centering
\includegraphics[width=0.48\textwidth]{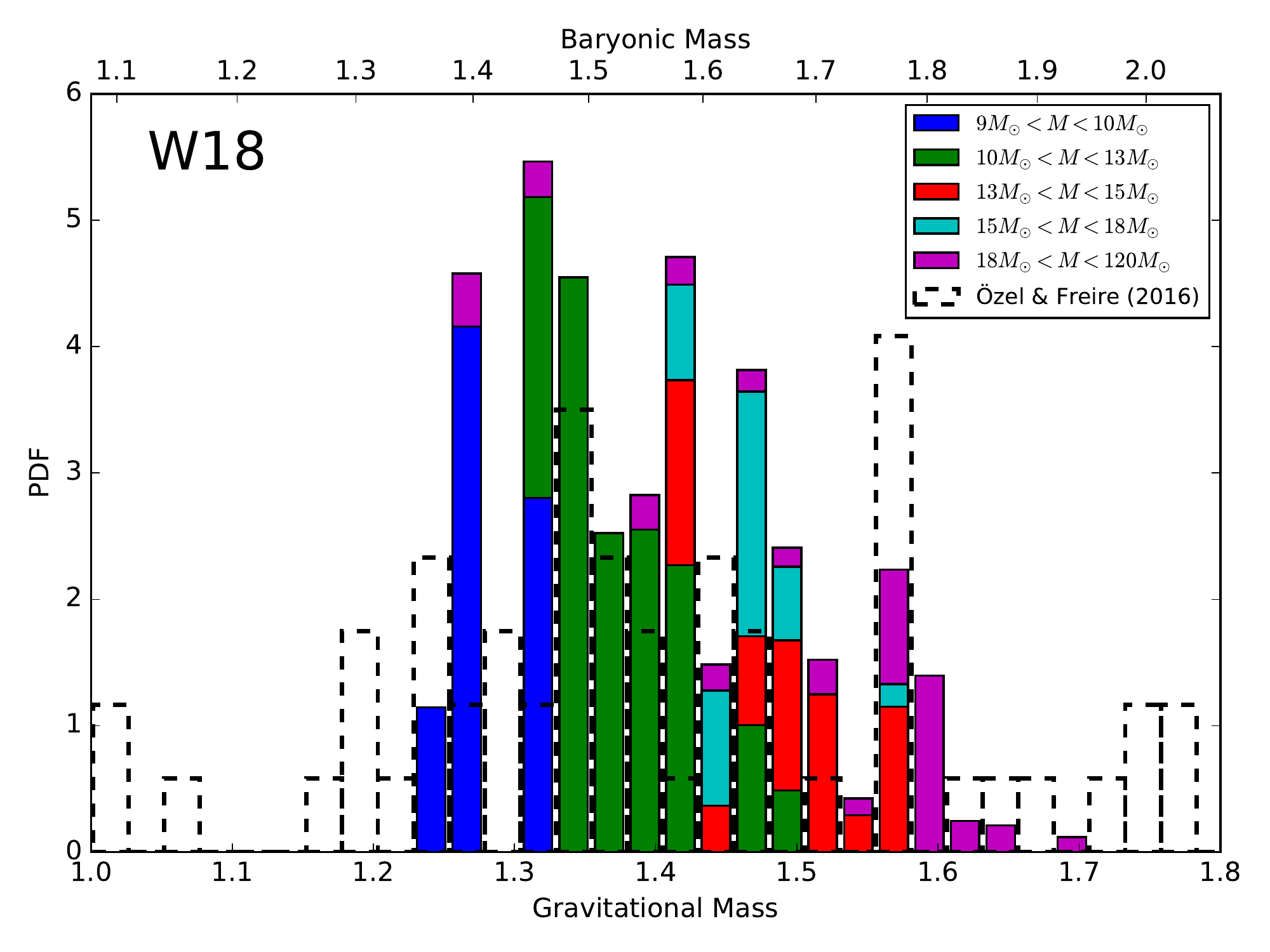}
\includegraphics[width=0.48\textwidth]{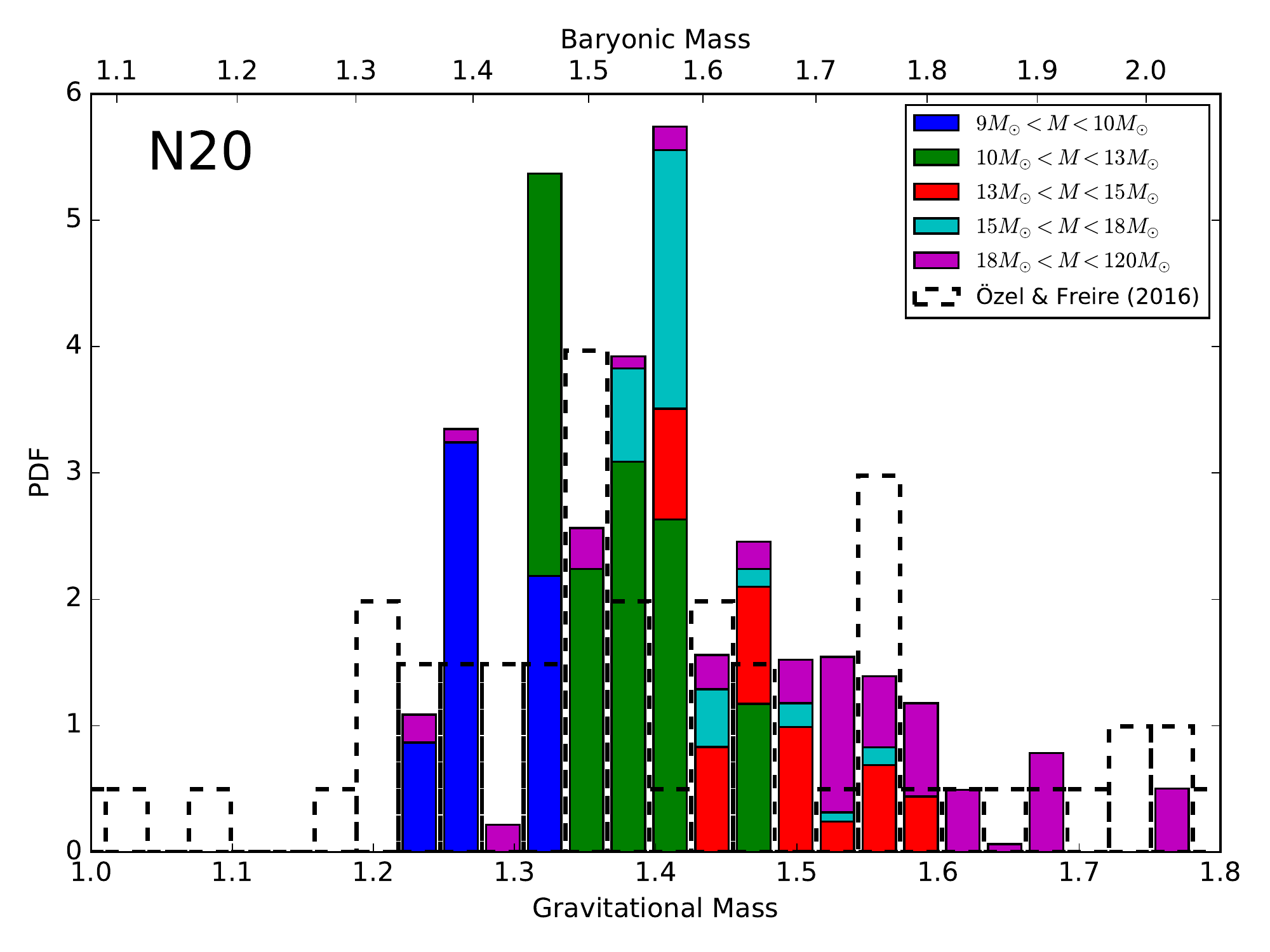}
\caption{Distributions of neutron star masses for the explosions 
         calculated using P-HOTB compared with the observational 
         data from \citet{Oze12} and \citet{Oze16}. Several 
         neutron stars with masses greater than the maximum mass 
         plotted here have been observed and the lightest 
         observational masses have large error bars. In the top 
         figure, the W18 calibration is used, and in the bottom, 
         the N20. Both distributions show some weak evidence for 
         bimodality around 1.25 and 1.4 \Msun. Results have been 
         color coded to show the main sequence masses contributing 
         to each neutron star mass bin. Note that this is not a 
         direct comparison to the observations, as the measured 
         values may already include accretion (i.e., not birth 
         masses) and have not been properly weighted based on 
         their measurement uncertainties and the relative 
         contributions of different classes of neutron stars to 
         the total population.
         \lFig{nstarmass}}
\end{figure}

An average neutron star mass can also be generated from a IMF-weighted
sampling of the successful explosions. The results for four different
calibrations of the central engine for SN 1987A, each including the
normalization to the Crab for stars below 13 \Msun, are given for a
Salpeter IMF in \Tab{stats}. These values are not very sensitive to
the central engine used for SN 1987A, but are determined more by
presupernova properties. All lie close to the (error-weighted)
observational means given by \citet{Lat12} of 1.368 \Msun \ for x-ray
and optical binaries, 1.402 \Msun \ for neutron star binaries, and
1.369 \Msun \ for neutron star - white dwarf binaries.  They are not
far from the average neutron star mass found by \citet{Sch10}, $1.325
\pm 0.056 \Msun$.  These new theoretical values are closer to the
observations than the results from \citet{Bro13}.

\Tab{stats} also gives the average baryonic mass of the remnants of
successful explosions, $\bar M_b$, the average supernova kinetic
energy at infinity, $\bar E$, and the range of average $^{56}$Ni
masses, all as calculated in P-HOTB. The lower bound ignores the
contribution from the neutrino-powered wind while the upper bound
assumes all of the wind not in the form of $\alpha$-particles is
$^{56}$Ni. Based on post-processing with KEPLER this may be an
overestimate of the actual $^{56}$Ni production by 10 to 20\%.
The column labelled SN\% is the IMF-weighted fraction of
all stars studied that blew up and left neutron star remnants.  One
minus this fraction is the percentage that became black holes. The
last column gives the numerical fraction of all successful explosions
that have main sequence masses above 12 \Msun, 20 \Msun, and 30
\Msun. The small value above 20 \Msun\ compares much better than past
surveys with the results of \citet{Sma09,Sma15}, who place an upper
limit on observational supernovae of about 18 \Msun.  The successful
explosions above 30 \Msun \ are all Type Ib or Ic.

\begin{deluxetable*}{r r@{$\times10^{51}$} r r r r r r r r r r}
\tablecaption{Integrated Statistics (see \Sect{massiveyield} for descriptions; all masses in \Msun)}
\tablehead{\colhead{Cal.} & \colhead{$\overline{E}$ (erg)} & \colhead{$\overline{M_{b}}$} & \colhead{$\overline{M_{g}}$} & \colhead{Lower $\overline{M_{\mathrm{BH}}}$} & \colhead{Upper $\overline{M_{\mathrm{BH}}}$} & \colhead{$\overline{M_{\rm{Ni}},l}$} & \colhead{$\overline{M_{\rm{Ni}},u}$} & \colhead{SN\%} & \colhead{($>12$)} & \colhead{($>20$)} & \colhead{($>30$)}}
\startdata
W15.0 & 0.68 &\ 1.55 & 1.40 & 8.40 & 13.3 & 0.040 & 0.049 & 66 & 47 & 8 & 2\\
W18.0 & 0.72 &\ 1.56 & 1.40 & 9.05 & 13.6 & 0.043 & 0.053 & 67 & 48 & 9 & 2 \\
W20.0 & 0.65 &\ 1.54 & 1.38 & 7.69 & 13.2 & 0.036 & 0.044 & 55 & 37 & 3 & 0\\
N20.0 & 0.81 &\ 1.56 & 1.41 & 9.23 & 13.8 & 0.047 & 0.062 & 74 & 52 & 13 & 5\\
\enddata
\lTab{stats}
\end{deluxetable*}

\subsection{Black Holes}
\lSect{bh}

Stellar collapses that fail to create a strong outward moving shock
after 3--15\,s in P-HOTB (with variations due to the
progenitor-dependent mass-accretion rate) are assumed to form black
holes. In the absence of substantial rotation, it is assumed that the
rest of the core of helium and heavy elements collapses into that
hole. The fate of the hydrogen envelope is less clear.  In the more
massive stars, above about 30 \Msun, the envelope will already have
been lost to a wind. In the lighter stars, all of which are red
supergiants, the envelope is very tenuously bound. Typical net binding
energies are $\sim$10$^{47}$erg. Any small core disturbance prior to
explosion \citep{Shi14} or envelope instability \citep{Smi14} could
lead to its ejection in many cases. Even if the envelope is still in
place when the iron core collapses, the sudden loss of mass from the
core as neutrinos can lead to the unbinding of the envelope
\citep{Nad80,Lov13}.

\Fig{mass_budget} shows the masses ejected and neutron star remnant
masses for the successful explosions using the Z9.6 and N20 central
engines. For those stars that made black holes, the helium core and
envelope masses are indicated and, for all stars, the mass loss to
winds before star death is indicated. A few stars made black holes by
fallback and are also shown. \Fig{bhmass} shows the distribution of
black hole masses under two assumptions: a) that only the helium core
accretes, and b) that the entire presupernova star falls into the
black hole. A distribution of IMF-weighted black hole frequency,
calculated just as it was for neutron stars, is given in
\Tab{stats}. No subtraction has been made for the mass lost to
neutrinos, that is the gravitational mass has been taken equal to the
baryonic mass. It is expected that a proto-neutron star will form in
all cases and radiate neutrinos until collapsing inside its event
horizon. The amount of emission before trapped surface formation is
uncertain, but unlikely to exceed the binding energy of the maximum
mass neutron star, about 0.3 \Msun\ \citep{Oco11,Ste13}.

Assuming the entire collapse of any black-hole forming star, including
its hydrogen envelope, gives an upper bound to the mass of the black
hole formed. This limiting case is not in good agreement with the
existing measurements
\citep{Wik14} \footnote{http://stellarcollapse.org/bhmasses; retrieved
  6 Sep. 2015 \label{BH}}, in terms of range and frequency of observed masses. The helium core mass seems a better
indicator \citep{Zha08,Koc14,Koc15}.

In addition to their production by stars that fail to launch a
successful outgoing shock, black holes can also be made in successful
explosions that experience a large amount of fallback. Only a few
cases of this were found in the present survey, and the resulting
black hole masses were always significantly less than the helium core
mass. They were made in some of the most massive stars that exploded.
The weakest central engine, W20, did not produce any black holes by
fallback.  Cases that might have had large fallback failed to explode in
the first place.  The W15 series yielded only one black hole with a
mass of 4.7 \Msun \ produced in a star that on the main sequence was
60 \Msun. Series W18 produced black holes by fallback at 27.2 and 27.3
\Msun \ with masses of 3.2 and 6.2 \Msun. The strongest two engines
S19.8 and N20 gave a few such cases at a slightly higher mass,
resulting into black holes in the range 4.1 to 7.3 \Msun.  In all
cases, the black hole mass was substantially less than the helium core
mass, which ranged from 9.2 to 10.2 \Msun.


\begin{figure}[h]
\centering
\includegraphics[width=\columnwidth]{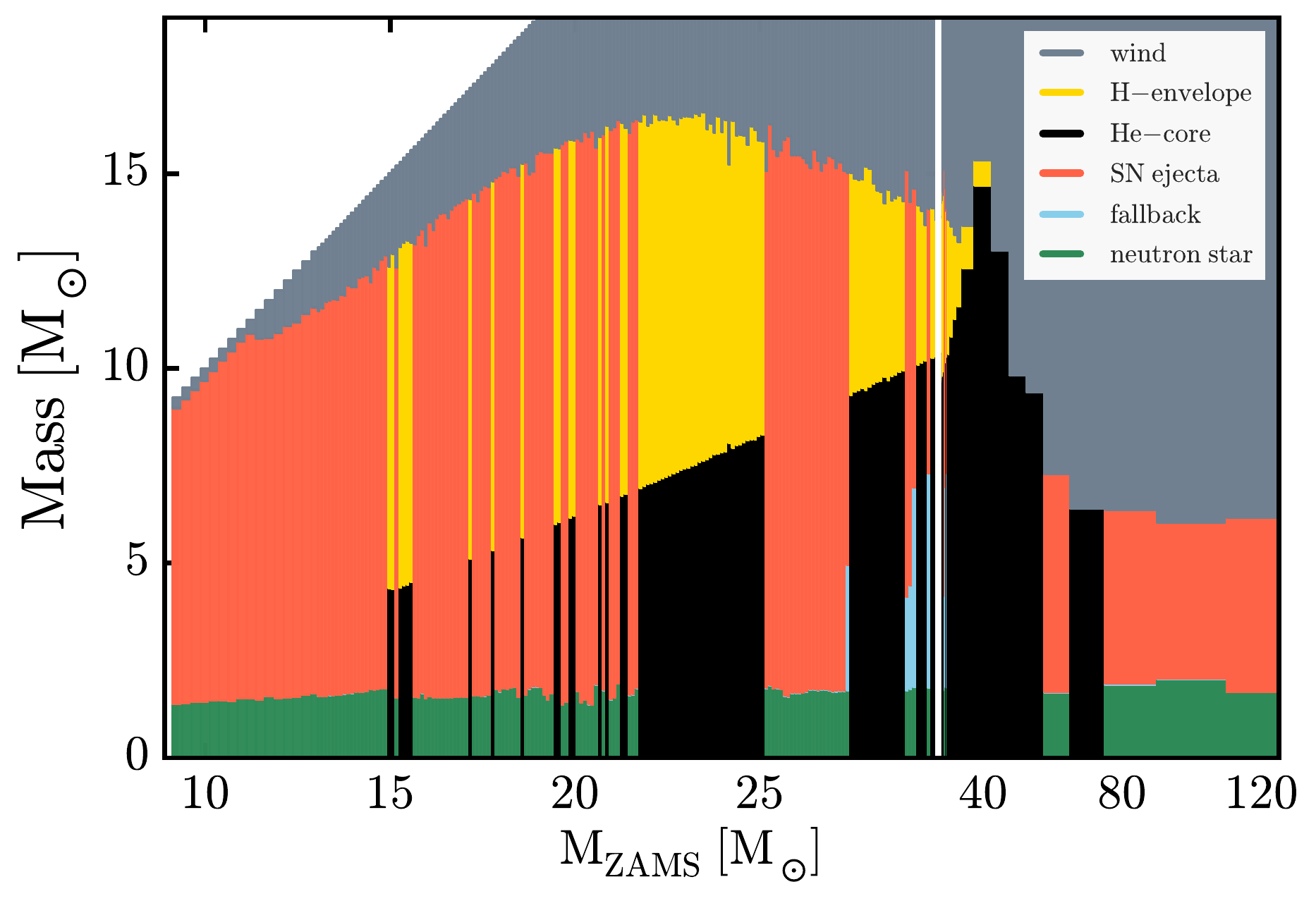}
\caption{The mass 'budget' for each star based on results using the
  Z9.6 and N20 engines. Grey shows mass lost to winds. For the
  successfully exploded models, the compact remnant mass is shown in
  green. For a few models that experienced fallback, the fallback mass
  is shown in blue. Except near 30 \Msun, fallback is negligible.  The
  helium-cores and the hydrogen-envelopes of the ``failed'' explosions
  are shown in black and yellow respectively. The resulting black hole
  mass from an implosion will most likely include the full
  presupernova star (black plus yellow), or just the helium core
  (black).  Results using the W18 engine are qualitatively similar,
  with fewer explosions and fewer cases with significant fallback.
  \lFig{mass_budget}}
\end{figure}

This tendency of neutrino-powered models to either explode robustly or
not at all has been noted previously, and naturally accounts for a
substantial mass gap between the heaviest neutron stars and the
typical black hole mass \citep{Ugl12}. In any successful explosion of
a quite massive star (i.e., above 12 \Msun), a few hundredths of a
solar mass of photodisintegrated matter reassembles yielding a lower
bound to the explosion energy of a few $\times 10^{50}$ erg
\citep[e.g.,][]{Sch06}. On the other hand, the ejection of the
hydrogen envelope and collapse of the entire helium core requires that
the final kinetic energy at infinity be less than about 10$^{50}$ erg
\citep{Lov15}. Given that the observations favor the implosion of the
helium core, but not of the entire star (\Fig{bhmass}), it seems that
another mechanism is at work. One natural explanation is that the
hydrogen envelope is ejected during the collapse of these massive
``failures'' by the Nadyozhin-Lovegrove effect
\citep{Nad80,Lov13}. The loss in binding energy due to neutrino
emission of the proto-neutron star launches a weak shock that ejects the
loosely bound envelope. If so, faint, red supernovae may be a
diagnostic of typical black hole formation in massive stars
\citep{Lov13,Koc14,Lov15}.


\begin{figure}
\centering
\includegraphics[width=0.48\textwidth]{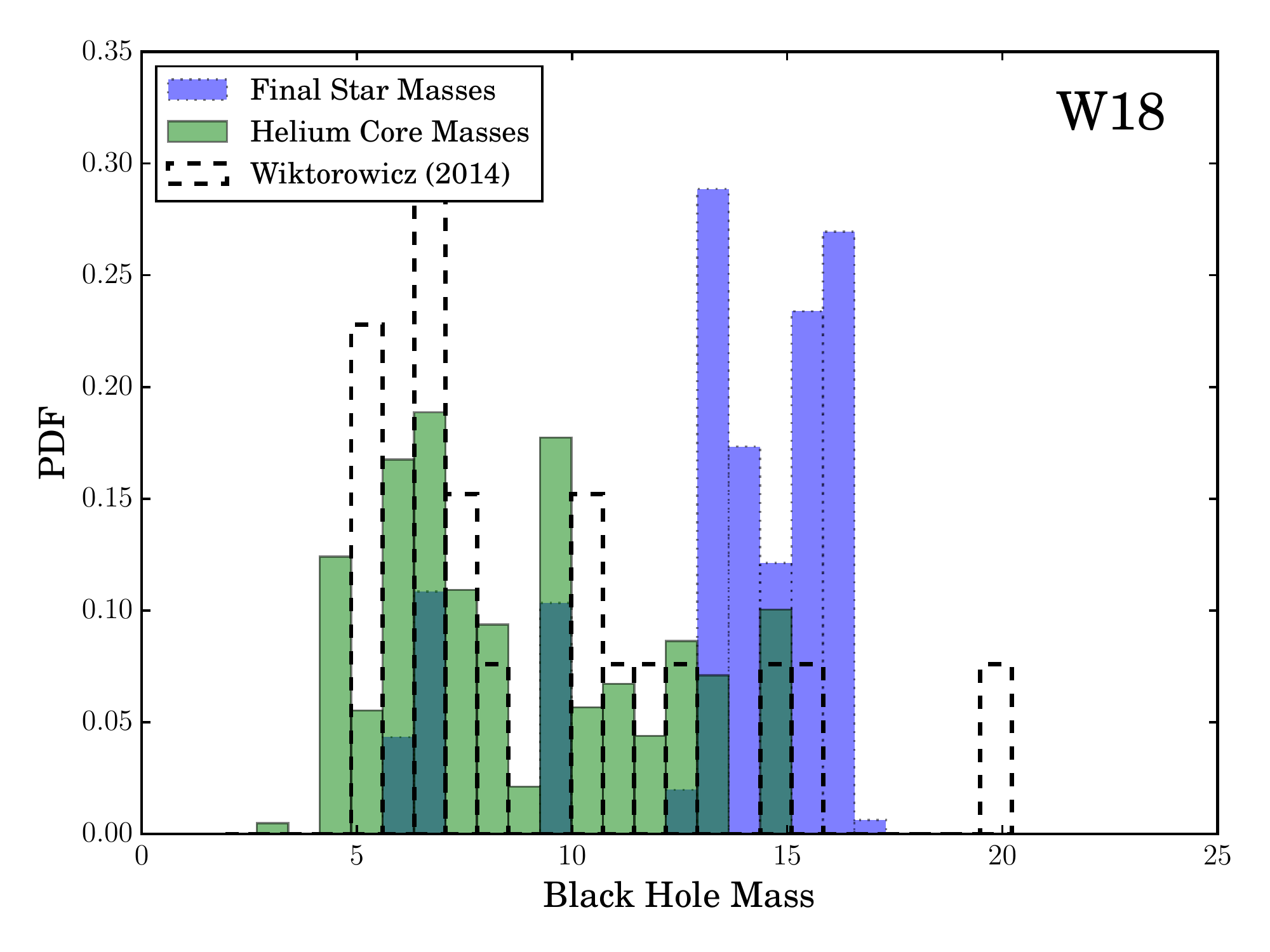}
\includegraphics[width=0.48\textwidth]{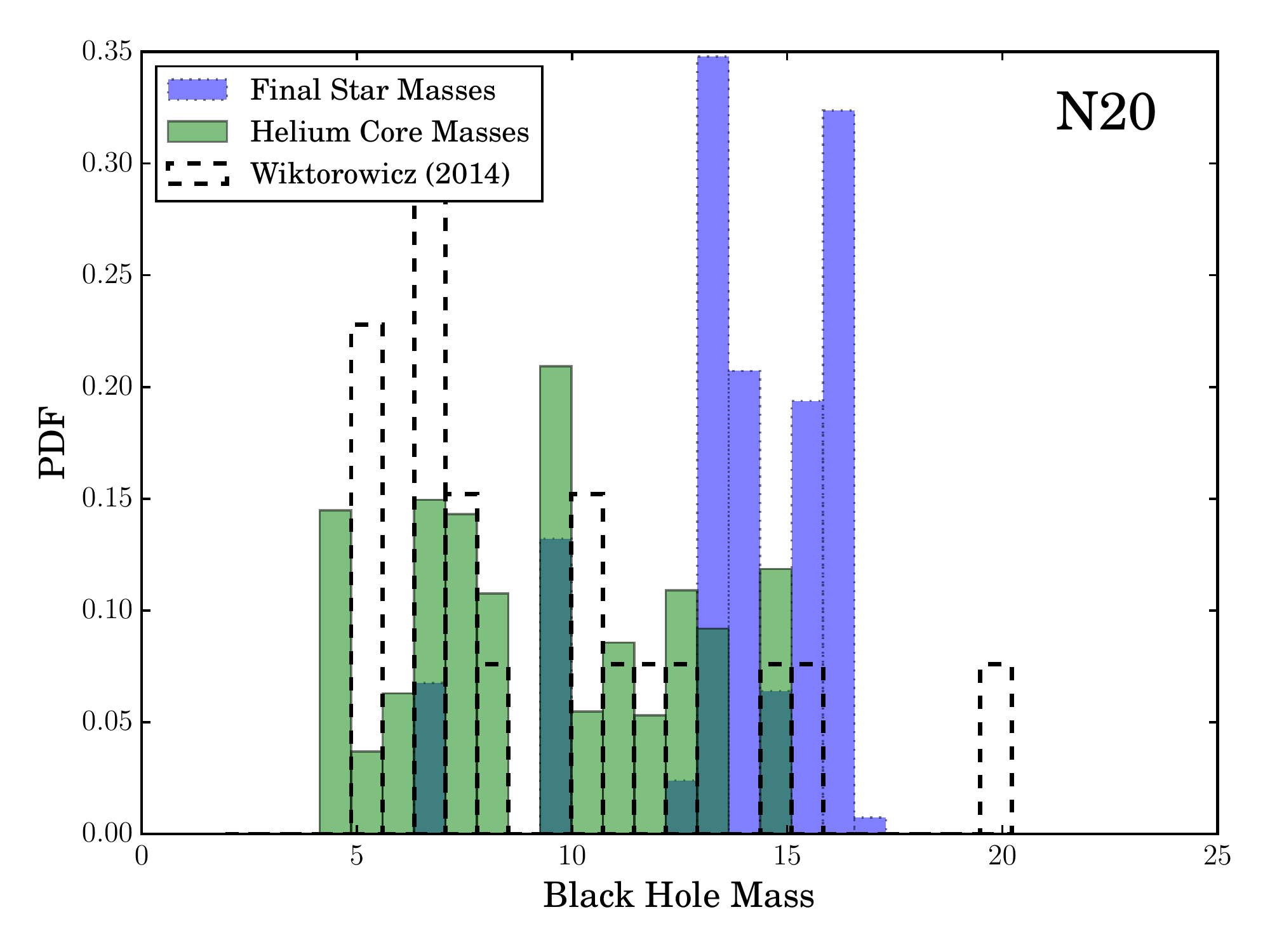}
\caption{Distributions of black hole masses for the explosions 
         calculated using P-HOTB compared with the observational 
         data from \citet{Wik14}\textsuperscript{\ref{BH}} in 
         gray. Theoretical results are shown based on two 
         assumptions: 1) that only the helium core implodes 
         (green); or 2) that the whole presupernova star implodes 
         (blue). Observations are more consistent with just the 
         helium core imploding. As in \Fig{nstarmass}, this is 
         not a direct comparison to the observations.
         \lFig{bhmass}}
\end{figure}


\section{Nucleosynthesis}
\lSect{nucleo}

Detailed isotopic nucleosynthesis, from hydrogen to bismuth, was
calculated using the KEPLER code for all the models presented in this
paper, and they are included in the electronic edition in a tar.gz 
package. Sample output for two supernovae with main sequence masses 
14.9 \Msun \ and 25.2 \Msun \ in \Tab{yieldtab} gives the ejected 
masses in solar masses in both the presupernova winds and the 
explosions using the W18 engine.

Rather than discuss the yields of individual stars, however, this
section gives and discusses a summary of the nucleosynthesis for the
two main explosion series, N20 and W18, averaged over an initial mass
function. To give an indication of the variation with mass, separate
results are given for three mass ranges: 9 -- 12 \Msun; 12 -- 30
\Msun; and 30 -- 120 \Msun. These ranges are chosen to represent the
qualitatively different nature (\Sect{solgrid}) of supernova
progenitors with very compact core structures (first group);
supernovae that have actually made most of the elements (second
group); and events where mass loss or an uncertain initial mass
function plays a key role (third group). At the end, we also summarize
the total nucleosynthesis (9 -- 120 \Msun) and consider the effect of
adding a SN Ia component to pick up deficiencies in iron-group and
intermediate-mass element production (\Sect{combined}).

To proceed, the yields of each star, $M_i$, are separated into two
categories: those species produced in the winds of the star during its
presupernova evolution, and those ejected during the supernova itself
(only wind for the imploded stars). For each star, these
yields, in solar masses, are multiplied by the fractional area under a
Salpeter \citep{Sal55} initial mass function (IMF) that describes the
number of stars in each bin between $M_{i}$ and $M_{i+1}$.  The area
has been normalized so that the total area under the IMF curve from 9
to 120 \Msun is 1. Summing the results across a mass range provides
the average ejecta, in solar masses per massive star within that
range.

A characteristic mass fraction is then computed which is the ratio of
the IMF-weighted ejected mass of a given element (or isotope) produced
in a mass range divided by the total mass of all species ejected. The
computation may or may not include the wind; both cases are
considered. The production factor is then the ratio of this mass
fraction to the corresponding mass fraction in the sun \citep{Lod03}.

Defined in this way, a production factor of 1 implies no change in the
mass fraction of an isotope from the star or mass range considered
compared to the presupernova progenitor. What came in is what goes
out. The net production of a given species is proportional to 1 -
P. There is also a change in all species since some mass stays behind
in the bound remnant and is lost.  This is not accounted for in the
definition used here for P.  If half of the mass of the star is lost
to the remnant, but the mass ejected from the star is all of exactly
solar composition, the production factor by our definition is still 1,
but the {\sl net} production would be this factor times $(M_{\rm ZAMS}
- M_{\rm rem})/M_{\rm rem} = 1/2$.  Values of remnant masses are given
in \Tab{eni_Z9.6} and \Tab{eni_N20_W18}. As presently designed, the
production factor can only be less than 1 if the species is destroyed
by nuclear reactions (e.g., deuterium, beryllium).

For purposes of presentation, all production factors are divided by
that of $^{16}$O so that the production factor of $^{16}$O, an
abundant element known to be produced in massive stars, is one. To
retain information about the unnormalized production factor, a line is
given in each figure at the reciprocal of the unnormalized production
factor of $^{16}$O. Points lying on this line have no net production,
i.e, have an unnormalized production factor of 1.

In order to demonstrate the dependence on uncertain mass loss rates,
especially in the heavier stars, some yields are presented with and
without mass loss. The ``without mass loss results'' are an artificial
construct because we have not carried out a self-consistent evolution
and explosion of constant mass stars. The explosion of such stars
would be quite different. For example, the more massive stars would
all make black holes. The less massive ones might explode and eject
the same matter the wind would have removed. Here, the mass lost in
the wind is simply discarded.  The effect of removing it shows its
importance.

In those calculations that include the wind, the mass loss of all
stars has been added, including those that end up making black
holes. All stars have winds; only those that explode have supernova
ejecta. The tricky issue of how to treat the hydrogen envelope of
stars that form black holes, but are presumed (not calculated) to eject
their residual envelopes is deferred to a later section (\Sect{26al}).

Because of the lower bound on the mass, 9 \Msun, no contribution from
lighter stars is included here, so it is to be expected that species
like the heavy $s$-process, $^{13}$C, $^{14}$N etc. which are known to
be efficiently produced in AGB stars, will be under-represented. Any
$r$-process from merging neutron stars or the neutrino wind is also
missing, as are the products of cosmic ray spallation.

Some interesting statistics relevant to nucleosynthesis are given in
\Tab{rangestats}.  For the W18 and N20 series, the typical supernova
mass, based only upon those stars that actually exploded, is given
along with the fraction of the total ejecta that come from each mass
range.

\begin{deluxetable}{lccc}
\tablecaption{Supernova Mass Statistics}
\tablehead{\colhead{Mass Range}                  &
           \colhead{Engine(s)}                   & 
           \colhead{$\overline{\rm M_{\rm SN}}$} & 
           \colhead{f (no wind) }
           }
\startdata
$\leq$12 & Z9.6+W18 & 10.25 & 0.19 (0.40)  \\
12-30    &    -     & 14.5  & 0.37 (0.54)  \\
30-120   &    -     & 60.0  & $\:\:$0.44 (0.010) \\
9-120    &    -     & 12.0  & 1     \\
$\leq$12 & Z9.6+N20 & 10.25 & 0.18 (0.40)  \\
12-30    &    -     & 15.2  & 0.39 (0.58)  \\
30-120   &    -     & 80.0  & $\:\:$0.43 (0.026) \\
9-120    &    -     & 12.25 & 1
\enddata
\lTab{rangestats}
\end{deluxetable}

\subsection{9 to 12 \Msun}
\lSect{9to12nucleo}

\begin{figure}[h]
	\centering
	\includegraphics[width=0.9\columnwidth]{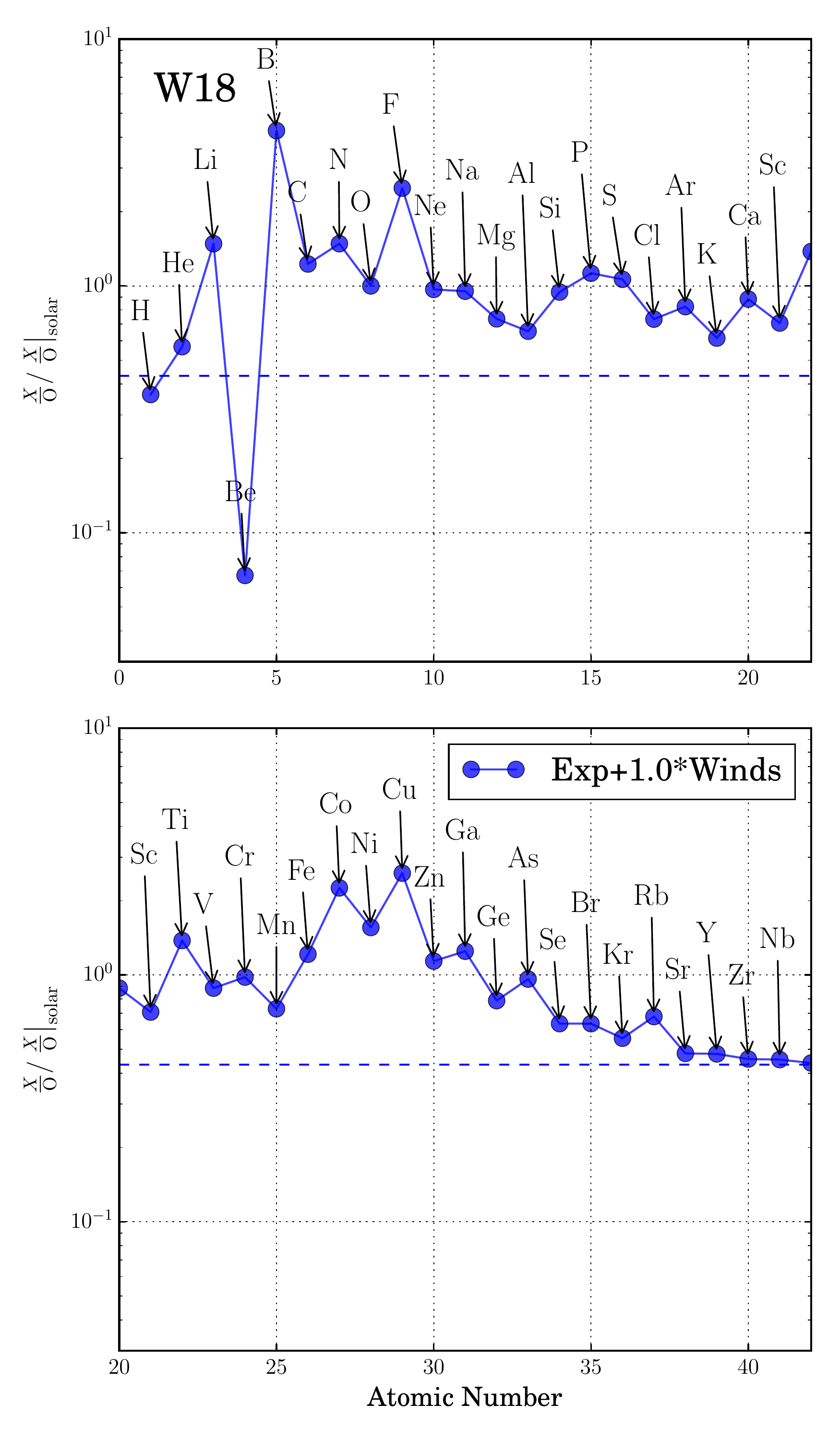}
	\caption{\label{fig:low} Nucleosynthesis in low mass
          supernovae. The IMF-averaged yields of just the explosions
          from 9 --12 \Msun \ are given for the elements from hydrogen
          through niobium. The calculations actually included elements
          up to bismuth (Z = 83), but no appreciable production
          occurred for the heavier elements.  Production factors
          (defined in the text) have been normalized to that of
          $^{16}$O, which is 2.37, by dividing them all by this
          factor, making the oxygen production unity. The dashed line
          at (2.37)$^{-1}$ = 0.43 is thus a line of no net change.
          Elements below the dashed line are destroyed in the star;
          those above it experience net production. The relative large
          yields of iron group elements, which are nearly solar
          compared to oxygen, is a consequence of the low oxygen yield
          in these light supernovae, and does not characterize heavier
          stars where more nucleosynthesis happens. The $s$-process
          just above the iron group is significantly
          underproduced. Note, however, the large yields of Li and B
          produced by the neutrino process.  Fluorine is also
          overproduced by a combination of neutrinos and helium
          burning. Stellar winds of all models have been included.}
\end{figure}

While roughly half of all observed supernovae lie within this range,
(\Tab{stats}), their contribution to the overall stellar
nucleosynthesis is relatively small. \Fig{low} shows reasonably good
agreement with solar system elemental abundances for elements heavier
than beryllium (Z = 4) all the way through the iron group, but this
can be misleading since a lot of the yield was in the star to begin
with in solar proportions.  The actual production factor for oxygen,
before renormalization, is 2.37, so a value of 1./2.37 = 0.422 in the
figure indicates no net change in the abundance due to stellar
nucleosynthesis. This is apparent for the heaviest elements plotted,
around Z = 40. To the extent that other elements have the same
production factor as oxygen, their abundance is also about 2.37 times
greater than what the star had, external to its final gravitationally
bound remnant, at birth.

\Fig{low} reveals some interesting systematics for low mass
supernovae though. First, the abundances {\sl are} roughly solar, even for
the iron group. Some elements, Co, Ni, and Cu are even
overproduced. This is surprising given that only a minor fraction of
iron-group elements is expected to come from massive stars
\citep[e.g.][]{Tim95,Cay04}. More should come from Type Ia supernovae
(SN Ia). Their apparent large production here really just reflects
the fact that these stars make very little oxygen.  As we shall see
later, massive stars altogether make only about one-quarter of solar
iron. Still it is an interesting prediction that iron to oxygen in the
remnants of low mass supernova remnants may be roughly solar.

By the same token, the productions of carbon and nitrogen is not
generally attributed to massive stars, but they do have slightly
super-solar proportions to oxygen in this mass range. Nitrogen, which
is made in the hydrogen envelope by the CNO cycle operating in a large
mass on primordial C and O, seems large here because oxygen, made by
helium burning in a relatively thin shell is small. Overall, we shall
find that nitrogen, at least nowadays, is probably not a massive star
product, though the component coming from massive stars is
appreciable. The case of carbon is less clear and depends on mass loss
rates and the initial mass function.

Also quite abundant in \Fig{low} are lithium, boron, and fluorine. 
$^{7}$Li and $^{11}$B are made by the neutrino process 
\citep{Woo90}. The assumed temperature of the $\mu$ and $\tau$ 
neutrinos here was 6 MeV. The large production of $^{11}$B reflects 
the large carbon abundance in these lower mass stars, and the 
proximity of the carbon shell to the collapsing iron core which emits 
the neutrinos.  $^7$Li is made by the spallation of $^4$He by 
$\mu$- and $\tau$-neutrinos by the reaction sequence 
$^4$He($\nu,\nu$'n)$^3$He($\alpha,\gamma)^7$Be. Its large 
production (again compared with a small oxygen abundance) reflects 
the proximity of the helium shell to the core and the importance of 
the alpha-rich freeze out in the bottom-most layers ejected. Fluorine 
is made by a combination of the neutrino irradiation of neon and 
helium burning \citep{Mey93,Woo95}, with the former making most 
of the $^{19}$F above 15 \Msun, but the latter dominating for lower 
mass supernovae \citep[see also][]{Sie15}, and slightly dominant 
overall. The synthesis of $^{19}$F by neutrinos is primary, however, 
since the target is $^{20}$Ne, while the synthesis by helium burning 
depends upon the abundance of $^{14}$N and neutrons and is 
therefore secondary. The production of fluorine by helium burning will 
thus be reduced in lower metallicity stars.

The low production of Be reflects its destruction in massive stars,
$^9$Be and $^{10}$B are thought to be produced by cosmic ray
spallation.


\subsection{12 to 30 \Msun}
\lSect{12to30nucleo}

Stars from 12 to 30 \Msun \ are responsible for most of the
nucleosynthesis that happens in supernovae. \Fig{mid} shows the
production factors for this range. The intermediate mass elements are
made in roughly solar relative proportions compared with oxygen,
though the low yield of calcium is a concern That the iron-peak
elements are deficient is not surprising. Most of iron comes from SN
Ia. Some species in the iron group, especially Co, Ni, and Cu, are
produced here in roughly solar proportions. All are made in the
alpha-rich freeze out: $^{58}$Ni as itself, and $^{59}$Co and
$^{63,65}$Cu as $^{59}$Cu, $^{63}$Ga, and $^{65}$Ge
respectively. Production of these species is sensitive to the
treatment of the inner boundary and the neutrino reactions that might
go on there changing the value of $Y_e$. The calculated production
here is thus more uncertain than for other species.

The production of Li, B, C, N, and F have all declined greatly from
the large values seen for the lower mass supernovae. This is mostly to
do with the larger oxygen production to which these yields are now
normalized, but also to the lower carbon abundance and larger radii of
the carbon and helium shells.

Of particular note in this mass range is a significant underproduction
of $s$-process elements, especially Sc and the elements just above the
iron group. This problem will persist in the final integrated
nucleosynthesis from all masses and is a challenge for the present
models and approach. Part of this deficiency may be due to an
uncertain reaction rate for $^{22}$Ne($\alpha,$n)$^{25}$Mg
(\Sect{sproc}).


\begin{figure}[ht]
	\centering
	\includegraphics[width=0.9\columnwidth]{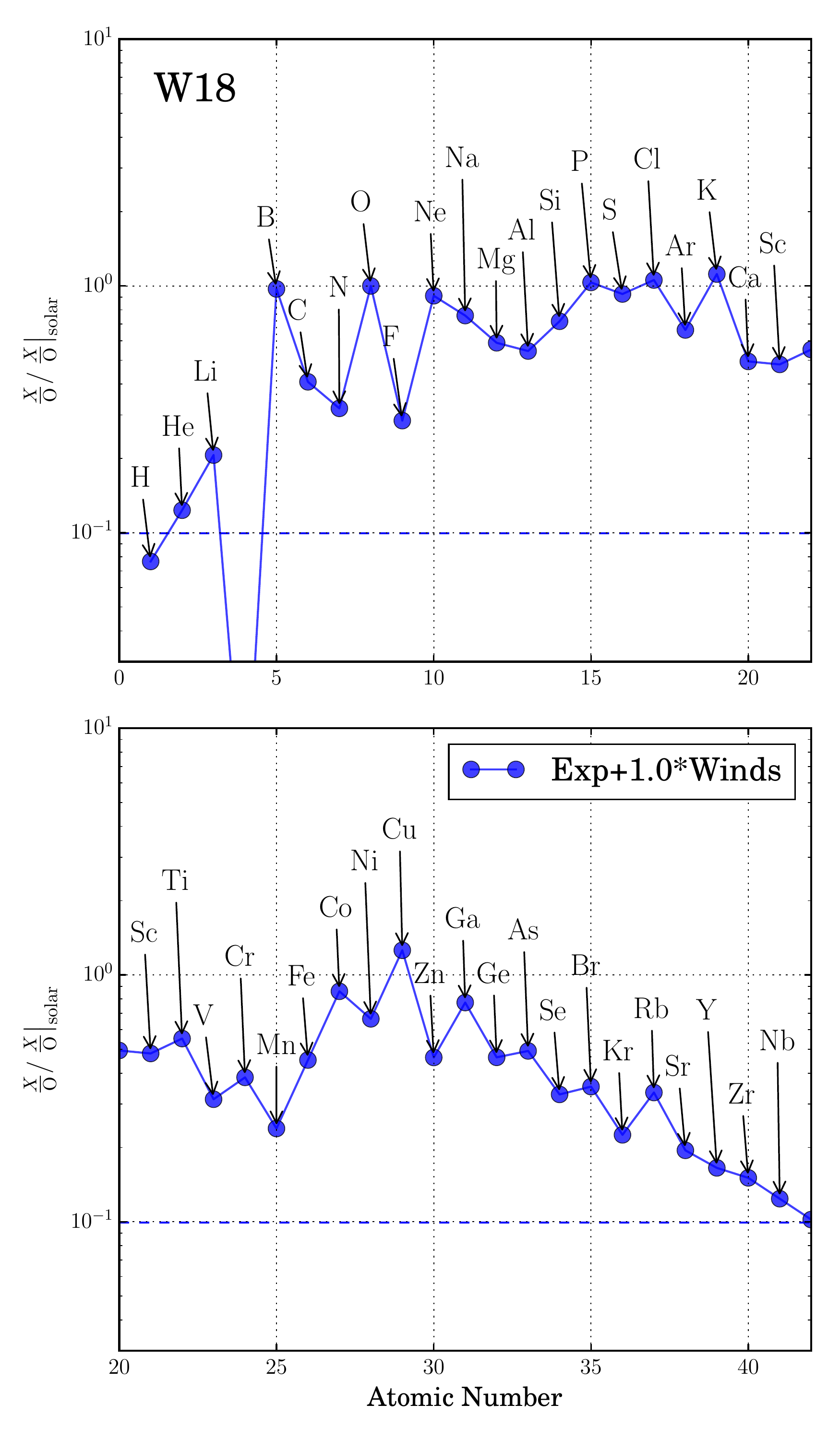}
	\caption{Similar to \Fig{low}, the IMF-averaged production of
          elements in supernovae from 12 to 30 \Msun \ is shown using
          the W18 calibration for the central engine. This is the
          range of masses where most supernova nucleosynthesis occurs.
          The production factor of oxygen before normalization was
          10. Production factors above the dashed line at 0.10 thus
          indicate net nucleosynthesis in the stars considered.  The
          production of iron-group elements is substantially lower than
          in \Fig{low}, but  will be supplemented by SN Ia. The
          production of light $s$-process elements is substantially
          greater, but still inadequate to explain their solar
          abundances. Production of Li and B by the neutrino process
          is diminished, but still significant, especially for
          B. Contributions to nucleosynthesis by the winds of all
          stars, including those that became black holes is included.}
        \lFig{mid}
\end{figure}

\subsection{30 to 120 \Msun}
\lSect{30to120nucleo}

Most stars above 30 \Msun \ in the present study become black holes,
and their remnants absorb most of the core-processed elements.  Their
winds will escape prior to collapse however, and contribute to
nucleosynthesis, especially of the lighter elements like CNO. For 
the assumed mass loss rates, some of the heaviest stars in this range 
(i.e. from 60 \Msun\ upward) also explode, but only because winds have 
stripped the core of the star down to a manageable size. Such stars, 
having carbon-oxygen cores comparable to stars in the 12 -- 39 \Msun\ range will contribute similar ucleosynthesis, but more enhanced in 
the products of helium burning.


\begin{figure}[ht]
	\centering
	\includegraphics[width=0.9\columnwidth]{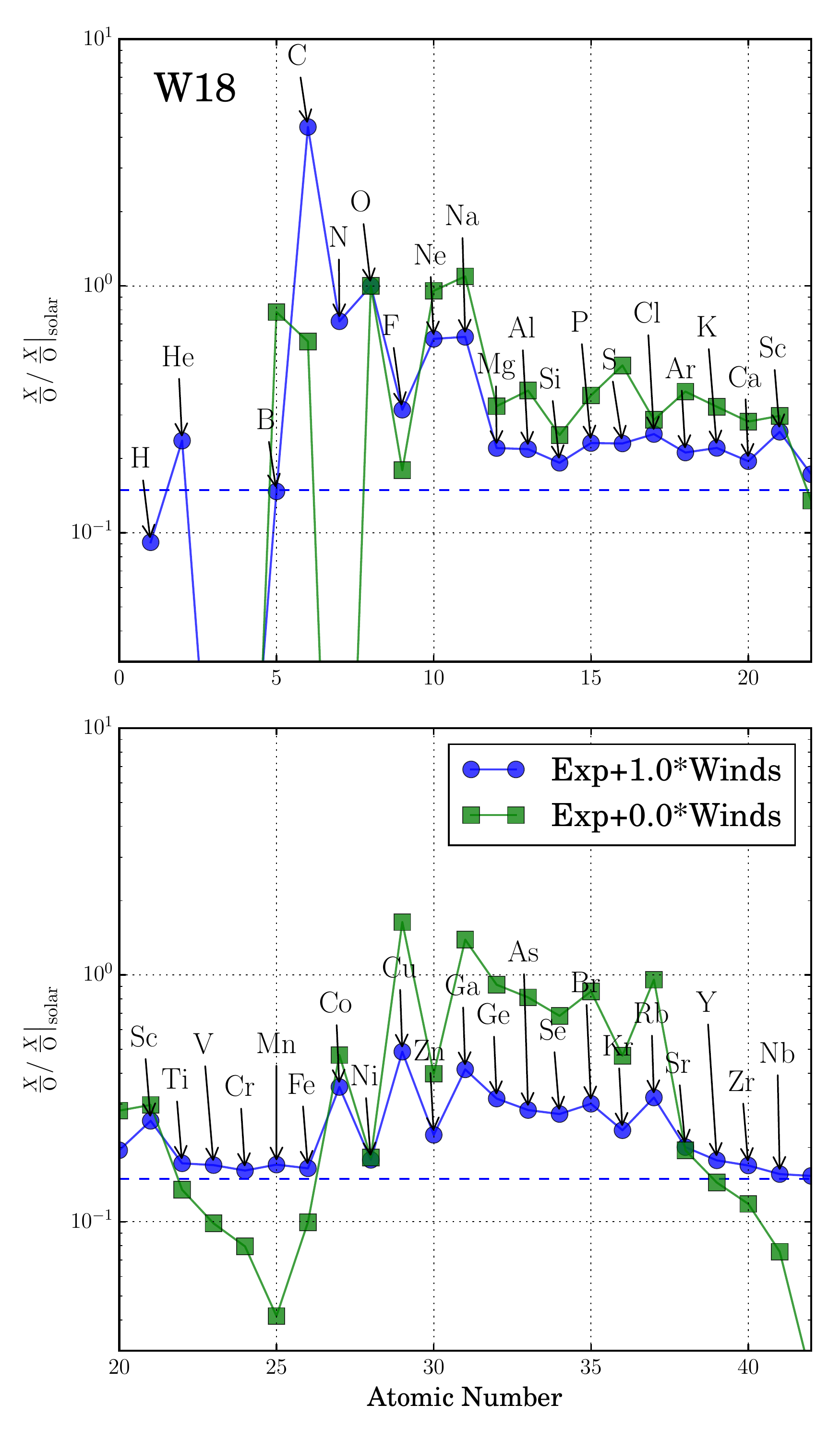}
	\caption{Similar to \Fig{low}, the IMF-averaged production of
          the elements is shown for stars from 30 to 120 \Msun, using
          the W18 calibration for the central engine. Most, though not
          all stars in this very massive range become black holes, so
          their nucleosynthesis is dominated by the winds that they
          eject before dying.  Production factors are normalized to
          oxygen. If winds are included, net production of an element
          occurs in this mass range if its average production factor
          is above 0.15 (blue dashed line).  Also shown are the
          production factors if the wind is discarded. Since oxygen
          was being appreciably synthesized in the winds, neglecting
          them results in an overall renormalization of all production
          factors.  The cut-off for the net production without winds
          falls to 0.010, off the scale of the plot.  Including winds,
          these stars make so much carbon, nitrogen and oxygen that
          they can be responsible for little else.}  \lFig{high}
\end{figure}

\Fig{high} shows that the winds contain excesses of $^{4}\rm{He}$,
both from the envelopes of these massive stars that have been enriched
by hydrogen burning and convective dredge up, and from the winds of
Wolf-Rayet stars that have lost their envelopes altogether. This
production significantly augments the primordial helium in the
original star from the Big Bang and previous generations of stars. The
nuclei $^{12,13}\rm{C}$, $^{14}\rm{N}$, and $^{16,18}\rm{O}$ are also
significantly enriched, but $^{15}$N and $^{17}$O are not.
$^{22}\rm{Ne}$ is greatly enhanced owing to Wolf-Rayet winds. This
nucleus comes from two alpha captures on $^{14}\mathrm{N}$ and is
abundant in the helium shell from convective helium burning. The
s-process up to Rb is also produced in the winds of these very massive
stars, but its production is still inadequate to explain its abundance
in the sun when the lighter stars are folded in. Intermediate mass
elements are underproduced compared with the large oxygen synthesis
in the winds.

\subsection{Integrated Yields from Massive Stars 9 -- 120 \Msun}
\lSect{massiveyield}

The integrated production factors from massive stars of all masses
above 9 \Msun \ are shown in \Fig{full} summed over isotopes. The
individual isotopic abundances are given in \Fig{fulliso}. The integrated
production factor is calculated from a combined weighting of the three
mass ranges just discussed. If $P_{\rm low}$ is the production factor
for stars 9 -- 12 \Msun; $P_{\rm mid}$, for stars 12 -- 30 \Msun; and
$P_{\rm high}$, for stars 30 -- 120 \Msun, the total production
factor, $P_{\rm tot}$ is
\begin{equation}
P_{\rm tot} \ = \ f_1 P_{\rm low} + f_2 P_{\rm mid} + f_3 P_{\rm high}
\end{equation}
where $f_i$ is given in \Tab{rangestats}.  Because $P_{\rm low}$ is
generally low and $P_{\rm high}$ is low except for species produced in
the winds like CNO, the total production for most elements is dominated by $ P_{\rm mid}$.


\begin{figure}[h]
	\centering
	\includegraphics[width=0.43\textwidth]{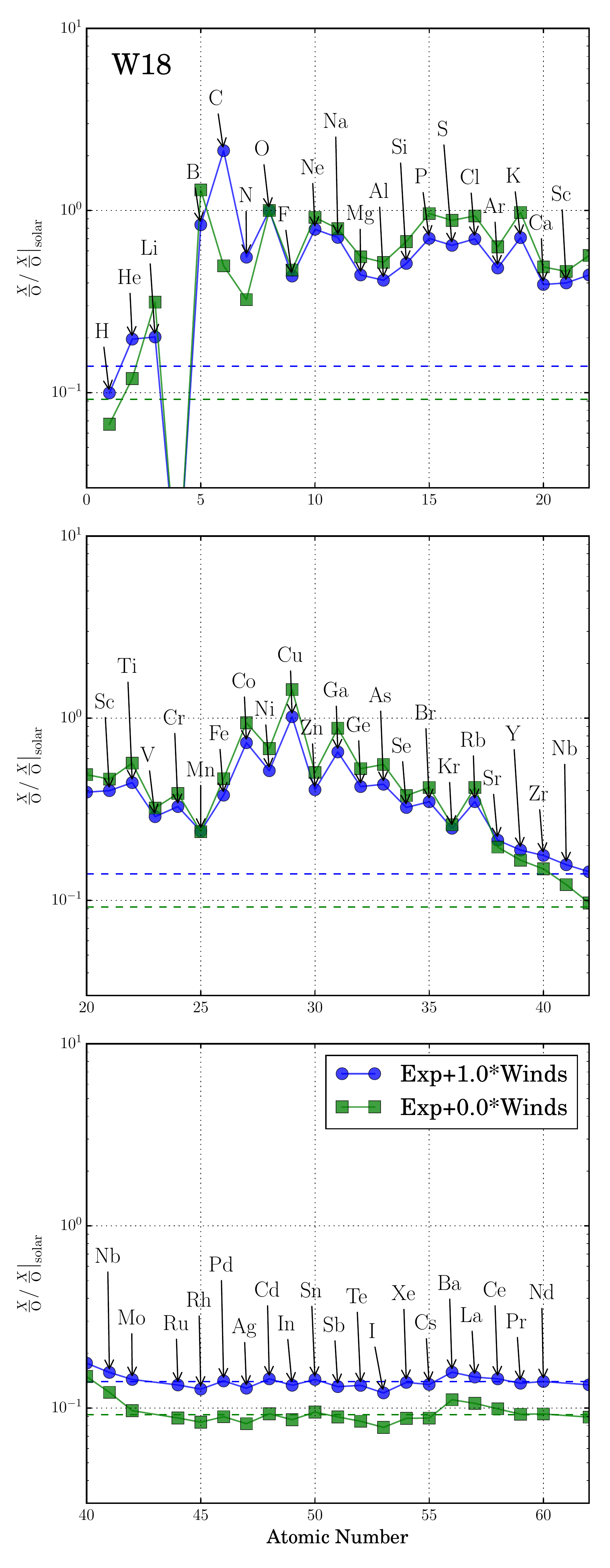}
	\caption{Integrated nucleosynthesis of the elements for all
          stars considered (9 -- 120 \Msun) using the W18 central
          engine. The results have been normalized to oxygen and the
          dashed line show net production occurring for production
          factors bigger than 0.14, if winds are included, 0.092 if
          they are not.  The displacement of the overall normalization
          is due to the production of oxygen itself in winds.  While
          individual isotopes may be affected (\Fig{fulliso}), no net
          nucleosynthesis occurs for elements above Z = 40.}
        \lFig{full}
\end{figure}

The probable nucleosynthetic sites for producing the various isotopes
heavier than lithium have been summarized in Table III of
\citet{Woo02} and our discussion follows that general delineation and
omits references given there.  We also include the full integrated
yield of the N20 series in \Fig{n20iso}.  The primary difference is
that the N20 model produces a bit more $^{16}$O, which can be seen by
comparing the baseline solar abundance of the heavy isotopes.


\begin{figure}[ht]
\centering
\includegraphics[width=\columnwidth]{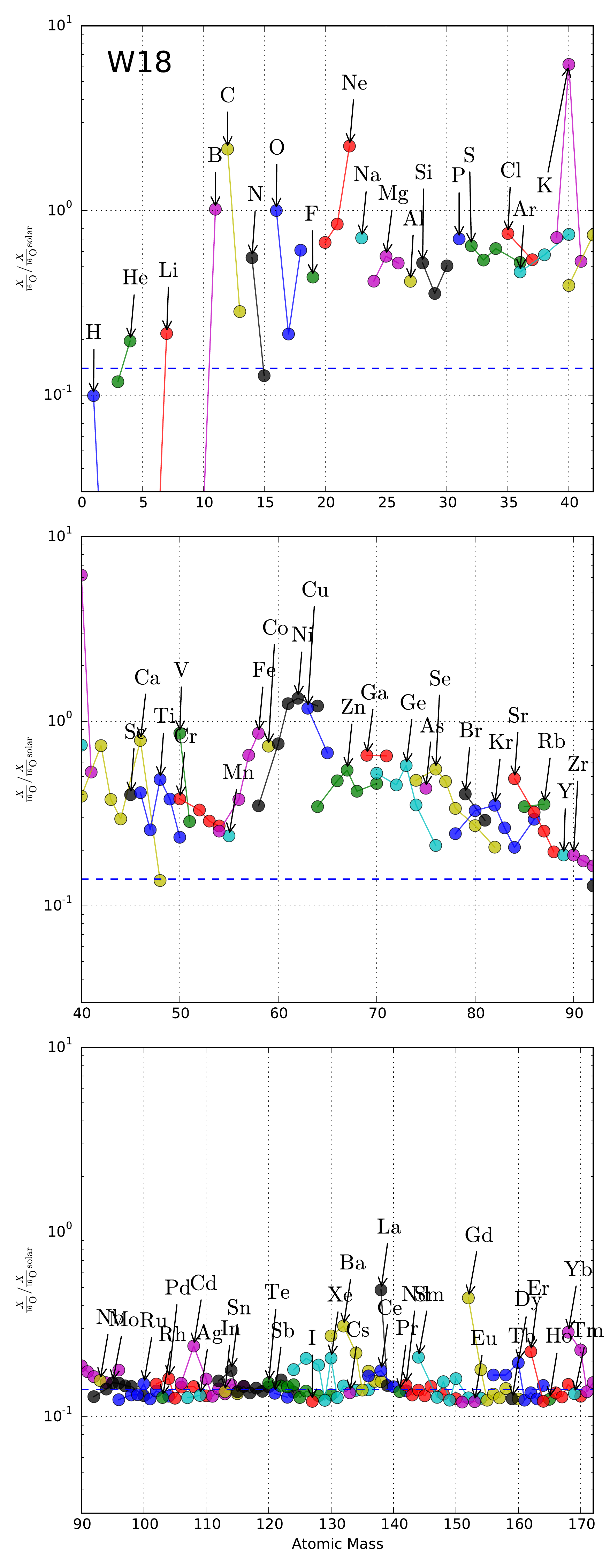}
\caption{\lFig{fulliso} The same yields as in \Fig{full} for the W18
  central engine including the wind, but now given for the individual
  isotopes of each element.}
\end{figure}


\begin{figure}[ht]
  \centering
  \includegraphics[width=\columnwidth]{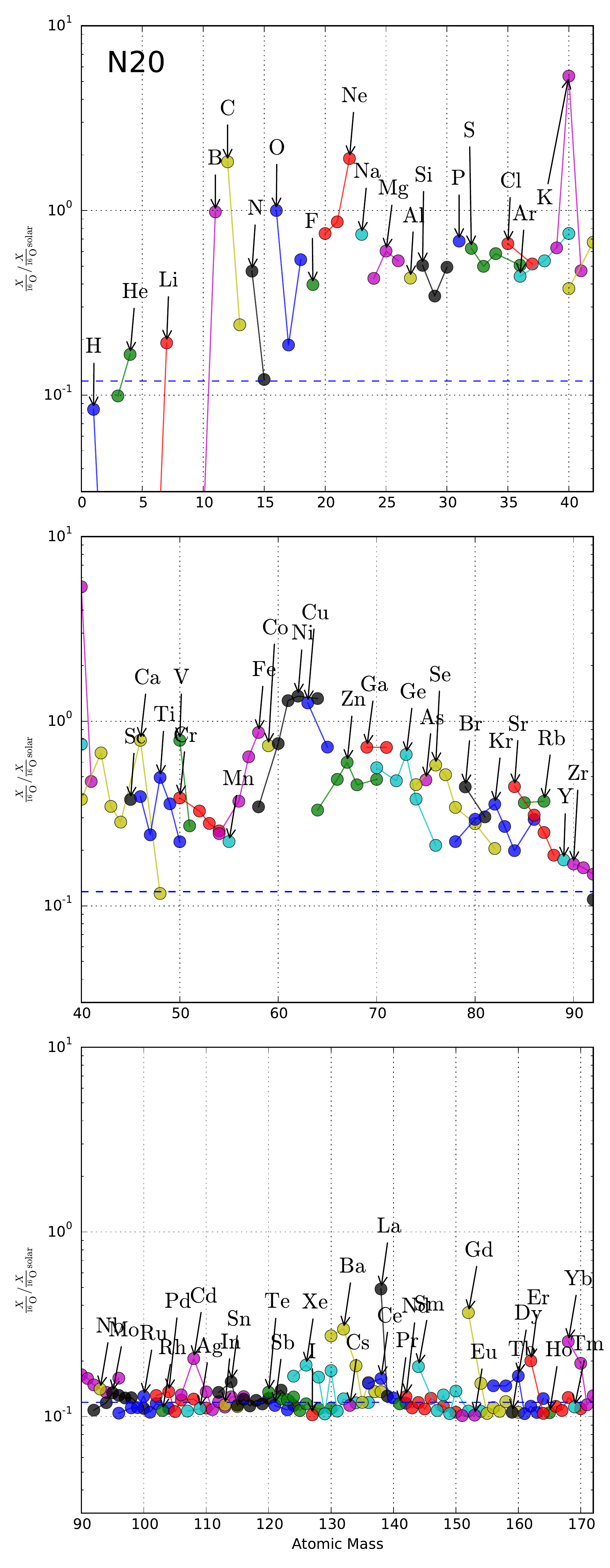}
  \caption{\lFig{n20iso} Similar to \Fig{fulliso}, this figure gives
    results using the N20 central engine instead of W18. There are no
    significant differences.}
\end{figure}

Additionally, one could vary the power of the initial mass function.
Lowering the power from that of a Salpeter IMF of $-2.35$ to a value
as low as $-2.7$ affects the results in a similar way to a reduction
in the contributions of mass loss rates. The value of $-2.7$ was
chosen as being a low value within the known uncertainties
\citep[e.g.][]{Sca86,Cha03}. The results in \Fig{fullimf} show that a
steeper IMF acts in the same way as reducing the contributions from
the winds of very heavy stars, but with a smaller effect.


\begin{figure}[ht]
\centering
\includegraphics[width=\columnwidth]{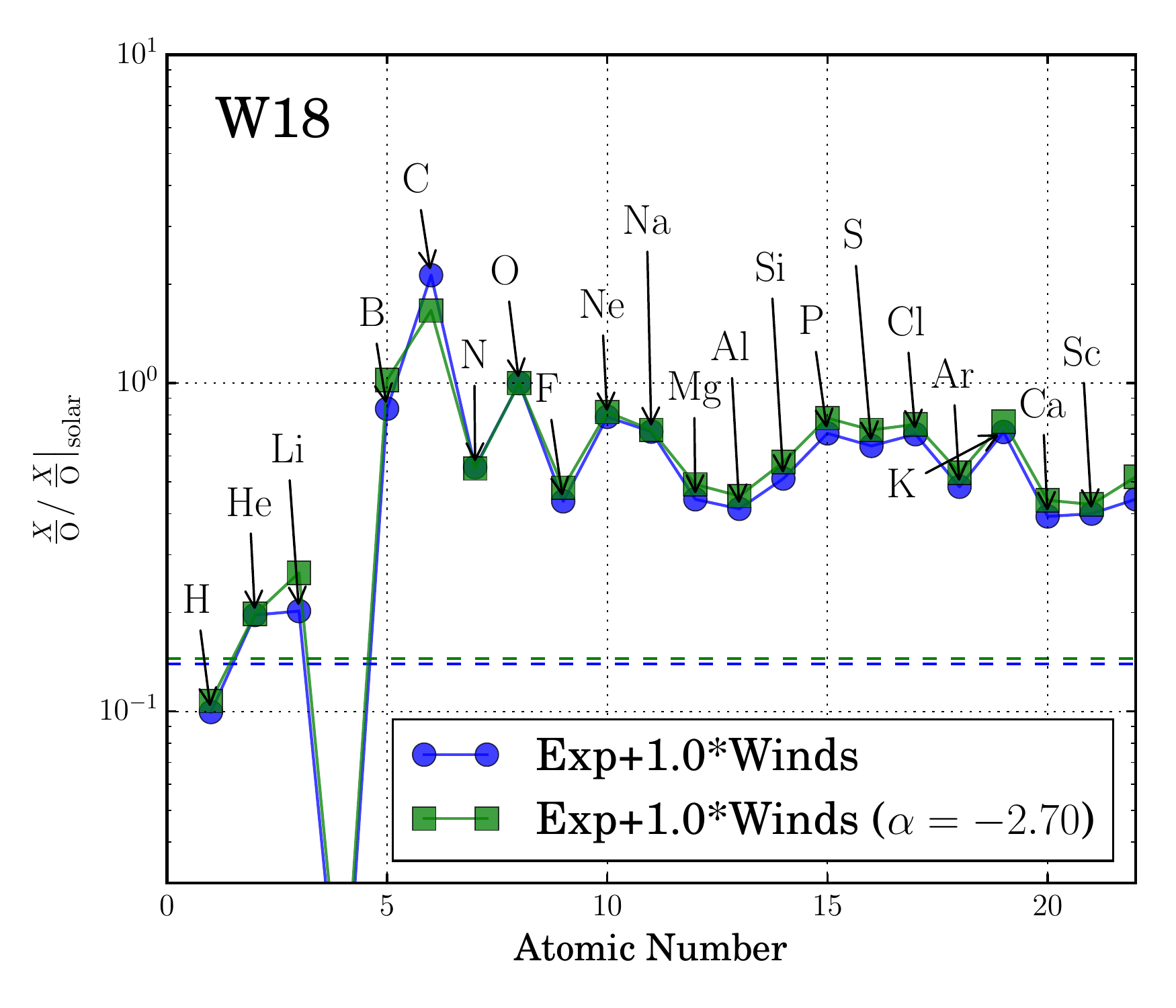}
  \caption{Similar to \Fig{full}, but using a steeper
    mass dependence for the IMF. Here the results of using a Salpeter
    IMF with a power-law dependence of $-2.35$ are compared to those
    obtained using a slope of $-2.7$. The dashed lines here indicate
    net production for both ensembles at 0.14.}  \lFig{fullimf}
\end{figure}

\subsubsection{Light elements - lithium through aluminum}
\lSect{zeq5to13}

While perhaps not widely recognized, massive stars can make an
important, though not dominant contribution to the abundance of $^7$Li
(though not $^6$Li). The $^7$Li is made mostly by the neutrino
process, which spalls helium in the deeper layers to make $^{3}$He
which combines with $^4$He to make $^7$Li (after a decay). Some
$^{3}$He is also made in the hydrogenic envelope. The total
contribution to each is about 20\% of their solar value. $^6$Li,
$^9$Be, and $^{10}$B must be made elsewhere, perhaps by cosmic ray
spallation.

$^{11}$B, on the other hand, is abundantly produced by the neutrino
process. Spallation of carbon by $\mu$ and $\tau$-neutrinos produces
the isotope in the carbon shell. Without neutrinos, $^{11}$B would be
greatly underproduced. Its abundance in the sun thus provides a
useful thermometer for measuring the typical temperature of the neutrinos
(though their spectrum is varies significantly from thermal). The
assumed temperature of the neutrinos here was 6 MeV. A slightly
smaller value would have sufficed, but larger values are ruled out.

Carbon, perhaps surprisingly is adequately produced in massive stars,
but the yield is very sensitive to the treatment of mass loss and, to
a lesser extent, the IMF (\Fig{fullimf}).  Too much carbon is made for
the mass loss rates assumed here \citep[see also][]{Mae92}, but a mass 
loss rate half as large is certainly allowed and would produce carbon 
nicely. In the unrealistic case of no mass loss, only about one-third 
of carbon would be made in massive stars. Care should be taken though 
because folding the yields from a population of solar metallicity 
stars with an IMF is very approximate compared with a full model for 
galactic chemical evolution. Lower metallicity massive stars would 
presumably have diminished winds and would make less carbon. The 
isotope $^{13}$C is always underproduced in massive stars and is 
presumably made in the winds and planetary nebulae of lower mass stars.

Nitrogen too is underproduced and needs to be mostly made
elsewhere. Presumably $^{14}$N is made by the CNO cycle in lower mass
stars and $^{15}$N is made in classical novae by the hot CNO cycle.

Oxygen is the normalization point and both $^{16}$O and $^{18}$O are
copiously produced, both by helium burning. The yield of $^{18}$O is
sensitive to the metallicity, but that of $^{16}$O is not directly,
though it is sensitive by way of the mass loss. $^{17}$O is not made
due to its efficient destruction by
$^{17}$O(p,$\alpha$)$^{14}$N. Perhaps it too is made in classical
novae.

Fluorine is somewhat underproduced in the overall ensemble, despite
having a very substantial production by the neutrino process and
helium burning in low mass stars.

The abundance of neon is uncertain in the sun, but $^{20}$Ne and
$^{21}$Ne are well produced and $^{22}$Ne is overproduced. The
synthesis of $^{22}$Ne depends on the initial metallicity of the star,
however, since CNO is converted into $^{22}$Ne during helium
burning. Lower metal stars would make less and the abundance of
$^{22}$Ne might come down in a full study of galactic chemical
evolution. The yield of $^{22}$Ne is also sensitive to mass loss, more
so than $^{20}$Ne and $^{21}$Ne, and lower metallicity would imply
lower mass loss.

All isotopes of sodium, magnesium, and aluminium are produced
reasonably well by carbon and neon burning. Magnesium and aluminium
are a little deficient, however, and this is the beginning of a trend
that persists through the intermediate mass elements. It could reflect
a systematic underestimate of the CO core size due to the neglect of 
rotation or inadequate overshoot mixing, since these isotopes will be 
produced in greater abundances in larger CO cores, but increasing the 
core size might make the stars harder to blow up. Or it may indicate 
an incompleteness in the present approach. Many of the stars that 
made black holes here would have contributed to the intermediate mass 
elements. Perhaps rotation or other multi-dimensional effects on the 
explosion play a role?

\subsubsection{Intermediate-mass elements - silicon through scandium}
\lSect{zeq14to21}

The major isotopes of the intermediate-mass elements from silicon
through calcium are consistently co-produced in solar
proportions. The total mass of these, however, is about 12\% of that
of $^{16}$O, which is only half the solar value of 22\%. A similar
underproduction was seen by \citet{Woo07} (their Fig 8), so it is not
solely a consequence of the new approach. Part of the difference might
be picked up by SN Ia which, aside from being prolific sources of
iron, can also produce a significant amount of intermediate-mass elements
\citep[][and \Sect{combined}]{Iwa99}. Still the systematic
underproduction of so many species generally attributed to massive
stars is troubling. Better agreement between intermediate mass
element and oxygen productions existed in earlier studies
\citep{Tim95,Woo02} which used a larger solar abundance for $^{16}$O.

Several isotopes with particularly anomalous production in this mass
range warrant mention. $^{40}$K is greatly overproduced compared even
with the abundance in the zero age sun. The difference presumably
reflects the lengthy time in which decay occurred between the last
typical supernova and the sun's birth.  $^{44}$Ca is underproduced in
massive stars. Presumably it is made by sub-Chandrasekhar mass models
for SN Ia (\Sect{combined}). $^{48}$Ca along with several neutron-rich
iron-group nuclei like $^{50}$Ti and $^{54}$Cr, is presumably made in
a neutron-rich nuclear statistic equilibrium as might exist in a rare
variety of Type Ia supernova igniting at high density.  $^{45}$Sc is
due to the $s$-process and its underproduction is a portent of
problems to come (\Sect{sproc}).

\subsubsection{The Iron Group}
\lSect{fegroup}

The iron yields here were calibrated to be the maximum calculated in
P-HOTB (\Tab{eni_Z9.6} and \Tab{eni_N20_W18}), where it was assumed that the
neutrino wind makes an appreciable contribution. This was one of the
agreements forced upon the KEPLER recalculation (\Fig{ni_W18}).

As expected, even taking this upper bound, the iron group is severely
underproduced in massive stars, since most of the iron in the sun has
been made by SN Ia. In the next section, we shall consider the
consequences of combining both varieties of supernovae. The ratio of
the mass of new iron made here as $^{56}$Ni to new oxygen (neglecting
the initial iron and oxygen in the star because it had solar
metallicity) is proportional to the ratio of (P-1) for the two
species, where P here is the unnormalized production factor
(\Sect{nucleo}).  When normalized to the solar mass fractions, the
fraction of solar iron made in core-collapse supernovae is
\begin{equation}
F_{\rm Fe} = \frac{P_{\rm Fe} -1}{P_{\rm Ox} -1}. 
\end{equation}
For the series W18 $P_{\rm Fe}$ = 2.17 and $P_{\rm Ox}$ = 7.16; for
the N20 series $P_{\rm Fe}$ = 3.09 and $P_{\rm Ox}$ = 8.38. Both sets
of numbers include the low mass contributions from the Z9.6 series.  In
both cases the implied iron production is 28\%. Given the way this was
calculated, using yields normalized to the maximum production in
\Tab{stats}, this is probably an upper bound, though not by much.

While SN Ia make most of the iron, it is noteworthy that massive stars
do contribute appreciably to many species in the iron group. $^{50}$V
is well produced by carbon burning and $^{58}$Fe by the
$s$-process. Cobalt, copper and the nickel isotopes are well produced
and, in the case of $^{62}$Ni, actually overproduced. All three
elements are made mostly by the $\alpha$-rich freeze out (as
$^{58}$Ni, $^{59}$Cu, and $^{60,61,62}$Zn) and are thus most sensitive
to the treatment of the inner boundary in KEPLER and the explosion
itself of P-HOTB. The abundances of $^{58}$Ni and $^{62}$Zn, which
makes $^{62}$Ni, are sensitive to the neutron excess in the innermost
zones. In KEPLER, appreciable electron capture occurs as these zones
fall to very high density and are heated to high temperature by the
shock. In P-HOTB the neutron excess in these deepest zones can be
changed by the neutrinos flowing through, but this has been treated
approximately.  As a result, the composition of the neutrino-powered
wind has not been accurately determined in either code. A higher value
of $Y_e$ in those zones would reduce the synthesis of both $^{58}$Ni
and $^{62}$Ni, so their production here must be regarded as
uncertain. An appreciable, though lesser, contribution to $^{62}$Ni
also comes from the $s$-process in the helium shell

Also noteworthy is the underproduction of $^{55}$Mn in massive
stars. This species is made mostly as $^{55}$Co in a normal
($\alpha$-deficient) freeze out from nuclear statistical equilibrium,
and its synthesis is not so uncertain as that of nickel which is made
deeper in.

\subsubsection{Heavier elements}
\lSect{sproc}

In contrast to previous studies \citep[e.g.][]{Woo07}, the production
of species above the iron group falls off rapidly. This is problematic
given that the synthesis of $s$-process isotopes up to about A = 90 is
often attributed to massive stars \citep{Kap11}. The problem is
exacerbated by the fact that these are secondary elements and so are
expected to be even more underproduced in stars of lower
metallicity. If anything, one would like to produce an excess of
isotopes in this metallicity range compared with solar values.

\Fig{full} shows that production in the mass range A = 65 -- 90 is 20
- 50\% of what is needed. Previous studies \citep{Woo02,Woo07} using
the same code and stellar and nuclear physics showed a slight
overproduction in the same mass range. The change is a direct
consequence of fewer massive stars exploding \citep{Bro13}. A larger
value for the $^{22}$Ne($\alpha$,n)$^{25}$Mg reaction rate might help
here. For calibration, the present study used reaction rates at $3
\times 10^8$ K of $2.58 \times 10^{-11}$ Mole$^{-1}$ s$^{-1}$ for
$^{22}$Ne($\alpha,$n)$^{25}$Mg and $8.13 \times 10^{-12}$ Mole$^{-1}$
s$^{-1}$ for $^{22}$Ne($\alpha,\gamma)^{26}$Mg, the same as used for
the last 15 years. \citet{Lon12} recently suggested a best value of
$3.36 \times 10^{-11}$ Mole$^{-1}$ s$^{-1}$ (range 2.74 to 4.15
$\times 10^{-11}$ for $^{22}$Ne($\alpha,$n)$^{25}$Mg and $1.13 \times
10^{-11}$ Mole$^{-1}$ s$^{-1}$ (range 0.932 to $1.38 \times 10^{-11}$)
for $^{22}$Ne($\alpha,\gamma)^{26}$Mg. Use of more modern rates would
thus improve the agreement slightly, but would probably not eliminate
the discrepancy. An improved treatment of overshoot mixing might also 
make a difference \citep{Cos06}. Future studies to further study this 
discrepancy are planned. For the time being, we take the deficient 
production of the light $s$-process as the strongest indicator yet 
that something may be awry in our present 1D modelling. The 
possibility that a substantial fraction of heavy element production 
occurs in stars that blow up because of rotation or other 
multi-dimensional effects must be considered.

Accompanying the underproduction of light $s$-process elements is a
notable underproduction of the heavy $p$-process, again compared with
past results. Some have suggested that the light $p$-process could
also be made in the neutrino-powered wind
\citep{Hof96,Fro06,Wan11b,Wan11c}.

\subsubsection{$^{26}$Al and $^{60}$Fe}
\lSect{26al}

The synthesis processes and sites of the long-lived radioactivities 
$^{26}$Al and $^{60}$Fe and their importance to gamma-line astronomy 
has been extensively discussed in the literature \citep{Tim95b,Lim06a,Lim06b}. Here, using the W18 calibration, the 
IMF-averaged ejection masses for $^{26}$Al and $^{60}$Fe are, 
respectively, $2.80\times10^{-5}$ and $2.70\times10^{-5}$ solar 
masses per typical massive star. Using the N20 calibration instead 
gives $3.63\times10^{-5}$ and $3.20\times10^{-5}$ solar masses per 
star. Gamma-ray line observations \citep{Wan07} imply a number ratio 
of $^{60}\mathrm{Fe}$ to $^{26}\mathrm{Al}$ in the interstellar medium 
of 0.148, which implies a mass ratio of 0.34. The ratios of 0.97 and 
0.88 from our two calibrations thus exceed observations almost by a 
factor of three. Still, this is an improvement from \citet{Woo07} 
where the ratio, using the same nuclear physics as here, was 1.8. 
This factor-of-two decrease reflects both the trapping of substantial 
$^{60}$Fe in stars that no longer explode and weaker typical explosion
energies. \citet{Lim06a} have discussed other uncertainties that might 
bring the models and observations into better agreement, including a 
modified initial mass function, mass loss, modified cross sections, 
and overshoot mixing.

As in \citet{Woo07}, we also note that alternate, equally justifiable
choices for key nuclear reaction rates, especially for
$^{26}$Al(n,p)$^{26}$Mg, $^{26}$Al(n,$\alpha)^{23}$Na, and
$^{59,60}$Fe(n,$\gamma)^{60,61}$Fe, might bring this ratio in our
models down by an additional factor of two, hence within observational
errors. This possibility will be explored in a future work.
Interestingly the production ratio for $^{60}\mathrm{Fe}$ to
$^{26}\mathrm{Al}$ is sensitive to stellar mass. Averaging over just
the explosions from 9 to 12 \Msun, the ratio of masses ejected is
close to 2.6 for both the W18 and N20 calibrations, while for heavier
stars it is 0.65 (\Tab{isotopes}). 

We have also examined the potential effect of ejecting the $^{26}$Al
that resides in the envelopes of the stars that implode to become
black holes. While not lost to winds, even weak explosions may eject
these loosely bound envelopes while allowing the helium core to
collapse to a singularity \citep{Nad80,Lov13}.  For these, we summed
the total mass of $^{26}$Al still in the hydrogen envelopes of the
stars that collapsed and added these to the yields of the stars prior
to the integration. We find only a very minor improvement to the total
integrated yield of aluminium, of about 7\%.  This helps to reduce the
underproduction of $^{26}$Al, but not appreciably.

Additional information on $^{60}$Fe production is available from the
$^{60}$Fe/$^{56}$Fe ratio observed in cosmic rays \citep{Isr15}. There
the ratio is thought to reflect chiefly the synthesis of both species
in massive stars, diluted with some fraction of unprocessed material
prior to acceleration. Observations of cosmic rays near the earth give
a mass ratio that, when propagated back to the source, is
$^{60}$Fe/$^{56}$Fe = 0.80 $\pm$ 0.30 $\times 10^{-4}$ by mass. Our
IMF average of the mass ratio for $^{60}$Fe/$^{56}$Fe ejected in both
sets of explosions and winds is near $7 \times 10^{-4}$. Assuming a
standard dilution factor of 4 to one for normal ISM to massive star
ejecta \citep{Hig03,Isr15} gives an expected ratio in the local cosmic
rays of $1.4 \times 10^{-4}$, in reasonable agreement with the
observations. Again though, a reduction of a factor of about 2 in
$^{60}$Fe would be desirable. This ratio too is sensitive to the mass
range of supernovae that are sampled (\Tab{isotopes}) and is larger in
lower mass stars and somewhat smaller in heavier ones. Note that the
iron in \Tab{isotopes} is expressed in solar masses per massive star
death, while that in \Tab{stats} is per supernova.

\begin{deluxetable}{llccc}
\tablecaption{Selected Isotopes Production in \Msun}
\tablehead{\colhead{Mass Range}          &
           \colhead{Engine(s)}           & 
           \colhead{$^{26}$Al/10$^{-5}$} & 
           \colhead{$^{60}$Fe/10$^{-5}$} & 
           \colhead{$^{56}$Fe} }
\startdata
9-120    & Z9.6+W18 & 2.80 & 2.70 & 0.038 \\
9-120    & Z9.6+N20 & 3.63 & 3.20 & 0.052 \\
$\leq$12 & Z9.6     & 1.36 & 2.73 & 0.026 \\
$>$12    & W18      & 3.17 & 2.66 & 0.053 \\
$>$12    & N20      & 4.22 & 3.28 & $\:$0.066 
\enddata
\lTab{isotopes}
\end{deluxetable}

\subsection{A Type Ia Supernova Contribution}
\lSect{combined}

\subsubsection{Nucleosynthesis}



\begin{figure}[t]
	\centering
	\includegraphics[width=0.45\textwidth]{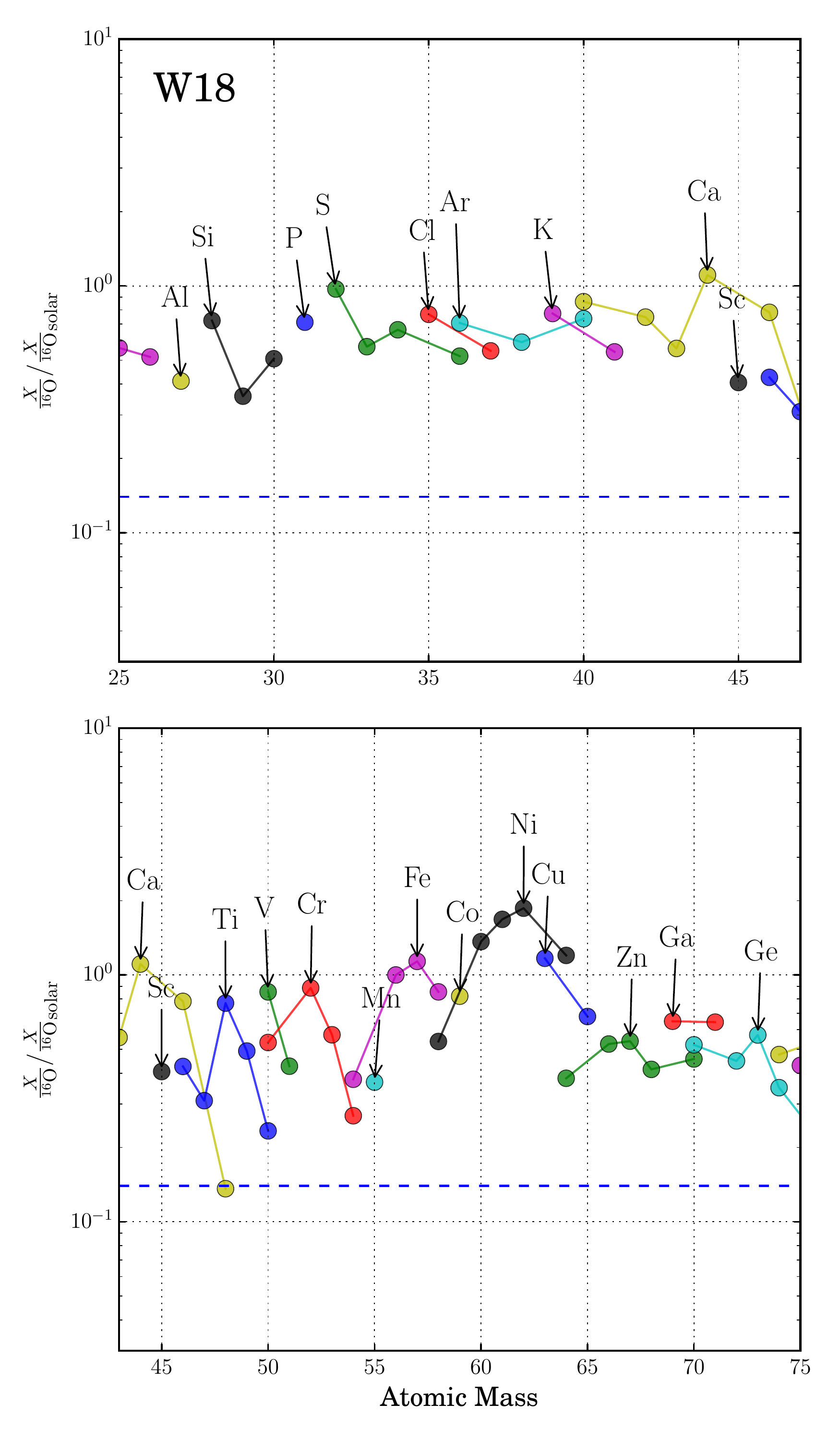}
	\caption{Similar to \Fig{fulliso}, this figure includes
          contributions from all 9--120 \Msun \ supernovae and their
          winds, but also an additional component from SN Ia. Added to
          the average yield in \Fig{fulliso} is that of a typical
          sub-Chandrasekhar mass SN Ia \citep{Woo11} with a sufficient
          quantity of iron to make its solar abundance relative to
          oxygen. Not only does this variety of SN Ia make iron, but
          also raises the production of $^{44}$Ca to its full solar
          value and increases the abundances of Si, S, Ar, and Ca so
          that they are closer to solar. Note, however, the
          underproduction of Mn/Fe.  The dashed line for net
          production here is at 0.14.  } \lFig{full_ia_nuc}
\end{figure}


\begin{figure}[t]
  \centering
  \includegraphics[width=0.45\textwidth]{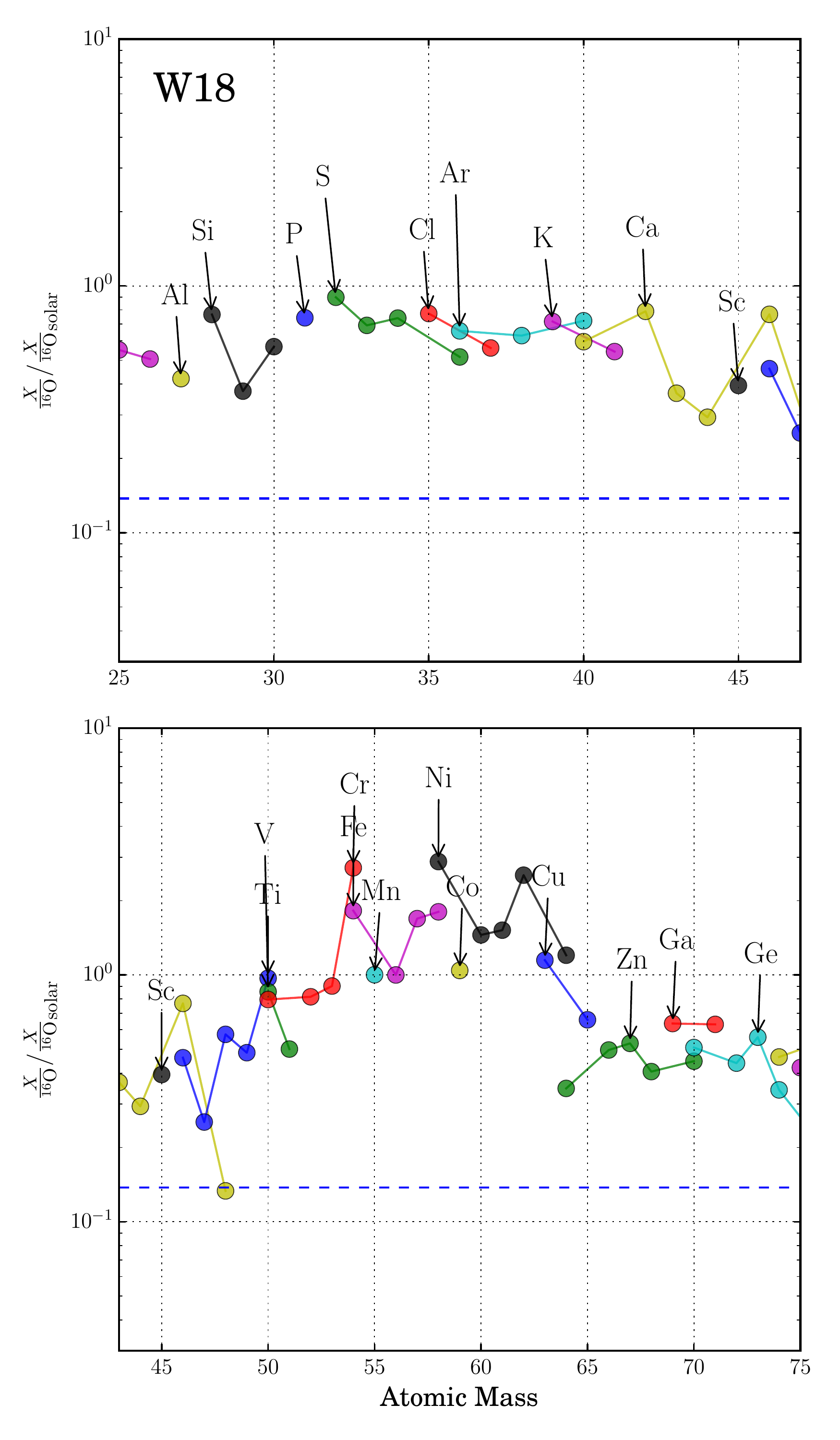}
  \caption{Similar to \Fig{full_ia_nuc}, but including the yields of a
    Chandrasekhar mass model for SN Ia (Nomoto's W7) rather than the
    sub-Chandrasekhar model.  The dashed line for net production here
    is at 0.14.  Manganese is much improved, but $^{44}$Ca is lost}
  \lFig{full_w7_nuc}
\end{figure}

The correct model or models for Type Ia supernovae is currently under much
debate. In terms of nucleosynthesis, however, the models segregate
into two general categories: explosions near the Chandrasekhar mass by
a combination of deflagration and detonation, and prompt detonations in
sub-Chandrasekhar mass white dwarfs.  The requirement that any
successful model makes $\sim$0.6 \Msun \ of $^{56}$Ni in order to
explain the light curve and $\sim$0.2 \Msun \ of intermediate mass
elements to explain the spectrum determines the acceptable bulk
nucleosynthesis, so that the models differ only in detail.  Here, as a
representative example of the Chandrasekhar mass model we use the
nucleosynthesis from Nomoto's highly successful ``W7'' model
\citep{Iwa99}. For the sub-Chandrasekhar mass model we use Model 10HC
of \citet{Woo11}. This detonation of a 1.0 \Msun \ carbon-oxygen white
dwarf capped by 0.0445 \Msun\ of accreted helium produced 0.636 \Msun
\ of $^{56}$Ni in an explosion with final kinetic energy $1.2 \times
10^{51}$ erg. The contribution from each SN Ia model was normalized so
as to produce a solar proportion of $^{56}$Fe/$^{16}$O when combined
with the integrated yield of massive stars from 9 to 120
\Msun. 

The combined nucleosynthesis is given in \Fig{full_ia_nuc} for a
sub-Chandrasekhar white dwarf and in \Fig{full_w7_nuc} for the
Chandrasekhar mass model. Most of the intermediate mass elements are
unchanged, though some of these isotopes do change.  Not surprisingly,
the introduction of Type Ia supernovae improves the overall fit
especially for the iron group. The sub-Chandrasekhar mass model
greatly improves the production of $^{44}$Ca (made by helium
detonation as $^{44}$Ti) and seems to be required if only because of
that unique contribution. The Chandrasekhar mass model, on the other
hand, does a much better job of producing manganese, suggesting that
this component may also be needed \citep[see
  also][]{Sei13a,Yam15}. The Chandrasekhar mass model overproduces
nickel however. This is a known deficiency of Model W7 having to do
with excess electron capture at high density. It might be circumvented
in more modern multi-dimensional models \citep[e.g.][]{Sei13b}
especially if the white dwarf expands appreciably during the
deflagration stage prior to detonating.

\subsubsection{Implications for Supernova Rates}
\lSect{sn1arate}

As noted previously (\Sect{fegroup}), our massive star models are 
responsible for producing about 28\% of the iron in the sun. The 
average SN Ia makes very nearly 0.6 \Msun. That being the case, 
and assuming that the typical successful core collapse supernova 
(mostly SN IIp in the present study) makes 0.04 to 0.05 \Msun\ of 
new iron (\Tab{stats}), the implied ratio of SN Ia event rates to 
SN IIp is 1/6 to 1/5. If core-collapse supernovae made a little less 
iron, as seems likely, then the SN Ia rate would need to be a bit 
bigger. This ratio is consistent with observations for spiral 
galaxies similar to the Milky Way \citep{Li11}.

We can also use these abundances, particularly the fact that each
massive star above 9 \Msun\ produces an average of 0.57 \Msun\ of
$^{16}\rm{O}$, to infer a rate of Type II supernova (assuming stars
below 9 \Msun produce little oxygen in their evolution).  From the
solar abundance of $^{16}$O of $6.6\times10^{-3}$ \citep{Lod03}, we
find that there must have been 0.012 massive stars per solar mass of
Population I material. Assuming that 66\% of these stars become
supernovae, as we found in our models, we find that integrated over
galactic time, there must be 0.0076 core collapse supernovae (visible
or invisible) per solar mass of Pop I material in order to explain the
present solar abundance. This seems a reasonable, albeit very
approximate number, especially if the supernova rate was different in
the past.


\begin{figure}[h]
\centering
\includegraphics[width=.48\textwidth]{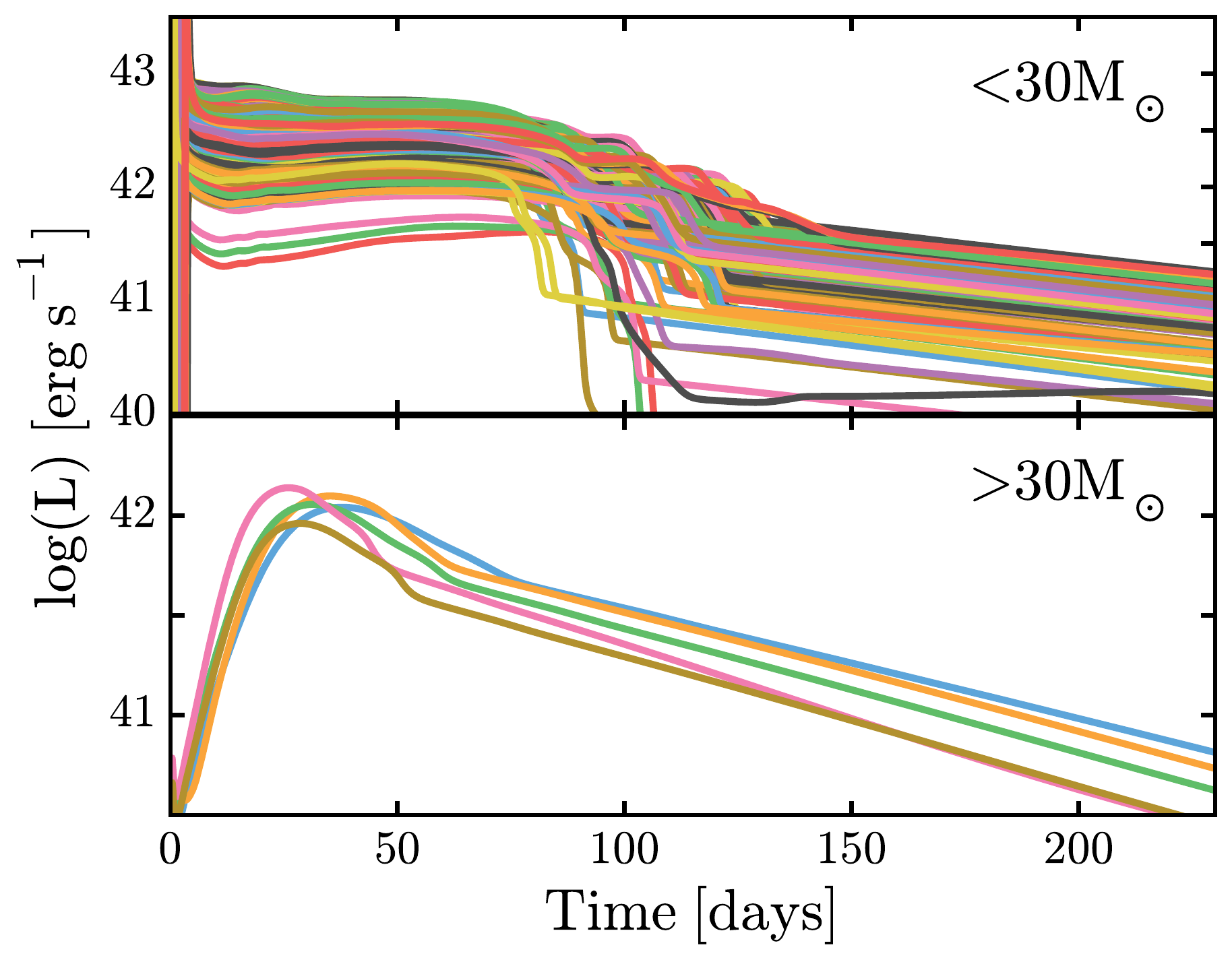}
\caption{For each successful explosion calculation, the bolometric
  light curve was followed for 230 days. Lighter models (top) that
  retained a good fraction of their envelopes were SN IIp, while
  heavier models (bottom) which lost most or all of their
  envelopes were SN Ib or Ic. Both panels show all models that were
  successfully exploded by Z9.6, W18 and N20 calibrators.  \lFig{lcs}}
\end{figure}


\section{Light Curves}
\lSect{lite}

The KEPLER code incorporates flux-limited radiative diffusion and thus
is capable of calculating approximate bolometric light curves for
supernovae of all types. The light curve calculation includes both
contributions from the diffusion of energy deposited by the shock wave
and from the decay of radioactive $^{56}$Ni and $^{56}$Co.  Light
curves were calculated as described in \citet{Woo07} using a full
Saha-solver for 19 elements, hydrogen to nickel, to obtain the
electron density, and adopting a lower bound to the opacity of
$10^{-5}$ cm$^2$ g$^{-1}$.  The full model set corresponding to the
successful explosions using the central engines Z9.6, W18 and N20 was
post-processed to obtain the luminosity evolution until about 230
days, which was well past the plateau and peak for all models. For the
mass loss prescription used, explosions below 40 \Msun, i.e., most
successful explosions, produced Type IIp supernovae with a distinct
plateau phase. The heavier stars gave rise to Type Ib or Ic
supernovae. \Fig{lcs} illustrates the diversity of outcomes from the
full model set. In this section, we discuss the systematics of these
results and how they compare with simple analytic scalings in the
literature. Some scalings of particular interest to observers are
summarized in \Sect{system}. All of our light curve data are 
included in the electronic edition in a tar.gz. package.

Since the light curve results from W18 and N20 engines are qualitatively 
similar, in the following analysis, only the N20 series is considered, supplemented by Z9.6 below 12 \Msun.

\subsection{Type IIp}
\lSect{type2} 

Since all of the models that retained their hydrogen envelope were, at
death, red supergiants, most of the supernovae modeled here were of
Type IIp. A wide diversity of durations and luminosities on the
plateau is expected observationally \citep{Arc12}, and was found owing
to the variable presupernova radius, explosion energy, envelope mass,
$^{56}$Ni ejected, and mixing. Luminosities on the plateau ranged
across at least a decade in luminosity (10$^{41.5}$ - 10$^{43}$ erg
s$^{-1}$), with the most luminous events coming from the energetic
explosions of stars with exceptionally large radii, touching 10$^{14}$
cm. \Tab{eni_Z9.6} and \Tab{eni_N20_W18} also show a variable amount
of $^{56}$Ni ranging from 0.003 to 0.15 \Msun. The envelope masses
($M_i - M_{\alpha}$) and radii of the exploding models can be inferred
from \Tab{progenitor} and, for common events, are typically 7 -- 10
\Msun \ and 3 -- 8 $\times 10^{13}$ cm. The explosion energy varied
from 0.1 to 2 $\times 10^{51}$ erg (see \Fig{niandevscomp}), with a
typical value around $0.7 \times 10^{51}$ erg (\Tab{stats}).

Approximate analytic scaling rules have been derived for both the
luminosity on the plateau and its duration by \citet{Pop93} and
\citet{Kas09}. An important issue is how to measure these quantities,
both observationally and for the models. A variety of definitions is
used in the literature. The luminosity is not really constant on the
``plateau'', and the plateau's onset might be counted as beginning at
shock breakout; the cooling and recombination of hydrogen in the
outermost zone; or the time when the effective temperature first falls
below some value.  Even more uncertain, the end of the plateau might
be measured as when the envelope first combines; when the whole ejecta
first becomes optically transparent; or the beginning of the
radioactive tail, if there is one.  Here, it is assumed that the
bolometric luminosity is evaluated 50 days after shock breakout and
that shock breakout defines the beginning of the plateau. This is
clearly a lower bound to the actual commencement. For most models,
recombination begins about 3 days later, which is short compared with
the approximate 100 day duration of the plateau. All SN IIp light
curve plateaus were substantially longer than 50 days.

Two measures are used to bracket the end of the plateau. One is when
the photospheric radius recedes to less than one-half of its maximum
value. Empirically,  this corresponds roughly to when the
recombination front reaches the base of the hydrogen envelope (or the
density increase associated with that former boundary, if the
composition has been mixed). This is an operational lower bound to
the duration of the plateau. Later, after the internal energy of the
helium and heavy element core has diffused out, the whole star
recombines and the photosphere shrinks inside 10$^{14}$ cm. It is at
this point that any radioactive contribution reaches steady state with
the supernova luminosity and the ``tail'' of the light curve
begins. For the observer, these two limits thus correspond to the
first strong downward inflection of the light curve, and the onset of
the characteristic exponential decay on the tail.

All of our models are artificially mixed after explosive 
nucleosynthesis has ended, in order to account empirically for both 
the turbulent convection that goes on behind the shock during the
explosion and the Rayleigh-Taylor mixing following the reverse
shock. The mixing is calibrated, crudely, to SN 1987A
\citep{Woo88a,Kas09}. Because most of our models experienced little 
fall back, mixing is not very important to the nucleosynthesis 
calculated here.

\subsubsection{Analytic Scalings}
\lSect{popov}

The scalings of \citet{Pop93}, to which our luminosities and plateau
durations are compared, are:
\begin{equation}
\begin{aligned}
L &\propto E^{5/6} M^{-1/2} R^{2/3} \kappa^{-1/3} T_i^{4/3}, \\
t &\propto E^{-1/6} M^{1/2} R^{1/6} \kappa^{1/6} T_i^{-2/3}, \\
\end{aligned}
\lEq{popov93}
\end{equation}
where $E$ is the explosion energy; $M$, the envelope mass; $R$, the progenitor
radius; $\kappa$, the opacity;  and $T_i$, the ionization temperature,
which is the effective emission temperature divided by 2$^{1/4}$.
Adopting, from our models on the plateau, a typical effective temperature of
6300 K (Popov used 6000 K), and retaining his opacity of 0.34 cm$^2$
g$^{-1}$, one has the analytic results:
\begin{equation}
\begin{aligned}
L_{50} &= C_L \ E_{51}^{5/6} M_{10}^{-1/2} R_{0,500}^{2/3}
\quad \rm ergs\ s^{-1}, \\
t_{p,0} &= C_t \ E_{51}^{-1/6} M_{10}^{1/2} R_{0,500}^{1/6}\quad \rm days,
\end{aligned}
\lEq{popov}
\end{equation}
where $C_L$ and $C_t$ are $1.82\times10^{42}$ erg s$^{-1}$ and 96 days,
respectively. $L_{50}$ is the bolometric luminosity 50 days after
shock breakout; $t_{\rm p,0}$, the plateau duration without any
radioactive contribution; $ E_{51}$, the final kinetic energy of the
explosion in 10$^{51}$ erg units; $R_{0,500}$, the radius of the
presupernova star in units of 500 \Rsun; and $M_{10}$, the mass of
the presupernova hydrogen envelope in units of 10 \Msun.

For the bolometric luminosity, calibrated to our entire set of 181 type 
IIp models, 50 days after core collapse we find:
\begin{equation}
L_{50} \ = \ 1.85\times10^{42} \ E_{51}^{5/6} M_{10}^{-1/2} R_{0,500}^{2/3} \
\rm ergs\ s^{-1}.
\lEq{lumscale}
\end{equation}
This agrees extremely well with Popov's predicted scaling and is our
recommended value for this survey. The agreement with all the  supernova
models is excellent, as shown in \Fig{L50}.


\begin{figure}[h]
\centering
\includegraphics[width=.48\textwidth]{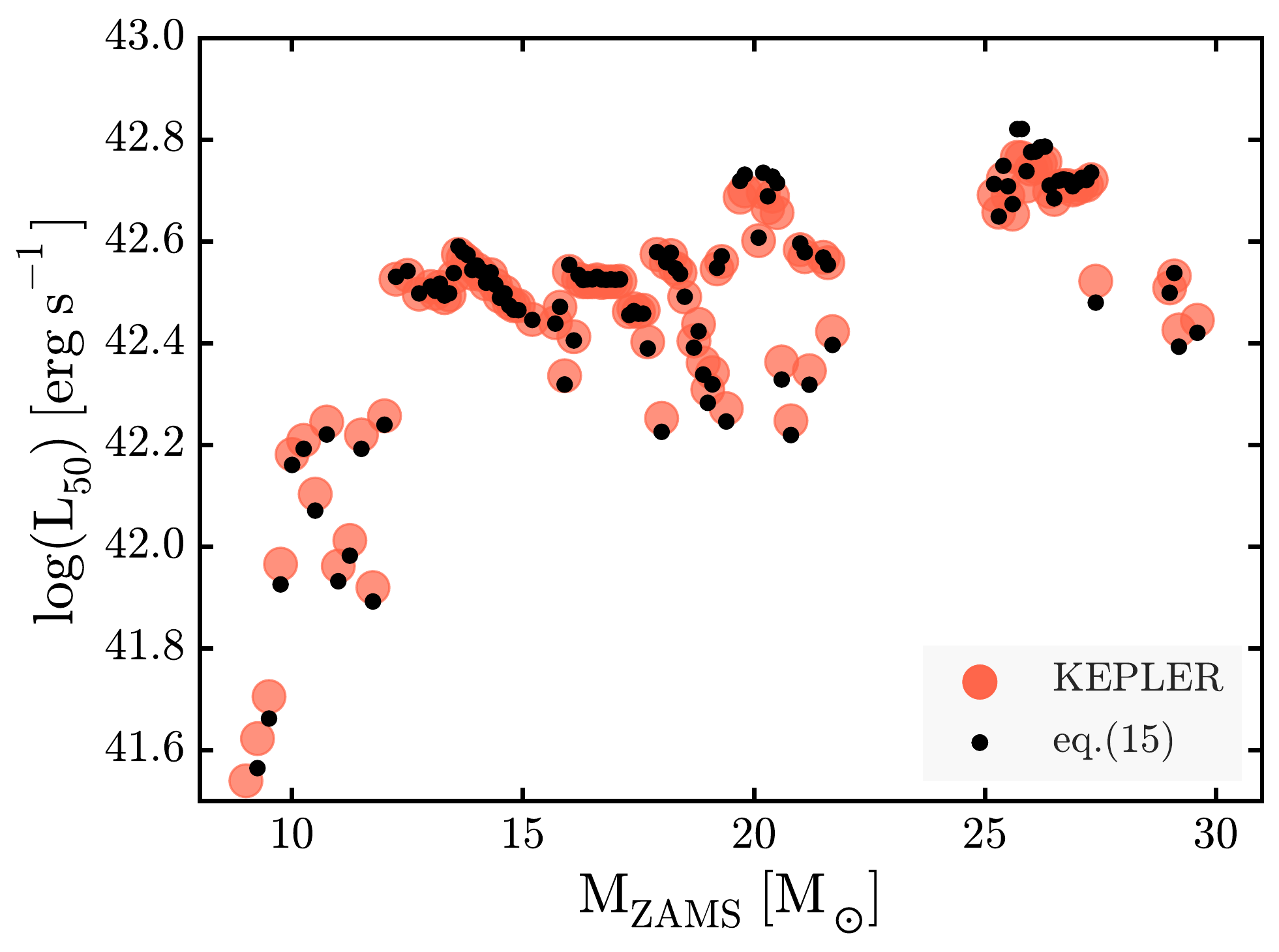}
\caption{Plateau luminosities 50 days post-breakout for the
  successfully exploded models of the Z9.6 and N20 series (red
  circles) are compared with \Eq{lumscale} (black circles). The
  agreement is striking. Results for the W18 series are not shown, but
  are very similar to those of N20.  \lFig{L50}}
\end{figure}

\citet{Kas09} gave similar scalings to \Eq{popov}, and calibrated
them to a small set of numerical models that used similar physics to
the present work:
\begin{equation}
\begin{aligned}
L_{50} &= 1.26 \times10^{42} \ E_{51}^{5/6} M_{10}^{-1/2}
R_{0,500}^{2/3} X_{\rm He,0.33}^{1} \quad \rm ergs\ s^{-1}, \\ 
t_{p,0} &= 122 \ E_{51}^{-1/4} M_{10}^{1/2} R_{0,500}^{1/6} X_{\rm
  He,0.33}^{-1/2}\quad \rm days, 
  \lEq{KW09}
\end{aligned}
\end{equation}
where $X_{\rm He}$ is the helium mass fraction in the envelope. Two
typographical errors are fixed here in Eq.(11) of \citet{Kas09}: (1)
$X_{\rm He}$ is normalized to 0.33, and (2) $t_{\rm p,0}\propto X_{\rm
  He,0.33}^{-1/2}$. In comparison to \Eq{popov93} and \Eq{popov}, this
formula assumes scaling with $X_{\rm He}$, which ranged from 0.30 to
0.53 in the models. Given that the helium mass fraction is typically
greater than 0.33, Kasen and Woosley's formula for the luminosity is
similar to Popov's, though slightly fainter. Their formula for the
plateau duration has been empirically adjusted to models (hence
E$^{-1/4}$ instead of E$^{-1/6}$) and is a bit longer, possibly due
  to differing definitions of what constitutes the plateau.

\subsubsection{Plateau Duration and Recombination Time Scale}
\lSect{tplateau}

As alluded to earlier, the duration of the plateau is more ambiguously
defined than the luminosity on a particular day. It is also more
sensitive to corrections for mixing and radioactivity. The original
scalings of \citet{Pop93} accounted for neither.  To compare with
\Eq{popov} and determine our own estimate of $t_{\rm p,0}$, the
plateau duration without radioactivity, the light curves of 26 models
from the Z9.6 and N20 series with masses from 9 \Msun \ to 27
\Msun\ were recalculated by assuming no $^{56}$Ni was produced in the
explosion. Two possible durations of the plateau were estimated for each
corresponding to the time between shock breakout and when the photospheric
radius shrinks to a) 50\% of its maximum and b) 10$^{14}$ cm. The top
panel of \Fig{tp0_and_frad} compares this range from the models with Popov's
expression - \Eq{popov}.


\begin{figure}[h]
	\centering
	\includegraphics[width=\columnwidth]{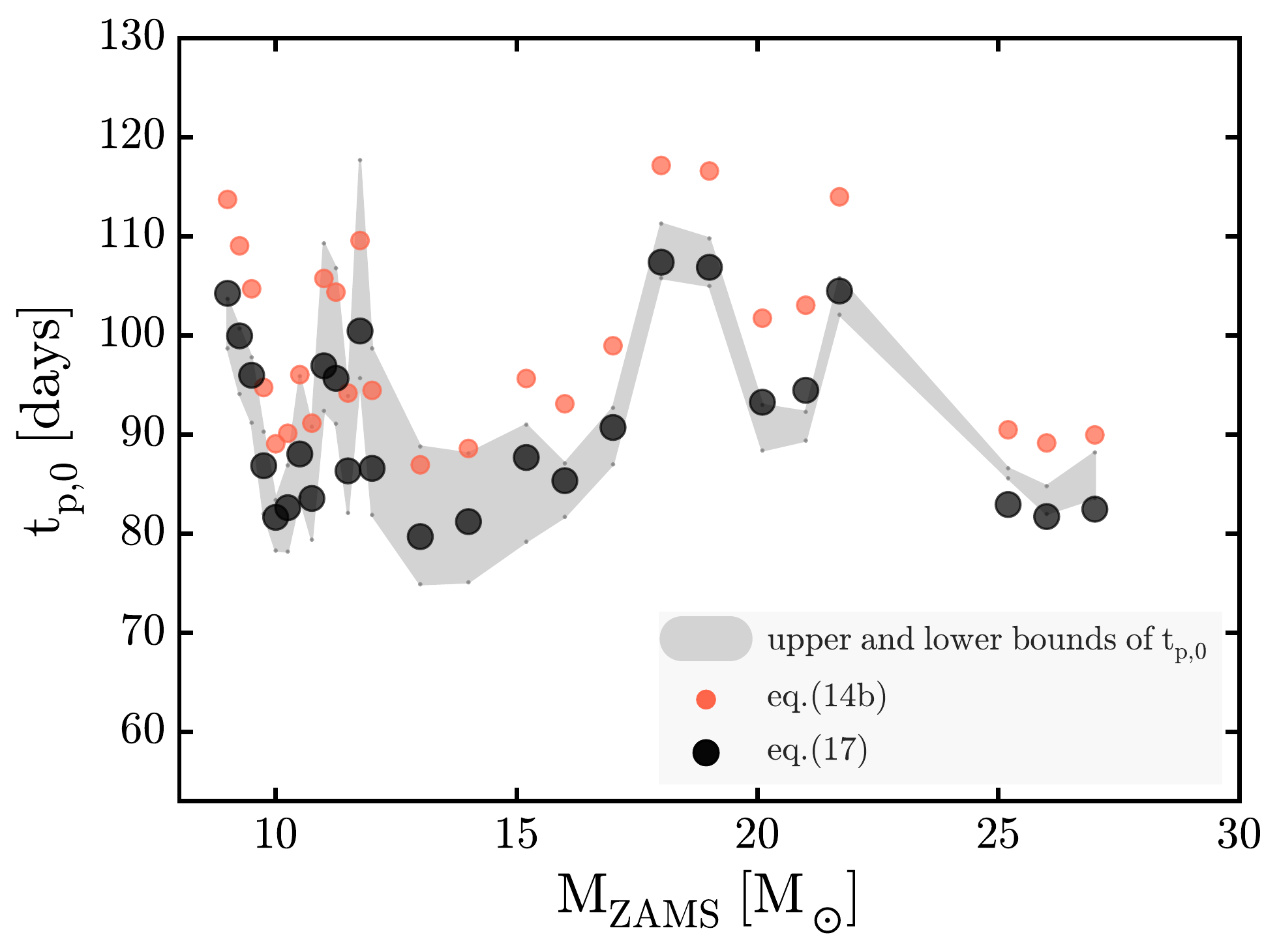}
	\includegraphics[width=\columnwidth]{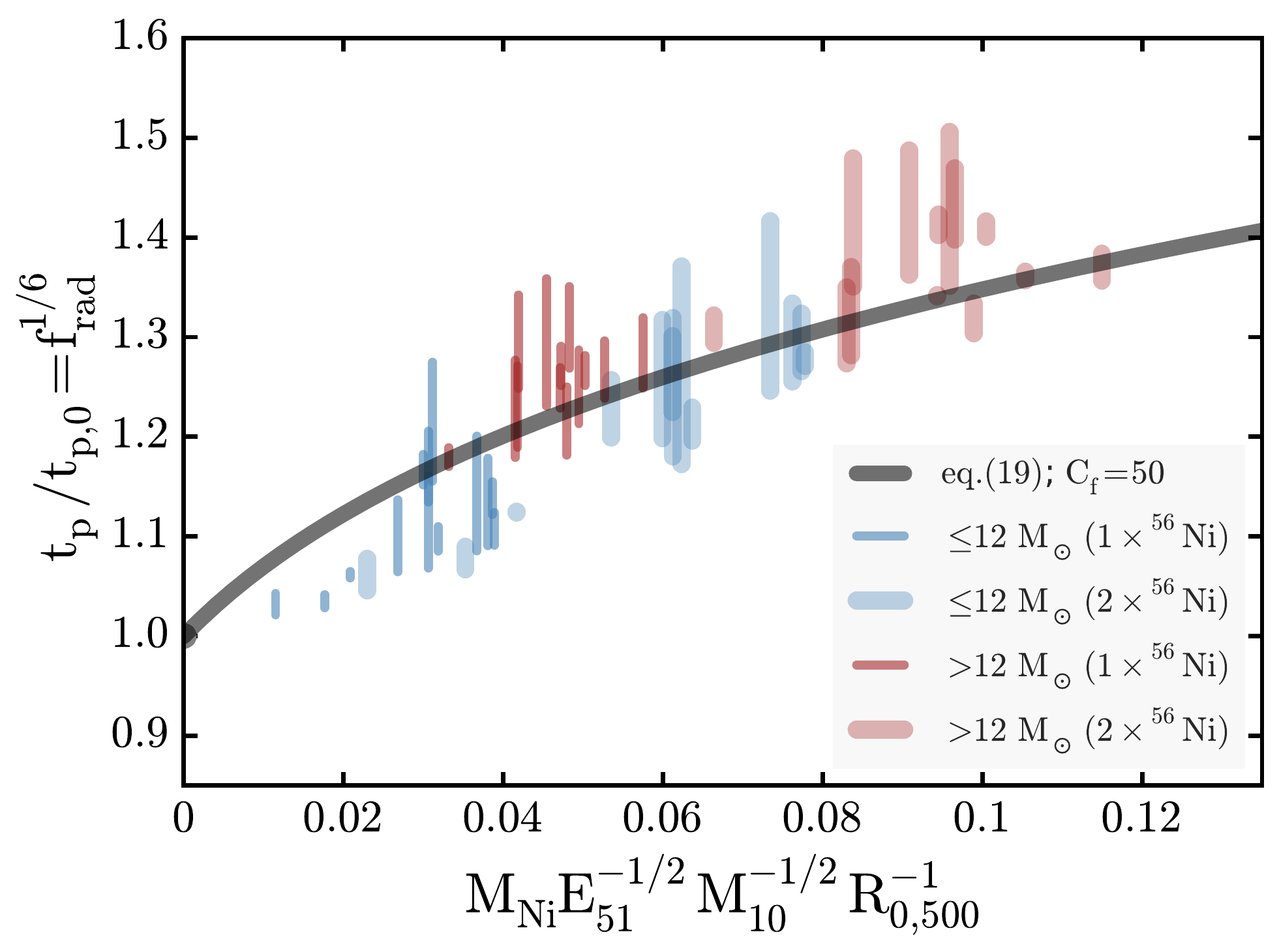}
	\caption{Top: the plateau durations from a set of 26 models
          calculated assuming no radioactive decay are shown in
          comparison with Popov's unmodified scaling law (\Eq{popov})
          and a renormalized version. The upper and lower bounds on
          $t_{\rm p,0}$ from the models (see text) give the gray
          shaded region. The \Eq{popov} with $C_t = 96$ days
          systematically overestimates the measured durations by about
          to 10\%. The same scaling normalized to $C_t$=88 days
          produces better agreement.  Bottom: the effect of
          radioactivity on the plateau duration is shown from the same
          set of models that included no radioactivity ($t_{p,0}$),
          with the default calculated yields of $^{56}$Ni ($t_p$), and
          twice that value ($t_p$). Models from progenitors with
          initial mass larger than about 12 \Msun\ follow the
          correction term derived in \Eq{frad} reasonably well, but
          smaller models are overestimated. The fit is improved when
          the constant in the brackets of \Eq{frad} $C_f$ is 50,
          i.e. the black curve is: $(1+50M_{\rm
            Ni}E_{51}^{-1/2}M_{10}^{-1/2}R_{0,500}^{-1})^{1/6}$.}
        \lFig{tp0_and_frad}
\end{figure}

The range between upper and lower bounds (shaded gray) is mostly due
to the energy that continues to diffuse out of the helium core before
the whole star recombines. Compared to this range, \Eq{popov}
overestimates the duration by about 10\%. Reducing the normalization
constant, $C_t$ in \Eq{popov} to 88 days gives a better fit to our
models. Thus for light curves with no radioactive contribution, we
recommend:
\begin{equation}
t_{\rm p,0} = 88E_{51}^{-1/6}M_{10}^{1/2}R_{0,500}^{1/6} \ {\rm days}. 
\lEq{timescale}
\end{equation}

As pointed out by \citet{Kas09}, the presence of radioactivity does
not substantially influence the luminosity during most of the plateau
because the diffusion time out of the core is long compared with the
recombination time. There are exceptions if the star makes more than
about 0.1 \Msun \ of $^{56}$Ni, or if the progenitor is not a red
supergiant (e.g., SN 1987A). The duration of the plateau is another
matter, however. Even 0.01 \Msun\ of $^{56}$Ni can appreciably
lengthen it. Radioactivity affects $t_p$ both by contributing to the
energy budget and by keeping the gas ionized longer so that its
opacity and diffusion time remain high.

\Fig{lcs_ni} shows light curves due to various amounts of nickel
masses for the lightest supernova studied, 9 \Msun , and a more
'typical' model of 15.2 \Msun.  For the lower mass model, $^{56}$Ni
masses over 0.01 \Msun \ are not physical, and the results shown are
purely for comparison. For a realistic amount of $^{56}$Ni, the
plateau of the 9.0 \Msun \ is not appreciably lengthened, but the 15.2
\Msun \ model is. Note also a bump at the end of the plateau for
both zero $^{56}$Ni models, but especially for the larger mass
star. This is the energy diffusing out of the helium core after the
hydrogen envelope has recombined. This effect is weaker in the smaller
model since the helium core is much smaller.


\begin{figure}[h]
\centering
\includegraphics[width=.48\textwidth]{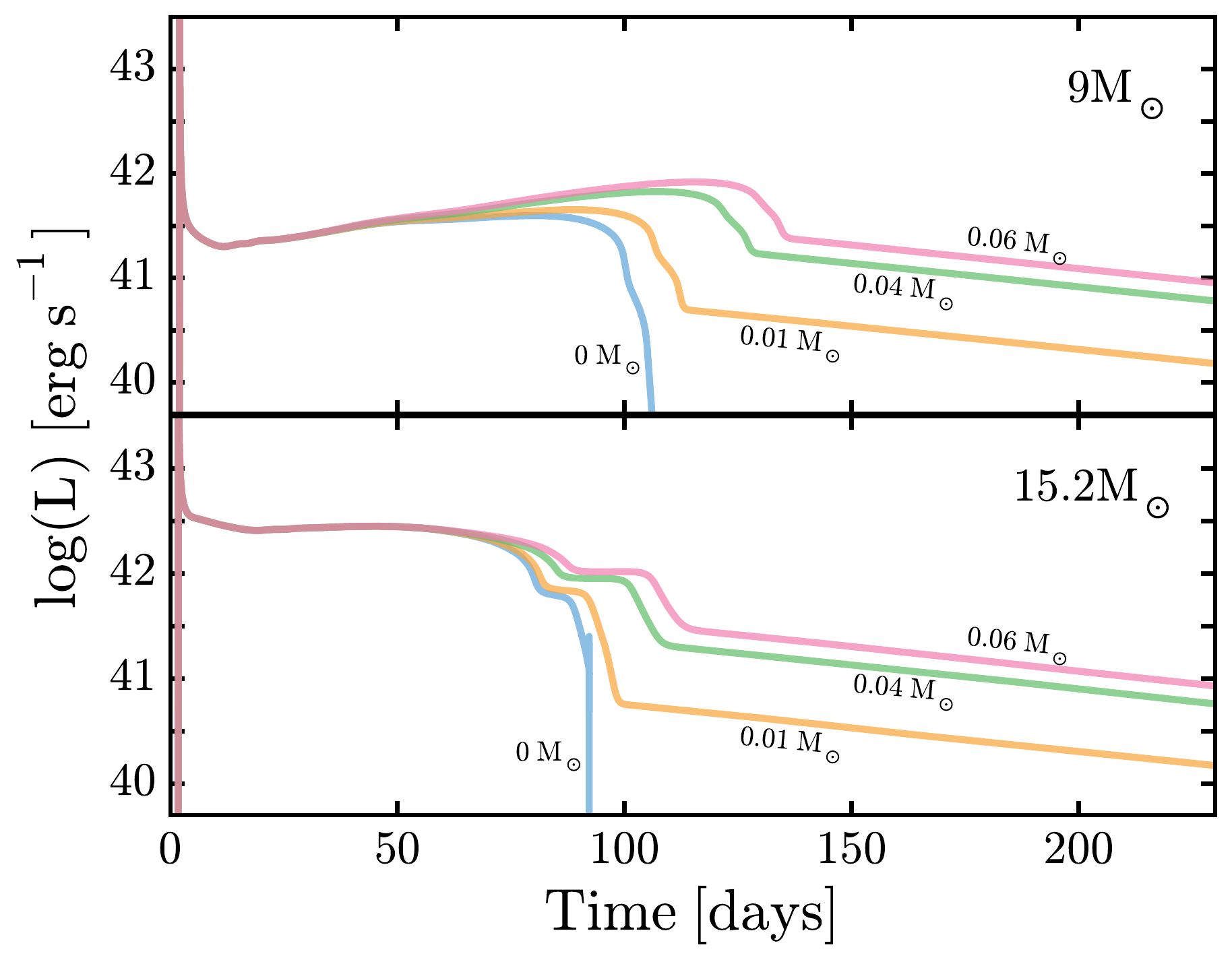}
\caption{The light curves due to various amounts of nickel masses are
  shown for the 9.0 \Msun\ model of the Z9.6 series (top) and the 15.2
  \Msun\ model from the N20 series. The KEPLER model gave 0.004 \Msun
  nickel for the 9.0 \Msun\ model and 0.070 \Msun\ for the 15.2
  \Msun\ model. Energy generation from the decay of these masses were
  multiplied by a constant to simulate the production of various
  amounts of $^{56}$Ni and the resulting curves are shown. $^{56}$Ni
  masses of more than 0.01 \Msun\ for the 9.0 \Msun \ model are
  unlikely, but shown for comparison.  \lFig{lcs_ni}}
\end{figure}

Following \citet{Kas09}, we adopt a correction factor to the internal
energy due to the decay of $^{56}$Co:
\begin{equation}
f_{\rm rad} \ = \ 1 + 24 \frac{M_{\rm Ni}}{E_{51}} \frac{t_{\rm Co}}{t_e}
\end{equation}
where $M_{\rm Ni}$ is the mass of $^{56}$Ni produced in the explosion
in units of \Msun\ ; $t_{\rm Co}$ is the mean life of $^{56}$Co, 113
days; $t_e$ is a characteristic expansion time given by the initial
radius of the star divided by a typical speed, $v = (2E/M_{\rm
  env})^{1/2}$. Two things worth noting in the derivation are that the
decay of $^{56}$Ni itself has no effect on $t_{\rm p}$ (this just
increases the expansion kinetic energy by a small amount and is
neglected), only $^{56}$Co does.  Second, the derivation implicitly
assumes complete mixing, that is the $^{56}$Co is distributed
homogeneously in the star.

In the derivation of the scaling shown in \Eq{timescale} two energies
appear that are taken, within a constant multiplier, to be the same,
the internal thermal energy, $E_T$, and the kinetic energy, $E_K$. The
scaling actually depends on $(E_T/E_K^2)^{1/6}$ or $E^{-1/6}$ if,
e.g., $E_T = E_K = E$. Since $f_{\rm rad}$ is assumed to only multiply
the internal energy, the correction factor enters in as the positive
one-sixth power. That is
\begin{equation}
\begin{aligned}
t_p &= t_{p,0} \times  f_{\rm rad}^{1/6}\\
    &= t_{p,0} \times (1 \ + \ C_f  \ M_{\rm Ni} E_{51}^{-1/2} M_{10}^{-1/2}
R_{0,500}^{-1})^{1/6}
\end{aligned}
\lEq{frad}
\end{equation}
where $C_f\sim 21$. This is similar to eq. (13) of \citet{Kas09},
but corrects two typographical errors: (1) the constant $C_f\approx$21 and (2)
$M$ is raised to the -1/2 power.

The bottom panel of \Fig{tp0_and_frad} shows the effect of this
correction factor on the plateau duration of models with
radioactivity. The same set of 26 models as used in the zero
radioactivity case were compared with those in which the calculated
amount of radioactivity was turned on. Additional models multiplied
the energy generation from radioactivity by two. For progenitor masses
above about 12 \Msun, the model results agree well with \Eq{frad},
though there are substantial differences at lower masses.  Better
agreement with our models is obtained if $C_f\sim50$ and that
is our suggested value.


\begin{figure}[h]
\centering
\includegraphics[width=.48\textwidth]{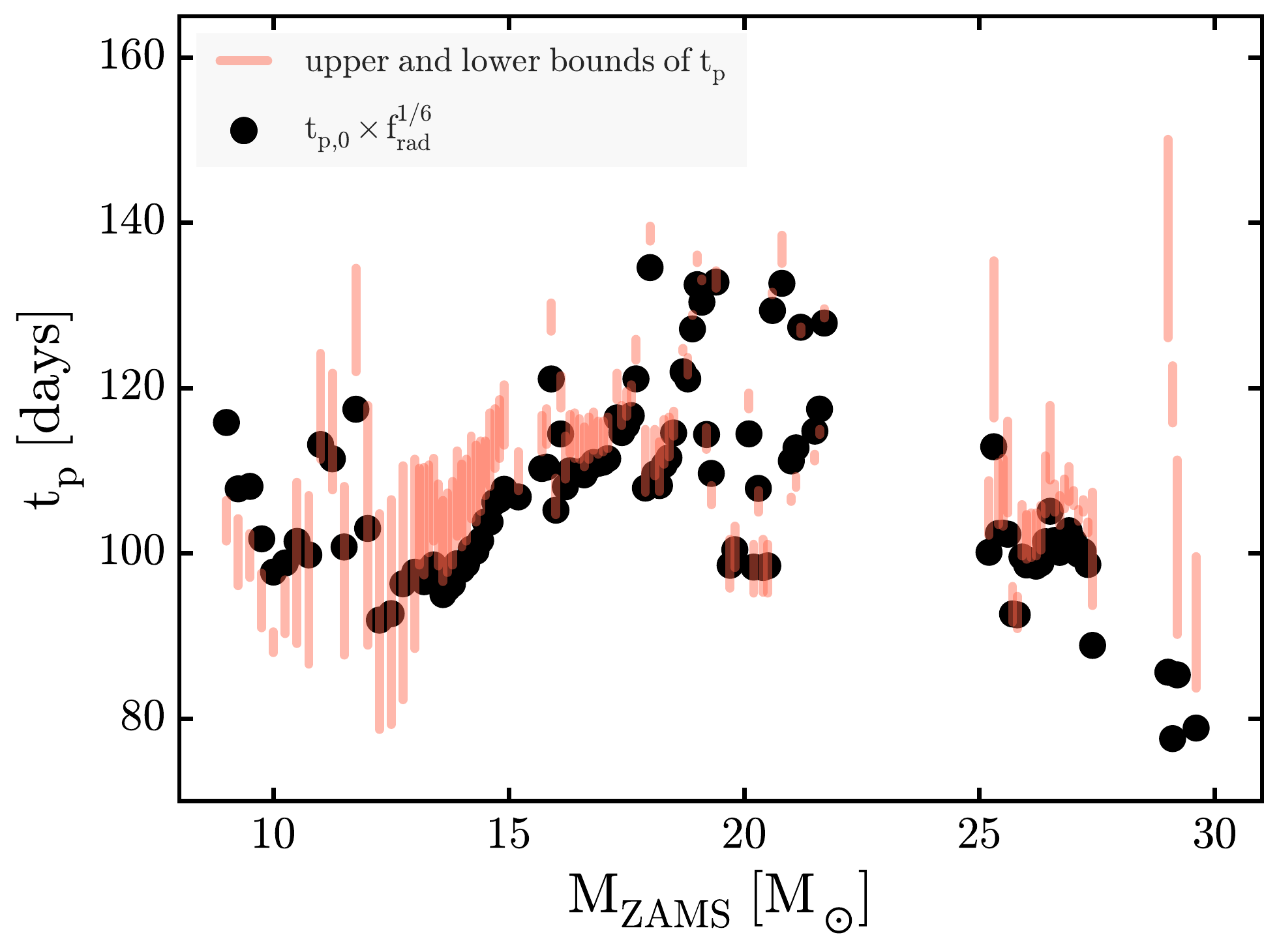}
\caption{The scaling laws that showed good agreement in
  \Fig{tp0_and_frad} are now compared against the standard explosion
  models with their $^{56}$Ni included. The models used are from the
  Z9.6 and N20 series. A red vertical bar represents a range of
  possible plateau durations calculated from the models assuming that
  1) the photospheric radius had receded to 50\% of its maximum value
  (lower bound); or 2) the whole star had recombined (upper
  bound). Black points result from multiplying \Eq{popov} with
  $C_t=88$ by the correction term, \Eq{frad}, with $C_f=50$.
  \lFig{tp}}
\end{figure}

The breakdown below 12 \Msun\ is not particularly surprising.  These
low mass supernovae have low energy ($\sim10^{50}$ erg), but also low
$^{56}$Ni mass ($< 0.01 \Msun$).  The combination $M_{\rm Ni}E^{-1/2}$
is still appreciable even though the mass of radioactivity itself is
very small. If the mass fraction of $^{56}$Ni in a given zone, after
mixing, is less than about 0.001, the decay energy, $6.4 \times
10^{16}$ erg g$^{-1}$ of $^{56}$Co, will, after time $t_{\rm Co}$, be
only a small fraction of the internal energy at recombination,
$\sim10^{15}E_{51}$ erg g$^{-1}$, and recombination will not be
greatly affected. For the very low mass explosions and mixing
prescription assumed, the cobalt abundance is below this limiting value
in all zones except at the very base of the hydrogen envelope.  In any
case the correction factor is small, less than 10\%, in these low
energy explosions.

Combining our best fit to the no-radioactivity models with this best
fit correction factor results in good agreement with the full set of
models with radioactivity included (\Fig{tp}). As expected, most of the
large deviations are for the lightest and heaviest stars. For small
stars, the scaling relation overestimates the plateau duration
because the correction term, $f_{\rm rad}$, is too big. At high
mass, there are cases with a large amount of fallback
(\Fig{mass_map}), and therefore no $^{56}$Ni production. Indeed, such
short plateaus in massive stars with normal luminosities (\Fig{L50})
may be an observable signature of black hole formation.

\subsubsection{Systematics}
\lSect{lcsystematic}

Given the generally good agreement of our SN IIp model characteristics
with analytic expressions, it is worth exploring whether these
systematics can be exploited to obtain insights into the masses and
energies of observed events. Another interesting question is whether
SN IIp, or some subset, can be used as standard candles.

Neglecting the radioactive correction, \Eq{frad}, the luminosity and
plateau duration are proportional to powers of the explosion energy,
presupernova radius, and envelope mass. These are not independent
variables. The radius of the presupernova star increases monotonically
with its main sequence mass, M$_{\rm ZAMS}$ and, to good approximation
in the mass range where most SN IIp occur (9 -- 20 \Msun), R$_0
\propto$ M$_{\rm ZAMS}^{3/4}$ (\Fig{menv_r}). Above 20 \Msun, the
hydrogen envelope mass of the presupernova star declines with M$_{\rm
  ZAMS}$ due to mass loss, but below 20 \Msun, it too has a monotonic
scaling, M$_{\rm env} \propto$ M$_{\rm ZAMS}^{2/3}$. The explosion
energy also tends to increase with mass because of the greater
efficiency of neutrino absorption in heavier stars and the larger
neutrino luminosity from accretion, but owing to the non-monotonic
behaviour of compactness, this is not as well-defined a
correlation. The compactness parameter does generally increase with
mass though, especially at low mass (\Fig{cp_all}), and the explosion
energy there is correlated with compactness (\Fig{niandevscomp}).  Given
these correlations, one can derive some simple relations among the
luminosity, energy, duration, and main sequence mass.


\begin{figure}[h]
\centering 
\includegraphics[width=.48\textwidth]{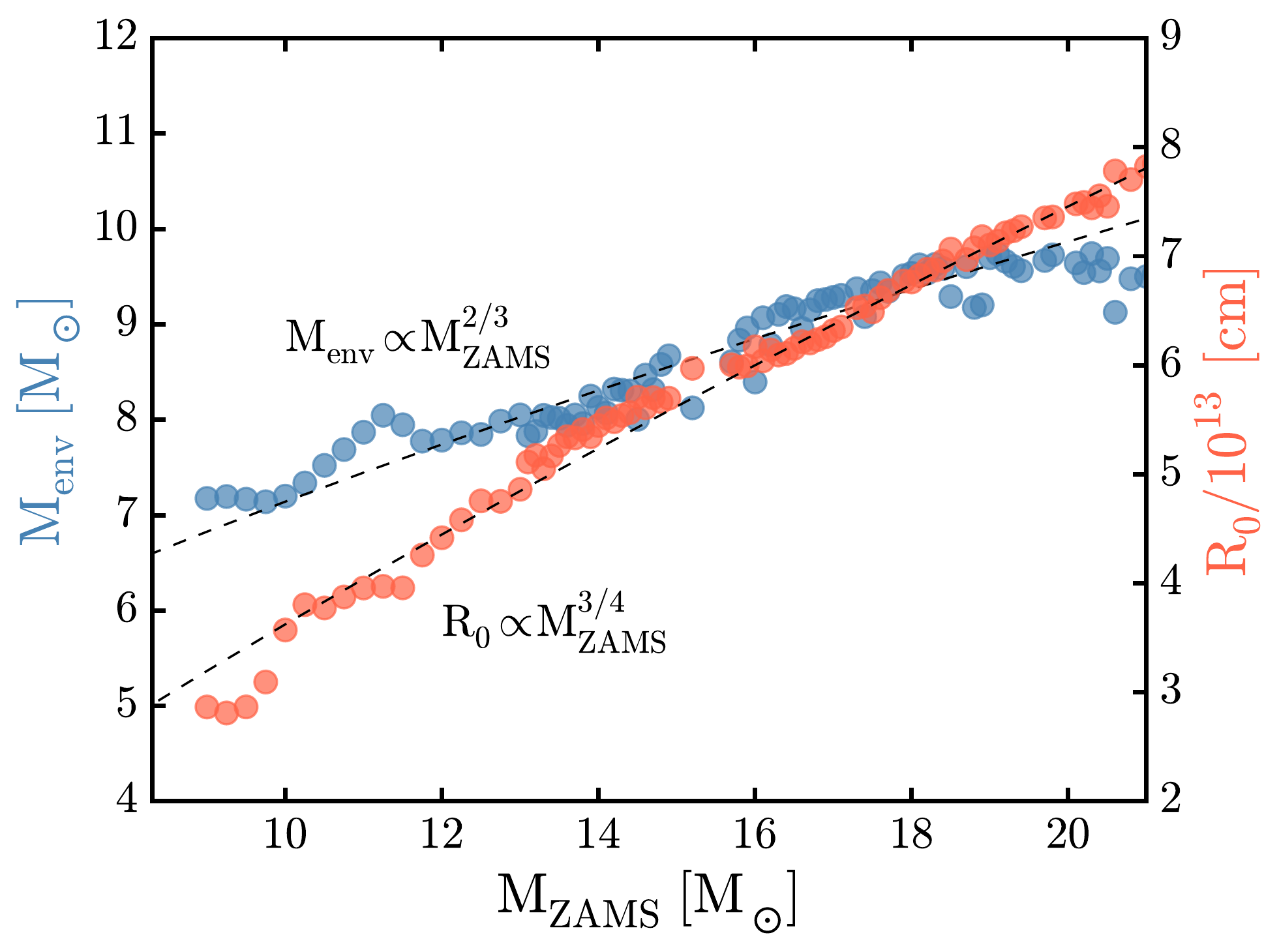}
\caption{The scaling of the envelope mass and radius with the initial
  mass for progenitors below 21 \Msun. The radius scales as $M_{\rm
    ZAMS}^{3/4}$, while the envelope mass scales approximately as
  $M_{\rm ZAMS}^{2/3}$. The envelope mass peaks near 20 \Msun\ and
  begins to decline at higher mass due to mass loss, so these scalings
  cannot be used outside the range shown.  Additional progenitor
  information is given in \Tab{progenitor}.  \lFig{menv_r}}
\end{figure}

First, consider the width-luminosity relation for SN IIp.
\Fig{WLR} shows $L_{50}$ vs $t_p$ with $t_p$ measured from breakout
until the onset of the radioactive tail. The results have been segregated 
by mass groups, and the explosions used the Z9.6 and N20 central engines. 
The scaling relations (\Eq{popov}) predict 
$L_{50} t^3_p \propto E_{51}^{1/3} M_{10} R_{0,500}^{7/6}$. From 12.25 
\Msun \ to 20 \Msun, roughly half of all supernovae, the energy and 
envelope mass change only a little and the radius increases by only 30\%, 
so crudely one expects $L_{50} \propto t_p^{-3}$, which is approximately what 
the figure shows. For stars with $t_p$ near 110 days, most supernovae 
have a standard bolometric luminosity between log L = 42.45 and 42.58.


\begin{figure}[h]
\centering
\includegraphics[width=.48\textwidth]{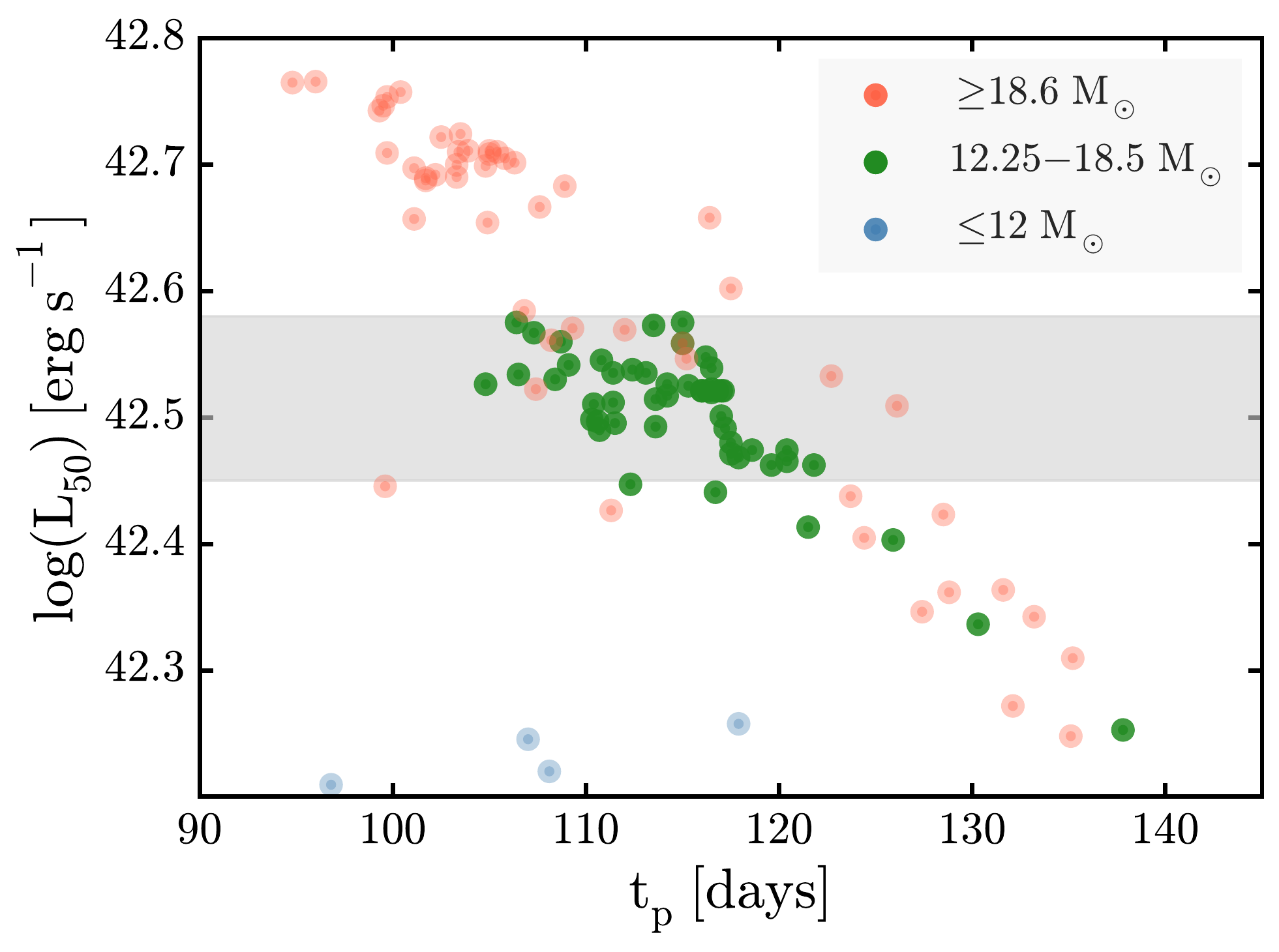}
\caption{The correlation between the plateau luminosities and
  durations is shown for the exploded models based on the Z9.6 and N20
  engines. The plateau duration here is an upper bound, the starting
  time of the radioactive tail when the full star has
  recombined. Green points, which represent 35\% of all supernovae,
  show a strong clustering. Above log L$_{50}$ = 42.3 erg s$^{-1}$, the
  luminosity is clearly anti-correlated with the plateau duration. The
  low mass models from series Z9.6 all have luminosities well below
  the gray box, and some of them have log L$_{50}$ below 42.2 erg
  s$^{-1}$ and fall on the plotted grid.  \lFig{WLR}}
\end{figure}

Unfortunately, the low mass supernovae, with their unusually low
energy explosions, contaminate an otherwise nice relation. These might
be selected against on the basis of their low photospheric speeds or
even by their colors, but that lies beyond the scope of the present
work. Our large model set will be available\textsuperscript{\ref{mpa_db}} 
and represents a resource from which information on dilution factors 
and spectral velocities, for example, could be extracted. These could 
be useful for the expanding photosphere method for distance 
determination \citep[e.g.][]{Eas96,Vin12}.

Next consider the kinetic energy of the explosion.  Can it be
determined just from observations of the light curve? \Fig{L50KE}
shows a well-defined relation between the bolometric luminosity on day
50 (after shock breakout) and the kinetic energy of the supernova.
This is expected on the basis of \Eq{lumscale} which shows the
luminosity scaling as $E_{51}^{5/6} M_{10}^{-1/2}
R_{0,500}^{2/3}$. From 9 to 20 \Msun, which includes about 95\% of
successful SN IIp explosions, the hydrogen envelope mass varies
between 7 and 10 \Msun, and is usually 8 or 9 \Msun. The radius varies
about a factor of 3, but enters as a 2/3 power. Moreover, the radius
scales as $R \propto M^{3/4}$, and bigger mass stars tend to have
bigger explosion energies. Thus the fact that the luminosity scales
roughly as the explosion energy is not surprising.

The figure shows though that just by measuring the bolometric
luminosity on the plateau one gets a good estimate of the total
explosion energy. Models above 22 \Msun \ lie off the main curve,
but there are very few such explosions.  These plots include only the
contributions of radioactivity and a point explosion, and not any
contribution from a magnetar or circumstellar interaction. An inferred
explosion energy above $2 \times 10^{51}$ erg would be suggestive of a
breakdown in this assumption.


\begin{figure}[h]
\centering 
\includegraphics[width=.48\textwidth]{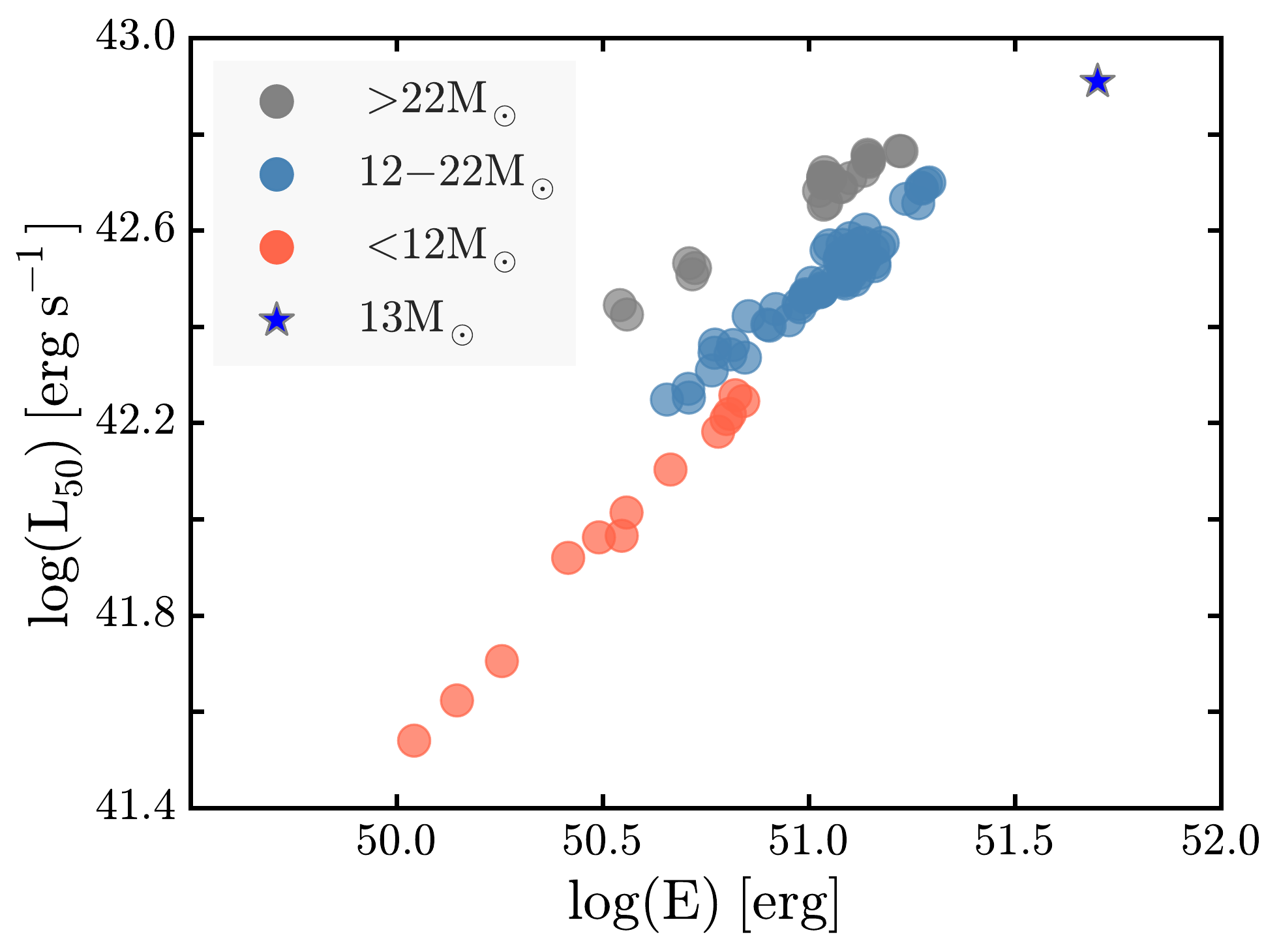}
\includegraphics[width=.48\textwidth]{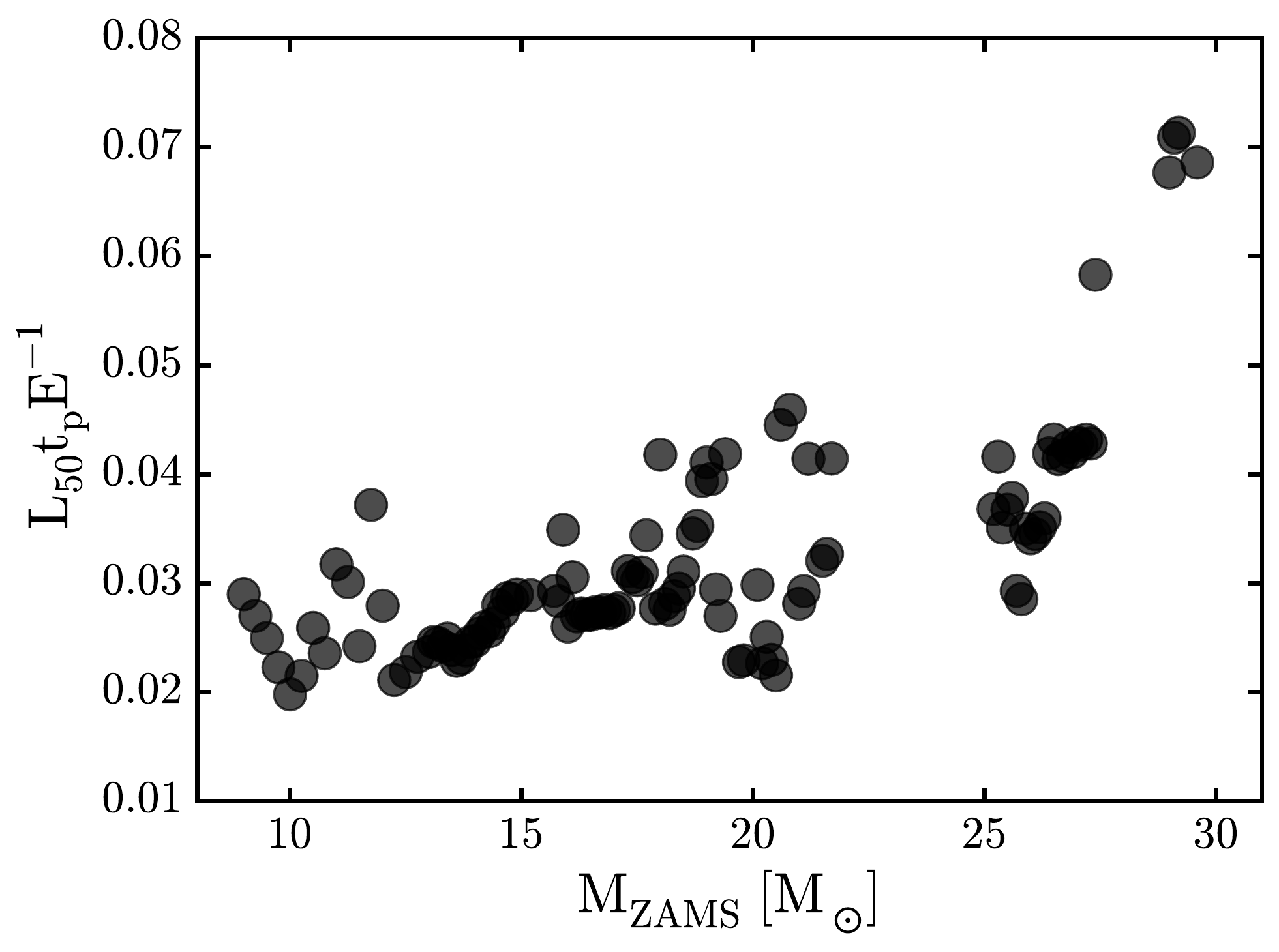}
\caption{Top: the correlation between plateau luminosities at 50 days
  and the explosion energies for Z9.6 and N20 engines. The 13
  \Msun\ model from the N20 engine has been re-calculated with the
  explosion energy scaled up to $5\times10^{51}$ ergs and is plotted
  as a single blue star. Explosions from stars smaller than 22 \Msun,
  which are most supernovae, show a strong positive correlation of
  luminosity with energy.  Bottom: the ratio of $\rm L_{50}t_p$ and
  explosion energy is shown as a function of initial progenitor
  mass. Integration of the light curve shows that $\rm L_{50}t_p$ is
  an accurate proxy for the total radiated energy. The typical SN IIp
  supernova radiates 2-4\% of its explosion energy as light.
  \lFig{L50KE}}
\end{figure}

Also shown in \Fig{L50KE} is the quantity $L_{50} t_p/E$ vs
progenitor initial mass. This is the ratio of the total power emitted 
by the supernova on its plateau to the kinetic energy of the explosion 
and is expected to scale as $E_{51}^{-1/3} R_{0,500}^{5/6}$.  
Since $E_{51}$ generally increases with mass and therefore with radius, 
the figure shows $L_{50} t_p /E$ is roughly constant.  The total radiated energy in light for most neutrino-powered supernovas is roughly 
1/35 of the kinetic energy of the explosion. The full range of this ratio
between 9 and 20 \Msun \ is 0.02 to .04. This gives another way of
estimating the energy of a neutrino-powered supernova from observables,
$L_{50}$ and $t_p$. 

Finally we consider whether the mass of the zero-age main sequence
(ZAMS) star can be determined from the supernova light curve.
\Fig{getmass} shows quantity $(L_{50} t_p)^{12/7}/E$, which should
scale as $R^{10/7}$, plotted against $M_{\rm ZAMS}$.  Since $R \propto
M_{ZAMS}^{3/4}$, this implies that $(L_{50} t_p)^{12/7}/E$ should
scale as $M_{\rm ZAMS}^{15/14}$ which is consistent with \Fig{getmass}. If
one observes $L_{50}$ and t$_p$ and determines the explosion energy
from \Fig{L50KE}, the ZAMS mass of a supernova in our data set is
roughly determined.


\begin{figure}[h]
\centering 
\includegraphics[width=.48\textwidth]{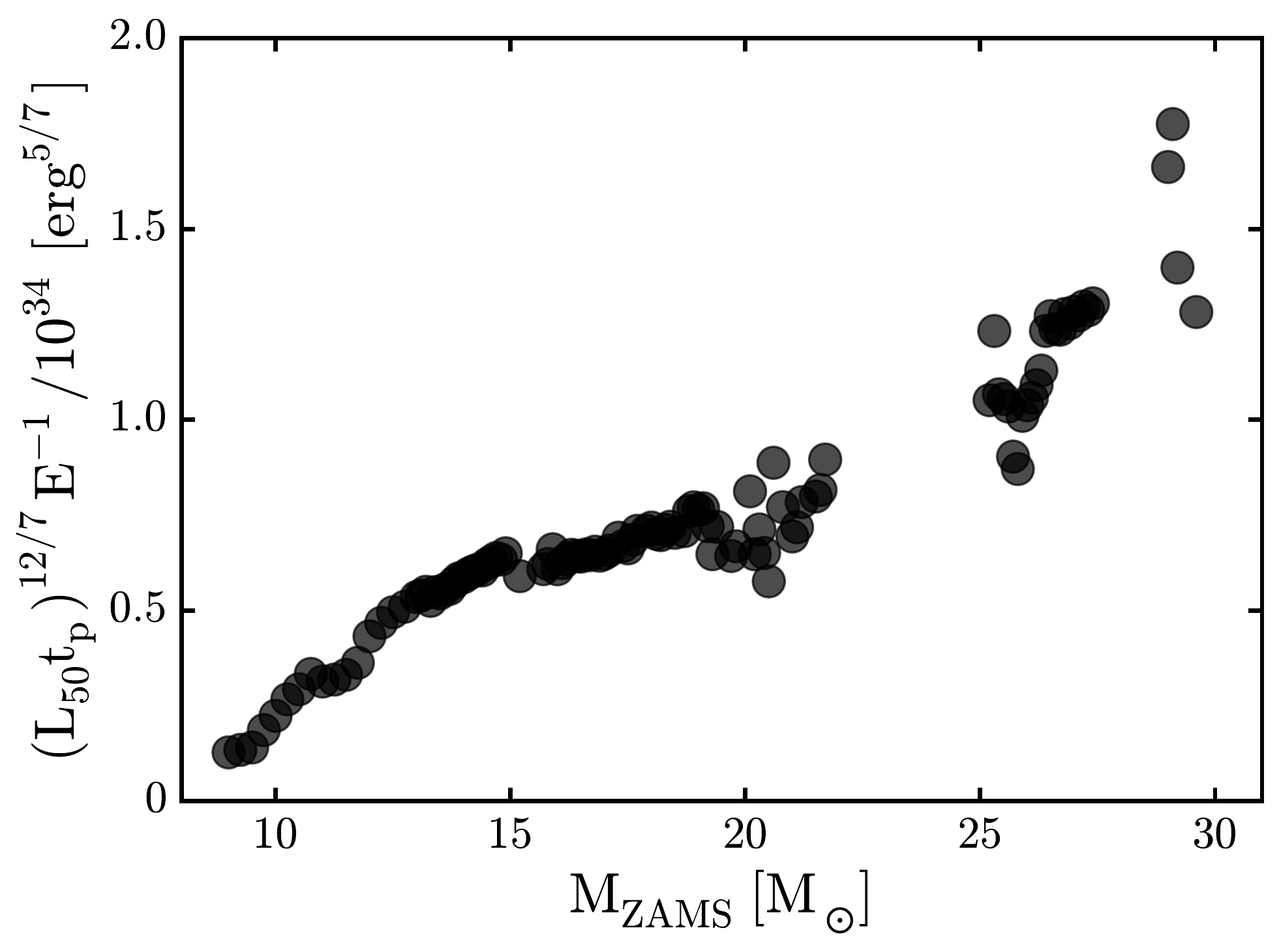}
\caption{The quantity $(L_{50}t_p)^{12/7}E^{-1}$ is plotted against
  the ZAMS mass of the progenitor shown for all exploded models of
  Z9.6 and N20 engines.  This correlation allows an approximate
  determination of the progenitor mass given measurements of plateau
  luminosity and duration along with an explosion energy estimated
  from \Fig{L50KE}.
  \lFig{getmass}}
\end{figure}

\subsection{Type Ib/c}
\lSect{type1} 

With the adopted mass loss prescription, stars with initial mass
greater than about 35 \Msun\ lose all of their hydrogen envelope
before dying. Successful explosions will then produce supernovae of
Type Ib and Ic. Models heavier than $\geq$45 \Msun\ have lost most 
of their helium envelope and all have surface $^{4}$He mass fractions 
less than 0.2 (\Tab{progenitor}), however, the exact distinction 
between Ib and Ic is as much dependent on mixing as on mass loss 
\citep{Des11}, and we shall not attempt to separate the two classes 
in this work. Use of the N20 engine resulted in the explosion of the 
60, 80, 100, 120 \Msun\ progenitors, while the W18 engine exploded 
only the 60 and 120\Msun\ progenitors. The resulting light curves 
are shown in the lower panel of \Fig{lcs}.

These are not common Ib's or Ic's. Their light curves are too broad
and faint. Presumably common Type Ib and Ic supernovae come from
mass-exchanging binary star systems and lower mass progenitors
\citep{Des11,Eld13}. The supernovae here are bigger and more slowly
expanding, but should exist in nature and might be sought as a
separate subclass of Type Ibc. Since the neutrino mechanism appears
inadequate to produce an explosion energy greater than about $2 \times
10^{51}$ erg \citep{Ugl12,Ert15}, these would be fainter and broader
than e.g., SN 2009ff \citep{Val11}. Conversely, events with much
greater energy than $2 \times 10^{51}$ erg would not be powered
exclusively by neutrinos.


\begin{figure}[h]
\centering
\includegraphics[width=.48\textwidth]{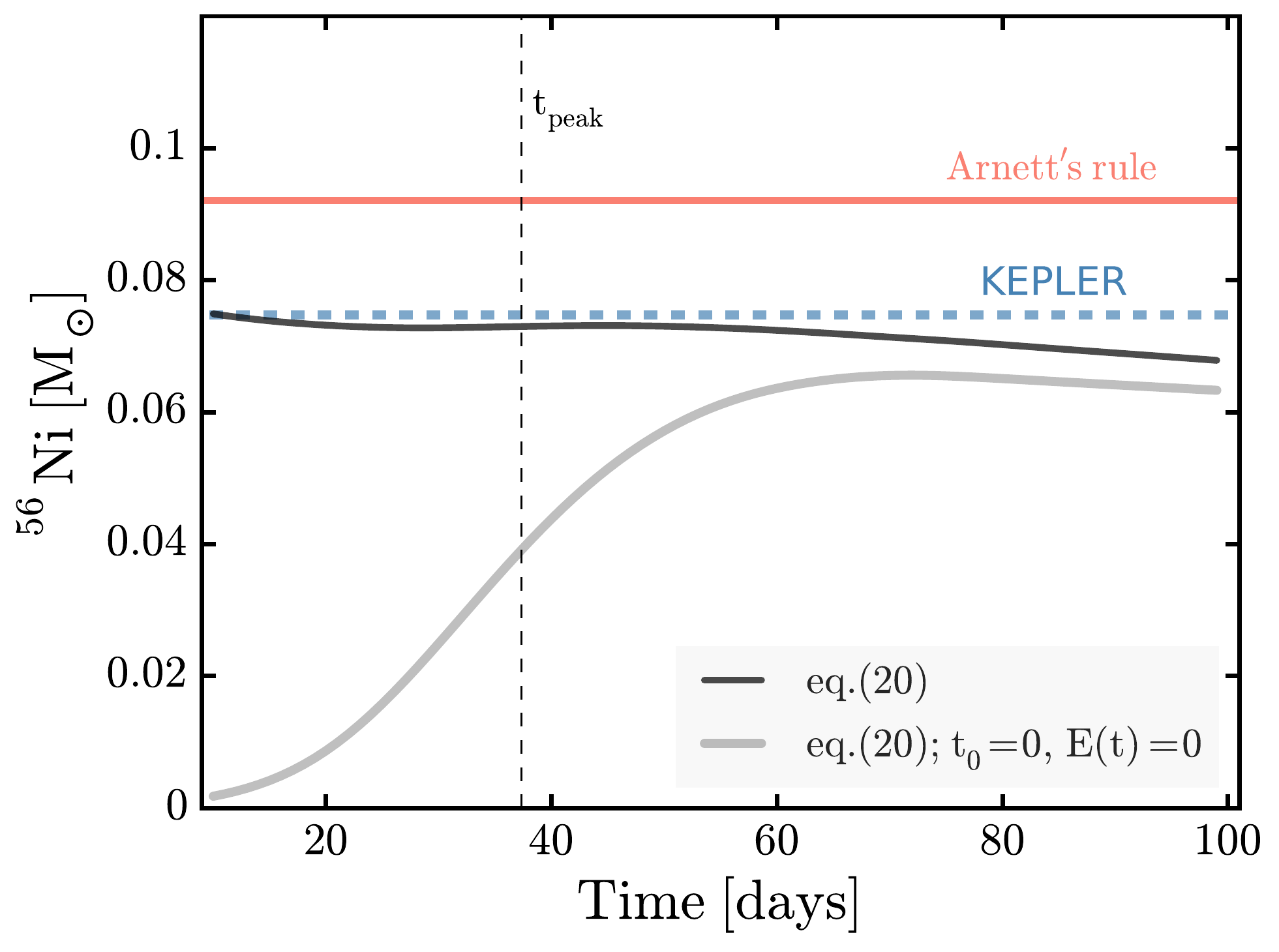}
\caption{Estimates of the $^{56}$Ni mass for the 80 \Msun \ progenitor
  exploded using the N20 engine. The KEPLER calculation synthesized
  0.074 \Msun\ of $^{56}$Ni. The energy conservation argument
  described in \Eq{katz} closely agrees with this value until late
  times when the leakage of gamma rays becomes significant. The
  special case of this argument presented in \citet{Katz13} for type
  Ia light curves, underestimates the nickel mass, but works
  better than the Arnett's rule.  \lFig{katz}}
\end{figure}

These Type Ibc light curves have peak luminosities larger than
predicted by 'Arnett's rule' \citep{Arn82}, which states that $L_{\rm
  peak} = L_{\rm decay}(\rm t_{\rm peak})$. Consequently,  the nickel mass
content cannot be accurately inferred in the same way as for Type Ia
supernovae. \citet{Des15} studied similar light curves and found that the
energy-conservation-based arguments of \citet{Katz13} gave better
agreement. Their argument follows from the first law of thermodynamics
and the assumption that the ejecta are freely coasting, radiation
dominated, and powered only by radioactivity. Then
\begin{equation}
tE(t)-t_0E(t_0) = \int_{t_0}^t t'L_{\rm dec.}(t')dt'-\int_{t_0}^t t'L_{\rm rad.}(t')dt'
\lEq{katz}
\end{equation}
where $L_{\rm rad.}$ is the radiated luminosity, and $E$ is the
internal energy of the ejecta. In the special case where $t_0 = 0$,
$t$ is sufficiently large such that $E(t) \approx 0$, and $L_{\rm
  rad.} = L_{\rm bolometric}$, the argument reduces to a connection
between an observable (bolometric luminosity) and a physical quantity
($^{56}$Ni).

\Fig{katz} illustrates this method applied to one of our models. The
approximation works until about 60 days post-explosion (black curve),
after which the leakage of gamma rays becomes significant. If one
assumes full trapping, \Eq{katz} will hold as long as the radiation
dominates the internal energy. Some deviation is expected because the
internal energy is not all in the form of radiation; an appreciable
fraction is in the electrons.

In the special case when $t_0=0$ and $E(t)=0$, \Eq{katz} expresses the
equality between the two integrals. At around 70 days the nickel mass
inferred from this relation peaks and reaches its closest point with
respect to that in the actual model (blue dashed line). After the peak, it
goes down again due to leakage, but it doesn't quite reach the 'true'
value because the internal energy is not small enough at that time. At
70 days the internal energy is roughly $2\times10^{47}$ ergs, only
about 3 times less than its value at 3 days. Though it underestimates
the nickel mass, this method is more accurate than 'Arnett's rule' for
all of our Ib and Ic models.

\subsection{Type IIL}
\lSect{other_lc}

None of the exploded models from the Z9.6, W18 and N20 engines showed
a clearly linear decline after the shock breakout. All of the
explosions from progenitors smaller than 30 \Msun\ showed a plateau of
roughly constant luminosity.  The shortest plateau, as measured by the 
lower bound, was 79 days for the 12.25 \Msun \ model with a 7.8 \Msun\ 
envelope. And the exploded model with the least amount of envelope 
mass, 29.6 \Msun\ model of N20 engine with 3.9 \Msun\ envelope, yielded 
84 days long plateau. Given that Type IIL should come from at least 
some stars in the 9 to  120 \Msun\ range, their absence warrants an 
explanation.

There are several possibilities. We may just have missed them
because of sparse mass sampling above 35 \Msun. This is unlikely. Type IIL
is expected to come from a star that has lost most of its hydrogen
envelope, but retains a large radius and, possibly, makes a lot of
$^{56}$Ni. Our model set indeed had 5 \Msun \ spacing from 35 to 60
\Msun, but none of those stars blew up.

Second, their absence may reflect a deficiency in presupernova
modelling. This is quite possible. Over the years, one of us has noted
a chronic problem calculating {\sl ab initio} models of very massive
stars with envelope masses below 1 \Msun \ using KEPLER. The envelopes
expand to such large radii (over 10$^{14}$ cm) that they begin to
recombine. Density and even pressure inversions develop near the
surface and the calculation cannot be carried further. Such was the
case with the 35 \Msun \ model studied here and in \citet{Woo07}. At
central helium depletion this star had an extended low density
envelope of about 1 \Msun\ with a radius of $8.3 \times 10^{13}$ cm
and a luminosity of $1.35 \times 10^{39}$ erg s$^{-1}$, only slightly
below the Eddington limit for its remaining mass, 15 \Msun. As the
star attempted to adjust its structure and ignite carbon, its
luminosity increased above $1.6 \times 10^{39}$ erg s$^{-1}$, and the
envelope become unstable. It is quite possible that at this point, the
envelope comes off. The central temperature was over $5 \times 10^8$ K
and it would not have been long before the star died. Fortunately
perhaps, there is no way neutrinos would blow up this model, so this
behaviour is not an issue for the present paper, but a similar fate
might befall a lower mass star in a binary system, or the KEPLER
calculation may just be wrong. More study is clearly needed.

Finally, it should be noted that the mass range where Type IIL might
have occurred, 30 - 40 \Msun, is also a mass range where the effects
of rotation are expected to be important to the death of the star
\citep{Heg05}.

Type IIL supernovae could thus be a consequence of collisions with
recently ejected envelopes, binary systems, jets, magnetars, stellar
evolution still to be done properly, or all five, and that makes them
interesting.

\section{Conclusions} 
\lSect{conclude}

The deaths of massive stars from 9.0 to 120 \Msun \ as supernovae have
been surveyed using an improved depiction of the explosion
physics. Models are one-dimensional, of a single metallicity - solar,
and do not include any effects of rotation. Nevertheless, they capture
many essential aspects of a general solution neglected in past
studies, including the explosion effects of a non-monotonic variation 
of the presupernova compactness (\Fig{cp_all}) with mass. The explosions 
are calibrated to reproduce the characteristics of two well-studied
supernovae: the Crab and SN 1987A. The Crab is believed to have been a
low mass supernova, near 10 \Msun, with a low explosion energy,
$\sim10^{50}$ erg that made very little iron. SN 1987A was a star near
18 \Msun \ that exploded with $1.3 \times 10^{51}$ erg and made 0.07
\Msun \ of $^{56}$Ni. These two calibration points anchor our
survey. Multiple models for the central engine for 87A were employed
and one for the Crab.

Calibrating to these two events, and using lessons learned from
previous 3D modelling efforts of core collapse, results in fiducial
``central engines'' that describe the evolution of only the inner 1.1
\Msun \ of the collapsing iron core (\Sect{coremodel}). These central
engines, or inner boundary conditions are characterized by just a few
physically descriptive parameters (\Tab{P-HOTBparams}). These inner
core evolution prescriptions are then used to follow the evolution of
200 presupernova stars (\Sect{solgrid}; \Tab{progenitor}) in the mass
range 9.0 -- 120 \Msun. Outside 1.1 \Msun, neutrino transport was
followed in an approximate, efficient way for each stellar model,
thus assuring a physical representation of the explosion that was
sensitive to the structure of individual presupernova stars. Above 12
\Msun, five central engines for various SN 1987A models were employed,
two of them extensively. From 9.0 to 12 \Msun, the properties of the
central engine were interpolated between Crab-like and 87A-like
behaviour. The Crab model itself was based on a 9.6 \Msun \ star whose
explosion has been previously studied in 3D.

As a result of these simulations, a large diversity of successful and
failed explosions was generated. Kinetic energies at infinity varied
from 0.11 to 2.03 $\times 10^{51}$ erg. The outcome of each
calculation depended upon the structure of the presupernova star and
central engine employed. All stars below 12 \Msun \ exploded, albeit
weakly.  Despite their small compactness parameters and ease of
explosion, their final energies were low because a smaller fraction of
the neutrino power radiated by the proto-neutron star was converted
into kinetic energy of expansion. The steep density gradient outside
the iron core reduced the absorption. Conversely, larger stars trapped
a larger proportion of the neutrino radiation and had a bigger
neutrino luminosity from accretion, but also had a larger binding
energy and ram pressure to overcome. Some exploded with variable
energy up to $2 \times 10^{51}$ erg; some not at all.

There was no single mass below which all stars exploded and above
which black holes formed, but rather there were islands of
``explodability'' in a sea of black hole formation (\Fig{mass_map}). A
similar result was found by \citet{Pej15b} suggesting that the outcome
of supernova explosions is more influenced by presupernova structure
than details of the central engine, provided that engine is
sufficiently powerful to explode many stars. Stars above 35
\Msun\ lost their envelopes and a few that experienced severe mass
loss as Wolf-Rayet stars exploded as Type Ibc supernovae, but most
became black holes. Typically 95\% of the stars that did explode were
SN IIp with masses less than 20 \Msun \ (\Tab{stats}). The median
supernova mass, neglecting any explosions below 9 \Msun, was 12
\Msun. Half were heavier and half lighter. The heavier half accounted
for most of the nucleosynthesis.

From the baryonic masses of the remnants, it was possible to create a
distribution of neutron star masses and an average. Both agreed very
well with observations (\Tab{stats} and \Fig{nstarmass}) with
typical neutron star gravitational masses near 1.40 \Msun. The
heaviest neutron star made was 1.77 \Msun \ and the lightest, 1.23
\Msun. While electron capture supernovae were not computed here, their
neutron stars should result from the collapse of Chandrasekhar mass
cores with $Y_e$ still close to 0.50. The neutron stars resulting from
our 9 and 10 \Msun \ models, which have experienced appreciable
electron capture during oxygen and silicon burning prior to collapse,
can actually be lighter.

For those stars that made black holes, which were most, but not all of
the stars above 20 \Msun \ and some below 20 \Msun, only a few
produced their black holes by fallback. Stars that in past surveys
produced black holes by fallback more frequently did not explode at
all in the present study.

The present calculations with P-HOTB also included a contribution to
the explosion energy from the neutrino-powered wind and from
recombination, of order 10$^{49}$ to 10$^{50}$ erg, that gave an extra
push to the inner zones during the first second or so after the shock
was launched. This helped to prevent their reimplosion.  Usually black
hole production involved the collapse of the full star, including its
hydrogen envelope. Other mechanisms are necessary and were invoked to
explain envelope ejection during black hole formation. If the full
presupernova star always collapsed the average black hole mass (IMF
weighted) was 13.3 -- 13.8 \Msun; if only the helium core collapsed,
it was 7.7 -- 9.2 \Msun \ (\Tab{stats}). The latter is in better
agreement with observations and suggests that black hole formation is
accompanied by weak explosions that eject only the hydrogen
envelope. These weak supernovae would have distinctive light curves
and colors. The gap between neutron stars and black hole masses 
from $\sim$2 to $\sim$4 \Msun\ suggested by observations was naturally obtained because of the lack of fallback supernovae that could fill 
this mass interval.

One important result of this study is the nucleosynthesis expected
from massive stars calculated using the more realistic depiction of
explosion. This was calculated using the KEPLER code and an amply
large nuclear reaction network complete up to the element bismuth. A
trajectory near the final mass cut in P-HOTB was taken as the inner
boundary for the KEPLER calculation, and the infall and time of
reversal were nearly the same. The velocity of the piston in KEPLER
was varied however, to give near perfect agreement in explosion energy
with the P-HOTB calculations and to reproduce the iron synthesis to
better than 10\% in most cases. In a few cases the latter required a 
slight adjustment of the piston mass.  Neutrino interactions, except 
for the neutrino process of nucleosynthesis, were neglected in the 
KEPLER studies and the neutrino-powered wind was not carried. In 
practice this meant the neglect of a possible $r$-process or 
$\nu p$-process component and some uncertainty in the nucleosynthesis 
of nickel, copper and zinc.

The contributions from various mass stars were then determined as well
as their overall IMF-weighted production factors. The lighter
supernovae below 12 \Msun, half of all supernovae numerically, make a
relatively small contribution to the overall nucleosynthesis, but are
important for some elements. Carbon, nitrogen, lithium, boron, and
fluorine as well as numerous species in the iron group were produced
in solar or super-solar proportions to oxygen (\Fig{low}). Besides
their contribution to Pop I nucleosynthesis in nature, these 
abundances are also of interest for the large fraction of supernova remnants they would characterize. The iron to oxygen ratio was 
nearly normal, despite the fact that overall, the massive stars
only account for a minor fraction of the iron in the sun 
(\Sect{fegroup} and \citet{Whe89,Mel02,Chi03}). Boron was unusually 
high owing to the operation of the neutrino process in a shell 
close to the neutron star with a large abundance of carbon.

Explosions from 12 to 30 \Msun \ are responsible for most of the
nucleosynthesis. The elemental production pattern (\Fig{mid}) is
consistently solar from boron to copper with the expected deficiency
in the iron group due to SN Ia production. Carbon, nitrogen, and
fluorine are also greatly reduced compared with the low mass
supernovae. In general, there is a tendency to slightly underproduce
the intermediate mass elements, Si, S, Ar, and Ca with respect to
oxygen. This may indicate a smaller than optimal abundance for oxygen
in the sun, or the lack of some other component, e.g., SN Ia.  The
$s$-process production, from Cu to Zr is less than in previous works
\citep[e.g.,][]{Woo07} and is problematic.

Above 30 \Msun, there were few explosions, but even stars that make
black holes still contribute their winds (\Fig{high}). Because mass
loss rates are uncertain, the nucleosynthesis of these stars must be
treated with caution. For standard assumptions, carbon and oxygen are
substantially produced in the winds of massive Wolf-Rayet stars, and
nitrogen, neon, and sodium also have important components. Indeed the
abundances of these light elements are so great that stars in this
mass range make little else when the other productions are normalized
to them.

Integrating over the entire full range from 9 to 120 \Msun, and
examining isotopes as well as elements, the nucleosynthesis bears
strong resemblance both to the solar pattern and to the earlier survey
by \citet{Woo07}, but with notable exceptions.  $^{12}$C and $^{22}$Ne
are slightly overproduced. This might be an indication that the mass
loss rates used here are high by about a factor of two, but the
$^{22}$Ne yield would also be reduced in stars with less than solar
metallicity, because of the smaller neutron excess. C and O production
in the winds of low metallicity stars would also be reduced. The
excesses would instead end up in black holes. A full galactic chemical
evolution model would need to be done to see if these overproductions
are problematic.  Fluorine is underproduced by a factor of two,
suggesting a possible contribution from lower mass stars. Radioactive
$^{40}$K is greatly overproduced, but this is not a problem since a
large uncertain fraction of the radioactive species would decay before
the sun was born. The iron group retains the underproduction seen in
the mid-mass explosions, but some copper and nickel are slightly
overproduced, possibly due to the neglect of neutrino interactions at
the base of the ejecta.  Also notable are the deficient productions of
$^{44}$Ca and $^{55}$Mn and of the $s$- and $p-$processes. 

In fact, above the iron group, the fit to solar abundances is
substantially poorer than in \citet{Woo07} which used essentially the
same stellar physics, but an inferior model for the explosion. The
underproductions are a consequence of many massive stars that were
once prolific sources of the $s$-process now imploding to black
holes. To a lesser extent, it also reflects the use of a rate for the
$^{22}$Ne($\alpha,$n)$^{25}$Mg reaction that is about $\sim$50\%
smaller than a recent reanalysis of experimental data. Given the
critical role of the light $s$-process as a diagnostic here, our
calculations should be repeated with a variable value for this rate to
test the sensitivity of the outcome. The heavy $p$-process is also
underproduced both compared with the sun and with the results from 
\citet{Woo07}. This again reflects a deficiency of massive star 
explosions with thick layers that experience explosive neon burning. 
The origin of the light $p$-process below A = 130 and specifically of 
the nucleus $^{92}$Mo continues to be a mystery.  In total, the nucleosynthesis would be improved if more massive stars blew up \citep{Bro13}.

More $^{26}$Al is produced relative to $^{60}$Fe than in the 2007
survey because some of the major producers of $^{60}$Fe now make black
holes. This is good news, but should be treated cautiously because the
same changes necessary to increase light $s$-process production might
also increase the yield of $^{60}$Fe. The production ratio from the
present survey for $^{60}$Fe/$^{26}$Al, averaged over a Salpeter IMF,
is, by mass, $\sim$0.90. The observed value is 0.35 $\pm$ 0.1. The 
\citet{Woo07} survey gave 1.8. Part of the remaining discrepancy probably has
to do with uncertain reaction rates.

As is well known, the underproduction of iron in massive stars is
actually desirable since most of iron comes from SN Ia. Here we find
28\% of the solar abundance relative to oxygen is made in massive
stars. Since both are primary elements whose synthesis is independent
of neutron excess, the same ratio would probably characterize low
metallicity stars. If the SN Ia rate is at least 1/6 to 1/5 of the SN
IIp rate then iron is made in solar proportions. Roughly comparable
contributions from both sub-Chandrasekhar mass models and the
traditional Chandrasekhar mass model are needed, however, to produce
both $^{44}$Ca and $^{55}$Mn. The sub-Chandrasekhar model also helps
boost the deficient production of the alpha-nuclei from silicon
through calcium.

The excellent agreement of $^{11}$B/$^{16}$O in the integrated sample
provides an accurate measure of the $\mu$- and $\tau$-neutrino
temperature during proto-neutron star evolution.

A total of 194 supernova light curves were calculated. The vast
majority of these were SN IIp but the successful explosions over 30
\Msun \ made a variety of SN Ibc. These are not the common SN Ibc
though, because they are too faint and broad. Presumably the common
events come from lower mass explosions stripped of their envelopes in
binary systems. The massive faint variety should exist in nature though,
and should be sought.

The 181 SN IIp models are a community resource that is freely available\textsuperscript{\ref{mpa_db}}, and they offer a rich opportunity 
for modelling the spectrum and colors of a large comprehensive sample 
with standard identical physics. Here we analysed (\Sect{popov}) the 
bolometric light curves using semi-analytic approximations from \citet{Pop93} and \citet{Kas09}. While the range in plateau 
luminosities spans one and a half orders of magnitude, its systematics 
are amazingly well represented by the analytic formulae (\Fig{L50}). 
This shows that the systematics are well understood and encapsulated 
in simple scaling relations, lending confidence to the proposition 
that, with additional constraints from the spectrum, colors, dilution 
factor, etc., SN IIp may be useful cosmological yardsticks.

\Fig{L50KE} shows a tight correlation between SN IIp luminosity on the
plateau and its total explosion energy. This is well worth checking with
an observational sample with  spectroscopically determined kinetic 
energies. It also gives a useful tool for finding ``unusual''
supernovae. None of our explosions exceeded $2.03 \times 10^{51}$ erg
and none were brighter than $10^{42.7}$ erg s$^{-1}$. A SN IIp
brighter than this, say 10$^{43}$ erg s$^{-1}$, then is not powered by
neutrinos. This gives a physical basis for discerning ``superluminous'' 
supernovae of type IIp. \Fig{L50KE} also shows that the fraction
of kinetic energy that comes out as light in a neutrino powered SN IIp
is roughly constant, $\sim 3 \pm 1$\% for the 95\% of SN IIp lighter than 
21 \Msun.

While the width luminosity relation for SN IIp is not nearly so tight
or ``calibratable'' as for SN Ia, typical SN IIp of a given plateau
duration have the same luminosity at day 50 to about $\pm25$\%
(\Fig{WLR}). This is not true, however, for a substantial sample of low
energy supernovae below 12 \Msun. These would need to be
spectroscopically distinguished on the basis of their low photospheric
speed.

With less precision, the mass of the supernova progenitor can also be
determined from measurements of the total radiated light (to good
approximation $L_{50} t_p$) and estimates of the explosion energy
(\Fig{getmass}). The figure is especially useful for distinguishing
low mass supernovae but loses utility in separating supernovae in the
mass range 13 to 21 \Msun \ because of the much slower variation of
integrated light in this range. In practice, the need to determine the
explosion energy accurately may limit the application.

The success of the present approach in fitting many observational
constraints, including neutron star masses, black hole masses,
supernova kinetic energies, light curves, and nucleosynthesis is
gratifying, and gives hope that it may be more broadly applied in the
future. The existence of a large set of models with controlled, well
understood physics will be useful for the statistical analysis of
large data sets in the future. There is surely room for improvement,
though. Nuclear reaction rates need updating and the effect of varying
the mass loss should be explored. Rotation can be included in the
presupernova models, and approximated in the explosion. A broader
range of metallicities can be explored. The core physics, currently 1D
and calibrated using just two observed events can be improved by
calibrating to a greater range of 3D simulations and observations.  The
survey can be done in 2D to capture essential aspects of the mixing,
and the neutrino wind included in the yields. Given the vast amount of
data expected in the near future from large dedicated transient
surveys, the effort is worth it and it will happen.

For now though we offer not only the results and analysis in this
paper, but all presupernova models, exploded models, nucleosynthetic 
yields and the light curves in the electronic edition, and 
also at the MPA-Garching archive\textsuperscript{\ref{mpa_db}}.

\section{Acknowledgements}

We thank Alex Heger for the data of his Z9.6 progenitor model, his
numerous contributions in developing the KEPLER code and many insights into the topics of this paper. The help of Feryal \"Ozel 
with comparing our neutron star masses with observational data 
and of Gabriel Mart{\'{\i}}nez-Pinedo with our analysis of $^{19}$F 
production is also appreciated. TE and HTJ thank L.~H\"udepohl, 
T.~Melson, M.~Ugliano and A.~Wongwathanarat for numerical 
support and R.~Hix and F.-K.~Thielemann for providing the NSE 
solver introduced into P-HOTB by K.~Kifonidis.  This work was 
supported by NASA (NNX14AH34G) and the UC Office of the 
President (12-LR-237070).  At Garching, funding by Deutsche 
Forschungsgemeinschaft through grant EXC~153 ``Excellence 
Cluster Universe'' and the European Research Council through grant 
ERC-AdG No.~341157-COCO2CASA is acknowledged.

\clearpage

\begin{deluxetable*}{ccccc}
\tablewidth{0pt}
\tablecaption{Explosion Results for the Z9.6 Engine}
\tablehead{
 & \colhead{M$_{\mathrm{Ni}}$ [$\Msun$]} & \colhead{E$_{\mathrm{expl.}}$ [B]} & \colhead{M$_{\mathrm{remnant}}$ [$\Msun$]} \\
\colhead{Progenitor} & \colhead{K.$|$P.} & \colhead{K.$|$P.} & \colhead{P.}}
\startdata
  9.0 & 0.004 $|$ 0.006 & 0.11 $|$ 0.11 & 1.35  \\ 
 9.25 & 0.003 $|$ 0.007 & 0.14 $|$ 0.14 & 1.39  \\ 
  9.5 & 0.006 $|$ 0.009 & 0.18 $|$ 0.18 & 1.41  \\ 
 9.75 & 0.014 $|$ 0.018 & 0.35 $|$ 0.35 & 1.45  \\ 
 10.0 & 0.027 $|$ 0.031 & 0.60 $|$ 0.60 & 1.45  \\ 
10.25 & 0.029 $|$ 0.034 & 0.63 $|$ 0.63 & 1.47  \\ 
 10.5 & 0.018 $|$ 0.027 & 0.46 $|$ 0.46 & 1.48  \\ 
10.75 & 0.031 $|$ 0.046 & 0.69 $|$ 0.69 & 1.47  \\ 
 11.0 & 0.018 $|$ 0.019 & 0.31 $|$ 0.31 & 1.54  \\ 
11.25 & 0.019 $|$ 0.022 & 0.36 $|$ 0.36 & 1.53  \\ 
 11.5 & 0.025 $|$ 0.038 & 0.64 $|$ 0.64 & 1.50  \\ 
11.75 & 0.018 $|$ 0.018 & 0.26 $|$ 0.26 & 1.59  \\ 
 12.0 & 0.034 $|$ 0.043 & 0.66 $|$ 0.66 & 1.53  \\ 
\enddata
\tablecomments{K = KEPLER and P = P-HOTB. The remnant masses are
  listed only for the P-HOTB calculations since they are almost always
  identical to that of KEPLER.}
\lTab{eni_Z9.6}
\end{deluxetable*}

\clearpage

\LongTables
\begin{deluxetable*}{ccccccccc}
\tablewidth{0pt}
\tablecaption{Explosion Results for the N20 and W18 Engines}
\tablehead{
 & \multicolumn{2}{c}{M$_{\mathrm{Ni}}$ [$\Msun$]} & \multicolumn{2}{c}{E$_{\mathrm{expl.}}$ [B]} & \multicolumn{2}{c}{M$_{\mathrm{remnant}}$ [$\Msun$]} \\
\colhead{Progenitor} & \colhead{N20} & \colhead{W18} & \colhead{N20} & \colhead{W18} & \colhead{N20} & \colhead{W18}\\
& \colhead{K.$|$P.} & \colhead{K.$|$P.} & \colhead{K.$|$P.} & \colhead{K.$|$P.} & \colhead{P.} & \colhead{P.}}
\startdata
12.25 & 0.055 $|$ 0.089 & 0.063 $|$ 0.086 & 1.44 $|$ 1.44 & 1.36 $|$ 1.36 & 1.56 & 1.56  \\ 
 12.5 & 0.059 $|$ 0.092 & 0.059 $|$ 0.088 & 1.44 $|$ 1.44 & 1.35 $|$ 1.35 & 1.58 & 1.58  \\ 
12.75 & 0.062 $|$ 0.087 & 0.060 $|$ 0.082 & 1.29 $|$ 1.29 & 1.20 $|$ 1.20 & 1.63 & 1.64  \\ 
 13.0 & 0.070 $|$ 0.094 & 0.065 $|$ 0.083 & 1.32 $|$ 1.32 & 1.18 $|$ 1.18 & 1.66 & 1.68  \\ 
 13.1 & 0.061 $|$ 0.086 & 0.058 $|$ 0.080 & 1.22 $|$ 1.22 & 1.11 $|$ 1.11 & 1.59 & 1.60  \\ 
 13.2 & 0.061 $|$ 0.088 & 0.058 $|$ 0.082 & 1.26 $|$ 1.26 & 1.14 $|$ 1.14 & 1.59 & 1.60  \\ 
 13.3 & 0.061 $|$ 0.086 & 0.059 $|$ 0.081 & 1.22 $|$ 1.22 & 1.12 $|$ 1.12 & 1.60 & 1.61  \\ 
 13.4 & 0.063 $|$ 0.086 & 0.061 $|$ 0.081 & 1.21 $|$ 1.21 & 1.12 $|$ 1.12 & 1.61 & 1.62  \\ 
 13.5 & 0.064 $|$ 0.093 & 0.062 $|$ 0.087 & 1.33 $|$ 1.33 & 1.23 $|$ 1.23 & 1.61 & 1.62  \\ 
 13.6 & 0.073 $|$ 0.104 & 0.070 $|$ 0.097 & 1.51 $|$ 1.51 & 1.38 $|$ 1.38 & 1.62 & 1.64  \\ 
 13.7 & 0.071 $|$ 0.103 & 0.068 $|$ 0.096 & 1.48 $|$ 1.48 & 1.35 $|$ 1.35 & 1.63 & 1.64  \\ 
 13.8 & 0.072 $|$ 0.101 & 0.069 $|$ 0.096 & 1.43 $|$ 1.43 & 1.33 $|$ 1.33 & 1.65 & 1.66  \\ 
 13.9 & 0.071 $|$ 0.097 & 0.069 $|$ 0.091 & 1.36 $|$ 1.36 & 1.27 $|$ 1.27 & 1.66 & 1.67  \\ 
 14.0 & 0.070 $|$ 0.097 & 0.068 $|$ 0.091 & 1.36 $|$ 1.36 & 1.27 $|$ 1.27 & 1.67 & 1.68  \\ 
 14.1 & 0.069 $|$ 0.094 & 0.067 $|$ 0.089 & 1.30 $|$ 1.30 & 1.24 $|$ 1.23 & 1.69 & 1.69  \\ 
 14.2 & 0.067 $|$ 0.090 & 0.066 $|$ 0.086 & 1.25 $|$ 1.25 & 1.20 $|$ 1.19 & 1.69 & 1.70  \\ 
 14.3 & 0.072 $|$ 0.096 & 0.068 $|$ 0.089 & 1.31 $|$ 1.31 & 1.21 $|$ 1.21 & 1.70 & 1.71  \\ 
 14.4 & 0.070 $|$ 0.090 & 0.069 $|$ 0.088 & 1.22 $|$ 1.22 & 1.19 $|$ 1.19 & 1.72 & 1.72  \\ 
 14.5 & 0.077 $|$ 0.089 & 0.077 $|$ 0.088 & 1.09 $|$ 1.09 & 1.07 $|$ 1.07 & 1.76 & 1.76  \\ 
 14.6 & 0.072 $|$ 0.090 & 0.071 $|$ 0.086 & 1.17 $|$ 1.17 & 1.13 $|$ 1.13 & 1.75 & 1.75  \\ 
 14.7 & 0.079 $|$ 0.089 & 0.078 $|$ 0.086 & 1.07 $|$ 1.07 & 1.01 $|$ 1.01 & 1.77 & 1.78  \\ 
 14.8 & 0.072 $|$ 0.085 & 0.071 $|$ 0.083 & 1.07 $|$ 1.07 & 1.05 $|$ 1.05 & 1.78 & 1.78  \\ 
 14.9 & 0.076 $|$ 0.088 & 0.075 $|$ 0.085 & 1.07 $|$ 1.07 & 1.04 $|$ 1.04 & 1.78 & 1.78  \\ 
 15.2 & 0.070 $|$ 0.082 & 0.071 $|$ 0.079 & 0.94 $|$ 0.94 & 0.83 $|$ 0.83 & 1.55 & 1.57  \\ 
 15.7 & 0.075 $|$ 0.086 & 0.075 $|$ 0.081 & 0.95 $|$ 0.95 & 0.81 $|$ 0.81 & 1.57 & 1.59  \\ 
 15.8 & 0.085 $|$ 0.097 & 0.074 $|$ 0.074 & 1.06 $|$ 1.06 & 0.65 $|$ 0.65 & 1.56 & 1.64  \\ 
 15.9 & 0.079 $|$ 0.079 &       $|$       & 0.70 $|$ 0.70 &      $|$      & 1.67 &       \\ 
 16.0 & 0.094 $|$ 0.110 & 0.075 $|$ 0.079 & 1.26 $|$ 1.26 & 0.78 $|$ 0.78 & 1.54 & 1.62  \\ 
 16.1 & 0.075 $|$ 0.084 & 0.078 $|$ 0.087 & 0.89 $|$ 0.89 & 0.97 $|$ 0.97 & 1.59 & 1.59  \\ 
 16.2 & 0.095 $|$ 0.111 & 0.076 $|$ 0.081 & 1.23 $|$ 1.23 & 0.79 $|$ 0.79 & 1.55 & 1.62  \\ 
 16.3 & 0.097 $|$ 0.113 & 0.078 $|$ 0.083 & 1.23 $|$ 1.23 & 0.80 $|$ 0.80 & 1.55 & 1.62  \\ 
 16.4 & 0.093 $|$ 0.110 & 0.075 $|$ 0.079 & 1.24 $|$ 1.24 & 0.76 $|$ 0.76 & 1.55 & 1.63  \\ 
 16.5 & 0.093 $|$ 0.110 & 0.075 $|$ 0.078 & 1.23 $|$ 1.23 & 0.75 $|$ 0.75 & 1.55 & 1.63  \\ 
 16.6 & 0.097 $|$ 0.112 & 0.078 $|$ 0.082 & 1.22 $|$ 1.22 & 0.78 $|$ 0.78 & 1.56 & 1.63  \\ 
 16.7 & 0.096 $|$ 0.111 & 0.077 $|$ 0.081 & 1.23 $|$ 1.22 & 0.78 $|$ 0.78 & 1.55 & 1.62  \\ 
 16.8 & 0.096 $|$ 0.112 & 0.078 $|$ 0.082 & 1.22 $|$ 1.22 & 0.77 $|$ 0.77 & 1.57 & 1.64  \\ 
 16.9 & 0.095 $|$ 0.111 & 0.077 $|$ 0.080 & 1.22 $|$ 1.22 & 0.75 $|$ 0.75 & 1.57 & 1.64  \\ 
 17.0 & 0.093 $|$ 0.110 & 0.075 $|$ 0.078 & 1.21 $|$ 1.21 & 0.73 $|$ 0.73 & 1.57 & 1.65  \\ 
 17.1 & 0.095 $|$ 0.111 & 0.077 $|$ 0.079 & 1.21 $|$ 1.21 & 0.74 $|$ 0.74 & 1.58 & 1.66  \\ 
 17.3 & 0.089 $|$ 0.097 & 0.078 $|$ 0.076 & 0.98 $|$ 0.98 & 0.65 $|$ 0.65 & 1.61 & 1.67  \\ 
 17.4 & 0.088 $|$ 0.097 & 0.077 $|$ 0.075 & 0.98 $|$ 0.98 & 0.64 $|$ 0.64 & 1.60 & 1.67  \\ 
 17.5 & 0.086 $|$ 0.096 & 0.075 $|$ 0.074 & 0.99 $|$ 0.99 & 0.63 $|$ 0.63 & 1.59 & 1.66  \\ 
 17.6 & 0.088 $|$ 0.097 & 0.077 $|$ 0.075 & 0.98 $|$ 0.98 & 0.63 $|$ 0.63 & 1.60 & 1.67  \\ 
 17.7 & 0.085 $|$ 0.090 &       $|$       & 0.80 $|$ 0.80 &      $|$      & 1.63 &       \\ 
 17.9 & 0.079 $|$ 0.101 & 0.077 $|$ 0.097 & 1.36 $|$ 1.35 & 1.27 $|$ 1.27 & 1.77 & 1.78  \\ 
 18.0 & 0.081 $|$ 0.070 &       $|$       & 0.51 $|$ 0.51 &      $|$      & 1.72 &       \\ 
 18.1 & 0.078 $|$ 0.097 & 0.077 $|$ 0.093 & 1.28 $|$ 1.28 & 1.22 $|$ 1.22 & 1.79 & 1.79  \\ 
 18.2 & 0.078 $|$ 0.099 & 0.076 $|$ 0.095 & 1.33 $|$ 1.33 & 1.25 $|$ 1.25 & 1.77 & 1.77  \\ 
 18.3 & 0.079 $|$ 0.095 & 0.077 $|$ 0.091 & 1.23 $|$ 1.23 & 1.14 $|$ 1.14 & 1.81 & 1.82  \\ 
 18.4 & 0.079 $|$ 0.094 & 0.077 $|$ 0.090 & 1.19 $|$ 1.18 & 1.10 $|$ 1.10 & 1.82 & 1.83  \\ 
 18.5 & 0.083 $|$ 0.097 & 0.071 $|$ 0.072 & 1.01 $|$ 1.01 & 0.62 $|$ 0.62 & 1.57 & 1.64  \\ 
 18.7 & 0.078 $|$ 0.084 &       $|$       & 0.79 $|$ 0.79 &      $|$      & 1.63 &       \\ 
 18.8 & 0.095 $|$ 0.095 &       $|$       & 0.83 $|$ 0.83 &      $|$      & 1.78 &       \\ 
 18.9 & 0.089 $|$ 0.081 &       $|$       & 0.65 $|$ 0.65 &      $|$      & 1.83 &       \\ 
 19.0 & 0.084 $|$ 0.073 &       $|$       & 0.58 $|$ 0.58 &      $|$      & 1.83 &       \\ 
 19.1 & 0.087 $|$ 0.078 &       $|$       & 0.64 $|$ 0.64 &      $|$      & 1.82 &       \\ 
 19.2 & 0.096 $|$ 0.111 & 0.077 $|$ 0.079 & 1.19 $|$ 1.19 & 0.77 $|$ 0.77 & 1.62 & 1.70  \\ 
 19.3 & 0.075 $|$ 0.099 & 0.065 $|$ 0.080 & 1.26 $|$ 1.26 & 0.96 $|$ 0.96 & 1.50 & 1.54  \\ 
 19.4 & 0.071 $|$ 0.066 &       $|$       & 0.51 $|$ 0.51 &      $|$      & 1.67 &       \\ 
 19.7 & 0.058 $|$ 0.117 & 0.055 $|$ 0.107 & 1.88 $|$ 1.88 & 1.70 $|$ 1.70 & 1.37 & 1.38  \\ 
 19.8 & 0.075 $|$ 0.133 & 0.072 $|$ 0.124 & 1.96 $|$ 1.95 & 1.78 $|$ 1.78 & 1.44 & 1.46  \\ 
 20.1 & 0.122 $|$ 0.141 & 0.098 $|$ 0.102 & 1.37 $|$ 1.36 & 0.82 $|$ 0.82 & 1.72 & 1.82  \\ 
 20.2 & 0.062 $|$ 0.121 & 0.060 $|$ 0.112 & 1.92 $|$ 1.92 & 1.75 $|$ 1.75 & 1.43 & 1.44  \\ 
 20.3 & 0.112 $|$ 0.148 & 0.091 $|$ 0.115 & 1.72 $|$ 1.72 & 1.25 $|$ 1.25 & 1.50 & 1.57  \\ 
 20.4 & 0.057 $|$ 0.116 & 0.054 $|$ 0.107 & 1.87 $|$ 1.87 & 1.70 $|$ 1.70 & 1.38 & 1.40  \\ 
 20.5 & 0.053 $|$ 0.113 & 0.052 $|$ 0.104 & 1.84 $|$ 1.84 & 1.67 $|$ 1.67 & 1.37 & 1.38  \\ 
 20.6 & 0.086 $|$ 0.080 &       $|$       & 0.59 $|$ 0.59 &      $|$      & 1.90 &       \\ 
 20.8 & 0.056 $|$ 0.053 & 0.056 $|$ 0.054 & 0.45 $|$ 0.45 & 0.48 $|$ 0.48 & 1.75 & 1.74  \\ 
 21.0 & 0.086 $|$ 0.107 & 0.075 $|$ 0.087 & 1.26 $|$ 1.26 & 0.92 $|$ 0.92 & 1.51 & 1.56  \\ 
 21.1 & 0.087 $|$ 0.105 & 0.078 $|$ 0.088 & 1.21 $|$ 1.20 & 0.91 $|$ 0.91 & 1.55 & 1.60  \\ 
 21.2 & 0.060 $|$ 0.057 & 0.059 $|$ 0.056 & 0.59 $|$ 0.59 & 0.59 $|$ 0.59 & 1.92 & 1.92  \\ 
 21.5 & 0.102 $|$ 0.113 & 0.083 $|$ 0.083 & 1.12 $|$ 1.12 & 0.70 $|$ 0.70 & 1.62 & 1.70  \\ 
 21.6 & 0.106 $|$ 0.115 & 0.077 $|$ 0.068 & 1.10 $|$ 1.10 & 0.52 $|$ 0.52 & 1.64 & 1.76  \\ 
 21.7 & 0.094 $|$ 0.091 &       $|$       & 0.71 $|$ 0.71 &      $|$      & 1.80 &       \\ 
 25.2 & 0.110 $|$ 0.128 & 0.122 $|$ 0.142 & 1.19 $|$ 1.18 & 1.31 $|$ 1.31 & 1.80 & 1.77  \\ 
 25.3 & 0.138 $|$ 0.152 &       $|$       & 1.10 $|$ 1.10 &      $|$      & 1.88 &       \\ 
 25.4 & 0.127 $|$ 0.150 & 0.105 $|$ 0.116 & 1.36 $|$ 1.35 & 0.86 $|$ 0.86 & 1.79 & 1.88  \\ 
 25.5 & 0.113 $|$ 0.130 & 0.123 $|$ 0.143 & 1.20 $|$ 1.19 & 1.30 $|$ 1.30 & 1.80 & 1.77  \\ 
 25.6 & 0.088 $|$ 0.105 & 0.091 $|$ 0.108 & 1.09 $|$ 1.08 & 1.05 $|$ 1.05 & 1.78 & 1.78  \\ 
 25.7 & 0.071 $|$ 0.120 & 0.069 $|$ 0.111 & 1.66 $|$ 1.65 & 1.49 $|$ 1.48 & 1.60 & 1.61  \\ 
 25.8 & 0.068 $|$ 0.118 & 0.065 $|$ 0.109 & 1.68 $|$ 1.67 & 1.49 $|$ 1.49 & 1.58 & 1.59  \\ 
 25.9 & 0.108 $|$ 0.133 & 0.116 $|$ 0.139 & 1.26 $|$ 1.26 & 1.28 $|$ 1.28 & 1.68 & 1.67  \\ 
 26.0 & 0.122 $|$ 0.151 & 0.125 $|$ 0.152 & 1.40 $|$ 1.39 & 1.32 $|$ 1.31 & 1.67 & 1.68  \\ 
 26.1 & 0.129 $|$ 0.155 & 0.108 $|$ 0.119 & 1.40 $|$ 1.39 & 0.94 $|$ 0.94 & 1.68 & 1.76  \\ 
 26.2 & 0.130 $|$ 0.157 & 0.112 $|$ 0.125 & 1.39 $|$ 1.39 & 0.99 $|$ 0.99 & 1.69 & 1.76  \\ 
 26.3 & 0.146 $|$ 0.171 & 0.112 $|$ 0.118 & 1.39 $|$ 1.38 & 0.78 $|$ 0.78 & 1.71 & 1.82  \\ 
 26.4 & 0.128 $|$ 0.141 & 0.125 $|$ 0.134 & 1.09 $|$ 1.08 & 0.82 $|$ 0.82 & 1.77 & 1.84  \\ 
 26.5 & 0.125 $|$ 0.138 & 0.120 $|$ 0.128 & 1.06 $|$ 1.05 & 0.84 $|$ 0.84 & 1.77 & 1.82  \\ 
 26.6 & 0.131 $|$ 0.147 &       $|$       & 1.13 $|$ 1.12 &      $|$      & 1.75 &       \\ 
 26.7 & 0.131 $|$ 0.146 &       $|$       & 1.10 $|$ 1.10 &      $|$      & 1.76 &       \\ 
 26.8 & 0.137 $|$ 0.150 &       $|$       & 1.10 $|$ 1.10 &      $|$      & 1.76 &       \\ 
 26.9 & 0.135 $|$ 0.147 &       $|$       & 1.10 $|$ 1.10 &      $|$      & 1.75 &       \\ 
 27.0 & 0.129 $|$ 0.141 & 0.089 $|$ 0.092 & 1.09 $|$ 1.08 & 0.59 $|$ 0.59 & 1.72 & 1.85  \\ 
 27.1 & 0.134 $|$ 0.144 & 0.101 $|$ 0.102 & 1.08 $|$ 1.08 & 0.61 $|$ 0.61 & 1.73 & 1.86  \\ 
 27.2 & 0.127 $|$ 0.143 & 0.006 $|$ 0.000 & 1.08 $|$ 1.08 & 0.57 $|$ 0.56 & 1.73 & 3.19  \\ 
 27.3 & 0.133 $|$ 0.146 & 0.006 $|$ 0.000 & 1.09 $|$ 1.09 & 0.41 $|$ 0.41 & 1.73 & 6.24  \\ 
 27.4 & 0.003 $|$ 0.000 &       $|$       & 0.53 $|$ 0.53 &      $|$      & 4.96 &       \\ 
 29.0 & 0.005 $|$ 0.000 &       $|$       & 0.52 $|$ 0.52 &      $|$      & 4.14 &       \\ 
 29.1 & 0.004 $|$ 0.000 &       $|$       & 0.51 $|$ 0.51 &      $|$      & 4.43 &       \\ 
 29.2 & 0.001 $|$ 0.000 &       $|$       & 0.36 $|$ 0.36 &      $|$      & 6.94 &       \\ 
 29.6 & 0.001 $|$ 0.000 &       $|$       & 0.35 $|$ 0.35 &      $|$      & 7.30 &       \\ 
   60 & 0.079 $|$ 0.095 & 0.065 $|$ 0.066 & 0.92 $|$ 0.92 & 0.65 $|$ 0.65 & 1.70 & 1.80  \\
   80 & 0.078 $|$ 0.079 &       $|$       & 0.54 $|$ 0.54 &      $|$      & 1.91 &       \\
  100 & 0.074 $|$ 0.079 &       $|$       & 0.62 $|$ 0.62 &      $|$      & 2.03 &       \\ 
  120 & 0.079 $|$ 0.092 & 0.048 $|$ 0.066 & 1.04 $|$ 1.01 & 0.67 $|$ 0.67 & 1.70 & 1.76  \\ 
\enddata
\tablecomments{K = KEPLER and P = P-HOTB. The remnant masses are
  listed only for the P-HOTB calculations since they are almost always
  identical to that of KEPLER.}
\lTab{eni_N20_W18}
\end{deluxetable*}

\clearpage

\LongTables
\begin{deluxetable}{lrrrr}
\tablecaption{Yields in \Msun\ from the $14.9$ \Msun\ and $25.2$ \Msun\ \\
Models Using the W18 engine}
\tablehead{\colhead{Isotope} & \colhead{Ejecta} & \colhead{Winds} & \colhead{Ejecta} & \colhead{Winds}}
\startdata
$^{1}\mathrm{H}$    & 5.65E$+$00 & 1.42E$+$00 & 3.42E$+$00 & 6.69E$+$00 \\
$^{2}\mathrm{H}$    & 8.25E$-$09 & 8.40E$-$08 & 7.80E$-$09 & 1.97E$-$07 \\
$^{3}\mathrm{He}$   & 3.13E$-$04 & 1.11E$-$04 & 8.94E$-$05 & 4.92E$-$04 \\
$^{4}\mathrm{He}$   & 4.03E$+$00 & 5.61E$-$01 & 4.43E$+$00 & 3.29E$+$00 \\
$^{6}\mathrm{Li}$   & 1.56E$-$11 & 5.81E$-$11 & 1.33E$-$11 & 9.74E$-$11 \\
$^{7}\mathrm{Li}$   & 3.24E$-$07 & 1.68E$-$09 & 2.54E$-$07 & 2.74E$-$09 \\
$^{9}\mathrm{Be}$   & 4.53E$-$11 & 6.14E$-$11 & 2.19E$-$11 & 9.59E$-$11 \\
$^{10}\mathrm{B}$   & 2.19E$-$09 & 6.02E$-$10 & 1.63E$-$09 & 1.80E$-$09 \\
$^{11}\mathrm{B}$   & 8.36E$-$07 & 2.38E$-$09 & 1.61E$-$06 & 5.87E$-$09 \\
$^{12}\mathrm{C}$   & 1.53E$-$01 & 4.12E$-$03 & 3.71E$-$01 & 1.91E$-$02 \\
$^{13}\mathrm{C}$   & 8.09E$-$04 & 1.93E$-$04 & 4.98E$-$04 & 7.46E$-$04 \\
$^{14}\mathrm{N}$   & 3.22E$-$02 & 2.66E$-$03 & 4.04E$-$02 & 2.23E$-$02 \\
$^{15}\mathrm{N}$   & 1.53E$-$04 & 3.94E$-$06 & 1.02E$-$04 & 1.83E$-$05 \\
$^{16}\mathrm{O}$   & 7.78E$-$01 & 1.30E$-$02 & 3.53E$+$00 & 5.78E$-$02 \\
$^{17}\mathrm{O}$   & 7.30E$-$05 & 6.61E$-$06 & 6.28E$-$05 & 2.93E$-$05 \\
$^{18}\mathrm{O}$   & 1.83E$-$03 & 2.72E$-$05 & 6.96E$-$04 & 1.20E$-$04 \\
$^{19}\mathrm{F}$   & 2.28E$-$05 & 9.15E$-$07 & 2.38E$-$05 & 3.94E$-$06 \\
$^{20}\mathrm{Ne}$  & 1.14E$-$01 & 2.35E$-$03 & 3.21E$-$01 & 1.18E$-$02 \\
$^{21}\mathrm{Ne}$  & 5.80E$-$04 & 6.34E$-$06 & 9.90E$-$04 & 4.31E$-$05 \\
$^{22}\mathrm{Ne}$  & 8.42E$-$03 & 1.84E$-$04 & 2.53E$-$02 & 8.07E$-$04 \\
$^{23}\mathrm{Na}$  & 1.82E$-$03 & 8.37E$-$05 & 1.14E$-$02 & 5.60E$-$04 \\
$^{24}\mathrm{Mg}$  & 4.38E$-$02 & 1.13E$-$03 & 6.83E$-$02 & 5.72E$-$03 \\
$^{25}\mathrm{Mg}$  & 6.85E$-$03 & 1.48E$-$04 & 1.60E$-$02 & 6.68E$-$04 \\
$^{26}\mathrm{Mg}$  & 6.02E$-$03 & 1.74E$-$04 & 1.53E$-$02 & 9.50E$-$04 \\
$^{27}\mathrm{Al}$  & 4.31E$-$03 & 1.33E$-$04 & 1.15E$-$02 & 6.78E$-$04 \\
$^{28}\mathrm{Si}$  & 5.71E$-$02 & 1.52E$-$03 & 4.19E$-$01 & 7.64E$-$03 \\
$^{29}\mathrm{Si}$  & 2.46E$-$03 & 7.97E$-$05 & 7.03E$-$03 & 4.02E$-$04 \\
$^{30}\mathrm{Si}$  & 2.99E$-$03 & 5.44E$-$05 & 7.53E$-$03 & 2.74E$-$04 \\
$^{31}\mathrm{P}$   & 7.25E$-$04 & 1.52E$-$05 & 6.72E$-$03 & 7.69E$-$05 \\
$^{32}\mathrm{S}$   & 3.09E$-$02 & 7.95E$-$04 & 2.51E$-$01 & 4.01E$-$03 \\
$^{33}\mathrm{S}$   & 1.39E$-$04 & 6.46E$-$06 & 2.55E$-$03 & 3.26E$-$05 \\
$^{34}\mathrm{S}$   & 1.27E$-$03 & 3.73E$-$05 & 2.23E$-$02 & 1.88E$-$04 \\
$^{36}\mathrm{S}$   & 4.05E$-$06 & 1.61E$-$07 & 2.25E$-$05 & 8.10E$-$07 \\
$^{35}\mathrm{Cl}$  & 1.17E$-$04 & 8.15E$-$06 & 6.91E$-$03 & 4.11E$-$05 \\
$^{37}\mathrm{Cl}$  & 5.56E$-$05 & 2.75E$-$06 & 5.22E$-$04 & 1.39E$-$05 \\
$^{36}\mathrm{Ar}$  & 4.79E$-$03 & 1.83E$-$04 & 3.53E$-$02 & 9.24E$-$04 \\
$^{38}\mathrm{Ar}$  & 5.56E$-$04 & 3.51E$-$05 & 1.59E$-$02 & 1.77E$-$04 \\
$^{40}\mathrm{Ar}$  & 2.40E$-$06 & 5.64E$-$08 & 8.22E$-$06 & 2.85E$-$07 \\
$^{39}\mathrm{K}$   & 8.84E$-$05 & 7.89E$-$06 & 4.45E$-$03 & 3.98E$-$05 \\
$^{40}\mathrm{K}$   & 7.78E$-$07 & 1.25E$-$08 & 4.67E$-$05 & 6.35E$-$08 \\
$^{41}\mathrm{K}$   & 7.28E$-$06 & 5.98E$-$07 & 1.31E$-$04 & 3.02E$-$06 \\
$^{40}\mathrm{Ca}$  & 3.78E$-$03 & 1.43E$-$04 & 1.27E$-$02 & 7.22E$-$04 \\
$^{42}\mathrm{Ca}$  & 1.52E$-$05 & 1.00E$-$06 & 6.38E$-$04 & 5.06E$-$06 \\
$^{43}\mathrm{Ca}$  & 3.27E$-$06 & 2.15E$-$07 & 3.10E$-$05 & 1.08E$-$06 \\
$^{44}\mathrm{Ca}$  & 5.69E$-$05 & 3.39E$-$06 & 1.00E$-$04 & 1.71E$-$05 \\
$^{46}\mathrm{Ca}$  & 3.87E$-$07 & 6.75E$-$09 & 4.33E$-$07 & 3.40E$-$08 \\
$^{48}\mathrm{Ca}$  & 1.75E$-$06 & 3.33E$-$07 & 1.71E$-$06 & 1.68E$-$06 \\
$^{45}\mathrm{Sc}$  & 1.67E$-$06 & 9.03E$-$08 & 1.09E$-$05 & 4.56E$-$07 \\
$^{46}\mathrm{Ti}$  & 7.73E$-$06 & 5.40E$-$07 & 9.35E$-$05 & 2.72E$-$06 \\
$^{47}\mathrm{Ti}$  & 6.22E$-$06 & 4.96E$-$07 & 1.98E$-$05 & 2.50E$-$06 \\
$^{48}\mathrm{Ti}$  & 1.79E$-$04 & 5.04E$-$06 & 2.87E$-$04 & 2.54E$-$05 \\
$^{49}\mathrm{Ti}$  & 9.97E$-$06 & 3.77E$-$07 & 2.09E$-$05 & 1.90E$-$06 \\
$^{50}\mathrm{Ti}$  & 4.23E$-$06 & 3.69E$-$07 & 1.12E$-$05 & 1.86E$-$06 \\
$^{50}\mathrm{V}$   & 8.82E$-$08 & 2.11E$-$09 & 1.01E$-$06 & 1.06E$-$08 \\
$^{51}\mathrm{V}$   & 1.79E$-$05 & 8.61E$-$07 & 3.45E$-$05 & 4.34E$-$06 \\
$^{50}\mathrm{Cr}$  & 4.95E$-$05 & 1.64E$-$06 & 1.13E$-$04 & 8.27E$-$06 \\
$^{52}\mathrm{Cr}$  & 9.99E$-$04 & 3.29E$-$05 & 1.46E$-$03 & 1.66E$-$04 \\
$^{53}\mathrm{Cr}$  & 1.00E$-$04 & 3.79E$-$06 & 1.55E$-$04 & 1.91E$-$05 \\
$^{54}\mathrm{Cr}$  & 1.36E$-$05 & 9.63E$-$07 & 3.22E$-$05 & 4.86E$-$06 \\
$^{55}\mathrm{Mn}$  & 4.98E$-$04 & 2.95E$-$05 & 7.05E$-$04 & 1.49E$-$04 \\
$^{54}\mathrm{Fe}$  & 3.21E$-$03 & 1.55E$-$04 & 5.21E$-$03 & 7.83E$-$04 \\
$^{56}\mathrm{Fe}$  & 8.34E$-$02 & 2.53E$-$03 & 1.23E$-$01 & 1.28E$-$02 \\
$^{57}\mathrm{Fe}$  & 3.12E$-$03 & 5.94E$-$05 & 4.65E$-$03 & 3.00E$-$04 \\
$^{58}\mathrm{Fe}$  & 4.54E$-$04 & 8.03E$-$06 & 1.32E$-$03 & 4.05E$-$05 \\
$^{59}\mathrm{Co}$  & 4.04E$-$04 & 8.05E$-$06 & 9.19E$-$04 & 4.06E$-$05 \\
$^{58}\mathrm{Ni}$  & 2.51E$-$03 & 1.11E$-$04 & 3.81E$-$03 & 5.59E$-$04 \\
$^{60}\mathrm{Ni}$  & 2.70E$-$03 & 4.42E$-$05 & 5.29E$-$03 & 2.23E$-$04 \\
$^{61}\mathrm{Ni}$  & 1.79E$-$04 & 1.95E$-$06 & 5.05E$-$04 & 9.84E$-$06 \\
$^{62}\mathrm{Ni}$  & 6.44E$-$04 & 6.32E$-$06 & 1.66E$-$03 & 3.19E$-$05 \\
$^{64}\mathrm{Ni}$  & 8.78E$-$05 & 1.66E$-$06 & 3.99E$-$04 & 8.37E$-$06 \\
$^{63}\mathrm{Cu}$  & 9.24E$-$05 & 1.35E$-$06 & 3.12E$-$04 & 6.81E$-$06 \\
$^{65}\mathrm{Cu}$  & 2.35E$-$05 & 6.20E$-$07 & 1.31E$-$04 & 3.13E$-$06 \\
$^{64}\mathrm{Zn}$  & 7.53E$-$05 & 2.25E$-$06 & 1.38E$-$04 & 1.13E$-$05 \\
$^{66}\mathrm{Zn}$  & 3.34E$-$05 & 1.33E$-$06 & 2.31E$-$04 & 6.68E$-$06 \\
$^{67}\mathrm{Zn}$  & 5.55E$-$06 & 1.98E$-$07 & 2.61E$-$05 & 9.98E$-$07 \\
$^{68}\mathrm{Zn}$  & 1.74E$-$05 & 9.17E$-$07 & 1.21E$-$04 & 4.63E$-$06 \\
$^{70}\mathrm{Zn}$  & 1.44E$-$06 & 3.13E$-$08 & 6.45E$-$07 & 1.58E$-$07 \\
$^{69}\mathrm{Ga}$  & 2.52E$-$06 & 8.75E$-$08 & 2.17E$-$05 & 4.41E$-$07 \\
$^{71}\mathrm{Ga}$  & 1.70E$-$06 & 5.98E$-$08 & 1.25E$-$05 & 3.02E$-$07 \\
$^{70}\mathrm{Ge}$  & 1.64E$-$06 & 1.05E$-$07 & 3.18E$-$05 & 5.31E$-$07 \\
$^{72}\mathrm{Ge}$  & 2.22E$-$06 & 1.41E$-$07 & 2.93E$-$05 & 7.12E$-$07 \\
$^{73}\mathrm{Ge}$  & 9.87E$-$07 & 3.99E$-$08 & 5.44E$-$06 & 2.01E$-$07 \\
$^{74}\mathrm{Ge}$  & 2.83E$-$06 & 1.88E$-$07 & 1.60E$-$05 & 9.49E$-$07 \\
$^{76}\mathrm{Ge}$  & 6.65E$-$07 & 4.01E$-$08 & 3.43E$-$07 & 2.02E$-$07 \\
$^{75}\mathrm{As}$  & 7.39E$-$07 & 2.69E$-$08 & 3.82E$-$06 & 1.36E$-$07 \\
$^{74}\mathrm{Se}$  & 4.96E$-$08 & 2.53E$-$09 & 1.21E$-$06 & 1.28E$-$08 \\
$^{76}\mathrm{Se}$  & 4.21E$-$07 & 2.75E$-$08 & 1.03E$-$05 & 1.39E$-$07 \\
$^{77}\mathrm{Se}$  & 4.69E$-$07 & 2.27E$-$08 & 2.99E$-$06 & 1.14E$-$07 \\
$^{78}\mathrm{Se}$  & 7.18E$-$07 & 7.17E$-$08 & 9.32E$-$06 & 3.62E$-$07 \\
$^{80}\mathrm{Se}$  & 1.53E$-$06 & 1.53E$-$07 & 8.07E$-$06 & 7.74E$-$07 \\
$^{82}\mathrm{Se}$  & 3.94E$-$07 & 2.77E$-$08 & 1.64E$-$07 & 1.40E$-$07 \\
$^{79}\mathrm{Br}$  & 3.94E$-$07 & 2.67E$-$08 & 2.65E$-$06 & 1.35E$-$07 \\
$^{81}\mathrm{Br}$  & 3.56E$-$07 & 2.65E$-$08 & 2.22E$-$06 & 1.34E$-$07 \\
$^{78}\mathrm{Kr}$  & 6.25E$-$09 & 9.15E$-$10 & 1.05E$-$07 & 4.62E$-$09 \\
$^{80}\mathrm{Kr}$  & 6.16E$-$08 & 6.00E$-$09 & 1.66E$-$06 & 3.03E$-$08 \\
$^{82}\mathrm{Kr}$  & 2.76E$-$07 & 3.09E$-$08 & 5.69E$-$06 & 1.56E$-$07 \\
$^{83}\mathrm{Kr}$  & 3.33E$-$07 & 3.11E$-$08 & 1.64E$-$06 & 1.57E$-$07 \\
$^{84}\mathrm{Kr}$  & 1.11E$-$06 & 1.55E$-$07 & 5.85E$-$06 & 7.81E$-$07 \\
$^{86}\mathrm{Kr}$  & 7.23E$-$07 & 4.80E$-$08 & 1.38E$-$06 & 2.42E$-$07 \\
$^{85}\mathrm{Rb}$  & 3.57E$-$07 & 2.37E$-$08 & 1.52E$-$06 & 1.20E$-$07 \\
$^{87}\mathrm{Rb}$  & 1.27E$-$07 & 9.96E$-$09 & 1.00E$-$06 & 5.02E$-$08 \\
$^{84}\mathrm{Sr}$  & 3.95E$-$09 & 6.46E$-$10 & 1.60E$-$07 & 3.26E$-$09 \\
$^{86}\mathrm{Sr}$  & 8.84E$-$08 & 1.17E$-$08 & 2.12E$-$06 & 5.91E$-$08 \\
$^{87}\mathrm{Sr}$  & 5.67E$-$08 & 8.29E$-$09 & 8.19E$-$07 & 4.18E$-$08 \\
$^{88}\mathrm{Sr}$  & 8.10E$-$07 & 1.00E$-$07 & 4.34E$-$06 & 5.06E$-$07 \\
$^{89}\mathrm{Y}$   & 1.80E$-$07 & 2.41E$-$08 & 1.02E$-$06 & 1.21E$-$07 \\
$^{90}\mathrm{Zr}$  & 1.95E$-$07 & 3.07E$-$08 & 2.03E$-$06 & 1.55E$-$07 \\
$^{91}\mathrm{Zr}$  & 4.83E$-$08 & 6.81E$-$09 & 1.33E$-$07 & 3.43E$-$08 \\
$^{92}\mathrm{Zr}$  & 7.05E$-$08 & 1.05E$-$08 & 1.42E$-$07 & 5.30E$-$08 \\
$^{94}\mathrm{Zr}$  & 6.99E$-$08 & 1.09E$-$08 & 1.04E$-$07 & 5.48E$-$08 \\
$^{96}\mathrm{Zr}$  & 1.64E$-$08 & 1.79E$-$09 & 1.48E$-$08 & 9.01E$-$09 \\
$^{93}\mathrm{Nb}$  & 2.74E$-$08 & 4.11E$-$09 & 4.34E$-$08 & 2.08E$-$08 \\
$^{92}\mathrm{Mo}$  & 1.06E$-$08 & 2.09E$-$09 & 1.82E$-$08 & 1.05E$-$08 \\
$^{94}\mathrm{Mo}$  & 6.94E$-$09 & 1.33E$-$09 & 1.10E$-$08 & 6.70E$-$09 \\
$^{95}\mathrm{Mo}$  & 1.51E$-$08 & 2.31E$-$09 & 1.81E$-$08 & 1.16E$-$08 \\
$^{96}\mathrm{Mo}$  & 1.30E$-$08 & 2.45E$-$09 & 2.25E$-$08 & 1.23E$-$08 \\
$^{97}\mathrm{Mo}$  & 8.35E$-$09 & 1.42E$-$09 & 9.51E$-$09 & 7.15E$-$09 \\
$^{98}\mathrm{Mo}$  & 1.98E$-$08 & 3.61E$-$09 & 2.59E$-$08 & 1.82E$-$08 \\
$^{100}\mathrm{Mo}$ & 7.82E$-$09 & 1.47E$-$09 & 6.79E$-$09 & 7.43E$-$09 \\
$^{96}\mathrm{Ru}$  & 3.04E$-$09 & 5.94E$-$10 & 3.68E$-$09 & 3.00E$-$09 \\
$^{98}\mathrm{Ru}$  & 1.17E$-$09 & 2.05E$-$10 & 4.63E$-$09 & 1.03E$-$09 \\
$^{99}\mathrm{Ru}$  & 7.09E$-$09 & 1.41E$-$09 & 7.88E$-$09 & 7.10E$-$09 \\
$^{100}\mathrm{Ru}$ & 7.60E$-$09 & 1.40E$-$09 & 1.31E$-$08 & 7.08E$-$09 \\
$^{101}\mathrm{Ru}$ & 9.32E$-$09 & 1.92E$-$09 & 8.46E$-$09 & 9.69E$-$09 \\
$^{102}\mathrm{Ru}$ & 1.90E$-$08 & 3.59E$-$09 & 2.04E$-$08 & 1.81E$-$08 \\
$^{104}\mathrm{Ru}$ & 1.17E$-$08 & 2.17E$-$09 & 1.04E$-$08 & 1.09E$-$08 \\
$^{103}\mathrm{Rh}$ & 1.10E$-$08 & 2.25E$-$09 & 1.12E$-$08 & 1.13E$-$08 \\
$^{102}\mathrm{Pd}$ & 8.01E$-$10 & 8.75E$-$11 & 5.29E$-$09 & 4.41E$-$10 \\
$^{104}\mathrm{Pd}$ & 5.43E$-$09 & 9.78E$-$10 & 1.05E$-$08 & 4.93E$-$09 \\
$^{105}\mathrm{Pd}$ & 9.59E$-$09 & 1.97E$-$09 & 8.67E$-$09 & 9.95E$-$09 \\
$^{106}\mathrm{Pd}$ & 1.32E$-$08 & 2.45E$-$09 & 1.54E$-$08 & 1.23E$-$08 \\
$^{108}\mathrm{Pd}$ & 1.30E$-$08 & 2.41E$-$09 & 1.46E$-$08 & 1.21E$-$08 \\
$^{110}\mathrm{Pd}$ & 6.15E$-$09 & 1.08E$-$09 & 5.15E$-$09 & 5.47E$-$09 \\
$^{107}\mathrm{Ag}$ & 7.80E$-$09 & 1.60E$-$09 & 7.49E$-$09 & 8.07E$-$09 \\
$^{109}\mathrm{Ag}$ & 7.44E$-$09 & 1.51E$-$09 & 7.59E$-$09 & 7.63E$-$09 \\
$^{106}\mathrm{Cd}$ & 1.28E$-$09 & 1.23E$-$10 & 4.65E$-$09 & 6.22E$-$10 \\
$^{108}\mathrm{Cd}$ & 9.70E$-$10 & 8.93E$-$11 & 8.48E$-$09 & 4.51E$-$10 \\
$^{110}\mathrm{Cd}$ & 7.27E$-$09 & 1.28E$-$09 & 1.32E$-$08 & 6.45E$-$09 \\
$^{111}\mathrm{Cd}$ & 6.51E$-$09 & 1.32E$-$09 & 6.31E$-$09 & 6.67E$-$09 \\
$^{112}\mathrm{Cd}$ & 1.33E$-$08 & 2.51E$-$09 & 1.47E$-$08 & 1.27E$-$08 \\
$^{113}\mathrm{Cd}$ & 6.41E$-$09 & 1.29E$-$09 & 6.03E$-$09 & 6.49E$-$09 \\
$^{114}\mathrm{Cd}$ & 1.72E$-$08 & 3.05E$-$09 & 2.00E$-$08 & 1.54E$-$08 \\
$^{116}\mathrm{Cd}$ & 5.93E$-$09 & 8.11E$-$10 & 5.99E$-$09 & 4.09E$-$09 \\
$^{113}\mathrm{In}$ & 3.02E$-$10 & 5.18E$-$11 & 2.42E$-$09 & 2.61E$-$10 \\
$^{115}\mathrm{In}$ & 5.94E$-$09 & 1.17E$-$09 & 6.06E$-$09 & 5.89E$-$09 \\
$^{112}\mathrm{Sn}$ & 2.42E$-$09 & 2.39E$-$10 & 1.78E$-$08 & 1.21E$-$09 \\
$^{114}\mathrm{Sn}$ & 1.47E$-$09 & 1.65E$-$10 & 3.01E$-$08 & 8.30E$-$10 \\
$^{115}\mathrm{Sn}$ & 4.30E$-$10 & 8.53E$-$11 & 2.87E$-$09 & 4.30E$-$10 \\
$^{116}\mathrm{Sn}$ & 1.94E$-$08 & 3.69E$-$09 & 3.13E$-$08 & 1.86E$-$08 \\
$^{117}\mathrm{Sn}$ & 9.96E$-$09 & 1.97E$-$09 & 1.04E$-$08 & 9.92E$-$09 \\
$^{118}\mathrm{Sn}$ & 3.42E$-$08 & 6.26E$-$09 & 4.23E$-$08 & 3.16E$-$08 \\
$^{119}\mathrm{Sn}$ & 1.18E$-$08 & 2.25E$-$09 & 1.25E$-$08 & 1.13E$-$08 \\
$^{120}\mathrm{Sn}$ & 5.02E$-$08 & 8.57E$-$09 & 6.10E$-$08 & 4.32E$-$08 \\
$^{122}\mathrm{Sn}$ & 1.10E$-$08 & 1.24E$-$09 & 9.69E$-$09 & 6.25E$-$09 \\
$^{124}\mathrm{Sn}$ & 8.90E$-$09 & 1.57E$-$09 & 6.68E$-$09 & 7.94E$-$09 \\
$^{121}\mathrm{Sb}$ & 7.34E$-$09 & 1.34E$-$09 & 7.83E$-$09 & 6.74E$-$09 \\
$^{123}\mathrm{Sb}$ & 5.43E$-$09 & 1.02E$-$09 & 4.75E$-$09 & 5.13E$-$09 \\
$^{120}\mathrm{Te}$ & 3.24E$-$10 & 3.23E$-$11 & 1.62E$-$09 & 1.63E$-$10 \\
$^{122}\mathrm{Te}$ & 4.93E$-$09 & 8.97E$-$10 & 7.30E$-$09 & 4.53E$-$09 \\
$^{123}\mathrm{Te}$ & 1.69E$-$09 & 3.15E$-$10 & 2.15E$-$09 & 1.59E$-$09 \\
$^{124}\mathrm{Te}$ & 9.12E$-$09 & 1.69E$-$09 & 1.27E$-$08 & 8.52E$-$09 \\
$^{125}\mathrm{Te}$ & 1.27E$-$08 & 2.53E$-$09 & 1.26E$-$08 & 1.28E$-$08 \\
$^{126}\mathrm{Te}$ & 3.53E$-$08 & 6.75E$-$09 & 3.78E$-$08 & 3.40E$-$08 \\
$^{128}\mathrm{Te}$ & 5.80E$-$08 & 1.15E$-$08 & 4.82E$-$08 & 5.78E$-$08 \\
$^{130}\mathrm{Te}$ & 6.30E$-$08 & 1.24E$-$08 & 5.08E$-$08 & 6.27E$-$08 \\
$^{127}\mathrm{I}$  & 3.51E$-$08 & 7.43E$-$09 & 2.98E$-$08 & 3.75E$-$08 \\
$^{124}\mathrm{Xe}$ & 7.74E$-$10 & 5.06E$-$11 & 6.79E$-$09 & 2.55E$-$10 \\
$^{126}\mathrm{Xe}$ & 6.53E$-$10 & 4.46E$-$11 & 1.11E$-$08 & 2.25E$-$10 \\
$^{128}\mathrm{Xe}$ & 6.68E$-$09 & 9.01E$-$10 & 1.52E$-$08 & 4.55E$-$09 \\
$^{129}\mathrm{Xe}$ & 5.35E$-$08 & 1.12E$-$08 & 4.44E$-$08 & 5.65E$-$08 \\
$^{130}\mathrm{Xe}$ & 1.20E$-$08 & 1.80E$-$09 & 1.85E$-$08 & 9.09E$-$09 \\
$^{131}\mathrm{Xe}$ & 4.41E$-$08 & 9.03E$-$09 & 4.06E$-$08 & 4.56E$-$08 \\
$^{132}\mathrm{Xe}$ & 5.93E$-$08 & 1.10E$-$08 & 6.25E$-$08 & 5.56E$-$08 \\
$^{134}\mathrm{Xe}$ & 2.54E$-$08 & 4.09E$-$09 & 2.08E$-$08 & 2.07E$-$08 \\
$^{136}\mathrm{Xe}$ & 1.91E$-$08 & 3.39E$-$09 & 1.76E$-$08 & 1.71E$-$08 \\
$^{133}\mathrm{Cs}$ & 1.52E$-$08 & 2.87E$-$09 & 1.68E$-$08 & 1.45E$-$08 \\
$^{130}\mathrm{Ba}$ & 1.07E$-$09 & 3.51E$-$11 & 1.28E$-$08 & 1.77E$-$10 \\
$^{132}\mathrm{Ba}$ & 7.81E$-$10 & 3.41E$-$11 & 1.30E$-$08 & 1.72E$-$10 \\
$^{134}\mathrm{Ba}$ & 5.59E$-$09 & 8.27E$-$10 & 1.37E$-$08 & 4.17E$-$09 \\
$^{135}\mathrm{Ba}$ & 1.28E$-$08 & 2.27E$-$09 & 1.42E$-$08 & 1.14E$-$08 \\
$^{136}\mathrm{Ba}$ & 1.68E$-$08 & 2.73E$-$09 & 3.08E$-$08 & 1.38E$-$08 \\
$^{137}\mathrm{Ba}$ & 2.53E$-$08 & 3.93E$-$09 & 3.43E$-$08 & 1.98E$-$08 \\
$^{138}\mathrm{Ba}$ & 1.65E$-$07 & 2.53E$-$08 & 2.63E$-$07 & 1.28E$-$07 \\
$^{138}\mathrm{La}$ & 1.40E$-$10 & 3.25E$-$12 & 3.16E$-$10 & 1.64E$-$11 \\
$^{139}\mathrm{La}$ & 2.21E$-$08 & 3.59E$-$09 & 3.23E$-$08 & 1.81E$-$08 \\
$^{136}\mathrm{Ce}$ & 1.81E$-$10 & 1.73E$-$11 & 2.50E$-$09 & 8.74E$-$11 \\
$^{138}\mathrm{Ce}$ & 2.27E$-$10 & 2.37E$-$11 & 2.11E$-$09 & 1.20E$-$10 \\
$^{140}\mathrm{Ce}$ & 4.91E$-$08 & 8.49E$-$09 & 6.99E$-$08 & 4.28E$-$08 \\
$^{142}\mathrm{Ce}$ & 5.89E$-$09 & 1.08E$-$09 & 5.74E$-$09 & 5.46E$-$09 \\
$^{141}\mathrm{Pr}$ & 7.68E$-$09 & 1.44E$-$09 & 1.11E$-$08 & 7.25E$-$09 \\
$^{142}\mathrm{Nd}$ & 1.06E$-$08 & 1.89E$-$09 & 2.16E$-$08 & 9.54E$-$09 \\
$^{143}\mathrm{Nd}$ & 4.39E$-$09 & 8.45E$-$10 & 4.73E$-$09 & 4.26E$-$09 \\
$^{144}\mathrm{Nd}$ & 9.14E$-$09 & 1.68E$-$09 & 9.91E$-$09 & 8.48E$-$09 \\
$^{145}\mathrm{Nd}$ & 2.96E$-$09 & 5.90E$-$10 & 3.32E$-$09 & 2.98E$-$09 \\
$^{146}\mathrm{Nd}$ & 6.90E$-$09 & 1.23E$-$09 & 7.54E$-$09 & 6.20E$-$09 \\
$^{148}\mathrm{Nd}$ & 2.21E$-$09 & 4.18E$-$10 & 2.08E$-$09 & 2.11E$-$09 \\
$^{150}\mathrm{Nd}$ & 2.09E$-$09 & 4.14E$-$10 & 1.62E$-$09 & 2.09E$-$09 \\
$^{144}\mathrm{Sm}$ & 8.52E$-$10 & 6.60E$-$11 & 1.81E$-$08 & 3.33E$-$10 \\
$^{147}\mathrm{Sm}$ & 1.76E$-$09 & 3.39E$-$10 & 1.58E$-$09 & 1.71E$-$09 \\
$^{148}\mathrm{Sm}$ & 1.34E$-$09 & 2.49E$-$10 & 1.54E$-$09 & 1.26E$-$09 \\
$^{149}\mathrm{Sm}$ & 1.49E$-$09 & 3.07E$-$10 & 1.25E$-$09 & 1.55E$-$09 \\
$^{150}\mathrm{Sm}$ & 9.40E$-$10 & 1.66E$-$10 & 1.17E$-$09 & 8.36E$-$10 \\
$^{152}\mathrm{Sm}$ & 3.16E$-$09 & 6.06E$-$10 & 2.71E$-$09 & 3.06E$-$09 \\
$^{154}\mathrm{Sm}$ & 2.70E$-$09 & 5.22E$-$10 & 2.27E$-$09 & 2.63E$-$09 \\
$^{151}\mathrm{Eu}$ & 1.95E$-$09 & 4.04E$-$10 & 1.56E$-$09 & 2.03E$-$09 \\
$^{153}\mathrm{Eu}$ & 2.11E$-$09 & 4.46E$-$10 & 1.76E$-$09 & 2.25E$-$09 \\
$^{152}\mathrm{Gd}$ & 7.06E$-$11 & 5.98E$-$12 & 6.21E$-$11 & 3.05E$-$11 \\
$^{154}\mathrm{Gd}$ & 4.20E$-$10 & 6.54E$-$11 & 4.24E$-$10 & 3.31E$-$10 \\
$^{155}\mathrm{Gd}$ & 2.17E$-$09 & 4.48E$-$10 & 1.89E$-$09 & 2.26E$-$09 \\
$^{156}\mathrm{Gd}$ & 3.20E$-$09 & 6.22E$-$10 & 2.83E$-$09 & 3.14E$-$09 \\
$^{157}\mathrm{Gd}$ & 2.34E$-$09 & 4.80E$-$10 & 2.09E$-$09 & 2.42E$-$09 \\
$^{158}\mathrm{Gd}$ & 4.05E$-$09 & 7.65E$-$10 & 3.90E$-$09 & 3.86E$-$09 \\
$^{160}\mathrm{Gd}$ & 3.35E$-$09 & 6.82E$-$10 & 2.80E$-$09 & 3.44E$-$09 \\
$^{159}\mathrm{Tb}$ & 2.64E$-$09 & 5.52E$-$10 & 2.33E$-$09 & 2.78E$-$09 \\
$^{156}\mathrm{Dy}$ & 1.32E$-$11 & 1.98E$-$12 & 8.72E$-$11 & 9.97E$-$12 \\
$^{158}\mathrm{Dy}$ & 2.41E$-$11 & 3.43E$-$12 & 9.67E$-$11 & 1.73E$-$11 \\
$^{160}\mathrm{Dy}$ & 5.52E$-$10 & 8.49E$-$11 & 8.83E$-$10 & 4.29E$-$10 \\
$^{161}\mathrm{Dy}$ & 3.28E$-$09 & 6.91E$-$10 & 2.80E$-$09 & 3.48E$-$09 \\
$^{162}\mathrm{Dy}$ & 4.79E$-$09 & 9.37E$-$10 & 4.23E$-$09 & 4.73E$-$09 \\
$^{163}\mathrm{Dy}$ & 4.39E$-$09 & 9.21E$-$10 & 3.75E$-$09 & 4.65E$-$09 \\
$^{164}\mathrm{Dy}$ & 5.58E$-$09 & 1.05E$-$09 & 5.44E$-$09 & 5.29E$-$09 \\
$^{165}\mathrm{Ho}$ & 4.16E$-$09 & 8.71E$-$10 & 3.51E$-$09 & 4.39E$-$09 \\
$^{162}\mathrm{Er}$ & 3.88E$-$11 & 3.33E$-$12 & 1.95E$-$10 & 1.68E$-$11 \\
$^{164}\mathrm{Er}$ & 2.27E$-$10 & 3.95E$-$11 & 2.83E$-$10 & 1.99E$-$10 \\
$^{166}\mathrm{Er}$ & 4.30E$-$09 & 8.37E$-$10 & 3.89E$-$09 & 4.22E$-$09 \\
$^{167}\mathrm{Er}$ & 2.81E$-$09 & 5.74E$-$10 & 2.49E$-$09 & 2.90E$-$09 \\
$^{168}\mathrm{Er}$ & 3.77E$-$09 & 6.74E$-$10 & 3.51E$-$09 & 3.40E$-$09 \\
$^{170}\mathrm{Er}$ & 2.00E$-$09 & 3.81E$-$10 & 1.78E$-$09 & 1.92E$-$09 \\
$^{169}\mathrm{Tm}$ & 1.88E$-$09 & 3.67E$-$10 & 1.79E$-$09 & 1.85E$-$09 \\
$^{168}\mathrm{Yb}$ & 4.62E$-$11 & 3.19E$-$12 & 2.17E$-$10 & 1.61E$-$11 \\
$^{170}\mathrm{Yb}$ & 5.70E$-$10 & 7.53E$-$11 & 7.91E$-$10 & 3.80E$-$10 \\
$^{171}\mathrm{Yb}$ & 1.85E$-$09 & 3.55E$-$10 & 1.78E$-$09 & 1.79E$-$09 \\
$^{172}\mathrm{Yb}$ & 3.09E$-$09 & 5.48E$-$10 & 3.13E$-$09 & 2.76E$-$09 \\
$^{173}\mathrm{Yb}$ & 2.06E$-$09 & 4.07E$-$10 & 1.96E$-$09 & 2.06E$-$09 \\
$^{174}\mathrm{Yb}$ & 4.40E$-$09 & 8.07E$-$10 & 4.54E$-$09 & 4.07E$-$09 \\
$^{176}\mathrm{Yb}$ & 1.70E$-$09 & 3.27E$-$10 & 1.65E$-$09 & 1.65E$-$09 \\
$^{175}\mathrm{Lu}$ & 1.80E$-$09 & 3.57E$-$10 & 1.64E$-$09 & 1.80E$-$09 \\
$^{176}\mathrm{Lu}$ & 6.95E$-$11 & 1.04E$-$11 & 9.16E$-$11 & 5.28E$-$11 \\
$^{174}\mathrm{Hf}$ & 6.61E$-$11 & 2.81E$-$12 & 1.44E$-$10 & 1.42E$-$11 \\
$^{176}\mathrm{Hf}$ & 8.19E$-$10 & 9.13E$-$11 & 9.09E$-$10 & 4.61E$-$10 \\
$^{177}\mathrm{Hf}$ & 1.68E$-$09 & 3.29E$-$10 & 1.47E$-$09 & 1.66E$-$09 \\
$^{178}\mathrm{Hf}$ & 2.95E$-$09 & 4.84E$-$10 & 2.71E$-$09 & 2.44E$-$09 \\
$^{179}\mathrm{Hf}$ & 1.27E$-$09 & 2.43E$-$10 & 1.22E$-$09 & 1.23E$-$09 \\
$^{180}\mathrm{Hf}$ & 3.52E$-$09 & 6.30E$-$10 & 3.64E$-$09 & 3.18E$-$09 \\
$^{180}\mathrm{Ta}$ & 3.53E$-$12 & 2.65E$-$14 & 4.71E$-$12 & 1.34E$-$13 \\
$^{181}\mathrm{Ta}$ & 1.20E$-$09 & 2.23E$-$10 & 1.39E$-$09 & 1.12E$-$09 \\
$^{180}\mathrm{W}$  & 6.68E$-$11 & 1.62E$-$12 & 1.07E$-$10 & 8.15E$-$12 \\
$^{182}\mathrm{W}$  & 2.27E$-$09 & 3.61E$-$10 & 2.49E$-$09 & 1.82E$-$09 \\
$^{183}\mathrm{W}$  & 1.11E$-$09 & 1.96E$-$10 & 1.20E$-$09 & 9.90E$-$10 \\
$^{184}\mathrm{W}$  & 2.47E$-$09 & 4.23E$-$10 & 2.41E$-$09 & 2.14E$-$09 \\
$^{186}\mathrm{W}$  & 2.09E$-$09 & 3.95E$-$10 & 2.11E$-$09 & 1.99E$-$09 \\
$^{185}\mathrm{Re}$ & 1.09E$-$09 & 2.13E$-$10 & 1.05E$-$09 & 1.07E$-$09 \\
$^{187}\mathrm{Re}$ & 1.87E$-$09 & 3.89E$-$10 & 1.60E$-$09 & 1.96E$-$09 \\
$^{184}\mathrm{Os}$ & 4.24E$-$11 & 1.44E$-$12 & 1.03E$-$10 & 7.25E$-$12 \\
$^{186}\mathrm{Os}$ & 6.89E$-$10 & 1.17E$-$10 & 7.45E$-$10 & 5.90E$-$10 \\
$^{187}\mathrm{Os}$ & 4.72E$-$10 & 9.38E$-$11 & 4.34E$-$10 & 4.73E$-$10 \\
$^{188}\mathrm{Os}$ & 5.07E$-$09 & 9.88E$-$10 & 4.57E$-$09 & 4.98E$-$09 \\
$^{189}\mathrm{Os}$ & 5.66E$-$09 & 1.21E$-$09 & 4.47E$-$09 & 6.10E$-$09 \\
$^{190}\mathrm{Os}$ & 9.99E$-$09 & 1.98E$-$09 & 8.41E$-$09 & 9.98E$-$09 \\
$^{192}\mathrm{Os}$ & 1.49E$-$08 & 3.11E$-$09 & 1.17E$-$08 & 1.57E$-$08 \\
$^{191}\mathrm{Ir}$ & 1.25E$-$08 & 2.69E$-$09 & 9.72E$-$09 & 1.36E$-$08 \\
$^{193}\mathrm{Ir}$ & 2.15E$-$08 & 4.58E$-$09 & 1.70E$-$08 & 2.31E$-$08 \\
$^{190}\mathrm{Pt}$ & 2.12E$-$11 & 2.07E$-$12 & 4.20E$-$11 & 1.04E$-$11 \\
$^{192}\mathrm{Pt}$ & 1.10E$-$09 & 1.20E$-$10 & 1.25E$-$09 & 6.05E$-$10 \\
$^{194}\mathrm{Pt}$ & 2.56E$-$08 & 5.10E$-$09 & 2.16E$-$08 & 2.57E$-$08 \\
$^{195}\mathrm{Pt}$ & 2.50E$-$08 & 5.26E$-$09 & 2.08E$-$08 & 2.65E$-$08 \\
$^{196}\mathrm{Pt}$ & 2.09E$-$08 & 3.93E$-$09 & 1.90E$-$08 & 1.98E$-$08 \\
$^{198}\mathrm{Pt}$ & 5.99E$-$09 & 1.13E$-$09 & 4.79E$-$09 & 5.70E$-$09 \\
$^{197}\mathrm{Au}$ & 1.18E$-$08 & 2.27E$-$09 & 1.47E$-$08 & 1.14E$-$08 \\
$^{196}\mathrm{Hg}$ & 2.71E$-$09 & 7.25E$-$12 & 1.78E$-$08 & 3.65E$-$11 \\
$^{198}\mathrm{Hg}$ & 3.53E$-$09 & 4.78E$-$10 & 1.13E$-$08 & 2.41E$-$09 \\
$^{199}\mathrm{Hg}$ & 4.64E$-$09 & 8.15E$-$10 & 5.38E$-$09 & 4.11E$-$09 \\
$^{200}\mathrm{Hg}$ & 7.62E$-$09 & 1.12E$-$09 & 1.01E$-$08 & 5.64E$-$09 \\
$^{201}\mathrm{Hg}$ & 3.72E$-$09 & 6.42E$-$10 & 4.37E$-$09 & 3.24E$-$09 \\
$^{202}\mathrm{Hg}$ & 9.96E$-$09 & 1.46E$-$09 & 1.23E$-$08 & 7.37E$-$09 \\
$^{204}\mathrm{Hg}$ & 2.83E$-$09 & 3.39E$-$10 & 3.12E$-$09 & 1.71E$-$09 \\
$^{203}\mathrm{Tl}$ & 4.39E$-$09 & 6.50E$-$10 & 6.48E$-$09 & 3.28E$-$09 \\
$^{205}\mathrm{Tl}$ & 1.01E$-$08 & 1.56E$-$09 & 1.23E$-$08 & 7.89E$-$09 \\
$^{204}\mathrm{Pb}$ & 5.08E$-$09 & 7.73E$-$10 & 8.38E$-$09 & 3.90E$-$09 \\
$^{206}\mathrm{Pb}$ & 4.54E$-$08 & 7.27E$-$09 & 6.53E$-$08 & 3.67E$-$08 \\
$^{207}\mathrm{Pb}$ & 4.84E$-$08 & 8.09E$-$09 & 7.06E$-$08 & 4.08E$-$08 \\
$^{208}\mathrm{Pb}$ & 1.31E$-$07 & 2.33E$-$08 & 1.59E$-$07 & 1.17E$-$07 \\
$^{209}\mathrm{Bi}$ & 9.32E$-$09 & 1.70E$-$09 & 1.02E$-$08 & 8.59E$-$09 \\
$^{232}\mathrm{Th}$ & 7.19E$-$17 & 0.00E$+$00 & 0.00E$+$00 & 0.00E$+$00 \\
$^{235}\mathrm{U}$  & 1.04E$-$18 & 0.00E$+$00 & 0.00E$+$00 & 0.00E$+$00 \\
$^{238}\mathrm{U}$  & 0.00E$+$00 & 0.00E$+$00 & 0.00E$+$00 & 0.00E$+$00 \\
\enddata
$^{26}\mathrm{Al}$ & 3.16E$-$05 & 2.66E$-$09 & 7.78E$-$05 & 2.41E$-$06 \\
$^{60}\mathrm{Fe}$ & 4.70E$-$05 & 0.00E$+$00 & 3.58E$-$05 & 1.09E$-$16 \\
\enddata
\lTab{yieldtab}
\end{deluxetable}

\clearpage

%
%

\clearpage

\end{document}